\documentclass{msuphddissertation}
\usepackage{url}
\usepackage{textcomp}
\usepackage[T1]{fontenc}
\usepackage{ae,aecompl}
\usepackage{textcomp}
\usepackage{amsmath}
\usepackage[T1]{fontenc}
\usepackage{ae,aecompl}
\usepackage{graphicx}

\usepackage{simplewick}
\usepackage{footnote}
\usepackage{graphicx}
\usepackage{dcolumn}
\usepackage{bm}
\usepackage{stmaryrd}
\usepackage{amssymb}
\usepackage{amsmath}
\usepackage{bbm}
\usepackage{bbold}
\usepackage{multirow}

\newcommand{\s}{\hspace{2 pt}}
\newcommand{\matrixbb}[4]{\left(\hspace{-5 pt}\begin{tabular}{ c c } ${#1}$ & ${#2}$ \\ ${#3}$ & ${#4}$ \end{tabular}\hspace{-5 pt}\right)}

\newcommand{\matrixba}[2]{\left(\hspace{-5 pt}\begin{tabular}{ c } ${#1}$ \\ ${#2}$ \end{tabular}\hspace{-5 pt}\right)}

\newcommand{\matrixda}[4]{\left(\hspace{-5 pt}\begin{tabular}{ c } ${#1}$ \\ ${#2}$ \\ ${#3}$ \\ ${#4}$ \end{tabular}\hspace{-5 pt}\right)}

\newcommand{\ltr}{\llbracket}
\newcommand{\rtr}{\rrbracket}

\newcommand{\GEDOT}{G_{MN}^{(\text{5DOT})}}

\newcommand{\LRSf}[1]{\mathcal{L}_{#1}^{(\text{RS})}}
\newcommand{\LA}[1]{\overline{\mathcal{L}}_{A:#1}}
\newcommand{\LB}[1]{\overline{\mathcal{L}}_{B:#1}}
\newcommand{\LRSfeff}[1]{\mathcal{L}_{#1}^{(\text{RS,eff})}}
\newcommand{\LAeff}[1]{\mathcal{L}_{A:#1}^{(\text{eff})}}
\newcommand{\LBeff}[1]{\mathcal{L}_{B:#1}^{(\text{eff})}}

\newcommand{\kDD}{\kappa_{\text{4D}}}

\newcommand{\ahhh}[3]{a_{{#1}{#2}{#3}}}
\newcommand{\bhhh}[3]{b_{{#1}\hspace{1 pt}{#2}\hspace{1 pt}{#3}}}

\newcommand{\lBx}{{\mathpalette\lBxx\relax}}
\newcommand{\rBx}{{\mathpalette\rBxx\relax}}
\newcommand{\lBxx}[2]{\scalebox{2}[1]{${#1} [$}}
\newcommand{\rBxx}[2]{\scalebox{2}[1]{${#1} ]$}}

\newcommand{\lCx}{{\mathpalette\lCxx\relax}}
\newcommand{\rCx}{{\mathpalette\rCxx\relax}}
\newcommand{\lCxx}[2]{\scalebox{1.5}[1]{${#1} ($}}
\newcommand{\rCxx}[2]{\scalebox{1.5}[1]{${#1} )$}}

\newcommand{\bra}[1]{\langle{#1}|}
\newcommand{\ket}[1]{|{#1}\rangle}
\newcommand{\braket}[2]{\langle{#1}|{#2}\rangle}

\newcommand{\cbra}[1]{\lCx{#1}|}
\newcommand{\cket}[1]{|{#1}\rCx}
\newcommand{\cbracket}[2]{\lCx{#1}|{#2}\rCx}

\newcommand{\bracket}[2]{\langle{#1}|{#2}\rCx}
\newcommand{\sbra}[1]{\lBx{#1}|}
\newcommand{\sket}[1]{|{#1}\rBx}
\newcommand{\sbrasket}[2]{\lBx{#1}|{#2}\rBx}

\newcommand{\LD}{\pounds}

\newcommand{\vep}{\varepsilon}

\usepackage{tikz}
\usepackage{tikz-feynman}
\usepackage{contour}
\usetikzlibrary{arrows,decorations.pathmorphing}

\newcommand{\sDklrmn}[1]{\begin{tikzpicture}[scale={#1},photon/.style={decorate,decoration={snake,post length=1mm}}]
        \draw (0,4) -- (0.5,2);
        \draw[photon] (0,4) -- (0.5,2);
        \draw node[anchor=east] at (0,4) {\large $n_{1}$};
        \draw (0,0) -- (0.5,2);
        \draw[photon] (0,0) -- (0.5,2);
        \draw node[anchor=east] at (0,0) {\large $n_{2}$};
        \draw (0.5,2) -- (3.5,2);
        \draw (3.5,2) -- (4,4);
        \draw[photon] (3.5,2) -- (4,4);
        \draw node[anchor=west] at (4,4) {\large $n_{3}$};
        \draw (3.5,2) -- (4,0);
        \draw[photon] (3.5,2) -- (4,0);
        \draw node[anchor=west] at (4,0) {\large $n_{4}$};
      \node[circle,fill=darkgray,draw=black,inner sep=0pt,minimum size=0.1cm] at (0.5,2) {};
      \node[circle,fill=darkgray,draw=black,inner sep=0pt,minimum size=0.1cm] at (3.5,2) {};
      \draw node[anchor=south] at (2,2) {\large $r$};
\end{tikzpicture}}

\newcommand{\tDklrmn}[1]{\begin{tikzpicture}[scale={#1},photon/.style={decorate,decoration={snake,post length=1mm}}]
        \draw (0,4) -- (2,3.5);
        \draw[photon] (0,4) -- (2,3.5);
        \draw node[anchor=east] at (0,4) {\large $n_{1}$};
        \draw (0,0) -- (2,0.5);
        \draw[photon] (0,0) -- (2,0.5);
        \draw node[anchor=east] at (0,0) {\large $n_{2}$};
        \draw (2,0.5) -- (2,3.5);
        \draw (2,3.5) -- (4,4);
        \draw[photon] (2,3.5) -- (4,4);
        \draw node[anchor=west] at (4,4) {\large $n_{3}$};
        \draw (2,0.5) -- (4,0);
        \draw[photon] (2,0.5) -- (4,0);
        \draw node[anchor=west] at (4,0) {\large $n_{4}$};
      \node[circle,fill=darkgray,draw=black,inner sep=0pt,minimum size=0.1cm] at (2,3.5) {};
      \node[circle,fill=darkgray,draw=black,inner sep=0pt,minimum size=0.1cm] at (2,0.5) {};
      \draw node[anchor=east] at (2,2) {\large $r$};
\end{tikzpicture}}

\newcommand{\uDklrmn}[1]{\begin{tikzpicture}[scale={#1},photon/.style={decorate,decoration={snake,post length=1mm}}]
        \draw (0,4) -- (2,3.5);
        \draw[photon] (0,4) -- (2,3.5);
        \draw node[anchor=east] at (0,4) {\large $n_{1}$};
        \draw (0,0) -- (2,0.5);
        \draw[photon] (0,0) -- (2,0.5);
        \draw node[anchor=east] at (0,0) {\large $n_{2}$};
        \draw (2,0.5) -- (2,3.5);
        \draw (2,3.5) -- (4,0);
        \draw[photon] (2,3.5) -- (4,0);
        \draw node[anchor=west] at (4,0) {\large $n_{4}$};
        \draw (2,0.5) -- (4,4);
        \draw[photon] (2,0.5) -- (4,4);
        \draw node[anchor=west] at (4,4) {\large $n_{3}$};
      \node[circle,fill=darkgray,draw=black,inner sep=0pt,minimum size=0.1cm] at (2,3.5) {};
      \node[circle,fill=darkgray,draw=black,inner sep=0pt,minimum size=0.1cm] at (2,0.5) {};
      \draw node[anchor=east] at (2,2) {\large $r$};
\end{tikzpicture}}

\newcommand{\sDklzmn}[1]{\begin{tikzpicture}[scale={#1},photon/.style={decorate,decoration={snake,post length=1mm}}]
        \draw (0,4) -- (0.5,2);
        \draw[photon] (0,4) -- (0.5,2);
        \draw node[anchor=east] at (0,4) {\large $n_{1}$};
        \draw (0,0) -- (0.5,2);
        \draw[photon] (0,0) -- (0.5,2);
        \draw node[anchor=east] at (0,0) {\large $n_{2}$};
        \draw (0.5,2) -- (3.5,2);
        \draw[photon] (0.5,2) -- (3.5,2);
        \draw (3.5,2) -- (4,4);
        \draw[photon] (3.5,2) -- (4,4);
        \draw node[anchor=west] at (4,4) {\large $n_{3}$};
        \draw (3.5,2) -- (4,0);
        \draw[photon] (3.5,2) -- (4,0);
        \draw node[anchor=west] at (4,0) {\large $n_{4}$};
      \node[circle,fill=darkgray,draw=black,inner sep=0pt,minimum size=0.1cm] at (0.5,2) {};
      \node[circle,fill=darkgray,draw=black,inner sep=0pt,minimum size=0.1cm] at (3.5,2) {};
      \draw node[anchor=south] at (2,2) {\large $0$};
\end{tikzpicture}}

\newcommand{\tDklzmn}[1]{\begin{tikzpicture}[scale={#1},photon/.style={decorate,decoration={snake,post length=1mm}}]
        \draw (0,4) -- (2,3.5);
        \draw[photon] (0,4) -- (2,3.5);
        \draw node[anchor=east] at (0,4) {\large $n_{1}$};
        \draw (0,0) -- (2,0.5);
        \draw[photon] (0,0) -- (2,0.5);
        \draw node[anchor=east] at (0,0) {\large $n_{2}$};
        \draw (2,0.5) -- (2,3.5);
        \draw[photon] (2,0.5) -- (2,3.5);
        \draw (2,3.5) -- (4,4);
        \draw[photon] (2,3.5) -- (4,4);
        \draw node[anchor=west] at (4,4) {\large $n_{3}$};
        \draw (2,0.5) -- (4,0);
        \draw[photon] (2,0.5) -- (4,0);
        \draw node[anchor=west] at (4,0) {\large $n_{4}$};
      \node[circle,fill=darkgray,draw=black,inner sep=0pt,minimum size=0.1cm] at (2,3.5) {};
      \node[circle,fill=darkgray,draw=black,inner sep=0pt,minimum size=0.1cm] at (2,0.5) {};
      \draw node[anchor=east] at (2,2) {\large $0$};
\end{tikzpicture}}

\newcommand{\uDklzmn}[1]{\begin{tikzpicture}[scale={#1},photon/.style={decorate,decoration={snake,post length=1mm}}]
        \draw (0,4) -- (2,3.5);
        \draw[photon] (0,4) -- (2,3.5);
        \draw node[anchor=east] at (0,4) {\large $n_{1}$};
        \draw (0,0) -- (2,0.5);
        \draw[photon] (0,0) -- (2,0.5);
        \draw node[anchor=east] at (0,0) {\large $n_{2}$};
        \draw (2,0.5) -- (2,3.5);
        \draw[photon] (2,0.5) -- (2,3.5);
        \draw (2,3.5) -- (4,0);
        \draw[photon] (2,3.5) -- (4,0);
        \draw node[anchor=west] at (4,0) {\large $n_{4}$};
        \draw (2,0.5) -- (4,4);
        \draw[photon] (2,0.5) -- (4,4);
        \draw node[anchor=west] at (4,4) {\large $n_{3}$};
      \node[circle,fill=darkgray,draw=black,inner sep=0pt,minimum size=0.1cm] at (2,3.5) {};
      \node[circle,fill=darkgray,draw=black,inner sep=0pt,minimum size=0.1cm] at (2,0.5) {};
      \draw node[anchor=east] at (2,2) {\large $0$};
\end{tikzpicture}}

\newcommand{\sDkljmn}[1]{\begin{tikzpicture}[scale={#1},photon/.style={decorate,decoration={snake,post length=1mm}}]
        \draw (0,4) -- (0.5,2);
        \draw[photon] (0,4) -- (0.5,2);
        \draw node[anchor=east] at (0,4) {\large $n_{1}$};
        \draw (0,0) -- (0.5,2);
        \draw[photon] (0,0) -- (0.5,2);
        \draw node[anchor=east] at (0,0) {\large $n_{2}$};
        \draw (0.5,2) -- (3.5,2);
        \draw[photon] (0.5,2) -- (3.5,2);
        \draw (3.5,2) -- (4,4);
        \draw[photon] (3.5,2) -- (4,4);
        \draw node[anchor=west] at (4,4) {\large $n_{3}$};
        \draw (3.5,2) -- (4,0);
        \draw[photon] (3.5,2) -- (4,0);
        \draw node[anchor=west] at (4,0) {\large $n_{4}$};
      \node[circle,fill=darkgray,draw=black,inner sep=0pt,minimum size=0.1cm] at (0.5,2) {};
      \node[circle,fill=darkgray,draw=black,inner sep=0pt,minimum size=0.1cm] at (3.5,2) {};
      \draw node[anchor=south] at (2,2) {\large $j$};
\end{tikzpicture}}

\newcommand{\tDkljmn}[1]{\begin{tikzpicture}[scale={#1},photon/.style={decorate,decoration={snake,post length=1mm}}]
        \draw (0,4) -- (2,3.5);
        \draw[photon] (0,4) -- (2,3.5);
        \draw node[anchor=east] at (0,4) {\large $n_{1}$};
        \draw (0,0) -- (2,0.5);
        \draw[photon] (0,0) -- (2,0.5);
        \draw node[anchor=east] at (0,0) {\large $n_{2}$};
        \draw (2,0.5) -- (2,3.5);
        \draw[photon] (2,0.5) -- (2,3.5);
        \draw (2,3.5) -- (4,4);
        \draw[photon] (2,3.5) -- (4,4);
        \draw node[anchor=west] at (4,4) {\large $n_{3}$};
        \draw (2,0.5) -- (4,0);
        \draw[photon] (2,0.5) -- (4,0);
        \draw node[anchor=west] at (4,0) {\large $n_{4}$};
      \node[circle,fill=darkgray,draw=black,inner sep=0pt,minimum size=0.1cm] at (2,3.5) {};
      \node[circle,fill=darkgray,draw=black,inner sep=0pt,minimum size=0.1cm] at (2,0.5) {};
      \draw node[anchor=east] at (2,2) {\large $j$};
\end{tikzpicture}}

\newcommand{\uDkljmn}[1]{\begin{tikzpicture}[scale={#1},photon/.style={decorate,decoration={snake,post length=1mm}}]
        \draw (0,4) -- (2,3.5);
        \draw[photon] (0,4) -- (2,3.5);
        \draw node[anchor=east] at (0,4) {\large $n_{1}$};
        \draw (0,0) -- (2,0.5);
        \draw[photon] (0,0) -- (2,0.5);
        \draw node[anchor=east] at (0,0) {\large $n_{2}$};
        \draw (2,0.5) -- (2,3.5);
        \draw[photon] (2,0.5) -- (2,3.5);
        \draw (2,3.5) -- (4,0);
        \draw[photon] (2,3.5) -- (4,0);
        \draw node[anchor=west] at (4,0) {\large $n_{4}$};
        \draw (2,0.5) -- (4,4);
        \draw[photon] (2,0.5) -- (4,4);
        \draw node[anchor=west] at (4,4) {\large $n_{3}$};
      \node[circle,fill=darkgray,draw=black,inner sep=0pt,minimum size=0.1cm] at (2,3.5) {};
      \node[circle,fill=darkgray,draw=black,inner sep=0pt,minimum size=0.1cm] at (2,0.5) {};
      \draw node[anchor=east] at (2,2) {\large $j$};
\end{tikzpicture}}

\newcommand{\sgDklmn}[1]{\begin{tikzpicture}[scale={#1},photon/.style={decorate,decoration={snake,post length=1mm}}]
        \draw (0,4) -- (2,2);
        \draw[photon] (0,4) -- (2,2);
        \draw node[anchor=east] at (0,4) {\large $n_{1}$};
        \draw (0,0) -- (2,2);
        \draw[photon] (0,0) -- (2,2);
        \draw node[anchor=east] at (0,0) {\large $n_{2}$};
        \draw node[anchor=west] at (4,4) {\large $n_{3}$};
        \draw (2,2) -- (4,0);
        \draw[photon] (2,2) -- (4,0);
        \draw node[anchor=west] at (4,0) {\large $n_{4}$};
        \draw (2,2) -- (4,4);
        \draw[photon] (2,2) -- (4,4);
      \node[circle,fill=darkgray,draw=black,inner sep=0pt,minimum size=0.1cm] at (2,2) {};
      \node[circle,fill=darkgray,draw=black,inner sep=0pt,minimum size=0.1cm] at (2,2) {};
\end{tikzpicture}}

\newcommand{\totDklmn}[1]{\begin{tikzpicture}[scale={#1},photon/.style={decorate,decoration={snake,post length=1mm}}]
        \draw (0,4) -- (2,2);
        \draw[photon] (0,4) -- (2,2);
        \draw node[anchor=east] at (0,4) {\large $n_{1}$};
        \draw (0,0) -- (2,2);
        \draw[photon] (0,0) -- (2,2);
        \draw node[anchor=east] at (0,0) {\large $n_{2}$};
        \draw node[anchor=west] at (4,4) {\large $n_{3}$};
        \draw (2,2) -- (4,0);
        \draw[photon] (2,2) -- (4,0);
        \draw node[anchor=west] at (4,0) {\large $n_{4}$};
        \draw (2,2) -- (4,4);
        \draw[photon] (2,2) -- (4,4);
      \node[circle,fill=darkgray,draw=black,inner sep=0pt,minimum size=0.5cm] at (2,2) {};
\end{tikzpicture}}

\newcommand{\totDzzzz}[1]{\begin{tikzpicture}[scale={#1},photon/.style={decorate,decoration={snake,post length=1mm}}]
        \draw (0,4) -- (2,2);
        \draw[photon] (0,4) -- (2,2);
        \draw node[anchor=east] at (0,4) {\large $0$};
        \draw (0,0) -- (2,2);
        \draw[photon] (0,0) -- (2,2);
        \draw node[anchor=east] at (0,0) {\large $0$};
        \draw node[anchor=west] at (4,4) {\large $0$};
        \draw (2,2) -- (4,0);
        \draw[photon] (2,2) -- (4,0);
        \draw node[anchor=west] at (4,0) {\large $0$};
        \draw (2,2) -- (4,4);
        \draw[photon] (2,2) -- (4,4);
      \node[circle,fill=darkgray,draw=black,inner sep=0pt,minimum size=0.5cm] at (2,2) {};
\end{tikzpicture}}

\newcommand{\sDHHHHH}[1]{\begin{tikzpicture}[scale={#1},photon/.style={decorate,decoration={snake,post length=1mm}}]
        \draw (0,4) -- (0.5,2);
        \draw[photon] (0,4) -- (0.5,2);
        \draw node[anchor=east] at (0,4) {\large $H$};
        \draw (0,0) -- (0.5,2);
        \draw[photon] (0,0) -- (0.5,2);
        \draw node[anchor=east] at (0,0) {\large $H$};
        \draw (0.5,2) -- (3.5,2);
        \draw[photon] (0.5,2) -- (3.5,2);
        \draw (3.5,2) -- (4,4);
        \draw[photon] (3.5,2) -- (4,4);
        \draw node[anchor=west] at (4,4) {\large $H$};
        \draw (3.5,2) -- (4,0);
        \draw[photon] (3.5,2) -- (4,0);
        \draw node[anchor=west] at (4,0) {\large $H$};
      \node[circle,fill=darkgray,draw=black,inner sep=0pt,minimum size=0.1cm] at (0.5,2) {};
      \node[circle,fill=darkgray,draw=black,inner sep=0pt,minimum size=0.1cm] at (3.5,2) {};
      \draw node[anchor=south] at (2,2) {\large $H$};
\end{tikzpicture}}

\newcommand{\tDHHHHH}[1]{\begin{tikzpicture}[scale={#1},photon/.style={decorate,decoration={snake,post length=1mm}}]
        \draw (0,4) -- (2,3.5);
        \draw[photon] (0,4) -- (2,3.5);
        \draw node[anchor=east] at (0,4) {\large $H$};
        \draw (0,0) -- (2,0.5);
        \draw[photon] (0,0) -- (2,0.5);
        \draw node[anchor=east] at (0,0) {\large $H$};
        \draw (2,0.5) -- (2,3.5);
        \draw[photon] (2,0.5) -- (2,3.5);
        \draw (2,3.5) -- (4,4);
        \draw[photon] (2,3.5) -- (4,4);
        \draw node[anchor=west] at (4,4) {\large $H$};
        \draw (2,0.5) -- (4,0);
        \draw[photon] (2,0.5) -- (4,0);
        \draw node[anchor=west] at (4,0) {\large $H$};
      \node[circle,fill=darkgray,draw=black,inner sep=0pt,minimum size=0.1cm] at (2,3.5) {};
      \node[circle,fill=darkgray,draw=black,inner sep=0pt,minimum size=0.1cm] at (2,0.5) {};
      \draw node[anchor=east] at (2,2) {\large $H$};
\end{tikzpicture}}

\newcommand{\uDHHHHH}[1]{\begin{tikzpicture}[scale={#1},photon/.style={decorate,decoration={snake,post length=1mm}}]
        \draw (0,4) -- (2,3.5);
        \draw[photon] (0,4) -- (2,3.5);
        \draw node[anchor=east] at (0,4) {\large $H$};
        \draw (0,0) -- (2,0.5);
        \draw[photon] (0,0) -- (2,0.5);
        \draw node[anchor=east] at (0,0) {\large $H$};
        \draw (2,0.5) -- (2,3.5);
        \draw[photon] (2,0.5) -- (2,3.5);
        \draw (2,3.5) -- (4,0);
        \draw[photon] (2,3.5) -- (4,0);
        \draw node[anchor=west] at (4,0) {\large $H$};
        \draw (2,0.5) -- (4,4);
        \draw[photon] (2,0.5) -- (4,4);
        \draw node[anchor=west] at (4,4) {\large $H$};
      \node[circle,fill=darkgray,draw=black,inner sep=0pt,minimum size=0.1cm] at (2,3.5) {};
      \node[circle,fill=darkgray,draw=black,inner sep=0pt,minimum size=0.1cm] at (2,0.5) {};
      \draw node[anchor=east] at (2,2) {\large $H$};
\end{tikzpicture}}

\newcommand{\sgDHHHH}[1]{\begin{tikzpicture}[scale={#1},photon/.style={decorate,decoration={snake,post length=1mm}}]
        \draw (0,4) -- (2,2);
        \draw[photon] (0,4) -- (2,2);
        \draw node[anchor=east] at (0,4) {\large $H$};
        \draw (0,0) -- (2,2);
        \draw[photon] (0,0) -- (2,2);
        \draw node[anchor=east] at (0,0) {\large $H$};
        \draw node[anchor=west] at (4,4) {\large $H$};
        \draw (2,2) -- (4,0);
        \draw[photon] (2,2) -- (4,0);
        \draw node[anchor=west] at (4,0) {\large $H$};
        \draw (2,2) -- (4,4);
        \draw[photon] (2,2) -- (4,4);
      \node[circle,fill=darkgray,draw=black,inner sep=0pt,minimum size=0.1cm] at (2,2) {};
      \node[circle,fill=darkgray,draw=black,inner sep=0pt,minimum size=0.1cm] at (2,2) {};
\end{tikzpicture}}

\newcommand{\sDWWAWW}[1]{\begin{tikzpicture}[scale={#1},photon/.style={decorate,decoration={snake,post length=1mm}}]
        \draw (0,4) -- (0.5,2);
        \draw[photon] (0,4) -- (0.5,2);
        \draw node[anchor=east] at (0,4) {$W^{+}$};
        \draw (0,0) -- (0.5,2);
        \draw[photon] (0,0) -- (0.5,2);
        \draw node[anchor=east] at (0,0) {$W^{-}$};
        \draw[photon] (0.5,2) -- (3.5,2);
        \draw (3.5,2) -- (4,4);
        \draw[photon] (3.5,2) -- (4,4);
        \draw node[anchor=west] at (4,4) {$W^{+}$};
        \draw (3.5,2) -- (4,0);
        \draw[photon] (3.5,2) -- (4,0);
        \draw node[anchor=west] at (4,0) {$W^{-}$};
      \node[circle,fill=darkgray,draw=black,inner sep=0pt,minimum size=0.1cm] at (0.5,2) {};
      \node[circle,fill=darkgray,draw=black,inner sep=0pt,minimum size=0.1cm] at (3.5,2) {};
      \draw node[anchor=south] at (2,2) {$\gamma$};
\end{tikzpicture}}

\newcommand{\uDWWAWW}[1]{\begin{tikzpicture}[scale={#1},photon/.style={decorate,decoration={snake,post length=1mm}}]
        \draw (0,4) -- (2,3.5);
        \draw[photon] (0,4) -- (2,3.5);
        \draw node[anchor=east] at (0,4) {$W^{+}$};
        \draw (0,0) -- (2,0.5);
        \draw[photon] (0,0) -- (2,0.5);
        \draw node[anchor=east] at (0,0) {$W^{-}$};
        \draw[photon] (2,0.5) -- (2,3.5);
        \draw (2,3.5) -- (4,0);
        \draw[photon] (2,3.5) -- (4,0);
        \draw node[anchor=west] at (4,0) {$W^{+}$};
        \draw (2,0.5) -- (4,4);
        \draw[photon] (2,0.5) -- (4,4);
        \draw node[anchor=west] at (4,4) {$W^{-}$};
      \node[circle,fill=darkgray,draw=black,inner sep=0pt,minimum size=0.1cm] at (2,3.5) {};
      \node[circle,fill=darkgray,draw=black,inner sep=0pt,minimum size=0.1cm] at (2,0.5) {};
      \draw node[anchor=east] at (2,2) {$\gamma$};
\end{tikzpicture}}

\newcommand{\sDWWZWW}[1]{\begin{tikzpicture}[scale={#1},photon/.style={decorate,decoration={snake,post length=1mm}}]
        \draw (0,4) -- (0.5,2);
        \draw[photon] (0,4) -- (0.5,2);
        \draw node[anchor=east] at (0,4) {$W^{+}$};
        \draw (0,0) -- (0.5,2);
        \draw[photon] (0,0) -- (0.5,2);
        \draw node[anchor=east] at (0,0) {$W^{-}$};
        \draw[photon] (0.5,2) -- (3.5,2);
        \draw (3.5,2) -- (4,4);
        \draw[photon] (3.5,2) -- (4,4);
        \draw node[anchor=west] at (4,4) {$W^{+}$};
        \draw (3.5,2) -- (4,0);
        \draw[photon] (3.5,2) -- (4,0);
        \draw node[anchor=west] at (4,0) {$W^{-}$};
      \node[circle,fill=darkgray,draw=black,inner sep=0pt,minimum size=0.1cm] at (0.5,2) {};
      \node[circle,fill=darkgray,draw=black,inner sep=0pt,minimum size=0.1cm] at (3.5,2) {};
      \draw node[anchor=south] at (2,2) {$Z$};
\end{tikzpicture}}

\newcommand{\uDWWZWW}[1]{\begin{tikzpicture}[scale={#1},photon/.style={decorate,decoration={snake,post length=1mm}}]
        \draw (0,4) -- (2,3.5);
        \draw[photon] (0,4) -- (2,3.5);
        \draw node[anchor=east] at (0,4) {$W^{+}$};
        \draw (0,0) -- (2,0.5);
        \draw[photon] (0,0) -- (2,0.5);
        \draw node[anchor=east] at (0,0) {$W^{-}$};
        \draw[photon] (2,0.5) -- (2,3.5);
        \draw (2,3.5) -- (4,0);
        \draw[photon] (2,3.5) -- (4,0);
        \draw node[anchor=west] at (4,0) {$W^{+}$};
        \draw (2,0.5) -- (4,4);
        \draw[photon] (2,0.5) -- (4,4);
        \draw node[anchor=west] at (4,4) {$W^{-}$};
      \node[circle,fill=darkgray,draw=black,inner sep=0pt,minimum size=0.1cm] at (2,3.5) {};
      \node[circle,fill=darkgray,draw=black,inner sep=0pt,minimum size=0.1cm] at (2,0.5) {};
      \draw node[anchor=east] at (2,2) {$Z$};
\end{tikzpicture}}

\newcommand{\sDWWHWW}[1]{\begin{tikzpicture}[scale={#1},photon/.style={decorate,decoration={snake,post length=1mm}}]
        \draw (0,4) -- (0.5,2);
        \draw[photon] (0,4) -- (0.5,2);
        \draw node[anchor=east] at (0,4) {$W^{+}$};
        \draw (0,0) -- (0.5,2);
        \draw[photon] (0,0) -- (0.5,2);
        \draw node[anchor=east] at (0,0) {$W^{-}$};
        \draw (0.5,2) -- (3.5,2);
        \draw (3.5,2) -- (4,4);
        \draw[photon] (3.5,2) -- (4,4);
        \draw node[anchor=west] at (4,4) {$W^{+}$};
        \draw (3.5,2) -- (4,0);
        \draw[photon] (3.5,2) -- (4,0);
        \draw node[anchor=west] at (4,0) {$W^{-}$};
      \node[circle,fill=darkgray,draw=black,inner sep=0pt,minimum size=0.1cm] at (0.5,2) {};
      \node[circle,fill=darkgray,draw=black,inner sep=0pt,minimum size=0.1cm] at (3.5,2) {};
      \draw node[anchor=south] at (2,2) {$H$};
\end{tikzpicture}}

\newcommand{\uDWWHWW}[1]{\begin{tikzpicture}[scale={#1},photon/.style={decorate,decoration={snake,post length=1mm}}]
        \draw (0,4) -- (2,3.5);
        \draw[photon] (0,4) -- (2,3.5);
        \draw node[anchor=east] at (0,4) {$W^{+}$};
        \draw (0,0) -- (2,0.5);
        \draw[photon] (0,0) -- (2,0.5);
        \draw node[anchor=east] at (0,0) {$W^{-}$};
        \draw (2,0.5) -- (2,3.5);
        \draw[photon] (2,0.5) -- (2,3.5);
        \draw (2,3.5) -- (4,0);
        \draw[photon] (2,3.5) -- (4,0);
        \draw node[anchor=west] at (4,0) {$W^{+}$};
        \draw (2,0.5) -- (4,4);
        \draw node[anchor=west] at (4,4) {$W^{-}$};
      \node[circle,fill=darkgray,draw=black,inner sep=0pt,minimum size=0.1cm] at (2,3.5) {};
      \node[circle,fill=darkgray,draw=black,inner sep=0pt,minimum size=0.1cm] at (2,0.5) {};
      \draw node[anchor=east] at (2,2) {$H$};
\end{tikzpicture}}

\newcommand{\sgDWWWW}[1]{\begin{tikzpicture}[scale={#1},photon/.style={decorate,decoration={snake,post length=1mm}}]
        \draw (0,4) -- (2,2);
        \draw[photon] (0,4) -- (2,2);
        \draw node[anchor=east] at (0,4) {$W^{+}$};
        \draw (0,0) -- (2,2);
        \draw[photon] (0,0) -- (2,2);
        \draw node[anchor=east] at (0,0) {$W^{-}$};
        \draw node[anchor=west] at (4,4) {$W^{+}$};
        \draw (2,2) -- (4,0);
        \draw[photon] (2,2) -- (4,0);
        \draw node[anchor=west] at (4,0) {$W^{-}$};
        \draw (2,2) -- (4,4);
        \draw[photon] (2,2) -- (4,4);
      \node[circle,fill=darkgray,draw=black,inner sep=0pt,minimum size=0.1cm] at (2,2) {};
      \node[circle,fill=darkgray,draw=black,inner sep=0pt,minimum size=0.1cm] at (2,2) {};
\end{tikzpicture}}

\newcommand{\totDWWWW}[1]{\begin{tikzpicture}[scale={#1},photon/.style={decorate,decoration={snake,post length=1mm}}]
        \draw (0,4) -- (2,2);
        \draw[photon] (0,4) -- (2,2);
        \draw node[anchor=east] at (0,4) {$W^{+}$};
        \draw (0,0) -- (2,2);
        \draw[photon] (0,0) -- (2,2);
        \draw node[anchor=east] at (0,0) {$W^{-}$};
        \draw node[anchor=west] at (4,4) {$W^{+}$};
        \draw (2,2) -- (4,0);
        \draw[photon] (2,2) -- (4,0);
        \draw node[anchor=west] at (4,0) {$W^{-}$};
        \draw (2,2) -- (4,4);
        \draw[photon] (2,2) -- (4,4);
      \node[circle,fill=darkgray,draw=black,inner sep=0pt,minimum size=0.5cm] at (2,2) {};
\end{tikzpicture}}

\newcommand{\sDA}[1]{\begin{tikzpicture}[scale={#1},photon/.style={decorate,decoration={snake,post length=1mm}}]
        \draw (0,4) -- (0.5,2);
        \draw[photon] (0,4) -- (0.5,2);
        \draw node[anchor=east] at (0,4) {\large $1$};
        \draw (0,0) -- (0.5,2);
        \draw[photon] (0,0) -- (0.5,2);
        \draw node[anchor=east] at (0,0) {\large $4$};
        \draw (0.5,2) -- (3.5,2);
        \draw[photon] (0.5,2) -- (3.5,2);
        \draw (3.5,2) -- (4,4);
        \draw[photon] (3.5,2) -- (4,4);
        \draw node[anchor=west] at (4,4) {\large $2$};
        \draw (3.5,2) -- (4,0);
        \draw[photon] (3.5,2) -- (4,0);
        \draw node[anchor=west] at (4,0) {\large $3$};
      \node[circle,fill=darkgray,draw=black,inner sep=0pt,minimum size=0.1cm] at (0.5,2) {};
      \node[circle,fill=darkgray,draw=black,inner sep=0pt,minimum size=0.1cm] at (3.5,2) {};
      \draw node[anchor=south] at (2,2) {\large $5$};
\end{tikzpicture}}

\newcommand{\tDA}[1]{\begin{tikzpicture}[scale={#1},photon/.style={decorate,decoration={snake,post length=1mm}}]
        \draw (0,4) -- (2,3.5);
        \draw[photon] (0,4) -- (2,3.5);
        \draw node[anchor=east] at (0,4) {\large $1$};
        \draw (0,0) -- (2,0.5);
        \draw[photon] (0,0) -- (2,0.5);
        \draw node[anchor=east] at (0,0) {\large $4$};
        \draw (2,0.5) -- (2,3.5);
        \draw[photon] (2,0.5) -- (2,3.5);
        \draw (2,3.5) -- (4,4);
        \draw[photon] (2,3.5) -- (4,4);
        \draw node[anchor=west] at (4,4) {\large $2$};
        \draw (2,0.5) -- (4,0);
        \draw[photon] (2,0.5) -- (4,0);
        \draw node[anchor=west] at (4,0) {\large $3$};
      \node[circle,fill=darkgray,draw=black,inner sep=0pt,minimum size=0.1cm] at (2,3.5) {};
      \node[circle,fill=darkgray,draw=black,inner sep=0pt,minimum size=0.1cm] at (2,0.5) {};
      \draw node[anchor=east] at (2,2) {\large $1$};
\end{tikzpicture}}

\newcommand{\uDA}[1]{\begin{tikzpicture}[scale={#1},photon/.style={decorate,decoration={snake,post length=1mm}}]
        \draw (0,4) -- (2,3.5);
        \draw[photon] (0,4) -- (2,3.5);
        \draw node[anchor=east] at (0,4) {\large $1$};
        \draw (0,0) -- (2,0.5);
        \draw[photon] (0,0) -- (2,0.5);
        \draw node[anchor=east] at (0,0) {\large $4$};
        \draw (2,0.5) -- (2,3.5);
        \draw[photon] (2,0.5) -- (2,3.5);
        \draw (2,3.5) -- (4,0);
        \draw[photon] (2,3.5) -- (4,0);
        \draw node[anchor=west] at (4,0) {\large $2$};
        \draw (2,0.5) -- (4,4);
        \draw[photon] (2,0.5) -- (4,4);
        \draw node[anchor=west] at (4,4) {\large $3$};
      \node[circle,fill=darkgray,draw=black,inner sep=0pt,minimum size=0.1cm] at (2,3.5) {};
      \node[circle,fill=darkgray,draw=black,inner sep=0pt,minimum size=0.1cm] at (2,0.5) {};
      \draw node[anchor=east] at (2,2) {\large $2$};
\end{tikzpicture}}

\newcommand{\sgDA}[1]{\begin{tikzpicture}[scale={#1},photon/.style={decorate,decoration={snake,post length=1mm}}]
        \draw (0,4) -- (2,2);
        \draw[photon] (0,4) -- (2,2);
        \draw node[anchor=east] at (0,4) {\large $1$};
        \draw (0,0) -- (2,2);
        \draw[photon] (0,0) -- (2,2);
        \draw node[anchor=east] at (0,0) {\large $4$};
        \draw node[anchor=west] at (4,4) {\large $2$};
        \draw (2,2) -- (4,0);
        \draw[photon] (2,2) -- (4,0);
        \draw node[anchor=west] at (4,0) {\large $3$};
        \draw (2,2) -- (4,4);
        \draw[photon] (2,2) -- (4,4);
      \node[circle,fill=darkgray,draw=black,inner sep=0pt,minimum size=0.1cm] at (2,2) {};
      \node[circle,fill=darkgray,draw=black,inner sep=0pt,minimum size=0.1cm] at (2,2) {};
\end{tikzpicture}}

\newcommand{\propSc}[1]{\begin{tikzpicture}[scale={#1},photon/.style={decorate,decoration={snake,post length=1mm}}]
        \draw (0,0) -- (2,0);
\end{tikzpicture}}

\newcommand{\propGr}[1]{\begin{tikzpicture}[scale={#1},photon/.style={decorate,decoration={snake,post length=1mm}}]
        \draw (0,0) -- (2,0);
        \draw[photon] (0,0) -- (2,0);
        \draw node[anchor=east] at (0,0) {$\mu\nu$};
        \draw node[anchor=west] at (2,0) {$\rho\sigma$};
\end{tikzpicture}}

\newcommand{\propH}[1]{\begin{tikzpicture}[scale={#1},photon/.style={decorate,decoration={snake,post length=1mm}}]
        \draw (0,0) -- (2,0);
        \draw[photon] (0,0) -- (2,0);
        \draw node[anchor=east] at (0,0) {$MN$};
        \draw node[anchor=west] at (2,0) {$RS$};
        \draw node[anchor=south] at (1,0) {$H$};
\end{tikzpicture}}

\newcommand{\verhhh}[1]{\begin{tikzpicture}[scale={#1},photon/.style={decorate,decoration={snake,post length=1mm}}]
        \draw (0,4) -- (2,2);
        \draw[photon] (0,4) -- (2,2);
        \draw node[anchor=east] at (0,4) {\large $n_{1}$};
        \draw (0,0) -- (2,2);
        \draw[photon] (0,0) -- (2,2);
        \draw node[anchor=east] at (0,0) {\large $n_{2}$};
        \draw node[anchor=west] at (4,2) {\large $n_{3}$};
        \draw (2,2) -- (4,2);
        \draw[photon] (2,2) -- (4,2);
      \node[circle,fill=darkgray,draw=black,inner sep=0pt,minimum size=0.1cm] at (2,2) {};
      \node[circle,fill=darkgray,draw=black,inner sep=0pt,minimum size=0.1cm] at (2,2) {};
\end{tikzpicture}}

\newcommand{\verhhr}[1]{\begin{tikzpicture}[scale={#1},photon/.style={decorate,decoration={snake,post length=1mm}}]
        \draw (0,4) -- (2,2);
        \draw[photon] (0,4) -- (2,2);
        \draw node[anchor=east] at (0,4) {\large $n_{1}$};
        \draw (0,0) -- (2,2);
        \draw[photon] (0,0) -- (2,2);
        \draw node[anchor=east] at (0,0) {\large $n_{2}$};
        \draw node[anchor=west] at (4,2) {\large $r$};
        \draw (2,2) -- (4,2);
      \node[circle,fill=darkgray,draw=black,inner sep=0pt,minimum size=0.1cm] at (2,2) {};
      \node[circle,fill=darkgray,draw=black,inner sep=0pt,minimum size=0.1cm] at (2,2) {};
\end{tikzpicture}}

\newcommand{\verhhhh}[1]{\begin{tikzpicture}[scale={#1},photon/.style={decorate,decoration={snake,post length=1mm}}]
        \draw (0,4) -- (2,2);
        \draw[photon] (0,4) -- (2,2);
        \draw node[anchor=east] at (0,4) {\large $n_{1}$};
        \draw (0,0) -- (2,2);
        \draw[photon] (0,0) -- (2,2);
        \draw node[anchor=east] at (0,0) {\large $n_{2}$};
        \draw node[anchor=west] at (4,4) {\large $n_{3}$};
        \draw (2,2) -- (4,0);
        \draw[photon] (2,2) -- (4,0);
        \draw node[anchor=west] at (4,0) {\large $n_{4}$};
        \draw (2,2) -- (4,4);
        \draw[photon] (2,2) -- (4,4);
      \node[circle,fill=darkgray,draw=black,inner sep=0pt,minimum size=0.1cm] at (2,2) {};
      \node[circle,fill=darkgray,draw=black,inner sep=0pt,minimum size=0.1cm] at (2,2) {};
\end{tikzpicture}}

\newcommand{\verHHH}[1]{\begin{tikzpicture}[scale={#1},photon/.style={decorate,decoration={snake,post length=1mm}}]
        \draw (0,4) -- (2,2);
        \draw[photon] (0,4) -- (2,2);
        \draw node[anchor=east] at (0,4) {\large $H$};
        \draw (0,0) -- (2,2);
        \draw[photon] (0,0) -- (2,2);
        \draw node[anchor=east] at (0,0) {\large $H$};
        \draw node[anchor=west] at (4,2) {\large $H$};
        \draw (2,2) -- (4,2);
        \draw[photon] (2,2) -- (4,2);
      \node[circle,fill=darkgray,draw=black,inner sep=0pt,minimum size=0.1cm] at (2,2) {};
      \node[circle,fill=darkgray,draw=black,inner sep=0pt,minimum size=0.1cm] at (2,2) {};
\end{tikzpicture}}

\newcommand{\verHHHH}[1]{\begin{tikzpicture}[scale={#1},photon/.style={decorate,decoration={snake,post length=1mm}}]
        \draw (0,4) -- (2,2);
        \draw[photon] (0,4) -- (2,2);
        \draw node[anchor=east] at (0,4) {\large $H$};
        \draw (0,0) -- (2,2);
        \draw[photon] (0,0) -- (2,2);
        \draw node[anchor=east] at (0,0) {\large $H$};
        \draw node[anchor=west] at (4,4) {\large $H$};
        \draw (2,2) -- (4,0);
        \draw[photon] (2,2) -- (4,0);
        \draw node[anchor=west] at (4,0) {\large $H$};
        \draw (2,2) -- (4,4);
        \draw[photon] (2,2) -- (4,4);
      \node[circle,fill=darkgray,draw=black,inner sep=0pt,minimum size=0.1cm] at (2,2) {};
      \node[circle,fill=darkgray,draw=black,inner sep=0pt,minimum size=0.1cm] at (2,2) {};
\end{tikzpicture}}

\tikzstyle{arrow} = [thick,->,>=stealth]

\newcommand{\twototwoscattering}[1]{\begin{tikzpicture}
    \begin{feynman}[large]
        \vertex[blob] (c) at (0, 0) {};
        \vertex (n1) at (-1.5,1.5) {$1$};
        \vertex (a1) at (-1.7,1.2) {};
        \vertex (b1) at (-0.75,0.25) {};
        
        \vertex (n2) at (-1.5,-1.5) {$2$};
        \vertex (a2) at (-1.7,-1.2) {};
        \vertex (b2) at (-0.75,-0.25) {};
        
        \vertex (n3) at (1.5,1.5) {$3$};
        \vertex (a3) at (0.75,0.25) {};
        \vertex (b3) at (1.7,1.2) {};
        
        \vertex (n4) at (1.5,-1.5) {$4$};
        \vertex (a4) at (0.75,-0.25) {};
        \vertex (b4) at (1.7,-1.2) {};
        
        \diagram* {
            (n1) -- [photon] (c) -- [photon] (n3),
            (n2) -- [photon] (c) -- [photon] (n4),
            (a1) -- [arrow, edge label' = $p_{1}$] (b1),
            (a2) -- [arrow, edge label = $p_{2}$] (b2),
            (a3) -- [arrow, edge label' = $p_{3}$] (b3),
            (a4) -- [arrow, edge label = $p_{4}$] (b4)
        };
    \end{feynman}
\end{tikzpicture}}

\author{Dennis Foren}
\title{Scattering Amplitudes in Theories of Compactified Gravity}
\unit{Physics --- Doctor of Philosophy}

\begin{document}

\maketitlepage
\pagenumbering{roman}
\setcounter{page}{2}

\begin{abstract}
In this dissertation we discuss the properties of matrix elements describing the scattering of massive spin-2 particles in theories of compactified gravity. Our primary result is the calculation of 2-to-2 massive spin-2 Kaluza-Klein (KK) mode scattering matrix elements in the Randall-Sundrum 1 (RS1) model and the demonstration that those matrix elements grow no faster than $\mathcal{O}(s)$ irrespective of the KK mode numbers and helicities considered. Because this calculation requires summing infinitely-many spin-2 mediated diagrams which each diverge like $\mathcal{O}(s^{5})$, overall $\mathcal{O}(s)$ growth is only attained through cancellations between these diagrams. This in turn requires intricate cancellations between infinitely-many KK mode masses and couplings. We derive these sum rules, including their generalization to fully inelastic processes. We also consider these matrix elements in the five-dimensional orbifolded torus (5DOT) and large $kr_{c}$ limits, investigate the impact of including only finitely-many diagrams in the calculation (as measured via truncation error), and calculate the five-dimensional strong coupling scale $\Lambda_{\pi} \equiv M_{\text{Pl}}\, e^{-kr_{c}\pi}$ via the four-dimensional scattering calculation.
\end{abstract}

\begin{dedication}
\begin{center}
To those persons who act with empathy for others.
\end{center}
\end{dedication}

\begin{acknowledgment}
Dear reader,

Allow me a brief break from the royal ``we" to express my gratitude to you for taking the time to read this and to everyone who has supported me throughout my academic career. This dissertation marks the conclusion of an important chapter of my life, and is---in that sense---a selfish creation. Nonetheless, I have tried to imbue this text with material that might help others who aim to achieve similar goals. For those persons: I hope some fraction of this document is as useful to you as it has been to me.

I would be remiss to not acknowledge the immense privilege I had during this journey due to various aspects of my person (such as my race, gender expression, and being able-bodied). Because of this privilege, I was able to spend energy living life as I would like (researching, recharging), where others would have spent their energy overcoming unwarranted obstacles. If these barriers are to be removed from academia (and society more generally), action is required from those of us who do not encounter them. We must pay attention to and amplify the messages of affected persons, and act accordingly.

To my friends, my family, my advisors, and my committee for their time and energy along the way, as well as you and anyone else who takes the time and energy to read any amount of this document, from the depths of my being: thank you.\\

\begin{flushright}
{\it Wishing you the best, always,}\\

Dennis Foren

\end{flushright}

\end{acknowledgment}

\TOC
\LOT
\LOF

\pagenumbering{arabic}
\setcounter{page}{1}
\begin{doublespace}
\chapter{Introduction} \label{C - Introduction}

High-energy physics is the study of fundamental particles and their interactions. The success of modern high-energy physics is owed to the hard work of many experimental and theoretical physicists, including their development and application of quantum field theories. A quantum field theory (QFT) models each fundamental particle as an excitation of a field corresponding to that particle's species. Relativistic QFTs in particular combine the universal speed of light from special relativity (which provides well-defined meanings of particle mass and spin) with the probabilistic nature of reality that is intrinsic to quantum mechanics. With the help of a few additional features (the cluster decomposition principle, the LSZ reduction formula, etc.), high-energy physicists can calculate the probability that certain combinations of particles become other combinations of particles via scattering processes; knowing these probabilities allows the calculation of experimentally-relevant cross-sections and decay rates. However, before these probabilities can be calculated, the interested physicist must first calculate the Lorentz-invariant matrix element corresponding to the relevant scattering process, and to do that a physicist requires a Lagrangian.\footnote{We follow the standard high-energy convention of calling what is actually a ``Lagrangian density" (the integrand of an integral over spacetime) simply a ``Lagrangian" (which would otherwise be the integrand of an integral over time).}

Modern quantum field theory has streamlined the construction of model Lagrangians. In essence, a physicist decides on what matter particles and forces they would like included, chooses some interesting processes to investigate, and then puts together a Lagrangian that sums all terms consistent with that content which are relevant to those processes. Forces are typically included by declaring that the Lagrangian should have certain local symmetries, which then generate gauge bosons and their couplings to the matter particles. This is the way in which the champion of modern high-energy physics---the Standard Model (SM)---is constructed. The SM is presently our most accurate description of reality at subatomic scales, with high-energy experiments repeatedly confirming its predictions to increasingly high precision.

Prior to electroweak symmetry breaking (more on that in a moment), the Standard Model is an $\mathbf{SU(3)_{C}}\times \mathbf{SU(2)_{W}}\times \mathbf{U(1)_{Y}}$ gauge theory where
\begin{itemize}
    \item[$\bullet$] $\mathbf{SU(3)_{C}}$ generates the strong interaction and is gauged by eight gluons $G^{a}_{\mu}$,
    \item[$\bullet$] $\mathbf{SU(2)_{W}}$ generates the weak isospin interaction and is gauged by the triplet of vector bosons $\{W^{1}_{\mu},W^{2}_{\mu},W^{3}_{\mu}\}$, and
    \item[$\bullet$] $\mathbf{U(1)_{Y}}$ generates the weak hypercharge interaction and is gauged by the vector boson $B_{\mu}$.
\end{itemize}
The matter content of the theory (including each particle's mass, spin, and transformation behavior under the aforementioned local symmetry groups) is listed in Table \ref{Figure - PreEWSB SM}. The spin-$\tfrac{1}{2}$ quarks and leptons exhibit a generational structure (as emphasized by the subscript $m \in \{1,2,3\}$ on each field), the spin-$0$ Higgs doublet $\Phi$ does not, and all particles are massless. Everything changes when the electroweak gauge group $\mathbf{SU(2)_{W}}\times \mathbf{U(1)_{Y}}$ becomes spontaneously broken \cite{Glashow:1959wxa,Salam:1959zz,Weinberg:1967tq}.

\begin{table}
\bgroup
\def\arraystretch{1.5}
\begin{center}
{\bf The Matter Content of the Pre-EWSB Standard Model}\\
\text{ }\\
\begin{tabular}{ | c | c c | c | c | c | c | c |}
\hline & \multicolumn{2}{c|}{\bf Field Symbol} & {\bf Mass} & {\bf Spin} & $\mathbf{U(1)_Y}$   & $\mathbf{SU(2)_W}$ & $\mathbf{SU(3)_C}$ \\
 \hline
 \multirow{2}{*}{$\stackrel{\textbf{Left-Handed}}{\textbf{Quarks}}$} & \multirow{2}{*}{$\mathbf{q_{mL}}$} & $\mathbf{u_{mL}}$ &  \multirow{2}{*}{$0$} &  \multirow{2}{*}{$\tfrac{1}{2}$} &  \multirow{2}{*}{$+\tfrac{1}{3}$} & $+\tfrac{1}{2}$ &  \multirow{2}{*}{triplet} \\
 & & $\mathbf{d_{mL}}$ & &  & & $-\tfrac{1}{2}$ & \\
 \hline
 \multirow{2}{*}{$\stackrel{\textbf{Left-Handed}}{\textbf{Leptons}}$} & \multirow{2}{*}{$\mathbf{\ell_{mL}}$} & $\mathbf{\nu_{mL}}$ &  \multirow{2}{*}{$0$} &  \multirow{2}{*}{$\tfrac{1}{2}$} & $ \multirow{2}{*}{-1}$ & $+\tfrac{1}{2}$ &  \multirow{2}{*}{singlet}\\
 & & $\mathbf{e_{mL}}$ &  & & & $-\tfrac{1}{2}$ & \\
 \hline \multirow{2}{*}{$\stackrel{\textbf{Right-Handed}}{\textbf{Quarks}}$} & & $\mathbf{u_{mR}}$ & 0 & $\tfrac{1}{2}$ & $+\tfrac{4}{3}$ & $0$ & triplet \\
 \cline{2-8} & & $\mathbf{d_{mR}}$ & 0 & $\tfrac{1}{2}$ & $-\tfrac{2}{3}$ & $0$ & triplet\\
 \hline \multirow{2}{*}{$\stackrel{\textbf{Right-Handed}}{\textbf{Leptons}}$} & & $\mathbf{\nu_{mR}}$ & 0 & $\tfrac{1}{2}$ & $0$ & $0$ & singlet \\
 \cline{2-8} & & $\mathbf{e_{mR}}$ & 0 & $\tfrac{1}{2}$ & $-2$ & $0$ & singlet\\
\hline \raisebox{-0.45\height}{$\textbf{Higgs}$} & \multirow{2}{*}{$\mathbf{\Phi}$} & $\mathbf{\phi^{+}}$ &  \multirow{2}{*}{$0$} &  \multirow{2}{*}{$0^*$} &  \multirow{2}{*}{$+1$} & $+\tfrac{1}{2}$ &  \multirow{2}{*}{singlet} \\
 $\raisebox{0.45\height}{\textbf{Doublet}}$ & & $\mathbf{\phi^{0}}$ & & & & $-\tfrac{1}{2}$ & \\
 \hline
\end{tabular}
\end{center}
\egroup
\caption{The matter content of the Standard Model prior to electroweak symmetry breaking (EWSB) including their masses, internal spins, and gauge transformation properties. Rows are organized as to indicate matter fields that are related by the weak gauge group $\mathbf{SU(2)_{W}}$, i.e. $q_{mL}$ labels a weak gauge doublet with $+1/2$ component $u_{mL}$ and $-1/2$ component $d_{mL}$. The index $m \in \{1,2,3\}$ labels the generation of a given quark ($q$, $u$, $d$) or lepton ($\ell$, $e$, $\nu$) field, while a subscript ``L" or ``R" indicates whether it has left or right-handed chirality. The pre-EWSB Standard Model also contains gauge bosons $B_{\mu}$, $\{W^{1}_{\mu},W^{2}_{\mu},W^{3}_{\mu}\}$, and $\{G_{\mu}^{1},\dots,G_{\mu}^{8}\}$ corresponding to the weak hypercharge $\mathbf{U(1)_{Y}}$, weak isospin $\mathbf{SU(2)_{W}}$, and strong $\mathbf{SU(3)_{C}}$ gauge groups respectively. The left- and right-handed neutrinos $\nu_{mL}$ and $\nu_{mR}$ are called active and inert neutrinos respectively based on their $\text{SU}(2)_{\text{W}}$ transformation properties (or lack thereof). Whether or not the inert neutrinos $\nu_{mR}$ exist is an open question.}
\label{Figure - PreEWSB SM}
\end{table}

The electroweak gauge group breaks because the Higgs doublet spontaneously acquires a vacuum expectation value (vev), $v_{\text{EW}} = 0.246\text{ TeV}$, thereby isolating the Higgs boson $H$ from the rest of the doublet at low energies. This causes the electroweak gauge groups $\mathbf{SU(2)_{W}}\times \mathbf{U(1)_{Y}}$ to spontaneously break down to the electromagnetic gauge group $\mathbf{U(1)_{Q}}$. When this happens, a superposition of the $W^{3}_{\mu}$ and $B_{\mu}$ bosons become the massless spin-1 photon $A_{\mu}$ that gauges $\mathbf{U(1)_{Q}}$, while (in unitary gauge) an orthogonal mixture absorbs a fraction of the remaining Higgs doublet and becomes the massive $Z$-boson $Z_{\mu}$. The other $\mathbf{SU(2)_{W}}$ gauge bosons $W^{1}_{\mu}$ and $W^{2}_{\mu}$ absorb the rest of the Higgs doublet to become the massive $W$-bosons $W^{\pm}_{\mu}$. Simultaneously, interactions between the Higgs doublet and the (massless) fermionic matter fields are turned into mass and mixing terms, ultimately resulting in newly massive fermionic matter. Overall, electroweak symmetry breaking causes the low-energy SM to become an $\mathbf{SU(3)_{C}}\times \mathbf{U(1)_{Q}}$ gauge theory, wherein
\begin{itemize}
    \item[$\bullet$] $\mathbf{SU(3)_{C}}$ still generates the strong interaction and is gauged by the gluons $G^{a}_{\mu}$, and
    \item[$\bullet$] $\mathbf{U(1)_{Q}}$ generates the electromagnetic interaction and is gauged by the photon $A_{\mu}$.
\end{itemize}
and the matter content is as listed in Table \ref{Figure - SM}. In this way, electroweak symmetry breaking simultaneously explains the masses of the electroweak gauge bosons, expresses the weak force in terms of a local symmetry group, and generates masses for the Standard Model matter particles. The possibility that a single mechanism (``the Higgs mechanism") could explain all of these features motivated physicists in the 1960's to hypothesize the existence of the Higgs boson \cite{PhysRevLett.13.321,PhysRevLett.13.508,PhysRevLett.13.585}. Its eventual experimental confirmation in 2012 by the ATLAS and CMS collaborations at CERN is among the most celebrated achievements of physics in the 21st century thus far \cite{Aad:2012tfa,Chatrchyan:2012ufa}.

\begin{table}
\bgroup
\def\arraystretch{1.5}
\begin{center}
{\bf The Matter Content of the Post-EWSB Standard Model}\\
\text{ }\\
\begin{tabular}{ | c | c | c | c | c | c | c | }
    \hline
    & {\bf Name} & {\bf Symbol} & {\bf Mass (GeV/$c^{2}$)} & {\bf Spin} & $\mathbf{SU(3)_C}$ & $\mathbf{U(1)_Q}$\\
    \hline
    \multirow{3}{*}{$\stackrel{\textbf{Up-Type}}{\textbf{Quarks}}$} &  up quark & $u$ & $2.3\times10^{-3}$ &  \multirow{3}{*}{$\tfrac{1}{2}\otimes \tfrac{1}{2}$} &  \multirow{3}{*}{triplet} &  \multirow{3}{*}{$+\tfrac{2}{3}$} \\
    & charm quark & $c$ & $1.28$ & & & \\
    & top & $t$ & $173$ & & &\\
    \hline
    \multirow{3}{*}{$\stackrel{\textbf{Down-Type}}{\textbf{Quarks}}$} &  down quark & $d$ & $4.7\times 10^{-3}$ &  \multirow{3}{*}{$\tfrac{1}{2}\otimes \tfrac{1}{2}$} &  \multirow{3}{*}{triplet} &  \multirow{3}{*}{$-\tfrac{1}{3}$} \\
    & strange quark & $s$ & $9.5\times 10^{-2}$ & & & \\
    & bottom quark & $b$ & $4.18$ & & &\\
    \hline
    \multirow{3}{*}{$\stackrel{\textbf{Neutral}}{\textbf{Leptons}}$} &  neutrino $1$ & $\nu_{1}$ & ? &  \multirow{3}{*}{?} &  \multirow{3}{*}{singlet} &  \multirow{3}{*}{$0$} \\
    & neutrino $2$ & $\nu_{2}$ & ? & & & \\
    & neutrino $3$ & $\nu_{3}$ & ? & & &\\
    \hline
    \multirow{3}{*}{$\stackrel{\textbf{Charged}}{\textbf{Leptons}}$} &  electron & $e$ & $5.11\times 10^{-4}$ &  \multirow{3}{*}{$\tfrac{1}{2}\otimes \tfrac{1}{2}$} &  \multirow{3}{*}{singlet} &  \multirow{3}{*}{$-1$} \\
    & muon & $\mu$ & $0.106$ & & & \\
    & tauon & $\tau$ & $1.78$ & & &\\
    \hline
     & $\textbf{Higgs boson}$ & $H$ & $125$ & $0$ & singlet & $0$\\
    \hline
     & $\textbf{Z boson}$ & $Z$ & $91.2$ & $1$ & singlet & $0$\\
    \hline
    & $\textbf{W boson(s)}$ & $W^{+}$ & $80.3$ & $1^*$ & singlet & $+1$\\
    \hline
\end{tabular}
\end{center}
\egroup
\caption{The matter content of the Standard Model after electroweak symmetry breaking (EWSB) including their masses, internal spins, and gauge transformation properties \cite{PhysRevD.98.030001}. Rows group together matter fields that are related by generational structure. The Standard Model also contains the photon $A_{\mu}$ and the gluons $\{G_{\mu}^{1},\dots,G_{\mu}^{8}\}$ which are the gauge bosons corresponding to the electromagnetic $\mathbf{U(1)_{Q}}$ and strong $\mathbf{SU(3)_{C}}$ gauge groups respectively. The precise nature of the masses and spin structure of the neutrinos is an open question. The neutrino mass eigenstates $\nu_{1}$, $\nu_{2}$, $\nu_{3}$ are often reorganized via superposition into weak isospin eigenstates $\nu_{e}$, $\nu_{\mu}$, $\nu_{\tau}$ called the electron, muon, and tauon neutrinos respectively, which reconstruct the pre-EWSB active neutrinos at the cost of no longer having definite mass.}
\label{Figure - SM}
\end{table}

The SM is so successful in its predictions of subatomic phenomena that nearly every physically-descriptive QFT investigated in the high-energy literature hypothesizes new particles simply as add-ons to the SM. Of course, despite all that the Standard Model can predict, many physical phenomena lie outside its reach. For example, the SM does not predict the natures of neutrinos or dark matter or dark energy, nor does it incorporate gravity. 

A limited version of gravity {\it can} be added to the SM by considering four-dimensional general relativity in the weak field limit. Doing so generates a particle description of gravity, wherein the gravitational force is mediated by a massless spin-2 particle called the graviton. However, this modification breaks down at the Planck scale (or mass) $M_{\text{Pl}} = 2.435 \times 10^{15}$ TeV, reflecting its inability to describe strong or intrinsically quantum gravitational phenomenon that occur at higher energies. Furthermore, this SM + gravity theory possesses a vast range of energy scales between the electroweak's $v_{\text{EW}}$ and gravity's $M_{\text{Pl}}$ across which there is no new physics. Although nothing prevents such a hierarchy of scales in principle, the large ratio between the energy scales $M_{\text{Pl}}/v_{\text{EW}} \sim 5\times 10^{16}$ is technically unnatural.\footnote{Naturalness has several technical definitions in high-energy physics, but a theory tends to be natural if all of its parameters are set to values with similar magnitudes. Because the Higgs boson is a scalar particle, its mass-squared $m_{H}^{2}$ receives quantum corrections proportional to the square of the largest scales in the theory. By including gravity, that largest scale is the Planck mass $M_{\text{Pl}}$, and $16^{2} = 256$ decimal places of cancellations are required to obtain the experimentally-measured mass $m_{H} = 125\text{ GeV} \sim v_{\text{EV}}$ instead of a Planck scale mass $m_{H} \sim M_{\text{Pl}}$. Thus, the theory parameters must be fine-tuned to ensure this cancellation, and the ratio $M_{\text{Pl}}/v_{\text{EH}}$ is technically unnatural in the Standard Model.}

For many decades, physicists have attempted to solve this ``hierarchy problem" by hypothesizing a physical mechanism that would naturally generate a large ratio of scales $M_{\text{Pl}}/v_{\text{EW}}$. For example, in 1999, Randall and Sundrum proposed a five-dimensional gravity theory that could reparameterize the hierarchy problem via the warping of a non-factorizable extra-dimensional spacetime geometry \cite{Randall:1999ee,Randall:1999vf}. This theory is the Randall-Sundrum 1 (RS1) model, and the focus of this dissertation.

Relative to the usual four-dimensional (4D) spacetime, the RS1 model adds a finite extra dimension of space with length $\pi r_{c}$ which is parameterized by a coordinate $y\in\{0,\pi r_{c}\}$, where $r_{c}$ is called the compactification radius. At either end of the dimension is a four-dimensional hypersurface called a brane, with the five-dimensional spacetime between the branes being called the bulk. Typically, the four-dimensional world as we know it (e.g. the matter content) is placed on one brane (the ``visible brane") and only gravity is allowed to freely propagate through the bulk. Extra-dimensional warping is achieved by the presence of a warp factor $\vep \equiv e^{-kr_{c}|\varphi|}$ in the RS1 spacetime metric, where $k$ is called the warping parameter and $\varphi \equiv y/r_{c} \in \{0,\pi\}$ is a unitless version of the extra-dimensional coordinate. This warp factor enters into other aspects of RS1 calculations. For example, a fundamental energy scale $\Lambda$ in the bulk can be warped down to $\Lambda \, e^{-kr_{c}\pi} $ for an observer on the visible brane. In particular, we can set $\Lambda \approx M_{\text{Pl}}$ and its warped value $\Lambda \, e^{-kr_{c}\pi} \approx v_{\text{EW}}$ by choosing $kr_{c} \approx 12$, such that the hierarchy problem has gone from trying to explain the large ratio $M_{\text{Pl}}/v_{\text{EW}} \sim 5\times 10^{16}$ to trying to explain the order-10 number $kr_{c}\sim 12$. Unfortunately, this warp factor is not universally beneficial: whereas strong \& quantum gravitational effects force 4D gravity to break down at $M_{\text{Pl}}$, the RS1 model breaks down at the scale $\Lambda_{\pi} \equiv M_{\text{Pl}}\, e^{-kr_{c}\pi}$ instead. Thus, if $kr_{c} \sim 12$ as motivated by the hierarchy problem, then $\Lambda_{\pi} \sim v_{\text{EW}}$, and the theory becomes strongly coupled at LHC-relevant energy scales. As collider constraints confirm the Standard Model to increasingly high energies, $kr_{c}$ is driven necessarily lower, and the RS1 models creeps further away from a solution to the hierarchy problem. Nowadays, the RS1 model is utilized in relation to theoretical problems such as the AdS/CFT correspondence \cite{ArkaniHamed:2000ds,Rattazzi:2000hs} and as a model that generates phenomenologically-interesting massive spin-2 particles \cite{CMS:2016crm}.

Regardless of the specific value of $kr_{c}$ used, the size $\pi r_{c}$ of the extra-dimension is assumed small so that the five-dimensional (5D) nature of spacetime remains hidden at low energies (thereby explaining why we do not experience an extra spatial dimension in everyday life). In a sense, the relationship between the 5D RS1 spacetime and the usual 4D spacetime is similar to the relationship between a realistic sheet of paper (which has small but finite thickness) and its approximation as a two-dimensional plane. Because particles with sufficient energy can propagate throughout the full five-dimensional RS1 spacetime, the symmetry group relevant to high-energy particles is the 5D RS1 diffeomorphism group, which is gauged by the 5D RS1 graviton described by a 5D field $\hat{H}(x,y)$. At low energies, particles can no longer meaningfully probe the extra dimension, and the 5D RS1 diffeomorphism group is spontaneously broken down to a subgroup containing the usual 4D diffeomorphism group, which is gauged by the 4D graviton described by a 4D field $\hat{h}^{(0)}(x)$. In total, spontaneous symmetry breaking in the RS1 model results in the following 4D particle content:
\begin{itemize}
    \item[$\bullet$] the 4D graviton, $h^{(0)}$, a massless spin-2 particle
    \item[$\bullet$] the radion, $r^{(0)}$, a massless spin-0 particle
    \item[$\bullet$] KK modes, $h^{(n)}$ for $n\in\{1,2,\dots\}$, infinitely many massive spin-2 particles
\end{itemize}
in a process called Kaluza-Klein (KK) decomposition. The value $n$ for a particular KK mode $h^{(n)}$ is called its KK number. The KK modes gain masses by absorbing degrees of freedom from the 5D RS1 graviton, which is reflected in the fact that a massive spin-2 particle in four dimensions and a massless 5D graviton both have five states. Because of its qualitative similarities to electroweak symmetry breaking and its use of a nontrivial background geometry to achieve spontaneous symmetry breaking, this has been referred to as a ``geometric Higgs mechanism" \cite{Chivukula:2004qh}. The radion $r^{(0)}$ is a massless spin-0 particle generated by disturbing the separation distance between the branes.

Due to their common origin in the RS1 model, the scattering of 4D gravitons and the scattering of massive KK modes are closely related. In particular, (as demonstrated in this dissertation) the high-energy growth of the matrix elements describing 4D graviton and KK mode scatterings are identical. Before describing how this is possible in the RS1 model, let us first describe an analogous calculation in a model with finitely many particles: the Standard Model. In this case, the intermediate vector bosons ($W^{\pm}$, $Z$) are special with respect to electroweak symmetry breaking (EWSB) because they are massive superpositions of the original $\mathbf{SU(2)_{W}}\times\mathbf{U(1)_{Y}}$ gauge bosons ($W^{1}$,$W^{2}$,$W^{3}$,$B$); this contrasts with the situation of the fermions and even the Higgs boson, although they also gain masses as a result of EWSB. The only superposition of $\mathbf{SU(2)_{W}}\times\mathbf{U(1)_{Y}}$ gauge bosons that remains massless is the photon ($\gamma$), which gauges the electromagnetic $\mathbf{U(1)_{Q}}$.

Because the photon has no cubic or quartic self-interactions, its center-of-momentum frame 2-to-2 tree-level scattering matrix element (hereafter referred to simply as ``matrix element" for brevity) vanishes identically: $\mathcal{M} = 0$. Let $E$ denote the incoming center-of-momentum energy of this process, so that the Mandelstam variable $s$ equals $E^{2}$. In terms of high-energy growth, the photon scattering matrix element (trivially) scales like $\mathcal{O}(s^{0})$. Another way in which we could have arrived at this same scaling is by combining the following facts:
\begin{itemize}
    \item[$\bullet$] A 4D matrix element must be unitless.
    \item[$\bullet$] There is no energy scale available to this process.
\end{itemize}
The latter point means that there are no quantities with which to cancel any powers of energy introduced by factors of $s$, and thus the only way for the matrix element to be consistent with the first point is to scale like $\mathcal{O}(s^{0})$ at high energies (which $\mathcal{M}=0$ does trivially, as previously mentioned). Diagrammatically, we write
\begin{align}
    \mathcal{M}_{\gamma\gamma\rightarrow\gamma\gamma} &\hspace{0 pt} = \hspace{0 pt} \raisebox{-0.45\height}{\totDklmn{0.25}} \hspace{25 pt} \sim \hspace{25 pt} \mathcal{O}(s^{0})
\end{align}

In contrast, the 2-to-2 scattering of massive spin-1 particles (such as the W-bosons) {\it does} have access to another energy scale: the particle's mass. For example, an external massive spin-1 particle with mass $m$, $4$-momentum $p$, and helicity $\lambda$ will enter a matrix element calculation with any one of three possible polarization vectors:
\begin{align}
    [\epsilon^{\mu}_{\pm 1}(p)] = \pm \dfrac{e^{\pm i \phi}}{\sqrt{2}} \matrixda{0}{-c_{\theta}c_{\phi} \pm is_{\phi}}{-c_{\theta}s_{\phi}\mp i c_{\phi}}{s_{\theta}} \hspace{35 pt} [\epsilon^{\mu}_{0}(p)] = \dfrac{1}{m} \matrixda{|\vec{p}|}{E\, c_{\phi} s_{\theta}}{E\, s_{\phi} s_{\theta}}{E\, c_{\theta}} = \dfrac{1}{m} \matrixba{|\vec{p}|}{E\,\hat{p}}
\end{align}
corresponding to helicities $\lambda = \pm1$ and $\lambda = 0$ respectively, where $(\phi,\theta)$ determines the $3$-direction of $\vec{p}$ in spherical coordinates and $(c_{x},s_{x})\equiv (\cos x,\sin x)$. The components of the helicity-zero polarization vector $\epsilon^{\mu}_{0}(p)$ diverge like $\mathcal{O}(E/m) = \mathcal{O}(\sqrt{s}/m)$ at high energies, which is only made possible by the existence of the mass $m$. A massless spin-1 particle such as the photon only has access to the helicity $\lambda=\pm1$ states, which are independent of mass and energy.

Because each massive spin-1 state has three helicity options, the external states in a 2-to-2 massive spin-1 scattering process can be in any one of $3^{4} = 81$ helicity combinations (although many of these are related to one another through crossing symmetry).  Because the helicity-zero polarization vector diverges most quickly in energy, it is perhaps unsurprising that the fastest growing matrix element is typically attained by setting all external helicities to zero. We will refer to such a process as a ``helicity-zero process." It is not unusual for a helicity-zero matrix element describing massive spin-1 scattering to grow as fast as $\mathcal{O}(s^{2})$ at high energies.

However, this is not what happens in the SM. Instead, the helicity-zero matrix element grows like $\mathcal{O}(s^{0})$:
\begin{align}
    \mathcal{M}_{WW\rightarrow WW} &\hspace{0 pt} = \hspace{0 pt} \raisebox{-0.45\height}{\totDWWWW{0.25}} \hspace{25 pt} \mathrel{\mathop{\sim}^{\text{helicity}}_{\text{zero}}} \hspace{25 pt} \mathcal{O}(s^{0}) \label{MWWtoWW}
\end{align}
Table \ref{Table - WW-to-WW Scattering} summarizes the various diagrams that sum to form this matrix element, including their individual high-energy behaviors. Several channels exhibit $\mathcal{O}(s^{2})$ growth, but cancellations occur when all diagrams are summed together which ultimately result in a net $\mathcal{O}(s^{0})$ growth, the same growth as the photon scattering matrix element.

\begin{table}[t]
    \bgroup
    \begin{center}
        \begin{tabular}{| c || c | c | c | c |}
             \hline
             & & & & \\[-0.75 em]
             $\mathcal{M}_{\small WW\rightarrow WW} =$ & $\mathcal{M}_{\text{c}}$ & $+\, \mathcal{M}_{H}$ & $+\, \mathcal{M}_{\gamma}$ & $+\, \mathcal{M}_{Z}$\\
             & & & & \\[-0.75 em]
             \hline
             & & & & \\[-0.75 em]
             {\bf Mediator:} & {-} & {Higgs} & {photon} & Z-boson\\[0.25 em]
             \hline
             & & & & \\[-0.75 em]
             & \multirow{3}{*}{\sgDWWWW{0.20}} & \sDWWHWW{0.20} & \sDWWAWW{0.20} & \sDWWZWW{0.20}\\
             {\bf Diagrams:} & & $+$ & $+$ & $+$ \\
             &  & \uDWWHWW{0.20} & \uDWWAWW{0.20} & \uDWWZWW{0.20}\\[0.25 em]
             \hline
             & & & & \\[-0.75 em]
             {\bf Helicity-Zero} & \multirow{3}{*}{$\sim\mathcal{O}(s^{2})$} & \multirow{3}{*}{$\sim\mathcal{O}(s)$} & \multirow{3}{*}{$\sim\mathcal{O}(s^{2})$} & \multirow{3}{*}{$\sim\mathcal{O}(s^{2})$} \\
             {\bf High-Energy} & & & & \\ 
             {\bf Scaling:} & & & & \\[0.25 em]
             \hline
        \end{tabular}
    \end{center}
    \egroup
\caption{The various diagrams that contribute to the tree-level matrix element for the 2-to-2 Standard Model scattering process $W^{+}W^{-}\rightarrow W^{+}W^{-}$ and their high-energy behaviors when all external helicities vanish. The tree-level matrix element $\mathcal{M}_{WW\rightarrow WW}$ from Eq. \eqref{MWWtoWW} is the sum of these diagrams. Because of cancellations between diagrams, $\mathcal{M}_{WW\rightarrow WW}$ scales like $\mathcal{O}(s^{0})$, just like the 2-to-2 photon scattering matrix element $\mathcal{M}_{\gamma\gamma\rightarrow\gamma\gamma}$.}
\label{Table - WW-to-WW Scattering}
\end{table}

The existence of cancellations which reduce $\mathcal{O}(s^{2})$ growth to $\mathcal{O}(s^{0})$ growth is not a coincidence: even though the electroweak gauge group $\mathbf{SU(2)_{W}}\times\mathbf{U(1)_{Y}}$ has been spontaneously broken down to the electromagnetic gauge group $\mathbf{U(1)_{Q}}$, this fundamental symmetry still protects the scattering behavior of the related gauge bosons. Thus, the overall high-energy growth of the matrix element describing 2-to-2 scattering of the W-bosons (which are superpositions of the $\mathbf{SU(2)_{W}}$ gauge bosons) matches that of the 2-to-2 scattering of photons (which gauge the remaining $\mathbf{U(1)_{Q}}$).

The main result of this dissertation is the demonstration that similar cancellations occur in the Randall-Sundrum 1 model. In this case, a nontrivial background geometry at low energies causes the 5D RS1 diffeomorphism group to be spontaneously broken down to a subgroup containing the 4D diffeomorphism group. This latter group is gauged by the usual massless graviton.

Unlike the case of photon scattering that we previously considered, 4D gravity has an implicit energy scale: the Planck mass $M_{\text{Pl}}$. This scale enters the graviton scattering matrix element via the 4D gravitational coupling $\kappa_{\text{4D}}\equiv 2/M_{\text{Pl}}$, of which two instances are present in any given tree-level diagram. In order to be unitless overall, the matrix element must contribute a factor of $s = E^{2}$ to compensate, and thus it grows like
\begin{align}
    \mathcal{M}_{00\rightarrow 00} &\hspace{0 pt} = \hspace{0 pt} \raisebox{-0.45\height}{\totDzzzz{0.25}} \hspace{25 pt} \sim \hspace{25 pt} \mathcal{O}(s)
\end{align}
at high energies. The label ``0" indicates the 4D graviton, $h^{(0)}$, each instance of which can have helicity $\lambda = \pm 2$.

If we instead consider tree-level 2-to-2 scattering of massive spin-2 particles (such as the RS1 KK modes), then each external state will be associated with any one of five possible polarization tensors, $\epsilon^{\mu\nu}_{\lambda}(p)$:
\begin{align}
    \epsilon_{\pm 2}^{\mu\nu}(p) &= \epsilon_{\pm1}^{\mu}(p)\, \epsilon_{\pm 1}^{\nu}(p)~,\\
    \epsilon_{\pm 1}^{\mu\nu}(p) &= \dfrac{1}{\sqrt{2}}\left[\epsilon_{\pm 1}^{\mu}(p)\,\epsilon_{0}^{\nu}(p) + \epsilon_{0}^{\mu}(p)\, \epsilon_{\pm 1}^{\nu}(p)\right]~\\
    \epsilon_{0}^{\mu\nu}(p) &= \dfrac{1}{\sqrt{6}}\bigg[\epsilon^{\mu}_{+1}(p)\, \epsilon^{\nu}_{-1}(p) + \epsilon^{\mu}_{-1}(p)\, \epsilon^{\nu}_{+1}(p) + 2\epsilon^{\mu}_{0}(p)\,\epsilon^{\nu}_{0}(p)\bigg]~,
\end{align}
where $\epsilon^{\mu}_{\lambda}(p)$ are the previously-defined spin-1 polarization vectors. As in the massive spin-1 case, the most divergent of these is the helicity-zero option, which grows like $\mathcal{O}(s/m^{2})$ at large energies. Massive spin-2 scattering matrix elements have $5^{4} = 625$ possible helicity combinations (many related to one another via crossing symmetry), but the helicity-zero combination is typically the most divergent, usually growing as fast as $\mathcal{O}(s^{5})$.

Keeping this in mind, consider the matrix element $\mathcal{M}_{n_{1}n_{2}\rightarrow n_{3}n_{4}}$ corresponding to the helicity-zero KK mode scattering process $h^{(n_{1})}h^{(n_{2})}\rightarrow h^{(n_{3})}h^{(n_{4})}$ where the KK numbers $n_{1}$, $n_{2}$, $n_{3}$, and $n_{4}$ are all nonzero. Table \ref{Table - KK-to-KK Scattering} summarizes the diagrams which sum to form $\mathcal{M}_{n_{1}n_{2}\rightarrow n_{3}n_{4}}$ and their high-energy behaviors when all external helicities vanish. As anticipated in the previous paragraph, nearly every diagram that contributes to this matrix element diverges like $\mathcal{O}(s^{5})$. However, this dissertation demonstrates explicitly that nontrivial cancellations occur between these infinitely-many diagrams such that the full matrix element diverges like $\mathcal{O}(s)$:
\begin{align}
    \mathcal{M}_{n_{1}n_{2}\rightarrow n_{3}n_{4}} &\hspace{0 pt} = \hspace{0 pt} \raisebox{-0.45\height}{\totDklmn{0.25}} \hspace{25 pt} \mathrel{\mathop{\sim}^{\text{helicity}}_{\text{zero}}} \hspace{25 pt} \mathcal{O}(s) \label{M12to34}
\end{align}
which is precisely the energy growth found in the 4D graviton scattering channel. The conceptual similarities between the Standard Model and RS1 model are summarized in Table \ref{Table - Analogy SM to RS1}, with our original results indicated in bold and red. We also demonstrate in this dissertation that the RS1 strong coupling scale $\Lambda_{\pi} = M_{\text{Pl}}\, e^{-kr_{c}\pi}$ can be calculated directly from the 4D effective RS1 model.

\begin{table}[t]
    \bgroup
    \begin{center}
        \begin{tabular}{| c || c | c | c | c |}
             \hline
             & & & & \\[-0.75 em]
             $\mathcal{M}_{n_{1}n_{2}\rightarrow n_{3}n_{4}} =$ & $\mathcal{M}_{\text{c}}$ & $+\, \mathcal{M}_{\text{r}}$ & $+\, \mathcal{M}_{0}$ & ${\displaystyle +\sum_{j>0}\mathcal{M}_{j}}$\\
             & & & & \\[-0.75 em]
             \hline
             & & & & \\[-0.75 em]
             \multirow{2}{*}{\bf Mediator:} & \multirow{2}{*}{-} & \multirow{2}{*}{radion} & \multirow{2}{*}{graviton} & massive spin-2\\[0.25 em]
             & & & & KK mode\\[0.25 em]
             \hline
             & & & & \\[-0.75 em]
             \multirow{7}{*}{\bf Diagrams:}& & \sDklrmn{0.25} & \sDklzmn{0.25} & \sDkljmn{0.25}\\
             & & $+$ & $+$ & $+$ \\
             & \sgDklmn{0.25} & \tDklrmn{0.25} & \tDklzmn{0.25} & \tDkljmn{0.25}\\
             & & $+$ & $+$ & $+$ \\
             & & \uDklrmn{0.25} & \uDklzmn{0.25} & \uDkljmn{0.25}\\[0.25 em]
             \hline
             & & & & \\[-0.75 em]
             {\bf Helicity-Zero} & \multirow{3}{*}{$\sim\mathcal{O}(s^{5})$} & \multirow{3}{*}{$\sim\mathcal{O}(s^{3})$} & \multirow{3}{*}{$\sim\mathcal{O}(s^{5})$} & \multirow{3}{*}{$\sim\mathcal{O}(s^{5})$} \\
             {\bf High-Energy} & & & & \\ 
             {\bf Scaling:} & & & & \\[0.25 em]
             \hline
        \end{tabular}
    \end{center}
    \egroup
\caption{The various diagrams that contribute to the tree-level matrix element for the 2-to-2 RS1 model scattering process $h^{(n_{1})}h^{(n_{2})}\rightarrow h^{(n_{3})}h^{(n_{4})}$ and their high-energy behaviors when all external helicities vanish. The tree-level matrix element $\mathcal{M}_{n_{1}n_{2}\rightarrow n_{3}n_{4}}$ from Eq. \eqref{M12to34} is the sum of these diagrams. Because of cancellations between diagrams, the overall matrix element $\mathcal{M}_{n_{1}n_{2}\rightarrow n_{3}n_{4}}$ scales like $\mathcal{O}(s)$, just like the 2-to-2 graviton scattering matrix element $\mathcal{M}_{00\rightarrow 00}$. The confirmation and detailed demonstration of these cancellations is a major result of this dissertation.}
\label{Table - KK-to-KK Scattering}
\end{table}

\begin{table}
\bgroup
\begin{center}
    \begin{tabular}{| c || c | c |}
        \hline
         & & \\[-0.75 em]
         & {\bf Standard Model} & {\bf Randall-Sundrum 1}\\[0.25 em]
         \hline
         \hline
         & & \\[-0.75 em]
         The fundamental symmetry group... & $\mathbf{SU(2)_{W}}\times \mathbf{U(1)_{Y}}$ & 5D diffeomorphisms\\[0.25 em]
         \hline
         & & \\[-0.75 em]
         ... w/ unitarity-violation scale... & N/A & $\Lambda_{\pi} = M_{\text{Pl}}\, e^{-kr_{c}\pi}$\\[0.25 em]
         \hline
         & & \\[-0.75 em]
         ... and gauged by the... & electroweak bosons & 5D RS1 graviton\\[0.25 em]
         \hline
         & & \\[-0.75 em]
         ... is spontaneously broken by... & the Higgs vev & background geometry\\[0.25 em]
         \hline
         & & \\[-0.75 em]
         ... to a new symmetry group... & $\mathbf{U(1)_{Q}}$ & 4D diffeomorphisms*\\[0.25 em]
         \hline
         & & \\[-0.75 em]
         ... gauged by the... & photon, $\gamma$ & 4D graviton, $h^{(0)}$\\[0.25 em]
         \hline
         & & \\[-0.75 em]
         ... resulting in a spin-0 state... & Higgs boson, $H$ & radion, $r^{(0)}$\\[0.25 em]
         \hline
         & & \\[-0.75 em]
         ... as well as massive states & $W$-bosons, $W^{\pm}$ & spin-2 KK modes, $h^{(n)}$\\[0.25 em]
         built from fund. gauge bosons... & and $Z$-boson, $Z$ & for $n\in\{1,2,\dots\}$\\[0.25 em]
         \hline
         \hline
         & & \\[-0.75 em]
         The $2$-to-$2$ gauge boson process... & $\gamma\gamma \rightarrow \gamma\gamma$ & $h^{(0)}h^{(0)}\rightarrow h^{(0)}h^{(0)}$\\[0.25 em]
         \hline
         & & \\[-0.75 em]
         ... has $\mathcal{M}$ w/ high-energy growth $\sim$ & $\mathcal{O}(s^{0})$ & $\mathcal{O}(s)$\\[0.25 em]
         \hline
         & & \\[-0.75 em]
         ... or, if naively given mass, ... & $\mathcal{O}(s^{2})$ & $\mathcal{O}(s^{5})$\\[0.25 em]
         \hline
         & & \\[-0.75 em]
         ... yet $2$-to-$2$ massive state process & \multirow{2}{*}{$W^{+}W^{-}\rightarrow W^{+}W^{-}$} & \multirow{2}{*}{$h^{(n_{1})}h^{(n_{2})}\rightarrow h^{(n_{3})}h^{(n_{4})}$}\\[0.25 em]
         where mass arises via sym. break... &  & \\[0.25 em]
         \hline
         & & \\[-0.75 em]
         ... has $\mathcal{M}$ w/ high-energy growth $\sim$ & $\mathcal{O}(s^{0})$ & \textcolor{red}{$\mathbf{O(s)}$}\\[0.25 em]
         \hline
         \hline
         & & \\[-0.75 em]
         Breaking the fund. symmetry by... & elim. $Z$ & KK tower truncation \\[0.25 em]
         \hline
         & & \\[-0.75 em]
         ... makes massive states scatter like & \multirow{2}{*}{$\mathcal{O}(s^{2})$} & \multirow{2}{*}{\textcolor{red}{$\mathbf{O(s^{5})}$}}  \\[0.25 em]
         naively-massive gauge bosons, $\mathcal{M}\sim$ & & \\[0.25 em]
         \hline
         \hline
         & & \\[-0.75 em]
         Breaking the fund. symmetry by... & elim. the Higgs & elim. the radion \\[0.25 em]
         \hline
         & & \\[-0.75 em]
         ... makes massive states scatter $\sim$ & $\mathcal{O}(s)$ & \textcolor{red}{$\mathbf{O(s^{3})}$} \\[0.25 em]
         \hline
    \end{tabular}
\end{center}
\egroup
\caption{The Standard Model (SM) and the Randall-Sundrum 1 (RS1) model share a chain of conceptual similarities with respect to the scattering of particles made massive by spontaneous symmetry breaking. The Mandelstam variable $s\equiv E^{2}$, where $E$ is the incoming center-of-momentum energy. Original results presented in this dissertation are indicated in bold and red. (* - Technically, the new symmetry group is the Cartan subgroup of the 5D diffeomorphisms that contains the 4D diffeomorphisms.)}
\label{Table - Analogy SM to RS1}
\end{table}

Additionally, in practice if we intend to perform a numerical calculation (as might be relevant to experimental applications of the RS1 model) then we must truncate the number of KK modes we include as intermediate states (e.g. replacing the sum $\sum_{j=0}^{+\infty}\mathcal{M}_{j}$ in the matrix element with $\sum_{j=0}^{N}\mathcal{M}_{j}$ for some integer $N$). Because the entire tower is required in order to cancel the leading $\mathcal{O}(s^{5})$ growth, truncating the KK tower too low can cause the matrix element to violate partial wave unitarity well below the strong coupling scale $\Lambda_{\pi}$.\footnote{Truncation of the KK tower corresponds to explicit breaking of the underlying gravitational symmetry group. \cite{PhysRevLett.52.14} details this type of symmetry breaking in the case of the five-dimensional torus.} Furthermore, because the radion contributes matrix elements with $\mathcal{O}(s^{3})$ growth, proper inclusion of the radion is also vital to avoiding partial wave unitarity constraints. The effect of KK tower truncation and inclusion of the radion on the accuracy of KK mode scattering matrix elements is also investigated in this dissertation.

The remainder of the dissertation details the original results published in \cite{Chivukula:2019rij,Chivukula:2019zkt,Chivukula:2020hvi}, as well as generalizing and elaborating on aspects of those calculations in ways that have not yet been submitted for publication. It is organized as follows:
\begin{itemize}
    \item[$\bullet$] Chapter 2 establishes definitions and conventions from 4D quantum field theory relevant to the dissertation. In the interest of acting as a useful resource, it also provides a detailed derivation of 2-to-2 partial wave unitarity constraints and helicity eigenstates from first principles.
    \item[$\bullet$] Chapter 3 calculates the 5D weak field expanded RS1 Lagrangian $\mathcal{L}_{\text{5D}}$ to quartic order in the 5D fields or (equivalently) second order in the 5D coupling $\kappa_{\text{5D}}$. We demonstrate that all terms containing factors of $(\partial_{\varphi}|\varphi|)$ or $(\partial^{2}_{\varphi}|\varphi|)$ are cancelled to all orders in $\kappa_{\text{5D}}$.
    \item[$\bullet$] Chapter 4 presents an original parameterization of the 4D effective RS1 Lagrangian which manifests as a ``5D-to-4D formula" and categorizes all RS1 couplings as either ``A-type" or ``B-type" depending on the associated derivative content of the interaction. Many relationships between RS1 couplings and masses are derived; these significantly generalize our existing published work and will be submitted for publication in a future paper.
    \item[$\bullet$] Chapter 5 demonstrates that the matrix element describing massive spin-2 KK mode scattering in the 5D orbifolded torus and RS1 models exhibits $\mathcal{O}(s)$ growth after cancellations of more divergent behavior. From cancellations in the helicity-zero elastic case ($h^{(n)}h^{(n)}\rightarrow h^{(n)}h^{(n)}$) we derive sum rules relating KK mode masses and couplings, all but one of which we prove analytically. The final sum rule is demonstrated numerically. The RS1 strong coupling scale $\Lambda_{\pi} = M_{\text{Pl}}\, e^{-kr_{c}\pi}$ is calculated numerically in the 4D effective RS1 model and the effect of KK tower truncation on matrix element accuracy is investigated. These important original results have been published across several papers \cite{Chivukula:2019rij,Chivukula:2019zkt,Chivukula:2020hvi}.
    \item[$\bullet$] Chapter 6 concludes by summarizing the original results presented in the dissertation as well as future projects we will be pursuing based on this work.
\end{itemize}

\chapter{2-to-2 Scattering and Helicity Eigenstates} \label{C - QFT}

\section{Chapter Summary}
This chapter establishes various definitions and conventions from four-dimensional (4D) quantum field theory which are relevant to this dissertation, e.g. that we use the `mostly-minus' Minkowski metric and all indices are raised/lowered with the Minkowski metric. It is written with the aim of providing a self-consistent collection of standard derivations which all use the same conventions. This is done under the belief that such a collection could be useful to other physicists. As such, many details and observations are intentionally included which are often skipped in standard resources. For physicists who are already familiar with 2-to-2 scattering calculations involving helicity eigenstates, much of this chapter can be skimmed without missing details vital to the remainder of this dissertation.

This chapter is organized as follows:
\begin{itemize}
    \item[$\bullet$] Section \ref{S - Classical} derives the Lorentz and Poincar\'{e} groups from the assumption that the speed of light is globally invariant between reference frames. Active forms for the Poincar\'{e} transformations (rotations, boosts, spacetime translations) and their generators ($\vec{J}$, $\vec{K}$, $P^{\mu}$) are provided in the 4-vector representation, and the commutation structure of the generators is derived. The section closes by deriving the Lorentz-invariant phase space.
    \item[$\bullet$] Section \ref{S- Quantum Promo} considers the infinite-dimensional unitary representations of the Poincar\'{e} group, then finite-dimensional non-unitary representations of the Lorentz group. Unitary Poincar\'{e} representations are attained by promoting the Poincar\'{e} generators to Hermitian operators and the corresponding Poincar\'{e} transformations to unitary operators. The helicity operator $\Lambda$ is introduced.
    \item[$\bullet$] Section \ref{S - External States and Matrix Elements} defines single-particle $4$-momentum external states, which are then combined to form multi-particle $4$-momentum external states. Special care is taken to consider multi-particle states involving identical particles. The $S$-matrix element is introduced and its relation to the matrix element $\mathcal{M}$ is mentioned.
    \item[$\bullet$] Section \ref{S - 2to2} describes 2-to-2 particle processes in detail, with emphasis on scattering in the center-of-momentum (COM) frame and parameterization via the Mandelstam variables. An equation for simplifying integrals over the $4$-momenta of two particles is derived and then applied to unitarity of the $S$-matrix in order to derive the optical theorem.
    \item[$\bullet$] Section \ref{Angular Momentum} summarizes the usual treatment of angular momentum in quantum mechanics including how angular momentum representations are combined, and defines the Wigner D-matrix.
    \item[$\bullet$] Section \ref{S - Helicity} considers single-particle helicity eigenstates, which are then combined to form multi-particle helicity eigenstates. Using the relationship between helicity eigenstates and angular momentum eigenstates, the matrix element is decomposed in an angular momentum basis as to define partial wave amplitudes. The elastic and inelastic partial wave unitarity constraints are derived.
    \item[$\bullet$] Section \ref{S - Fields} derives the spin-1 and spin-2 polarization structures. Various canonical quadratic Lagrangians are considered, and their corresponding propagators are listed.
\end{itemize}

\section{Poincar\'{e} Group: 4-Vector Representation} \label{S - Classical}

\subsection{Preserving the Speed of Light} \label{Classical - Minkowski}

At the heart of modern relativity theory lies an axiom with far-reaching consequences: no matter how different the reference frames of two observers, they will agree that a wavepacket of light travels at a speed $c$. This defines the aptly-named speed of light.

Every reference frame is characterized by a choice of coordinates, which presently means a unique continuous association of every point of reality with a time coordinate $ct = x^{0}$ and some spatial coordinates $\vec{x} = (x^{1},x^{2},x^{3})$. In such a reference frame, a wavepacket of light will travel along some curve $\vec{x}(t)$ through three-dimensional space and, according to the aforementioned axiom of relativity, do so at the speed of light, such that $c = |d\vec{x}/dt|$; however, it is worthwhile to recast this universal property as an equation relating differentials along the motion of the wavepacket:
\begin{align}
    c =\left| \dfrac{d\vec{x}}{dt} \right| \hspace{35 pt} \implies \hspace{35 pt} c |dt| =|d\vec{x}| \hspace{35 pt} \implies \hspace{35 pt} c^{2} |dt|^{2} - |d\vec{x}|^{2} =0
\end{align}
This latter form is useful because it treats the space and time coordinates equivalently, with the speed of light amounting to a conversion from time duration units to length units. According to relativity theory, although an observer in a different inertial reference frame with different coordinates $(ct^{\prime},\vec{x}^{\,\prime})$ will measure that same wavepacket as traveling along a different trajectory $\vec{x}^{\,\prime}(t^{\prime})$, they will still find that its speed  $|d\vec{x}^{\,\prime}/dt|$ equals $c$ at every point along its path, or equivalently
\begin{align}
    c^{2} |dt^{\prime}|^{2} - |d\vec{x}^{\,\prime}|^{2} = 0 \label{speedoflightconstant}
\end{align}
This invariance greatly restricts the structure of reality. Imagine flooding reality with wavepackets of light that propagate in all directions and at every point of time and space. By the axiom of relativity, an observer in any other reference frame must also agree that every wavepacket in this vast network travels at the speed of light, even if their own perception of spacetime is wildly different. This puts a tight constraint on the local structure of reality itself, and requires that space and time must be woven together into a unified manifold of spacetime.

Consider the possible $4$-velocities $v^{\mu}\equiv (v^{0},\vec{v}) = (v^{0},v^{1},v^{2},v^{3})$ of a trajectory passing through a certain spacetime point. If the trajectory describes the motion of a wavepacket of light as in Eq. \eqref{speedoflightconstant}, then the $4$-velocity is light-like: $v^{2} \equiv v\cdot v= 0$, where
\begin{align}
    (v\cdot v) \equiv \eta_{\mu\nu} v^{\mu} v^{\nu} \equiv \sum_{\mu,\nu = 0}^{3} \eta_{\mu\nu} v^{\mu} v^{\nu} \label{Eq2point3}
\end{align}
and $\eta_{\mu\nu}$ is the Minkowski metric
\begin{align}
    [\eta_{\mu\nu}] &= \text{Diag}(+1,-1,-1,-1) \equiv \left(\hspace{-5 pt}\begin{tabular}{ c c c c}
    ${+1}$ & ${0}$ & ${0}$ & ${0}$\\
    ${0}$ & ${-1}$ & ${0}$ & ${0}$\\
    ${0}$ & ${0}$ & ${-1}$ & ${0}$\\
    ${0}$ & ${0}$ & ${0}$ & ${-1}$
    \end{tabular}\hspace{-5 pt}\right)
\end{align}
when expressed as a matrix with those components; this square bracket notation will be used throughout this dissertation. The metric is symmetric by construction ($\eta_{\mu\nu} = \eta_{\nu\mu}$) and we define it in the ``mostly-minus" convention, i.e. it has one $+1$ eigenvalue and three $-1$ eigenvalues corresponding to temporal and spatial information respectively. Note that the first equality in Eq. \eqref{Eq2point3} makes use of the Einstein summation convention, wherein repeated indices indicate sums over the corresponding index ranges; the Einstein summation convention will be used throughout the remainder of this dissertation as well. Using the Minkowski metric, we can rewrite and generalize Eq. \eqref{speedoflightconstant} as to define the invariant spacetime interval $ds^{2}$ associated with a generic (not necessarily light-like) infinitesimal spacetime displacement $dX$:
\begin{align}
    ds^{2} \equiv \eta_{\mu\nu}\, dX^{\mu}\, dX^{\nu}
\end{align}
This is termed ``invariant" for reasons that will be detailed shortly.

For the sake of performing calculations, it is vital to generalize the above language to include generic $4$-vectors, e.g. objects of the form $a = (a^{0},a^{1},a^{2},a^{3})$ for which $(a\cdot a)$ does not necessarily vanish. Through the Minkowski metric $\eta$, a generic $4$-vector $a^{\mu}$ implies a related $4$-covector $a_{\mu}$
\begin{align}
    a_{\mu} = (a_{0},a_{1},a_{2},a_{3}) \equiv \eta_{\mu\nu} a^{\nu} = (a^{0},-a^{1},-a^{2},-a^{3})
\end{align}
and for generic $4$-vectors $a$ and $b$ the previous inner product generalizes to
\begin{align}
    (a\cdot b) = \eta_{\mu\nu} a^{\mu} b^{\nu} = a_{\mu} b^{\mu} = a^{0} b^{0} - a^{1} b^{1} - a^{2} b^{2} - a^{3} b^{3}
\end{align}
where $(a\cdot a)$ is called the magnitude of $a$. Sometimes we will break a $4$-vector $a$ into its temporal $a^{0}$ and spatial $a^{i}$ components, the latter of which comprise a $3$-vector $\vec{a} = (a^{1},a^{2},a^{3})$. $3$-vectors are defined with the usual $3$-vector inner product, i.e.
\begin{align}
    \vec{a}\cdot \vec{b} = a^{i} b^{i} = a^{1}b^{1} + a^{2} b^{2} + a^{3}b^{3}
\end{align}
such that the $4$-vector inner product equals
\begin{align}
    a\cdot b = a^{0} b^{0} - \vec{a}\cdot \vec{b}
\end{align}
To avoid confusion, four-dimensional (4D) spacetime indices are labeled via lowercase Greek letters ($\mu$, $\nu$, $\rho$, ...) with $\mu \in\{0,1,2,3\}$, whereas three-dimensional (3D) spatial indices are labeled via lowercase Latin letters ($i$, $j$, $k$, ...) with $i \in\{1,2,3\}$. In the next chapter, we consider five-dimensional (5D) spacetime indices, which are labeled via uppercase Latin letters ($M$, $N$, $R$, ...) with $M\in\{0,1,2,3,5\}$. The $3$-vector components will sometimes be relabeled to make contact with the usual $(x,y,z)$-rectilinear $3$-space coordinates, in which case $a_{x} \equiv a^{1}$, $a_{y} \equiv a^{2}$, and $a_{z} \equiv a^{3}$. As above, we will use the 4D Minkowski metric for raising and lowering four-dimensional indices, whereas in the next chapter we will raise and lower five-dimensional indices with the 5D Minkowski metric $[\eta_{MN}] \equiv \text{Diag}(+1,-1,-1,-1,-1)$.

Returning to the invariance of the speed of light, consider the classification of all invertible linear transformations $\lambda$ that preserve light-like magnitudes:
\begin{align}
    v\cdot v = \eta_{\mu\nu} v^{\mu}v^{\nu} = 0\hspace{35 pt}\implies\hspace{35 pt} (\lambda v)\cdot(\lambda v) = \eta_{\mu\nu} (\lambda v)^{\mu} (\lambda v)^{\nu} = 0
\end{align}
As previously mentioned, demanding invariance of this inner product for all light-like $4$-vectors is a significant constraint. By expressing a generic $4$-vector as a sum of light-like $4$-vectors, it can be demonstrated that preserving light-like inner products necessarily implies the preservation of {\it all} inner products between $4$-vectors up to an overall rescaling. That is,
\begin{align}
    (\lambda a)\cdot (\lambda b) = \Omega\,(a\cdot b)
\end{align}
for a positive real number $\Omega$ (negative values of $\Omega$ are excluded because they would change the temporal dimension into a spatial dimension and vice-versa). Therefore, the linear transformation $\lambda$ decomposes into the composition of a dilation by an amount $\sqrt{\Omega}$ and a Lorentz transformation $\Lambda$ like so:
\begin{align}
    \lambda = \sqrt{\Omega} \Lambda
\end{align}
where $|\det\Lambda| = 1$ characterizes the Lorentz transformation. The dilation simply scales our time duration and length units by an equal amount $\sqrt{\Omega}$. Because we are interested in comparing reference frames that differ beyond a choice of units, we set $\Omega = 1$ so that $\lambda = \Lambda$, and we from here on restrict our attention to Lorentz transformations.

Lorentz transformations preserve $4$-vector magnitude, and therefore magnitudes can be classified in a frame-independent way: given a $4$-vector $a$, it is said to be space-like, light-like, or time-like if its magnitude is less than, equal to, or greater than $0$ respectively. These names are inspired by considering a spacetime displacement $\ell^{\mu}$ from the origin. If its magnitude vanishes $(\ell\cdot\ell) = 0$, then it is a displacement that could be traversed by a wavepacket of light. Meanwhile, a pure spatial displacement $\ell = (0,\vec{\ell})$ yields a negative magnitude $(\ell\cdot\ell) = -\vec{\ell}\cdot\vec{\ell} < 0$, and a pure temporal displacement $\ell = (\ell^{0},\vec{0}\,)$ yields a positive magnitude $(\ell\cdot\ell) = (\ell^{0})^{2} > 0$, and thus they are space-like and time-like respectively. A time-like (space-like) particle velocity corresponds to motion slower (faster) than the speed-of-light, and a trajectory is labeled space-, light-, or time-like if every $4$-velocity along that trajectory is also space-, light-, or time-like respectively.

A Lorentz $4$-vector is any $4$-vector ($4$-velocity or otherwise) that transforms under a Lorentz transformation in the way previously described: that is, the Lorentz $4$-vector $v^{\mu}$ goes to another Lorentz $4$-vector $\overline{v}^{\mu}$ after a Lorentz transformation $\Lambda$, where
\begin{align}
    \overline{v}^{\mu} \equiv {\Lambda^{\mu}}_{\nu} v^{\nu}
\end{align}
An index such as $\nu$ in $v^{\nu}$ which is transformed by contraction with ${\Lambda^{\mu}}_{\nu}$ under a Lorentz transformation $\Lambda$ is called a contravariant index. Because $(\Lambda a)\cdot (\Lambda b) = (a\cdot b)$ for all 4-vectors $a$ and $b$, Lorentz transformations preserve the metric in the following sense,
\begin{align}
    {\Lambda^{\rho}}_{\mu} {\Lambda^{\sigma}}_{\nu} \eta_{\rho\sigma} = \eta_{\mu\nu} \label{LorentzDef}
\end{align}
Furthermore, the Lorentz transformations define a group under composition (i.e. ${(\Lambda_{1})^{\mu}}_{\nu} {(\Lambda_{2})^{\nu}}_{\rho} = {(\Lambda_{3})^{\mu}}_{\rho}$), with a transformation $\Lambda$ related to its inverse $\Lambda^{-1}$ according to
\begin{align}
    {(\Lambda^{-1})^{\mu}}_{\nu} = {\Lambda_{\nu}}^{\mu}
\end{align}
because
\begin{align}
    {\Lambda_{\nu}}^{\mu} {\Lambda^{\nu}}_{\rho} = \left[\eta^{\mu\tau}\eta_{\sigma\nu}  {\Lambda^{\sigma}}_{\tau}\right] {\Lambda^{\nu}}_{\rho}  = \eta^{\mu\tau} \left[ {\Lambda^{\sigma}}_{\tau} {\Lambda^{\nu}}_{\rho} \eta_{\sigma\nu}\right] = \eta^{\mu\tau} \eta_{\tau \rho} = \eta^{\mu}_{\rho}
\end{align}
and $[\eta^{\mu}_{\rho}] = \text{Diag}(+1,+1,+1,+1) = \mathbbm{1}$. Thus, we refer to the collection of all Lorentz transformations as the Lorentz group. 

The Lorentz group can be further divided into four distinct connected components based on the determinant and temporal-temporal component of each transformation $\Lambda$:
\begin{itemize}
    \item[$\bullet$] If $\det \Lambda = +1$ then $\Lambda$ is proper. Otherwise, $\det \Lambda = -1$ and $\Lambda$ is improper.
    \item[$\bullet$] If $\Lambda_{00} \geq 1$, then $\Lambda$ is orthochronous. Otherwise, $\Lambda_{00} \leq -1$, and $\Lambda$ is antichronous.
\end{itemize}
These different connected components can be mapped onto one-another via the discrete Lorentz transformations $P$ and $T$,
\begin{align}
[{P^{\mu}}_{\nu}] &=  \left(\hspace{-5 pt}\begin{tabular}{ c c c c}
    ${+1}$ & ${0}$ & ${0}$ & ${0}$\\
    ${0}$ & ${-1}$ & ${0}$ & ${0}$\\
    ${0}$ & ${0}$ & ${-1}$ & ${0}$\\
    ${0}$ & ${0}$ & ${0}$ & ${-1}$
    \end{tabular}\hspace{-5 pt}\right) \hspace{35 pt} [{T^{\mu}}_{\nu}] =  \left(\hspace{-5 pt}\begin{tabular}{ c c c c}
    ${-1}$ & ${0}$ & ${0}$ & ${0}$\\
    ${0}$ & ${+1}$ & ${0}$ & ${0}$\\
    ${0}$ & ${0}$ & ${+1}$ & ${0}$\\
    ${0}$ & ${0}$ & ${0}$ & ${+1}$
    \end{tabular}\hspace{-5 pt}\right)
\end{align}
and their combined action $PT = TP$, where $P$ and $T$ are called the parity-inversion and time-reversal transformations respectively. We are most concerned with proper orthochronous Lorentz transformations, which are continuously connected to the identity transformation and form a subgroup of the wider Lorentz group. In fact, we will use this group so often that we drop the ``proper orthochronous" descriptor from hereon: unless otherwise indicated, these are the transformations to which we refer when discussing the Lorentz group.

The transformation behavior of a Lorentz $4$-vector can be used to derive the transformation behaviors of other Lorentz tensors. For example, a Lorentz $4$-covector $v_{\mu}$ becomes another Lorentz $4$-covector $\overline{v}_{\mu}$ under the Lorentz transformation $\Lambda$ according to
\begin{align}
    v_{\mu} = \eta_{\mu\nu} v^{\nu} \hspace{10 pt}\mapsto\hspace{10 pt} \overline{v}_{\mu} = \eta_{\mu\nu} \overline{v}^{\nu} = \eta_{\mu\nu} {\Lambda^{\nu}}_{\rho} v^{\rho} = {\Lambda_{\mu}}^{\rho} v_{\rho} = {(\Lambda^{-1})^{\rho}}_{\mu} v_{\rho}
\end{align}
where we have used Lorentz invariance of the metric ($\overline{\eta}_{\mu\nu} = \eta_{\mu\nu}$). As illustrated by the above result, symbols that require both the inversion label ``-1" and Lorentz indices are cumbersome. This is not the only time the inversion label clutters notations that are otherwise useful to this dissertation, so we will instead write inverses with a tilde, e.g. $\tilde{\Lambda}{{}^{\mu}}_{\nu} \equiv {(\Lambda^{-1})^{\mu}}_{\nu}$. In this notation, the transformed Lorentz $4$-covector is more succinctly written as $\overline{v}_{\mu} = \tilde{\Lambda}{{}^{\rho}}_{\mu} v_{\rho}$. A Lorentz index $\nu$ that transforms via contraction with $\tilde{\Lambda}{{}^{\nu}}_{\mu}$ is called a covariant index.

More generally, a Lorentz tensor ${X^{\alpha_{1}\cdots \alpha_{a}}}_{\beta_{1}\cdots\beta_{b}}$ is an object with $a$ contravariant indices $\alpha_{1}$, ..., $\alpha_{a}$ and $b$ covariant indices $\beta_{1}$, ..., $\beta_{b}$ that transforms under a Lorentz transformation $\Lambda$ according to
\begin{align}
    {X^{\alpha_{1}\cdots \alpha_{a}}}_{\beta_{1}\cdots\beta_{b}} \hspace{15 pt} \mapsto \hspace{15 pt} {\Lambda^{\alpha_{1}}}_{\gamma_{1}} \cdots {\Lambda^{\alpha_{a}}}_{\gamma_{a}} \tilde{\Lambda}{{}^{\delta_{1}}}_{\beta_{1}} \cdots \tilde{\Lambda}{{}^{\delta_{b}}}_{\beta_{b}} {X^{\gamma_{1}\cdots \gamma_{a}}}_{\delta_{1}\cdots\delta_{b}}
\end{align}
A tensor that transforms according to this rule is said to transform covariantly under Lorentz transformations or, in fewer words, to be Lorentz covariant. By contracting Lorentz indices between Lorentz tensors, a new Lorentz tensor can be formed. In particular, if all raised Lorentz indices are contracted with lower Lorentz indices in a product of Lorentz tensors and vice-versa (and the collection possesses no other transformation properties with regards to Lorentz transformations) then a Lorentz scalar is formed. For example, the inner product $(v\cdot v) = v^{\mu}v_{\mu}$ is a Lorentz scalar, and is thereby invariant under Lorentz transformations. Each field theory Lagrangian (density) is also a Lorentz scalar. In the context of this dissertation, we consider Lagrangians constructed from multiple rank-2 tensors $\hat{h}^{(n)}_{\mu\nu}$ that correspond to spin-2 fields. These nicely contract together like links in a chain, and their contractions are so common that it is worthwhile to grant them a special notation. We define the `twice-squared bracket' notation as follows:
\begin{align}
    \ltr \hat{h}^{(n_{1})}\rtr_{\mu\nu} &\equiv \hat{h}^{(n_{1})}_{\mu\nu}\\
    \ltr \hat{h}^{(n_{1})}\hat{h}^{(n_{2})} \rtr_{\mu\sigma} &\equiv \hat{h}^{(n_{1})}_{\mu\nu} \, \eta^{\nu\rho} \,  \hat{h}^{(n_{2})}_{\rho\sigma}\\
    \ltr \hat{h}^{(n_{1})}\hat{h}^{(n_{2})}\hat{h}^{(n_{3})} \rtr_{\mu\upsilon} &\equiv \hat{h}^{(n_{1})}_{\mu\nu} \,  \eta^{\nu\rho} \,  \hat{h}^{(n_{2})}_{\rho\sigma} \, \eta^{\sigma\tau} \, \hat{h}^{(n_{2})}_{\tau\upsilon} 
\end{align}
and so on. When the field indices are entirely contracted to form a trace (such that the chain is closed into a loop), the external indices are omitted:
\begin{align}
    \ltr \hat{h}^{(n_{1})}\cdots\hat{h}^{(n_{H})} \rtr \equiv \ltr \hat{h}^{(n_{1})}\cdots\hat{h}^{(n_{H})} \rtr_{\alpha\beta} \, \eta^{\alpha\beta}
\end{align}
The operation of connecting two such chains via contraction is called concatenation, and the identity chain with respect to concatenation is
\begin{align}
    \ltr 1 \rtr_{\mu\nu} \equiv \eta_{\mu\nu}
\end{align}
from which $\ltr 1 \rtr = 4$. (If we were instead working in $X$-dimensions, then $\ltr 1 \rtr_{MN} \equiv \eta_{MN}$ and $\ltr 1 \rtr = X$).

Regarding its group structure, the Lorentz group possesses two Casimir invariants, which are used to define particle content in quantum field theory. The first is the mass, which is defined from the (assumedly not space-like) $4$-momentum $p^{\mu} = (E/c,\vec{p}\,)$ where the quantities $E\geq 0$ and $\vec{p}$ are the energy and $3$-momentum of a particle excitation respectively. The mass $m \geq 0$ is defined from the Einstein equation,
\begin{align}
    E^{2} = m^{2} c^{4} + \vec{p}^{\,2}c^{2}
\end{align}
which we typically express instead as the on-shell condition $p^{2} \equiv p^{\mu} p_{\mu}= m^{2}c^{2}$ (``on-shell" being shorthand for ``on mass shell"). The collection of light-like $4$-momenta related by Lorentz transformations form the light cone, a right cone in $p^{\mu}$-space oriented along the energy axis. In contrast, if a $4$-momentum is time-like, then the mass is nonzero, and the collection of $4$-momenta with equal mass form a hyperboloid in $p^{\mu}$-space called a mass hyperboloid. Every mass hyperboloid contains a rest frame $4$-momentum $(m,\vec{0}\,)$ wherein $|\vec{p}\,| = 0$. Any two $4$-momenta on the light-cone or on the same mass hyperboloid can be related via a Lorentz transformation. The mass additionally dictates the kind of trajectories along which a given particle can travel: massless particles travel along light-like trajectories at the speed of light, whereas massive particles travel along time-like trajectories at speeds slower than the speed of light.

Regarding the second Casimir invariant---the Pauli-Lubanski pseudovector---we will not dwell on it beyond asserting that it allows a massive (massless) particle to be assigned a Lorentz-invariant total spin (helicity). For instance, the second Casimir invariant is why an electron can be assigned a definite internal spin of $\tfrac{1}{2}$. We adopt the standard convention of referring to a massless particle with total helicity $s$ as being a spin-$s$ particle. When a massive particle is in its rest frame, its total angular momentum equals its total internal spin. Whereas a massive spin-$s$ particle has $(2s+1)$ available helicities $\lambda \in \{-s,-s+1,\dots,s\}$, a massless spin-$s$ particle has at most two, $\lambda \in \{-s,+s\}$.

The above considerations for $4$-momentum apply more generally to other Lorentz $4$-vectors as well: any two (nonzero) light-like or time-like $4$-vectors $v$ and $w$ having equal magnitude $(v\cdot v) = (w\cdot w) \geq 0$ and same temporal component sign $\text{sign}(w^{0}) = \text{sign}(v^{0})$ can be related by a Lorentz transformation. Meanwhile, any two space-like $4$-vectors $v$ and $w$ with equal magnitude $(v\cdot v) = (w\cdot w) < 0$ can be related by a Lorentz transformation, regardless if they disagree on the signs of their temporal components. The collection of all $4$-vectors related to a particular $4$-vector $v$ by (proper orthochronous) Lorentz transformations is called the Lorentz-invariant hypersurface generated by $v$. In this language, a mass hyperboloid (light cone) is the Lorentz-invariant hypersurface generated by a time-like (light-like) $4$-momentum $p$. Note that the Lorentz-invariant hypersurface generated by a nonzero $4$-vector is a three-dimensional manifold because the four components of the $4$-vectors on that hypersurface have only one continuous constraint (i.e. maintaining the same overall $4$-vector magnitude). The ``nonzero" descriptor in the previous statement is important because the $4$-vector origin $0^{\mu}$ is individually invariant under Lorentz transformations, such that the hypersurface it generates is the zero-dimensional set $\{0^{\mu}\}$.\footnote{This is one way to understand the lack of a rest frame $4$-momentum on the light cone: if we could somehow map the light-like $4$-momentum of a massless particle to $0^{\mu}$, then we could (using the inverse transformation) map $0^{\mu}$ back to a different light-like $4$-momentum, but this would contradict the invariance of $\{0^{\mu}\}$. Therefore, massless particles cannot be at rest in any reference frame (this is, of course, a restatement of the invariance of the speed of light).}

In addition to its group structure, the Lorentz group forms a six-dimensional manifold: consider the magnitude
\begin{align}
    a\cdot a = (a^{0})^{2} - (a^{1})^{2} - (a^{2})^{2} - (a^{3})^{2}
\end{align}
of a $4$-vector $a^{\mu}$ for which all components are nonzero. A generic Lorentz transformation can alter any of these components but must ultimately preserve this magnitude. In particular, suppose a transformation alters one component slightly. Because all components of $a^{\mu}$ are assumedly nonzero, we can preserve the overall magnitude of $a$ by slightly increasing or decreasing a different component of $a$ by however much is necessary to accommodate the change of the first component. There are as many independent ways of performing this balancing trick as there are distinct pairs of components. Because $a$ has four components as a $4$-vector, there are six independent choices of component pairs. Furthermore, by chaining together the shifts of magnitude described by these six independent component pairs, we can form any (proper orthochronous) Lorentz transformation. Therefore, the Lorentz group is six-dimensional.

It is conventional to distinguish certain convenient Lorentz transformations:
\begin{itemize}
    \item[$\bullet$] Rotations are Lorentz transformations that leave the temporal 4-vector coordinate unchanged, and correspond to the usual collection of rotations in $3$-space. Their operation solely affects the $3$-vector part $\vec{a}$ of a $4$-vector $a$, and they form a closed subgroup of the Lorentz group. In the context of the aforementioned balancing trick, these transformations correspond to the ``space-space" mixing.
    \item[$\bullet$] Boosts are Lorentz transformations that leave a spatial $2$-plane unchanged, e.g. a boost along the $z$-axis will mix the $a^{0}$ and $a^{3}$ components of a $4$-vector, but leave the $a^{1}$ and $a^{2}$ components unchanged. Boosts do not form a closed subgroup of the Lorentz group. In the context of the aforementioned balancing trick, these transformations correspond to the ``time-space" mixing.
\end{itemize}
Any two $4$-vectors on the same Lorentz-invariant hypersurface can be related by a Lorentz transformation that combines rotations and boosts.

We arrived at the Minkowski metric $\eta$ by demanding that the speed of light be locally preserved between frames. If we now suppose the Minkowski metric describes spacetime globally as well (thereby ensuring we work in the realm of special relativity as opposed to general relativity), the trajectories $\vec{x}(t)$ of light-like wavepackets must be straight lines through $3$-space. That is, the wavepacket propagates such that at any time $t$ it is centered at $\vec{x}(t) = \vec{v} t + \vec{x}(0)$ for some initial position $\vec{x}(0)$ and velocity $|\vec{v}| = c$. If the $4$-velocity $(v^{0},\vec{v})$ transforms according to a Lorentz transformation
\begin{align}
    v^{\mu} \rightarrow {\Lambda^{\mu}}_{\nu} v^{\nu}
\end{align}
then the corresponding trajectory in $4$-space $x^{\mu} = (ct,\vec{x}(t))$ must transform according to the same Lorentz transformation plus a potential spacetime translation
\begin{align}
    x^{\mu} \rightarrow {\Lambda^{\mu}}_{\nu} x^{\nu} + \epsilon^{\mu} \label{PoincareTrajectory}
\end{align}
where $\epsilon^{\mu}$ is a generic $4$-vector. By once again considering a network of light-like wavepackets throughout spacetime, we can generalize this transformation behavior beyond a single trajectory and conclude that the coordinates of spacetime must generally transform according to Eq. \eqref{PoincareTrajectory}. The wider collection of transformations available to spacetime coordinates comprise the Poincar\'{e} group. Because $\epsilon^{\mu}$ has four real components and the Lorentz group is a six-dimensional manifold, the Poincar\'{e} group forms a ten-dimensional manifold.

The following subsections delve into more detail about specific transformations within the Poincar\'{e} group. To facilitate succinct expressions, we introduce unit $4$-vector basis elements,
\begin{align}
    v = v^{0} \hat{t} + v^{1} \hat{x} + v^{2} \hat{y} + v^{3} \hat{z}
\end{align}
where
\begin{align}
    [\hat{t}^{\mu}] = \matrixda{1}{0}{0}{0} \hspace{35 pt} [\hat{x}^{\mu}] = \matrixda{0}{1}{0}{0} \hspace{35 pt} [\hat{y}^{\mu}] = \matrixda{0}{0}{1}{0} \hspace{35 pt} [\hat{z}^{\mu}] = \matrixda{0}{0}{0}{1} \label{xyztDef}
\end{align}
For the same purpose, we also define abbreviations for the trigonometric and hyperbolic functions
\begin{align}
    c_{\alpha} \equiv \cos\alpha\hspace{35 pt}s_{\alpha} \equiv \sin\alpha\hspace{35 pt}ch_{\beta} \equiv \cosh\beta\hspace{35 pt}sh_{\beta} \equiv \sinh\beta
\end{align}
and utilize natural units for the remainder of this dissertation: $c = \hbar = 1$.

\subsection{Active vs. Passive Transformations} \label{Classical - Active vs Passive}
In order to quantify Lorentz and Poincar\'{e} transformations, we must decide whether to consider them as active or passive transformations. As to clarify the nuances of these perspectives, let us briefly restrict our attention to spacetime translations.

Consider a continuous function $\phi(x)$ of real numbers over spacetime that is sharply peaked at some spacetime point $x = X$ relative to an observer at the spacetime origin. Further suppose we want to describe this same distribution as instead having a peak at $X+a$ for some $4$-vector $a$ relative to that observer. We might use an active or passive transformation to achieve this: the active transformation shifts the entire distribution by an amount $a$ relative to the coordinate system, whereas the passive transformation instead keeps the distribution as-is and moves the observer (and the spacetime origin with them) by an amount $-a$. Because they ultimately describe the same physical reality---namely, that the peak is now at $X+a$ relative to the observer---these different transformations must be physically equivalent. More generally, an active Poincar\'{e} transformation $\mathcal{P}(\Lambda,a)$ on the distribution corresponds to a passive Poincar\'{e} transformation $\mathcal{P}(\Lambda,a)^{-1}$ on the observer and their coordinates.

When a transformation is used to switch between reference frames, it is typically written in the passive interpretation: in this interpretation, reality is fixed, and we are merely swapping between observers who have their own coordinate systems for observing that reality. However, the preceding discussion points out that we could equally well use active transformations as long as we are careful to invert the intended operation. Because we intend to eventually apply {\it active} transformations to quantum mechanical states, our discussion of the Lorentz group in the upcoming subsections is written in the active interpretation, even when those transformations are used to switch between reference frames. For example, our rotation operator $R_{z}(\alpha)$ corresponds to rotating {\it the physical system} by an angle $+\alpha$ about the $z$-axis, which is equivalent to rotating {\it the observer}  (and their coordinate system) by an angle $-\alpha$ about the $z$-axis. These are an active and passive transformation respectively.

That being said, there is an important transformation that we should always be cautious to interpret correctly: the time evolution transformation. An active time translation by an amount $\Delta t$ shifts our distribution $\phi(x)=\phi(t,\vec{x})$ to $\phi(t-\Delta t,\vec{x})$ and thereby ensures that a peak formerly at $X=(T,\vec{X})$ will subsequently occur at $X^{\prime} = (T+\Delta t,\vec{X})$. However, if we want to evolve the system in time by an amount $\Delta t$, we actually desire that $\phi(t,\vec{x})$ be mapped to $\phi(t+\Delta t,\vec{x})$. This can be achieved by either performing an active time translation by an amount $-\Delta t$ or (as it is usually expressed) performing a passive time translation by an amount $\Delta t$.

From here onward, the ``rotation", ``boost", ``translation", ``Lorentz", ``Poincar\'{e}", etc. transformations will be written as active transformations unless otherwise indicated, in contrast to the time evolution transformation, which (in the way just described) is always understood as a passive transformation.

\subsection{Rotations} \label{SS - Rotations}
For spatial coordinates, we utilize a standard right-handed $3$-space coordinate system (labeled such that $\hat{x}\times\hat{y} = \hat{z}$) and define our rotations using the right-hand rule. This means that, for example, an active rotation about the $z$-axis by an angle $\alpha$ on a generic 4-vector $x^{\mu}$ yields a new $4$-vector ${R_{z}(\alpha)^{\mu}}_{\nu} x^{\nu}$, where
\begin{align}
    [{R_{z}(\alpha)^{\mu}}_{\nu}] &= \left(\hspace{-5 pt}\begin{tabular}{ c c c c}
    ${1}$ & ${0}$ & ${0}$ & ${0}$\\
    ${0}$ & ${c_{\alpha}}$ & ${-s_{\alpha}}$ & ${0}$\\
    ${0}$ & ${s_{\alpha}}$ & ${c_{\alpha}}$ & ${0}$\\
    ${0}$ & ${0}$ & ${0}$ & ${1}$
    \end{tabular}\hspace{-5 pt}\right)
\end{align}
within which $c_{\alpha} \equiv \cos\alpha$ and $s_{\alpha} \equiv \sin\alpha$. Note that $R_{z}(\alpha)$ becomes the identity transformation when $\alpha = 0$. We can directly check that $R_{z}(\alpha)$ is a Lorentz transformation by considering how it (does not) affect the magnitude of a generic 4-vector $dX=(dt,dx,dy,dz)$:
\begin{align}
    \eta_{\mu\nu} \, \left[R_{z}(\alpha)\, dX\right]^{\mu} \, \left[R_{z}(\alpha)\, dX\right]^{\nu}&= dt^{2} - (c_{\alpha}\, dx - s_{\alpha}\, dy)^{2} - (s_{\alpha}\, dx + c_{\alpha}\, dy)^{2} - dz^{2}\\
    &= dt^{2} - (c_{\alpha}^{2} + s_{\alpha}^{2})dx^{2} - (c_{\alpha}^{2} + s_{\alpha}^{2})dy^{2} - dz^{2}\\
    &= dt^{2} - d\vec{x}^{\,2}\\
    &= \eta_{\mu\nu}\, dX^{\mu} \, dX^{\nu}
\end{align}
In principle, $R_{z}(\alpha)$ is an instantaneous mapping from one coordinate system to another. However, by taking $\alpha\rightarrow 0$, $R_{z}(\alpha)$ continuously goes to the identity (${[R_{z}(\alpha)^{\mu}}_{\nu}] \rightarrow [\eta^{\mu}_{\nu}] = [\delta_{\mu,\nu}]$), and thus (by reversing the direction of the limit) we can interpret a rotation $R_{z}(\alpha)$ as a continuous transformation that smoothly rotates $x^{\mu}$ to ${R_{z}(\alpha)^{\mu}}_{\nu}\,x^{\nu}$. In addition to being a nice conceptual feature, this continuity near the identity allows us to rewrite the rotation operator $R_{z}(\alpha)$ as the exponential of an angle-independent generator $J_{z}$:
\begin{align}
    [{R_{z}(\alpha)^{\mu}}_{\nu}] = \text{Exp}\bigg[\alpha [{(J_{z})^{\mu}}_{\nu}]\bigg] \equiv \sum_{n=0}^{+\infty} \dfrac{1}{n!}\bigg(\alpha [{(J_{z})^{\mu}}_{\nu}]\bigg)^{n}\hspace{25 pt}\text{where}\hspace{25 pt}J_{z} \equiv \left.\dfrac{\partial R_{z}(\alpha)}{\partial \alpha}\right|_{\alpha = 0}
\end{align}
from which we calculate
\begin{align}
    [{(J_{z})^{\mu}}_{\nu}] &= \left(\hspace{-5 pt}\begin{tabular}{ c c c c}
    ${0}$ & ${0}$ & ${0}$ & ${0}$\\
    ${0}$ & ${0}$ & ${-1}$ & ${0}$\\
    ${0}$ & ${+1}$ & ${0}$ & ${0}$\\
    ${0}$ & ${0}$ & ${0}$ & ${0}$
    \end{tabular}\hspace{-5 pt}\right) \label{Jz4V}
\end{align}
By having one index raised and another index lowered, we ensure that powers of $[{(J_{z})^{\mu}}_{\nu}]$ correctly reproduce a series of index contractions, e.g. $[{(J_{z})^{\mu}}_{\nu}]\, [{(J_{z})^{\nu}}_{\rho}] = [{(J_{z}^{2})^{\mu}}_{\rho}]$. For the rest of this chapter we will drop the index references on $[{(J_{z})^{\mu}}_{\nu}]$ and refer to it simply as $J_{z}$. Note that $J_{z}$ only leaves 4-vectors proportional to $(t,\hat{z})$ unchanged, which is consistent with $\hat{z}$ being the axis of the rotation generated by $J_{z}$. This same procedure can also be applied to rotations about the $x$- and $y$-axes, which have the rotation matrices,
\begin{align}
    [{R_{x}(\alpha)^{\mu}}_{\nu}] = \left(\hspace{-5 pt}\begin{tabular}{ c c c c}
    ${1}$ & ${0}$ & ${0}$ & ${0}$\\
    ${0}$ & ${1}$ & ${0}$ & ${0}$\\
    ${0}$ & ${0}$ & ${c_{\alpha}}$ & ${-s_{\alpha}}$\\
    ${0}$ & ${0}$ & ${s_{\alpha}}$ & ${c_{\alpha}}$
    \end{tabular}\hspace{-5 pt}\right)\hspace{35 pt}[{R_{y}(\alpha)^{\mu}}_{\nu}] = \left(\hspace{-5 pt}\begin{tabular}{ c c c c}
    ${1}$ & ${0}$ & ${0}$ & ${0}$\\
    ${0}$ & ${c_{\alpha}}$ & ${0}$ & ${s_{\alpha}}$\\
    ${0}$ & ${0}$ & ${1}$ & ${0}$\\
    ${0}$ & ${-s_{\alpha}}$ & ${0}$ & ${c_{\alpha}}$
    \end{tabular}\hspace{-5 pt}\right)
\end{align}
which can be expressed as exponentials $R_{x}(\alpha) = \text{Exp}[\alpha J_{x}]$ and $R_{y}(\alpha) = \text{Exp}[\alpha J_{y}]$, where
\begin{align}
    J_{x} = \left(\hspace{-5 pt}\begin{tabular}{ c c c c}
    ${0}$ & ${0}$ & ${0}$ & ${0}$\\
    ${0}$ & ${0}$ & ${0}$ & ${0}$\\
    ${0}$ & ${0}$ & ${0}$ & ${-1}$\\
    ${0}$ & ${0}$ & ${+1}$ & ${0}$
    \end{tabular}\hspace{-5 pt}\right)\hspace{35 pt}J_{y} = \left(\hspace{-5 pt}\begin{tabular}{ c c c c}
    ${0}$ & ${0}$ & ${0}$ & ${0}$\\
    ${0}$ & ${0}$ & ${0}$ & ${+1}$\\
    ${0}$ & ${0}$ & ${0}$ & ${0}$\\
    ${0}$ & ${-1}$ & ${0}$ & ${0}$
    \end{tabular}\hspace{-5 pt}\right) \label{JxJy4V}
\end{align}
are the corresponding generators. (These antisymmetric generators will be replaced by Hermitian operators when promoted to the analogous quantum mechanical description.) The generators $J_{i}$ have several convenient properties. For instance, they possess a closed commutator structure:
\begin{align}
    [J_{i},J_{j}] &= \epsilon_{ijk} J_{k}\hspace{35 pt}\implies\hspace{35 pt}\vec{J}\times\vec{J} = \vec{J} \label{Jcommutators}
\end{align}
where $i,j,k\in \{x,y,z\}$, $\vec{J}\equiv (J_{x},J_{y},J_{z})$, and $[A,B]\equiv AB - BA$. They can also be put into the combination
\begin{align}
\hspace{35 pt} \vec{J}^{\, 2} \equiv \vec{J}\cdot \vec{J} = \left(\hspace{-5 pt}\begin{tabular}{ c c c c}
    ${0}$ & ${0}$ & ${0}$ & ${0}$\\
    ${0}$ & ${-2}$ & ${0}$ & ${0}$\\
    ${0}$ & ${0}$ & ${-2}$ & ${0}$\\
    ${0}$ & ${0}$ & ${0}$ & ${-2}$
    \end{tabular}\hspace{-5 pt}\right) \equiv -2(\delta_{\mu,\nu} - \delta_{\mu,0}\delta_{\nu,0}) \label{J2in4V}
\end{align}
which commutes with every generator
\begin{align}
    [\vec{J},\vec{J}^{\, 2}] = 0
\end{align}
If a collection of three tensors $\{X_{x}, X_{y}, X_{z}\}$ happen to satisfy
\begin{align}
    [J_{i},X_{j}] = \epsilon_{ijk} X_{k} \label{3VectorRotationProperty}
\end{align}
where $i,j,k\in\{x,y,z\}$, then the collection transforms like a $3$-vector $\vec{X} \equiv (X_{x},X_{y},X_{z}) \equiv (X^{1}, X^{2}, X^{3})$ under rotations. In particular, via Eq. \eqref{Jcommutators}, $\vec{J}$ transforms as a proper $3$-vector under rotations, and so we can give the components of $\vec{J}$ legitimate $3$-vector indices: $\{J_{x},J_{y},J_{z}\} = \{J^{1},J^{2},J^{3}\}$. Because we use the mostly-minus metric convention, this means that, for example, $J_{x} = J^{1} = - J_{1}$. Exponentiating the generators together allows us to write a generic rotation matrix: a rotation $[R(\vec{\alpha})^{\mu}_{\nu}]$ around an axis $\hat{\alpha}$ by an angle $|\vec{\alpha}|$ equals
\begin{align}
    R(\vec{\alpha}) \equiv \text{Exp}[\vec{\alpha}\cdot \vec{J}] \label{Ralpha}
\end{align}
This is, of course, equivalent to a rotation by an angle $-|\vec{\alpha}|$ about $-\hat{\alpha}$ instead, if one so prefers.

As mentioned above, the rotation generator set $\{J_{x},J_{y},J_{z}\}$ is closed under the commutation bracket, Eq. \eqref{Jcommutators}. In fact, this specific commutation structure combined with the reality of the generators means they form the Lie algebra $\mathfrak{so}(3)$ and that the rotation group in three dimensions is the Lie group $\mathbf{SO(3)}$. $\mathbf{SO(3)}$---and its covering group, $\mathbf{SU(2)}$---is compact and thus admits finite-dimensional unitary representations (which we review in Section \ref{Angular Momentum}). The operator $\vec{J}^{\,2}$ (which we recall commutes with every generator) is the single Casimir operator belonging to $\mathfrak{so}(3)$. Like other Casimir operators, $\vec{J}^{\,2}$ is a geometric invariant that describes the dimensionalities of any invariant subgroups within a given representation of the rotation group. For example, although the $4$-vector representation above transforms under the rotation group in a well-defined way, it actually contains two distinct rotational behaviors which never mix under any rotation. This was hinted by the two distinct eigenvalues along the diagonal of $\vec{J}^{\,2}$ in Eq. \eqref{J2in4V}. It can also be identified directly from the transformation behavior of $4$-vectors if one knows what to search for: while the $3$-vector part $\vec{x}$ of a 4-vector $x^{\mu}$ is changed under any rotation in the usual way, its temporal component $x^{0}$ is left invariant, and so $x^{\mu}$ cleanly separates into $x^{0}$ and $\vec{x}$ as far as rotations are concerned. Regardless of how these invariant subspaces are derived, they correspond to spin-0 and spin-1 representations of the rotation group. In Subsection \ref{SS - Polarizations Derived}, we will use the spin-1 portion of the 4-vector representation to derive the canonical spin-1 and spin-2 polarizations.

The rotation $R(\vec{\alpha}\,)$ defined in Eq. \eqref{Ralpha} is only one of many ways of writing a generic rotation. Another (which is particularly useful for the purposes of this chapter) is the Euler angle parameterization. The Euler angles detail a sequence of rotations with which one can produce any orientation of a rigid body in $3$-space. They also happen to be a natural coordinate system for a symmetric top. Explicitly, we may write a generic rotation in terms of the Euler angles $\{\phi,\theta,\psi\}$ as
\begin{align}
    R(\phi,\theta,\psi) \equiv R_{z}(\phi) R_{y}(\theta) R_{z}(\psi) \label{EulerAngles}
\end{align}
where $\phi \in [0,2\pi)$, $\theta \in [0,\pi]$, and $\psi \in (-2\pi,0]$. When applied to a symmetric top which has been set to balance with its tip at the origin and with gravity pulling in the negative $\hat{z}$-direction, these angles correspond to the following motions:
\begin{itemize}
    \item[$\bullet$] $\psi$ describes the intrinsic rotation of the top about its own axis.
    \item[$\bullet$] $\theta$ describes nutation of the top, i.e. rotation of the top axis towards and away from the $z$-axis.
    \item[$\bullet$] $\phi$ describes precision of the top, i.e. rotation of the top axis about the $z$-axis.
\end{itemize}
In quantum mechanical problems where the relevant states are eigenkets of $z$-axis rotations but (necessarily) not of $x$- and $y$-axis rotations, the fact that Eq. \eqref{EulerAngles} begins and ends with $z$-axis rotations enables certain simplifications.

If the object we intend to rotate has no spatial extent beyond its axis of rotation (e.g. a symmetric top in the limit that it becomes a needle), then the intrinsic rotation angle $\psi$ has no physical effect and can be set to some conventional value. This will be relevant when we consider rotations of $3$-momenta, which can be rotated about their $3$-direction without affecting their value. Popular conventions include setting $\psi = -\phi$ and $\psi = 0$, of which we choose the former when such a choice is relevant. Setting the value of $\psi$ ensures that only two degrees of freedom remain, where the remaining angles correspond to the usual spherical coordinates $(\theta,\phi)$. For these cases, we define
\begin{align}
    R(\hat{p}) \equiv R(\phi,\theta) \equiv R(\phi,\theta,-\phi)
\end{align}
where $\hat{p}$ is the $3$-direction corresponding to $(\theta,\phi)$.

To phrase the previous point in a different way: any two $3$-vectors $\vec{v}$ and $\vec{w}$ which share the same magnitude $|\vec{v}|=|\vec{w}|$ are on the same rotation invariant hypersurface, and can be related via some choice of rotation. Because these hypersurfaces are $2$-spheres in $3$-space, we require only two degrees of freedom to parameterize the different $3$-vectors and, thus, the rotations relating them too. This is the language we use when discussing Lorentz transformations in the next subsection, after we derive the boost generators.

\subsection{Boosts}
An active boost along the $z$-axis with rapidity $\beta$ on a generic 4-vector $x^{\mu}$ yields a new $4$-vector ${B_{z}(\beta)^{\mu}}_{\nu} x^{\nu}$, where
\begin{align}
    [{B_{z}(\beta)^{\mu}}_{\nu}] &= \left(\hspace{-5 pt}\begin{tabular}{ c c c c}
    ${ch_{\beta}}$ & ${0}$ & ${0}$ & ${sh_{\beta}}$\\
    ${0}$ & ${1}$ & ${0}$ & ${0}$\\
    ${0}$ & ${0}$ & ${1}$ & ${0}$\\
    ${sh_{\beta}}$ & ${0}$ & ${0}$ & ${ch_{\beta}}$
    \end{tabular}\hspace{-5 pt}\right)
\end{align}
within which $ch_{\beta} \equiv \cosh\beta$ and $sh_{\beta} \equiv \sinh\beta$. When the rapidity vanishes ($\beta = 0$), $B_{z}(\beta)$ becomes the identity transformation. Like the rotations in the last subsection, we can directly check that $B_{z}(\beta)$ is a Lorentz transformation by considering its (lack of an) effect on the magnitude of a generic 4-vector $dX = (dt,dx,dy,dz)$:
\begin{align}
    \eta_{\mu\nu} \, \left[B_{z}(\beta)\, dX\right]^{\mu} \, \left[B_{z}(\beta)\, dX\right]^{\nu}&= (ch_{\beta} \, dt + sh_{\beta} \, dz)^{2} - dx^{2} - dy^{2} - (sh_{\beta} \, dt + ch_{\beta} \, dz)^{2}\\
    &= (ch_{\beta}^{2} - sh_{\beta}^{2})dt^{2} - dx^{2} - dy^{2} - (ch_{\beta}^{2} - sh_{\beta}^{2}) dz^{2} \\
    &= dt^{2} - d\vec{x}^{\,2}\\
    &= \eta_{\mu\nu} \, dX^{\mu}\, dX^{\nu}
\end{align}
Furthermore, because $B_{z}(\beta)$ is continuously connected to the identity, it can be interpreted as a smooth transformation (by evolving the rapidity from zero to $\beta$) and be expressed as an exponential of a rapidity-independent generator:
\begin{align}
    [{B_{z}(\beta)^{\mu}}_{\nu}] = \text{Exp}\bigg[\beta [{(K_{z})^{\mu}}_{\nu}]\bigg] \equiv \sum_{n=0}^{+\infty} \dfrac{1}{n!}\bigg(\beta [{(K_{z})^{\mu}}_{\nu}]\bigg)^{n}\hspace{25 pt}\text{ where }\hspace{25 pt}K_{z} \equiv \left.\dfrac{\partial B_{z}(\beta)}{\partial\beta}\right|_{\beta=0}
\end{align}
from which we calculate
\begin{align}
    [{(K_{z})^{\mu}}_{\nu}] &= \left(\hspace{-5 pt}\begin{tabular}{ c c c c}
    ${0}$ & ${0}$ & ${0}$ & ${+1}$\\
    ${0}$ & ${0}$ & ${0}$ & ${0}$\\
    ${0}$ & ${0}$ & ${0}$ & ${0}$\\
    ${+1}$ & ${0}$ & ${0}$ & ${0}$
    \end{tabular}\hspace{-5 pt}\right)
\end{align}
Again, we drop the index indicators and simply write $[{(K_{z})^{\mu}}_{\nu}]$ as $K_{z}$. Boosts along the $x$- and $y$-axes are defined similarly
\begin{align}
    [{B_{x}(\beta)^{\mu}}_{\nu}] &= \left(\hspace{-5 pt}\begin{tabular}{ c c c c}
    ${ch_{\beta}}$ & ${sh_{\beta}}$ & ${0}$ & ${0}$\\
    ${sh_{\beta}}$ & ${ch_{\beta}}$ & ${0}$ & ${0}$\\
    ${0}$ & ${0}$ & ${1}$ & ${0}$\\
    ${0}$ & ${0}$ & ${0}$ & ${1}$
    \end{tabular}\hspace{-5 pt}\right) \hspace{35 pt} [{B_{y}(\beta)^{\mu}}_{\nu}] &= \left(\hspace{-5 pt}\begin{tabular}{ c c c c}
    ${ch_{\beta}}$ & ${0}$ & ${sh_{\beta}}$ & ${0}$\\
    ${0}$ & ${1}$ & ${0}$ & ${0}$\\
    ${sh_{\beta}}$ & ${0}$ & ${ch_{\beta}}$ & ${0}$\\
    ${0}$ & ${0}$ & ${0}$ & ${1}$
    \end{tabular}\hspace{-5 pt}\right)
\end{align}
and can be expressed as exponentials $B_{x}(\beta) = \text{Exp}[\beta K_{x}]$ and $B_{y}(\beta) = \text{Exp}[\beta K_{y}]$, where
\begin{align}
    K_{x} &= \left(\hspace{-5 pt}\begin{tabular}{ c c c c}
    ${0}$ & ${+1}$ & ${0}$ & ${0}$\\
    ${+1}$ & ${0}$ & ${0}$ & ${0}$\\
    ${0}$ & ${0}$ & ${0}$ & ${0}$\\
    ${0}$ & ${0}$ & ${0}$ & ${0}$
    \end{tabular}\hspace{-5 pt}\right) \hspace{35 pt} K_{y} = \left(\hspace{-5 pt}\begin{tabular}{ c c c c}
    ${0}$ & ${0}$ & ${+1}$ & ${0}$\\
    ${0}$ & ${0}$ & ${0}$ & ${0}$\\
    ${+1}$ & ${0}$ & ${0}$ & ${0}$\\
    ${0}$ & ${0}$ & ${0}$ & ${0}$
    \end{tabular}\hspace{-5 pt}\right)
\end{align}
are the corresponding generators. Unlike the rotation generators, the boost generators are {\it not} closed with respect to the commutator bracket, and instead mix with the rotation generators:
\begin{align}
    [K_{i},K_{j}] &= -\epsilon_{ijk} J_{k} \label{BoostBoostComm}\\
    [J_{i},K_{j}] &= +\epsilon_{ijk} K_{k} \label{BoostAs3Vector}
\end{align}
where $i,j,k\in\{x,y,z\}$. By comparing Eq. \eqref{BoostAs3Vector} to Eq. \eqref{3VectorRotationProperty}, we note $\{K_{x},K_{y},K_{z}\}$ rotates like a proper $3$-vector under rotations, and label its components as such: $\vec{K}\equiv \{K_{x},K_{y},K_{z}\} = \{K^{1},K^{2},K^{3}\}$.

A generic boost $B(\vec{\beta}\,)$ along an axis $\hat{\beta}$ by an amount $\beta$ can be constructed from exponentiation of the generators:
\begin{align}
    B(\vec{\beta}\,) = \text{Exp}[\vec{\beta}\cdot\vec{K}] \label{Bbeta}
\end{align}
which leaves the $2$-plane perpendicular to $\vec{\beta}$ in $3$-space unchanged. In contrast to the rotation group, applying a sequence of boosts in the same direction to a $4$-vector never results in a return to the original $4$-vector. This reflects the fact that, unlike the rotation group, the Lorentz group is non-compact. Consequently, whereas the rotation group $\mathbf{SO(3)}$ admits finite-dimensional unitary representations, the Lorentz group $\mathbf{SO(1,3)}$ does not: its only unitary representations are infinite-dimensional, and any finite-dimensional representations are necessarily non-unitary. This will prove important in the following section, as well as when deriving polarization vectors and tensors in Subsection \ref{SS - Polarizations Derived}.

The rotation and boost generators together form the Lorentz generators $\{\vec{J},\vec{K}\}$, whose commutation structure defines the Lie algebra $\mathfrak{so}(1,3)$. They enumerate six independent degrees of freedom and can generate any (proper orthochronous) Lorentz transformation via exponentiation. Like in the case of a generic rotation, there are many ways to parameterize a generic Lorentz transformation. As one example, we may write a generic Lorentz transformation as a boost followed by a rotation:
\begin{align}
    {\Lambda(\vec{\alpha},\vec{\beta}\,)^{\mu}}_{\rho} \equiv {R(\vec{\alpha}\,)^{\mu}}_{\nu}\, {B(\vec{\beta}\,)^{\nu}}_{\rho}
\end{align}
which has six degrees of freedom $(\vec{\alpha},\vec{\beta}\,)$ as required. This is the parameterization we use for the remainder of this dissertation. Fewer parameters are required if we only seek to describe the Lorentz transformations that relate any two $4$-vectors on the same Lorentz-invariant hypersurface. In particular, if the $3$-vector part $\vec{v}$ of a light-like or time-like $4$-vector $v$ points in a direction $\hat{v}$, then we can obtain any other $4$-vector on the same Lorentz-invariant hypersurface by applying
\begin{align}
    {\Lambda(\vec{\alpha},\beta)^{\mu}}_{\rho} \equiv {R(\phi,\theta,-\phi)^{\mu}}_{\nu}\, {B(\beta\hat{v}\,)^{\nu}}_{\rho}
\end{align}
for some choice of $\phi$, $\theta$, and $\beta$. Explicitly, suppose the desired final $4$-vector is $w$. In this case, the boost takes the 3-magnitude $|\vec{v}\,|^{2}$ to $|\vec{w}\,|^{2}$, then the rotation redirects the $3$-vector $|\vec{w}\,|\,\hat{v}$ into the desired $3$-direction $\hat{w}$, and the final temporal component value $w^{0}$ is guaranteed to work out because the $4$-vector magnitude $(w\cdot w)=(v\cdot v)$ is invariant. Note that this specific collection of Lorentz transformations only has three degrees of freedom, consistent with the dimensionality of the light cone and mass hyperboloids.

In the next subsection, the translation generators are derived. Together, the translation and Lorentz generators form the Poincar\'{e} generators.

\subsection{Translations}
Although it is unnecessary to do so in more general contexts, it is advantageous for our current purposes to cast the translation operation as a matrix. To do so, we extend 4-vectors for the duration of this subsection to include a new auxiliary slot, e.g.
\begin{align}
    x^{\mu} \sim [x^{\mu}] = \matrixba{x^{\mu}}{1} = \left(\hspace{-5 pt}\begin{tabular}{ c }
    ${x^{0}}$\\
    ${x^{1}}$\\
    ${x^{2}}$\\
    ${x^{3}}$\\
    ${1}$
    \end{tabular}\hspace{-5 pt}\right)
\end{align}
and define a translation operator ${T(\epsilon)^{\nu}}_{\mu}$ as a $5\times5$ matrix,
\begin{align}
    [{T(\epsilon)^{\nu}}_{\mu}] = \matrixbb{\mathbbm{1}^{\nu}_{\mu}}{\epsilon^{\nu}}{0}{1}
\end{align}
where $\epsilon$ is a 4-vector, such that,
\begin{align}
    [{T(\epsilon)^{\nu}}_{\mu}x^{\mu}] = [{T(\epsilon)^{\nu}}_{\mu}]\, [x^{\mu}] = \matrixba{x^{\nu}+\epsilon^{\nu}}{1} = [x^{\nu} +\epsilon^{\nu}]
\end{align}
Like the previous transformations, the translation operator can be generated through exponentiation of certain translation generators $P^{\mu}$. However, let us be more careful about the signs in this exponentiation than we were in the rotation or boost cases. Specifically, to encourage Lorentz invariance, we would like to write the generators $P^{\mu}$ as a $4$-vector contracted with a generating parameter $\epsilon^{\mu}$, so that the exponentiation is of the form
\begin{align}
    \text{Exp}\bigg[\pm \bigg(\epsilon^{0} [{(P^{0})^{\mu}}_{\nu}] - \epsilon^{1} [{(P^{1})^{\mu}}_{\nu}] - \epsilon^{2} [{(P^{2})^{\mu}}_{\nu}] - \epsilon^{3} [{(P^{3})^{\mu}}_{\nu}]\bigg)\bigg]
\end{align}
where the overall sign of the exponent remains to be determined. The sign we ultimately choose is based on precedent: as written in Eqs. \eqref{Ralpha} and \eqref{Bbeta}, the exponents of the equivalent expressions for general rotations and boosts equal $+\vec{\alpha}\cdot\vec{J}$ and $+\vec{\beta}\cdot\vec{K}$ respectively. It would be nice if the $3$-vector part of the translation exponent equaled $+\vec{\epsilon}\cdot\vec{P}$ as well. Thus, we choose the lower sign.

Using this convention, the time-translation operator $H\equiv P^{0}$ is defined according to
\begin{align}
    [{T(\epsilon^{0}\hat{t})^{\nu}}_{\mu}] = \text{Exp}\bigg[-\epsilon^{0} [{H^{\mu}}_{\nu}]\bigg]\hspace{35 pt}\text{ where }\hspace{35 pt}H \equiv P^{0} = \left.\dfrac{\partial T(\epsilon^{0}\hat{t})}{\partial \epsilon^{0}}\right|_{\epsilon^{0}=0} \label{TranslationExponentiation}
\end{align}
from which we calculate
\begin{align}
    [{H^{\mu}}_{\nu}] = \matrixbb{\mathbb{0}^{\nu}_{\mu}}{-\hat{t}^{\nu}}{0}{0}  = \left(\hspace{-5 pt}\begin{tabular}{ c c c c c }
    ${0}$ & ${0}$ & ${0}$ & ${0}$ & ${-1}$\\
    ${0}$ & ${0}$ & ${0}$ & ${0}$ & ${0}$\\
    ${0}$ & ${0}$ & ${0}$ & ${0}$ & ${0}$\\
    ${0}$ & ${0}$ & ${0}$ & ${0}$ & ${0}$\\
    ${0}$ & ${0}$ & ${0}$ & ${0}$ & ${0}$
    \end{tabular}\hspace{-5 pt}\right)
\end{align}
As before, we drop the index indicators on the generators as we proceed. Like the above temporal translation, a pure spatial translation
\begin{align}
    [{T(\vec{\epsilon}\,)^{\nu}}_{\mu}] = \text{Exp}[\vec{\epsilon}\cdot \vec{P}]\hspace{35 pt}\text{ where }\hspace{35 pt}P^{i} = \left.\dfrac{\partial T(\vec{\epsilon}\,)}{\partial \epsilon^{i}}\right|_{\epsilon^{i}=0}
\end{align}
is accomplished via the space-translation generators $\{P_{x},P_{y},P_{z}\}$, which explicitly equal
\begin{align}
    P_{x} \equiv P^{1} =  \matrixbb{\mathbb{0}^{\nu}_{\mu}}{\hat{x}^{\nu}}{0}{0} \hspace{35 pt} P_{y} \equiv P^{2} =  \matrixbb{\mathbb{0}^{\nu}_{\mu}}{\hat{y}^{\nu}}{0}{0} \hspace{35 pt} P_{z} \equiv P^{3} =  \matrixbb{\mathbb{0}^{\nu}_{\mu}}{\hat{z}^{\nu}}{0}{0}
\end{align}
We will demonstrate that the $3$-vector indices $P^{i}$ are meaningful momentarily. By using the time and space-generators together, we may construct a spacetime translation by a generic $4$-vector $\epsilon^{\mu}$:
\begin{align}
    [T(\epsilon^{\mu})] = \text{Exp}\left[-(\epsilon\cdot P)\right]
\end{align}
where $\hat{x}^{\mu}$, $\hat{y}^{\mu}$, and $\hat{z}^{\mu}$ were defined in Eq. \eqref{xyztDef}. Every Poincar\'{e} transformation can be expressed as a combination of Lorentz transformations and spacetime translations, and thus we can now express all (proper orthochronous) Poincar\'{e} transformations as products of exponentiations of generators.

Combining the spacetime translation generators with the Lorentz generators yields the ten canonical Poincar\'{e} generators $\{P^{\mu},\vec{J},\vec{K}\}$, where the Lorentz generators have implicitly been extended to accommodate the $5\times5$ forms of the translations, e.g. given a Lorentz generator $G_{4\times4}$, a Poincar\'{e} generator $G_{5\times5}$ will have the same effect if defined as follows
\begin{align}
    G_{5\times5} \equiv \matrixbb{G_{4\times4}}{0}{0}{1}
\end{align}
We only distinguish the Poincar\'{e} generator $G_{5\times5}$ from the Lorentz generator $G_{4\times4}$ in the above definition. From here on, we just write $G$.

The commutators of the Poincar\'{e} generators equal, via explicit evaluation,
\begin{align}
    [J^{i},J^{j}] = +\epsilon_{ijk} J^{k} \hspace{35 pt} [J^{i},K^{j}] = +\epsilon_{ijk} K^{k} \hspace{35 pt} [K^{i},K^{j}] = -\epsilon_{ijk} J^{k} \label{PoincareGens1}
\end{align}
\begin{align}
    [H,J^{i}] = 0 \hspace{35 pt} [H,K^{i}] = +P^{i} \hspace{35 pt} [J^{i},P^{j}] = +\epsilon_{ijk} P^{k} \hspace{35 pt} [P^{i},K^{j}] = +H\delta_{i,j} \label{PoincareGens2}
\end{align}
\begin{align}
    [P^{\mu},P^{\nu}] = 0 \label{PoincareGens3}
\end{align}
where $i,j,k \in \{1,2,3\}$. The commutators of the form $[J^{i},\bullet]$ indicate that $\vec{P}$, $\vec{J}$, and $\vec{K}$ behave like $3$-vectors under rotations, such that their $3$-vector indices are meaningful. Although the Poincar\'{e} group does not preserve 4-vector magnitudes\footnote{For example, under a translation by a time-like 4-vector $\epsilon^{\mu}$, the origin $0^{\mu}$ (for which $(0\cdot 0) = 0$) is mapped to $\epsilon^{\mu}$ (for which $(\epsilon\cdot\epsilon) > 0$).}, Poincar\'{e} transformations of coordinates $x^{\mu}$ will preserve the magnitude of $4$-velocities (as well as $4$-momenta), as remarked in Subsection \ref{Classical - Minkowski}.

\subsection{Lorentz-Invariant Phase Space} \label{Classical - LIPS}
The preceding discussion detailed the Poincar\'{e} generators, quantities which can be exponentiated to yield Poincar\'{e} transformations. In the next section, we will promote these generators to quantum operators; however, before moving into the realm of quantum mechanics, it is useful to derive a Lorentz-invariant integral measure with which we will eventually normalize our quantum states.

Recall that a mass hyperboloid corresponding to a mass $m>0$ is a Lorentz-invariant hypersurface defined as the collection of $4$-momentum $p$ for which $E\equiv p^{0} >0$ and $p^{2} = (p\cdot p) = m^{2}$. An integral over a given mass hyperboloid is easily expressed as a $4$-momentum integral using these constraints
\begin{align}
    \int \dfrac{d^{4}p}{(2\pi)^{4}}\hspace{5 pt}(2\pi)\delta(p^{2} - m^{2})\hspace{5 pt}\theta(E)\hspace{5 pt}f(p) \label{4MomentumIntegral}
\end{align}
where $f$ is some function of the $4$-momentum, the Dirac delta function $\delta(p^{2}-m^{2})$ enforces $p^{2} = m^{2}$, and the Heaviside step function $\theta(E)$ enforces $E>0$. Note that this $4$-momentum integral is manifestly invariant under a Lorentz transformation $\Lambda$ so long as $f(p)$ is a Lorentz scalar, because $\theta(E)\mapsto \theta([\Lambda p]^{0})=\theta(E)$ and
\begin{align}
    d^{4}p\hspace{10 pt}\mapsto\hspace{10 pt} d^{4}(\Lambda p) = |\det\Lambda|\, d^{4}p = d^{4} p
\end{align}
\begin{align}
    \delta(p^{2} - m^{2}) \hspace{10 pt}\mapsto \hspace{10 pt} \delta\bigg((\Lambda p)^{2} - m^{2} \bigg) = \delta(p^{2} - m^{2})
\end{align}
Because the mass hyperboloid is a three-dimensional hypersurface, the goal of this subsection is to rewrite the $4$-momentum integral Eq. \eqref{4MomentumIntegral} as a $3$-momentum integral instead. We will first use the Dirac delta in order to eliminate the energy integral ($dE$ in the decomposition $d^{4}p = dE\, d^{3}\vec{p}\,$). However, the Dirac delta as written is not quite right for eliminating that integral, because it is of the form
\begin{align}
    \delta(p^{2} - m^{2}) = \delta(E^{2} - \vec{p}^{\,2} - m^{2})
\end{align}
instead of $\delta(E - E_{*})$ for some value $E_{*}$. To get it into this form, we reparameterize the Dirac delta using the following ``sum over zeros of the Dirac delta argument" property:
\begin{align}
    \delta(f(x)) = \sum_{x_{*}\text{ s.t. }f(x_{*}) = 0} \dfrac{\delta(x-x_{*})}{|f^{\prime}(x_{*})|}
\end{align}
where $f^{\prime}(x)$ denotes the derivative of $f$ with respect to its argument. Because
\begin{align}
    \dfrac{\partial}{\partial E}\bigg[ E^{2} - |\vec{p}\,|^{2} - m^{2} \bigg] = 2E
\end{align}
and $E^{2} - \vec{p}^{\,2} - m^{2} = 0$ when $E = \pm E_{\vec{p}} \equiv \pm \sqrt{m^{2} + \vec{p}^{\,2}}$,
\begin{align}
    \delta(p^{2} - m^{2}) = \dfrac{1}{2\sqrt{m^{2} + |\vec{p}\,|^{2}}}\bigg[\delta(E - E_{\vec{p}}) + \delta(E + E_{\vec{p}}) \bigg]
\end{align}
When we substitute this result into Eq. \eqref{4MomentumIntegral}, the Heaviside step function $\theta(E)$ causes the negative energy term---the term proportional to $\delta(E+E_{\vec{p}})$---to vanish, such that
\begin{align}
    \int \dfrac{d^{3}p}{(2\pi)^{3}} \hspace{5 pt}\dfrac{1}{2E_{\vec{p}}}\hspace{5 pt}f(E_{\vec{p}},\vec{p}\,)
\end{align}
which is a $3$-momentum integral as desired. Because the original integral is Lorentz invariant, this expression must be as well. The integration weight factor $d^{3}p/[(2\pi)^{3} 2E_{\vec{p}}]$ will occur frequently in definitions and calculations due to its Lorentz invariance.

When calculating quantities involving $n$ particles (labeled $1$, $2$, ..., $n$) with individual $4$-momenta $p_{i} = (E_{i},\vec{p}_{i})$ that are constrained to have some total $4$-momentum $P$ (but otherwise unconstrained), integrals of the form
\begin{align}
    \int \bigg[\prod_{i = 1}^{n} \dfrac{d^{3}p_{i}}{(2\pi)^{3}} \hspace{5 pt}\dfrac{1}{2E_{\vec{p}_{i}}}\bigg] \, \bigg[(2\pi)^{4}\delta^{4}\bigg(P - \sum_{i=1}^{n} p_{i} \bigg)\bigg]\, f(E_{\vec{p}_{1}},\vec{p}_{1},\cdots,E_{\vec{p}_{n}},\vec{p}_{n})
\end{align}
often occur, where the bracketed factors together form the $n$-particle Lorentz-invariant phase space element,
\begin{align}
    d\Pi_{n} &= \prod_{i=1}^{n} \bigg[\dfrac{d^{3}p_{i}}{(2\pi)^{3}} \dfrac{1}{2E_{\vec{p}_{i}}}\bigg] \, \bigg[(2\pi)^{4}\delta^{4}\bigg(P - \sum_{i=1}^{n} p_{i} \bigg)\bigg]
\end{align}
Although it was derived by considering massive particles, this expression is equally valid if any of the particles are massless. In the following section, we return to discussing the Poincar\'{e} generators, which will be promoted to quantum equivalents in preparation of defining external particle states with well-defined $4$-momentum and helicity.

\section{Poincar\'{e} Group: Quantum Promotion} \label{S- Quantum Promo}
\subsection{Quantum Mechanics} \label{SS - QM}
Demanding a universal speed of light motivated our investigation of the group of linear transformations that preserved $4$-vector inner products $p\cdot q = \eta_{\mu\nu} p^{\mu} q^{\nu}$. This led us to the Lorentz group, which combines rotations and boosts, and its generalization the Poincar\'{e} group, which additionally incorporates spacetime translations. The different transformations in the Poincar\'{e} group map between reference frames while globally preserving the speed of light.

In the present section, we extend these ideas to quantum mechanics. However, whereas our investigation of $4$-vector transformations was motivated by the frame independence of the speed of light, the promotion to quantum mechanics is motivated by the frame independence of experimental outcomes. To be concrete: while observers in inertial reference frames will disagree about their spacetime coordinates, once those differences are accounted for (by a Poincar\'{e} transformation) they should agree on---for example---how many heads or tails are measured in a sequence of coin flips. Consequently, so long as our experimental questions are phrased in frame-independent ways, the related experimental probabilities should be frame-independent as well.

A quantum mechanical state $\psi$ is described by a ket labeled $\ket{\psi}$. Two kets describe identical states if they differ at most by a phase, e.g. $\ket{\psi}$ and $e^{i\alpha}\ket{\psi}$ correspond to physically-indistinguishable systems for any real choice of $\alpha$. A complete set of kets spans a Hilbert space and is defined for a system by choosing a maximally-commuting set of observables, where those observables are described by self-adjoint operators ($A$ such that $A^{\dagger} = A$) whose eigenspectra encode the possible measured values of those observables. Despite there being technical differences between the two, we will use the descriptors ``self-adjoint" and ``Hermitian" interchangeably from here. Defining an orthonormality condition on a complete set of kets implies a complete set of bras $\bra{\psi}$ as well as a resolution of identity on the space. This defines an inner product between bras and kets which satisfies $\braket{\psi_{1}}{\psi_{2}}^{*} = \braket{\psi_{2}}{\psi_{1}}$ for any two kets $\ket{\psi_{1}}$ and $\ket{\psi_{2}}$.  The probability (or probability density) associated with measuring a state $\psi$ as another state $\psi^{\prime}$ is
\begin{align}
    \text{Prob}(\psi\rightarrow\psi^{\prime}) \equiv | \braket{\psi}{\psi^{\prime}}|^{2}
\end{align}
where it is assumed the kets are appropriately normalized.

A symmetry transformation $A$ on a Hilbert space is any transformation which preserves probabilities, i.e. if $\ket{\psi_{1}}$ and $\ket{\psi_{2}}$ are arbitrary kets in the Hilbert space and are transformed such that $\ket{\psi_{1}}\rightarrow \ket{A\psi_{1}} \equiv A\ket{\psi}$ and $\ket{\psi_{2}}\rightarrow \ket{A\psi_{2}} \equiv A\ket{\psi_{2}}$, then $A$ is a symmetry transformation if
\begin{align}
    |\braket{\psi_{1}}{\psi_{2}}|^{2} \hspace{10 pt}\mapsto\hspace{10 pt} |\braket{A\psi_{1}}{A\psi_{2}}|^{2} = |\bra{\psi_{1}}A^{\dagger}A \ket{\psi_{2}}|^{2} = |\braket{\psi_{1}}{\psi_{2}}|^{2}
\end{align}
Wigner's theorem states a symmetry transformation $A$ must either be unitary and linear,
\begin{align}
    \braket{A\psi_{1}}{A\psi_{2}} = \braket{\psi_{1}}{\psi_{2}}\hspace{20 pt}\text{ and }\hspace{20 pt}A\bigg[c_{1}\ket{\psi_{1}}+c_{2}\ket{\psi_{2}}\bigg] = c_{1}\ket{A\psi_{1}} + c_{2}\ket{A\psi_{2}}
\end{align}
or antiunitary and antilinear,
\begin{align}
    \braket{A\psi_{1}}{A\psi_{2}} = \braket{\psi_{1}}{\psi_{2}}^{*}\hspace{20 pt}\text{ and }\hspace{20 pt}A\bigg[c_{1}\ket{\psi_{1}}+c_{2}\ket{\psi_{2}}\bigg] = c_{1}^{*}\ket{A\psi_{1}} + c_{2}^{*}\ket{A\psi_{2}}
\end{align}
where $c_{1}$ and $c_{2}$ are complex numbers. If $A$ is unitary, then its inverse equals its Hermitian conjugate: $\tilde{A} = A^{\dagger}$, where we recall $\tilde{A}$ is an alternate notation for $A^{-1}$.

Suppose there exists a group of real transformations $\{A\}$ (like the $4$-vector representation of the Poincar\'{e} group) where each transformation is continuously connected to the identity such that each transformation $A$ can be expressed as exponentiations of real generators $G_{a}$
\begin{align}
    A(\xi) = \text{Exp}\bigg[\sum_{a}\xi_{a}G_{a}\bigg]
\end{align}
via real parameters $\xi_{a}$ and the generators satisfy some commutation relations
\begin{align}
    [G_{a},G_{b}] = \sum_{c} T_{abc} \, G_{c} \label{GeneratorCommutator}
\end{align}
for some real numbers $T_{abc}$. In the quantum theory, we can recreate the action of the set $\{A\}$ on our kets by mapping each transformation $A$ to a unitarity operator $\mathcal{U}[A]$ of the exponentiated form
\begin{align}
    \mathcal{U}[A(\xi)] =  \text{Exp}\bigg[-i\sum_{a}\xi_{a}\mathcal{H}[G_{a}]\bigg] \label{PromotingToUnitary}
\end{align}
where the Hermitian operators $\mathcal{H}[G_{a}]$ satisfy the commutation relations
\begin{align}
    \bigg[\mathcal{H}[G_{a}],\mathcal{H}[G_{b}]\bigg] = \sum_{c} i \, T_{abc} \, \mathcal{H}[G_{c}] \label{HermitianCommutator}
\end{align}
for those {\it same} real numbers $T_{abc}$. Heuristically, Eq. \eqref{GeneratorCommutator} goes to \eqref{HermitianCommutator} by replacing the generators $G_{a}$ with $-i\mathcal{H}[G_{a}]$. The operators $\mathcal{H}[G_{a}]$ are also called generators, although in this case they are generators of the unitary operators $\mathcal{U}[A]$. When context is sufficient (and to minimize clutter), we will simply write $\mathcal{H}[G_{a}]$ as $G_{a}$. If the original transformation is active (passive), then the resulting quantum operator will encode an active (passive) transformation as well. Recall that our generators from the previous section were derived in the active interpretation.

\subsection{Promoting the Poincar\'{e} Generators} \label{Quantum - Generators}
The spacetime coordinate transformations which globally preserve the speed of light comprise the Poincar\'{e} group, the generators of which were previously found to satisfy various commutation relations, Eqs. \eqref{PoincareGens1}-\eqref{PoincareGens3}. We now promote each of those generators to Hermitian operators as to create unitary representations of the corresponding Poincar\'{e} transformations, i.e. the matrices $\{P^{\mu}, J^{i}, K^{i}\}$ will be mapped to operators $\{\mathcal{H}[P^{\mu}], \mathcal{H}[J^{i}], \mathcal{H}[K^{i}]\}$. Note that this mapping is not unique: different particles within the same state often require different choices of Hermitian generators. However, whatever Hermitian generators we choose for a particular representation, they must satisfy the promoted version of the previously-derived commutation structure:
\begin{align}
    [J^{i},J^{j}] = +i\epsilon_{ijk} J^{k} \hspace{35 pt} [J^{i},K^{j}] = +i\epsilon_{ijk} K^{k} \hspace{35 pt} [K^{i},K^{j}] = -i\epsilon_{ijk} J^{k}
\end{align}
\begin{align}
    [H,J^{i}] = 0 \hspace{30 pt} [H,K^{i}] = +iP^{i} \hspace{30 pt} [J^{i},P^{j}] = +i\epsilon_{ijk} P^{k} \hspace{30 pt} [P^{i},K^{j}] = +iH\delta_{i,j}
\end{align}
\begin{align}
    [P^{\mu},P^{\nu}] = 0\label{QCommutator-PmuPnu}
\end{align}
where we have dropped the $\mathcal{H}$ label and have been cautious of the minus sign present in the exponentiation of the time-translation generator $H$ (as in Eq. \eqref{TranslationExponentiation}). The operator $H$ is the Hamiltonian, and an eigenket of $H$ with eigenvalue $E$ is said to have energy $E$. The operators $\vec{J}$ and $\vec{P}$ are the angular momentum and (linear) momentum operators respectively, and the rotation Casimir operator $\vec{J}^{\,2}$ is the total angular momentum operator.

Utilizing these generators, we obtain unitary operators that apply the effect of a generic rotation, boost, or translation to a ket:
\begin{align}
    \mathcal{U}[R(\vec{\alpha})] = \text{Exp}\bigg[-i\vec{\alpha}\cdot\vec{J}\,\bigg]\hspace{25 pt}\mathcal{U}[B(\vec{\beta})] = \text{Exp}\bigg[-i\vec{\beta}\cdot\vec{K}\bigg]\hspace{25 pt}\mathcal{U}[T(\epsilon)] = \text{Exp}\bigg[+i(\epsilon\cdot P)\bigg] \label{294}
\end{align}
Because the Lorentz group is non-compact, its (nontrivial) unitary representations are necessarily infinite dimensional. This is reflected in the kets that these operators act on, which (for example) might be labeled by a continuous parameter such as energy.

A vital feature of the quantum promotion is that it inadvertently expands the relevant spacetime symmetry group. For example, the rotation group $\mathbf{SO(3)}$ is not simply connected and thus possesses a distinct (simply connected) universal covering group, the Lie group $\mathbf{SU(2)}$. $\mathbf{SU(2)}$ is a double cover of $\mathbf{SO(3)}$, meaning each transformation in $\mathbf{SO(3)}$ is associated with two transformations in $\mathbf{SU(2)}$. However, the difference in connectedness amounts to differences in global structure, whereas locally---that is, near each group's identity element---$\mathbf{SU(2)}$ and $\mathbf{SO(3)}$ are identical. In other words, the Lie algebra $\mathfrak{so}(3)$ of the rotation group and the Lie algebra $\mathfrak{su}(2)$ of its covering group both have three generators $\{J^{1},J^{2},J^{3}\}$ that share identical commutation structures: $[J^{i},J^{j}] = +i\epsilon_{ijk} J^{k}$. Because the quantum operators are only restricted by the commutation relations in Eq. \eqref{HermitianCommutator}, we are able to represent $4$-vector rotations (elements of a representation of $\mathbf{SO(3)}$) as elements of unitary representations of $\mathbf{SU(2)}$ instead. Irreducible unitary representations of $\mathbf{SU(2)}$ are reviewed in Section \ref{Angular Momentum}. A similar phenomenon occurs in the wider Lorentz group, $\mathbf{SO(1,3)}$: the quantum theory uses irreducible representations of its covering group, $\mathbf{SL(2,\mathbb{C})}$, which is a double cover of $\mathbf{SO(1,3)}$.

As mentioned above, the time translation operator $H\equiv P^{0}$ is identified as the Hamiltonian, and yields a time evolution operator $\mathcal{U}(\Delta t)$,
\begin{align}
    \mathcal{U}(\Delta t) = \text{Exp}\bigg[-i\,(\Delta t)\, H\bigg]
\end{align}
Note the minus sign in the exponent relative to the time translation operator in Eq. \eqref{294}. This is consistent with the discussion in Subsection \ref{Classical - Active vs Passive}.

\subsection{The Square of the 4-Momentum Operator}
There are several important combinations of the generators relevant to our calculations. The first is square of the 4-momentum operator $P^{2} \equiv H^{2} - \vec{P}^{\,2}$, which is important because of its connection to particle mass. In particular, a state $\ket{\psi}$ has mass $M \geq 0$ if $P^{2} \ket{\psi} = M^{2} \ket{\psi}$. Because $P^{2}$ is a Casimir operator of the Poincar\'{e} group, all of the generators automatically commute with it. If a single-particle state is simultaneously an eigenstate of $P^{2}$ and $H$, then it is automatically also an eigenstate of the total $3$-momentum operator $\vec{P}^{\,2}$, and we can choose to label (and normalize) those states with either their energy or 3-momentum magnitude.

\subsection{The Helicity Operator} \label{SS - QM - Helicity Operator}
Another important operator formed by combining Poincar\'{e} generators is the helicity operator $\Lambda$. However, before we define the helicity operator, let us instead consider a related operator: the inner product $\vec{J}\cdot\vec{P} = J^{1} P^{1} + J^{2} P^{2} + J^{3}P^{3}$.

The operator $\vec{J}\cdot\vec{P}$ commutes with many of the Poincar\'{e} generators. For example, because $[AB,C] = [A,C]B + A[B,C]$ and $[P^{i},P^{j}] = 0$ (and recalling $P_{z} = P^{3}$),
\begin{align}
    [P_{z}, \vec{J}\cdot\vec{P}] &= \sum_{i=1}^{3} [P^{3},J^{i}] P^{i} + J^{i} [P^{3}, P^{i}]\\
    &= [P^{3},J^{1}]P^{1} + [P^{3},J^{2}]P^{2}\\
    &= iP^{2}P^{1} - iP^{1}P^{2}\\
    &= 0
\end{align}
such that $[P^{i},\vec{J}\cdot\vec{P}] = 0$ for all $i$ via cyclic symmetry, and thereby $[\vec{P}^{\,2},\vec{J}\cdot\vec{P}]=0$ as well. Similarly,
\begin{align}
    [J_{z}, \vec{J}\cdot\vec{P}] &= \sum_{i=1}^{3} [J^{3},J^{i}] P^{i} + J^{i} [J^{3}, P^{i}]\\
    &= [J^{3},J^{1}]P^{1} + J^{1}[J^{3},P^{1}] + [J^{3},J^{2}]P^{2} + J^{2}[J^{3},P^{2}]\\
    &= iJ^{2}P^{1} + iJ^{1}P^{2} - iJ^{1}P^{2} - iJ^{2} P^{1}\\
    &= 0
\end{align}
such that $[J^{i},\vec{J}\cdot\vec{P}] = 0$ via cyclic symmetry, and $[\vec{J}^{\,2},\vec{J}\cdot\vec{P}]=0$ as well. Finally, note that
\begin{align}
    [H,\vec{J}\cdot \vec{P}] &= \sum_{i=1}^{3} [H,J^{i}] P^{i} + J^{i} [H, P^{i}] = 0
\end{align}
vanishes too, because $[H,P^{i}] = [H,J^{i}] = 0$. Hence, in all, $\vec{J}\cdot\vec{P}$ commutes with $H$, $P^{i}$, $\vec{P}^{\,2}$, $J^{i}$, and $\vec{J}^{\,2}$.

Suppose we restrict our attention to eigenkets $\ket{E,M}$ of the Hamiltonian $H$ and total $4$-momentum operator $P^{\,2}$, which satisfy
\begin{align}
    H\ket{E,m} = E\ket{E,M} \hspace{35 pt} P^{\,2} \ket{E,m} = m^{2} \ket{E,m} \label{2107zz}
\end{align}
where $E > 0$ and $M \geq 0$ are the associated state energy and mass respectively. All of the single-particle states that we consider have well-defined energy and mass in this way. For these states, we define the helicity operator $\Lambda$ as
\begin{align}
    \Lambda \equiv \dfrac{\vec{J}\cdot \vec{P}}{\sqrt{E^{2} - M^{2}}} \label{HelicityDef}
\end{align}
which is pivotal to defining the external states relevant to this dissertation. Like the operator $\vec{J}\cdot\vec{P}$ to which it is proportional, $\Lambda$ commutes with $P^{i}$, $\vec{P}^{\,2}$, $J^{i}$, and $\vec{J}^{\,2}$.

When describing external single-particle states, we will consider the relation of two maximally-commuting sets of observable operators, both of which involve the helicity operator:
\begin{itemize}
    \item[$\bullet$] {\bf Option 1:} $P^{\mu}$, $\Lambda$
    \item[$\bullet$] {\bf Option 2:} $H$, $\vec{J}^{\,2}$, $J_{z}$, $\Lambda$
\end{itemize}
The single-particle states will also have definite masses and spins, and thus be eigenkets of the corresponding operators; however, because they are associated with Casimir operators of the Poincar\'{e} group, we can (and will) always include them in our maximally-commuting set. As such, we will not explicitly label our single-particle states with mass or spin after this point. Helicity eigenstates will be considered in more detail in Section \ref{S - Helicity}.

\subsection{Finite-Dimensional Lorentz Group Representations} \label{FiniteDimLorentzReps}
The preceding discussion concerned the construction of unitary representations of the Poincar\'{e} group. Because the Poincar\'{e} group is non-compact, its nontrivial unitary representations are necessarily infinite-dimensional, which is why most Poincar\'{e} kets end up labeled by continuous variables or indices with countably-infinite values. The same is true of the Lorentz group; however, the Lorentz group also admits finite-dimensional representations, albeit they are necessarily non-unitary. This subsection concerns the standard construction of irreducible finite-dimensional Lorentz representations, which include the usual Lorentz tensor fields (e.g. the spin-1 field $\hat{A}_{\mu}(x)$, the spin-2 field $\hat{h}_{\mu\nu}(x)$, and so-on).

We have actually already encountered such a representation: the $4$-vector representation defined in Section \ref{S - Classical}. Consider rewriting the previously-established 4-vector generators $\{(J^{i})_{\text{4-vector}},(K^{i})_{\text{4-vector}}\}$ so that they superficially resemble the generators we would obtain from the (unitary) quantum promotion procedure. That is, define generators $J^{i} = i(J^{i})_{\text{4-vector}}$ and $K^{i} = i(K^{i})_{\text{4-vector}}$ so that
\begin{align}
    J^{1} = \left(\hspace{-5 pt}\begin{tabular}{ c c c c}
    ${0}$ & ${0}$ & ${0}$ & ${0}$\\
    ${0}$ & ${0}$ & ${0}$ & ${0}$\\
    ${0}$ & ${0}$ & ${0}$ & ${-i}$\\
    ${0}$ & ${0}$ & ${+i}$ & ${0}$
    \end{tabular}\hspace{-5 pt}\right)\hspace{35 pt}J^{2} = \left(\hspace{-5 pt}\begin{tabular}{ c c c c}
    ${0}$ & ${0}$ & ${0}$ & ${0}$\\
    ${0}$ & ${0}$ & ${0}$ & ${+i}$\\
    ${0}$ & ${0}$ & ${0}$ & ${0}$\\
    ${0}$ & ${-i}$ & ${0}$ & ${0}$
    \end{tabular}\hspace{-5 pt}\right)\hspace{35 pt}J^{3} = \left(\hspace{-5 pt}\begin{tabular}{ c c c c}
    ${0}$ & ${0}$ & ${0}$ & ${0}$\\
    ${0}$ & ${0}$ & ${-i}$ & ${0}$\\
    ${0}$ & ${+i}$ & ${0}$ & ${0}$\\
    ${0}$ & ${0}$ & ${0}$ & ${0}$
    \end{tabular}\hspace{-5 pt}\right)
\end{align}
and
\begin{align}
    K^{1} = \left(\hspace{-5 pt}\begin{tabular}{ c c c c}
    ${0}$ & ${+i}$ & ${0}$ & ${0}$\\
    ${+i}$ & ${0}$ & ${0}$ & ${0}$\\
    ${0}$ & ${0}$ & ${0}$ & ${0}$\\
    ${0}$ & ${0}$ & ${0}$ & ${0}$
    \end{tabular}\hspace{-5 pt}\right)\hspace{35 pt}K^{2} = \left(\hspace{-5 pt}\begin{tabular}{ c c c c}
    ${0}$ & ${0}$ & ${+i}$ & ${0}$\\
    ${0}$ & ${0}$ & ${0}$ & ${0}$\\
    ${+i}$ & ${0}$ & ${0}$ & ${0}$\\
    ${0}$ & ${0}$ & ${0}$ & ${0}$
    \end{tabular}\hspace{-5 pt}\right)\hspace{35 pt}K^{3} = \left(\hspace{-5 pt}\begin{tabular}{ c c c c}
    ${0}$ & ${0}$ & ${0}$ & ${+i}$\\
    ${0}$ & ${0}$ & ${0}$ & ${0}$\\
    ${0}$ & ${0}$ & ${0}$ & ${0}$\\
    ${+i}$ & ${0}$ & ${0}$ & ${0}$
    \end{tabular}\hspace{-5 pt}\right)
\end{align}
with commutators
\begin{align}
    [J^{i},J^{j}]= +i\epsilon_{ijk} J^{k} \hspace{35 pt} [J^{i},K^{j}]= +i\epsilon_{ijk} K^{k} \hspace{35 pt} [K^{i},K^{j}]=-i\epsilon_{ijk} J^{k} \label{LorentzAlgebraCommutatorsAgain}
\end{align}
Using these, we may write a generic rotation and boost as $R(\vec{\alpha}) = \text{Exp}[-i\vec{\alpha}\cdot\vec{J}]$ and $B(\vec{\beta}) = \text{Exp}[-i\vec{\beta}\cdot\vec{K}]$ respectively. Note that although the rotation generators $\{J^{i}\}$ are Hermitian ($(J^{i})^{\intercal} = (J^{i})^{*}$), the boost generators $\{K^{i}\}$ are anti-Hermitian ($(K^{i})^{\intercal} = -(K^{i})^{*}$), thereby reinforcing that this representation is not unitary.

Thus far, we have done little of substance: we moved some factors of $i$ around in what otherwise remains the standard 4-vector Lorentz transformations. However, working with complex numbers does have its advantages. Consider the following complex linear combinations of rotation and boost generators:
\begin{align}
    \vec{\mathcal{A}} \equiv \dfrac{1}{2}(\vec{J} + i \vec{K}\,) \hspace{35 pt} \vec{\mathcal{B}} \equiv \dfrac{1}{2}(\vec{J} - i \vec{K}\,)
\end{align}
These combinations do not exist in the Lorentz algebra $\mathfrak{so}(1,3)$ (which only admits real linear combinations) but exist instead in the {\it complexified} Lorentz algebra $\mathfrak{so}(1,3)_{\mathbb{C}}$. Nonetheless, using the known commutators of $\{J^{i},K^{i}\}$, we may calculate the commutators of $\{\mathcal{A}^{i},\mathcal{B}^{i}\}$. Doing so, we find they equal
\begin{align}
    [\mathcal{A}^{i},\mathcal{A}^{j}] = +\epsilon_{ijk} \mathcal{A}^{k} \hspace{35 pt} [\mathcal{A}^{i},\mathcal{B}^{j}] = 0 \hspace{35 pt} [\mathcal{B}^{i},\mathcal{B}^{j}] = +\epsilon_{ijk} \mathcal{B}^{k} \label{AandBareSU2}
\end{align}
That is, not only do $\vec{\mathcal{A}}$ and $\vec{\mathcal{B}}$ decouple, but each individually satisfies the $\mathbf{SU(2)}$ commutation relations. Furthermore, they are Hermitian:
\begin{align}
    {\mathcal{A}}^{1} = \left(\hspace{-5 pt}\begin{tabular}{ c c c c}
    ${0}$ & ${-\tfrac{1}{2}}$ & ${0}$ & ${0}$\\
    ${-\tfrac{1}{2}}$ & ${0}$ & ${0}$ & ${0}$\\
    ${0}$ & ${0}$ & ${0}$ & ${-\tfrac{i}{2}}$\\
    ${0}$ & ${0}$ & ${+\tfrac{i}{2}}$ & ${0}$
    \end{tabular}\hspace{-5 pt}\right)\hspace{35 pt}{\mathcal{B}}^{1} = \left(\hspace{-5 pt}\begin{tabular}{ c c c c}
    ${0}$ & ${+\tfrac{1}{2}}$ & ${0}$ & ${0}$\\
    ${+\tfrac{1}{2}}$ & ${0}$ & ${0}$ & ${0}$\\
    ${0}$ & ${0}$ & ${0}$ & ${-\tfrac{i}{2}}$\\
    ${0}$ & ${0}$ & ${+\tfrac{i}{2}}$ & ${0}$
    \end{tabular}\hspace{-5 pt}\right)
\end{align}
\begin{align}
    {\mathcal{A}}^{2} = \left(\hspace{-5 pt}\begin{tabular}{ c c c c}
    ${0}$ & ${0}$ & ${-\tfrac{1}{2}}$ & ${0}$\\
    ${0}$ & ${0}$ & ${0}$ & ${+\tfrac{i}{2}}$\\
    ${-\tfrac{1}{2}}$ & ${0}$ & ${0}$ & ${0}$\\
    ${0}$ & ${-\tfrac{i}{2}}$ & ${0}$ & ${0}$
    \end{tabular}\hspace{-5 pt}\right)\hspace{35 pt}{\mathcal{B}}^{2} = \left(\hspace{-5 pt}\begin{tabular}{ c c c c}
    ${0}$ & ${0}$ & ${+\tfrac{1}{2}}$ & ${0}$\\
    ${0}$ & ${0}$ & ${0}$ & ${+\tfrac{i}{2}}$\\
    ${+\tfrac{1}{2}}$ & ${0}$ & ${0}$ & ${0}$\\
    ${0}$ & ${-\tfrac{i}{2}}$ & ${0}$ & ${0}$
    \end{tabular}\hspace{-5 pt}\right)
\end{align}
\begin{align}
    {\mathcal{A}}^{3} = \left(\hspace{-5 pt}\begin{tabular}{ c c c c}
    ${0}$ & ${0}$ & ${0}$ & ${-\tfrac{1}{2}}$\\
    ${0}$ & ${0}$ & ${-\tfrac{i}{2}}$ & ${0}$\\
    ${0}$ & ${+\tfrac{i}{2}}$ & ${0}$ & ${0}$\\
    ${-\tfrac{1}{2}}$ & ${0}$ & ${0}$ & ${0}$
    \end{tabular}\hspace{-5 pt}\right)\hspace{35 pt}{\mathcal{B}}^{3} = \left(\hspace{-5 pt}\begin{tabular}{ c c c c}
    ${0}$ & ${0}$ & ${0}$ & ${+\tfrac{1}{2}}$\\
    ${0}$ & ${0}$ & ${-\tfrac{i}{2}}$ & ${0}$\\
    ${0}$ & ${+\tfrac{i}{2}}$ & ${0}$ & ${0}$\\
    ${+\tfrac{1}{2}}$ & ${0}$ & ${0}$ & ${0}$
    \end{tabular}\hspace{-5 pt}\right)
\end{align}
which means the 4-vector transformations
\begin{align}
    \mathcal{U}[R_{A}(\vec{\theta}_{A})] \equiv \text{Exp}[-i\vec{\theta}_{A}\cdot\vec{\mathcal{A}}\,]\hspace{20 pt}\text{ and }\hspace{20 pt}\text{Exp}[-i\vec{\theta}_{B}\cdot\vec{\mathcal{B}}\,]
\end{align}
generated by $\vec{\mathcal{A}}$ and $\vec{\mathcal{B}}$ form unitary representations of $\mathbf{SU(2)}$ when the parameters $\vec{\theta}_{A}$ and $\vec{\theta}_{B}$ are real. The Casimir operators $\vec{\mathcal{A}}^{\,2}$ and $\vec{\mathcal{B}}^{\,2}$ of these $\mathbf{SU(2)}$ representations equal
\begin{align}
    \vec{\mathcal{A}}^{\,2} = \vec{\mathcal{B}}^{\,2} = \left(\hspace{-5 pt}\begin{tabular}{ c c c c}
    ${+\tfrac{3}{4}}$ & ${0}$ & ${0}$ & ${0}$\\
    ${0}$ & ${+\tfrac{3}{4}}$ & ${0}$ & ${0}$\\
    ${0}$ & ${0}$ & ${+\tfrac{3}{4}}$ & ${0}$\\
    ${0}$ & ${0}$ & ${0}$ & ${+\tfrac{3}{4}}$
    \end{tabular}\hspace{-5 pt}\right) \label{A2B234}
\end{align}
Note that the transformations $\mathcal{U}[R_{A}(\vec{\theta}_{A})]$ and $\mathcal{U}[R_{B}(\vec{\theta}_{B})]$ will typically map real 4-vectors to complex 4-vectors.

The commutators of $\{\mathcal{A}^{i},\mathcal{B}^{i}\}$ in Eq. \eqref{AandBareSU2} suggest that the complexified Lorentz algebra $\mathfrak{so}(1,3)_{\mathbb{C}}$ is isomorphic to two complexified copies of $\mathfrak{su}(2)$, i.e. $\mathfrak{so}(3,1)_{\mathbb{C}} \cong \mathfrak{su}(2)_{\mathbb{C}} \otimes \mathfrak{su}(2)_{\mathbb{C}}$.\footnote{Note that the degrees of freedom (DOF) work out: $\mathfrak{so}(1,3)$ has 6 real DOF, so $\mathfrak{so}(1,3)_{\mathbb{C}}$ has 6 complex DOF = 12 real DOF, whereas $\mathfrak{su}(2)$ has 3 real DOF, so $\mathfrak{su}(2)_{\mathbb{C}}$ has 3 complex DOF = 6 real DOF. Because $12 = 2\cdot 6$, all is well.} This isomorphism is correct, and enables a trick for finding all irreducible finite-dimensional representations of the Lorentz group. As will be reviewed in Subsection \ref{Finite Dim Angular Momentum Reps}, for each nonnegative half-integer $j$ there exists an irreducible $(2j+1)$-dimensional unitary representation of $\mathbf{SU(2)}$, wherein each state corresponds to a different $J_{z}$ eigenvalue $m \in \{-j,\cdots, j\}$ and all states are eigenstates of $\vec{J}^{\,2}$ with eigenvalue $j(j+1)$. Using this knowledge, the standard strategy for deriving irreducible finite-dimensional Lorentz representations is as follows:
\begin{enumerate}
    \item Choose two irreducible finite-dimensional unitary representations of $\mathbf{SU}(2)$, where one has $j=a$ and the other $j=b$ for some nonnegative half-integers $(a,b)$. Label the corresponding Lie algebras as $\mathfrak{su}(2)_{A}$ and $\mathfrak{su}(2)_{B}$, and their (Hermitian) generators as $\vec{\mathcal{A}}$ and $\vec{\mathcal{B}}$ respectively.
    \item Construct the complexifications of these algebras, $\mathfrak{su}(2)_{A,\mathbb{C}}$ and $\mathfrak{su}(2)_{B,\mathbb{C}}$, and form their direct product $\mathfrak{su}(2)_{A,\mathbb{C}}\otimes\mathfrak{su}(2)_{B,\mathbb{C}}$. In this new algebra, the collective generators $\{\mathcal{A}^{i},\mathcal{B}^{i}\}$ automatically satisfy the Eq. \eqref{AandBareSU2} commutators.
    \item Construct new operators $\vec{J}\equiv \vec{\mathcal{A}} + \vec{\mathcal{B}}$ and $\vec{K} \equiv -i(\vec{\mathcal{A}} - \vec{\mathcal{B}})$, which are necessarily Hermitian and anti-Hermitian respectively. These automatically satisfy the Eq. \eqref{LorentzAlgebraCommutatorsAgain} commutators, and thus correspond to the complexified Lorentz algebra $\mathfrak{so}(1,3)_{\mathbb{C}}$.
    \item Consider only the real linear combinations of $\{J^{i},K^{i}\}$, and thereby restrict their span to the real subalgebra of $\mathfrak{so}(1,3)_{\mathbb{C}}$, the usual Lorentz algebra $\mathfrak{so}(1,3)\cong \mathfrak{sl}(2,\mathbb{C})$.
    \item Exponentiate the operators $\{J^{i},K^{i}\}$ and thereby obtain an irreducible $(2a+1)(2b+1)$-dimensional representation of the covering group of the Lorentz group, $\mathbf{SL(2,\mathbb{C})}$. This is called the $(a,b)$ representation of the Lorentz group, or the $(a,b)$ Lorentz representation.
\end{enumerate}
Using this procedure, we can construct an irreducible finite-dimensional representation of the (covering group of the) Lorentz group for every half-integer pair $(a,b)$, and in doing so have accomplished the goal of this subsection.

Before moving on to the next section, let us reconsider the 4-vector representation from this more general perspective. In Eq. \eqref{A2B234}, we found that $\vec{\mathcal{A}}^{\, 2} = \vec{\mathcal{B}}^{\, 2} = +\tfrac{3}{4} \mathbbm{1}$, which indicates (because $\tfrac{3}{4} = \tfrac{1}{2}(\tfrac{1}{2}+1)$) that these representations of $\vec{\mathcal{A}}$ and $\vec{\mathcal{B}}$ correspond to $a=b=+\tfrac{1}{2}$. In other words, the 4-vector representation is the $(\tfrac{1}{2},\tfrac{1}{2})$ Lorentz representation. Although the 4-vector representation is irreducible as an $\mathbf{SL(2,\mathbb{C})}$ representation, it is reducible in terms of the rotation subgroup $\mathbf{SU(2)}$.  As will be discussed briefly in Subsection \ref{SS - Clebsch Gordan}, $\mathbf{SU(2)}$ representations can be combined to yield new $\mathbf{SU(2)}$ representations. In particular, if $\vec{J}_{1}$ and $\vec{J}_{2}$ are the respective spin-$j_{1}$ and spin-$j_{2}$ generators for unitary $\mathbf{SU}(2)$ representations, then the direct product of their algebras yields a sum of spin-$j$ representations, where $j\in\{|j_{1}-j_{2}|,\dots,j_{1}+j_{2}\}$. Because $\vec{J} = \vec{\mathcal{A}} +\vec{\mathcal{B}}$ is precisely of this form, the spin-$a$ and spin-$b$ representations implicit in the $(a,b)$ Lorentz representation combine to yield spin-$|a-b|$ through spin-$(a+b)$ representations. This means, for example, that the 4-vector representation $(\tfrac{1}{2},\tfrac{1}{2})$ encodes both spin-$0$ and spin-$1$ content. The spin-$1$ portion of the 4-vector representation yields precisely the canonical spin-$1$ polarization tensors, which we derive in Subsection \ref{SS - Polarizations Derived}.

\section{External States and Matrix Elements} \label{S - External States and Matrix Elements}
\subsection{Single-Particle States: Definite 4-Momentum} \label{External States - Single-Particle}

In quantum mechanics, the kets describing physical states are chosen to span the eigenvalues of certain Hermitian operators corresponding to observable quantities. Specifically, given a commuting set of observables $\{A_{1},\cdots ,A_{N}\}$ (so that $[A_{i},A_{j}]=0$ for any pair $A_{i},A_{j}$), we can form a complete set of kets $\{\ket{a_{1},\cdots, a_{n}}\}$ where
\begin{align}
    A_{i} \ket{a_{1},\cdots, a_{n}} = a_{i} \ket{a_{1}, \cdots, a_{n}}
\end{align}
for each $i\in\{1,\cdots,n\}$. Because each operator $A_{i}$ is Hermitian, each eigenvalue $a_{i}$ is real. The resulting collection of kets form a complete basis and are equipped with a convention-dependent orthonormalization condition.

For the duration of this chapter, we use (interaction picture) kets to describe the initial and final multi-particle states of scattering processes, each of which is built from direct products of single-particle states. Thus, we first focus on the construction of single-particle states. Following Wigner's classification \cite{Wigner:1939cj}, our single-particle states are ultimately chosen to be (infinite-dimensional) unitary irreducible representations of the Poincar\'{e} group which have definitive mass and total spin (or total helicity, if massless). For now, we will choose these states so that they only have well-defined $4$-momentum (helicity will be added in Section \ref{S - Helicity}). We can choose the components of $4$-momentum as quantum numbers because the $4$-momentum operators of Subsection $\ref{Quantum - Generators}$ form a commuting set ($[P^{\mu},P^{\nu}] = 0$ for all $\mu,\nu \in \{0,1,2,3\}$) and the $4$-momentum operators encode an observable. Because the energy eigenvalue $E$ associated with the Hamiltonian $H$ is constrained by the particle's mass $m$ to satisfy $E^{2} = m^{2} + \vec{p}^{\,2}$, we only label the kets with the $3$-momentum eigenvalues, i.e. as $\ket{\vec{p}\,} \equiv \ket{p_{x}\,,p_{y}\,,p_{z}}$. By definition, these satisfy
\begin{align}
    H\ket{\vec{p}\,} = \sqrt{m^{2} + |\vec{p}\,|^{2}} \, \ket{\vec{p}\,}\hspace{35 pt}\vec{P}\ket{\vec{p}\,} = \vec{p}\,\ket{\vec{p}\,}
\end{align}
where we recall that $H\equiv P^{0}$. Because $3$-momentum is a continuous degree of freedom, these kets are normalized by a Dirac delta, such that
\begin{align}
    \braket{\vec{p}\,}{\vec{p}^{\,\prime}} \propto \delta^{3}(\vec{p} - \vec{p}^{\,\prime})
\end{align}
up to some proportionality factor. The exact choice of this proportionality factor varies throughout the literature.  We motivate our particular choice via the Lorentz-invariant phase space element derived in Subsection \ref{Classical - LIPS}. Namely, we would like to normalize our kets such that we can resolve the identity on this space via an integral weighted by the Lorentz-invariant factor $d^{3}\vec{p}/[(2\pi)^{3} 2E_{\vec{p}}$:
\begin{align}
    \mathbbm{1} = \int \dfrac{d^{3}\vec{p}}{(2\pi)^{3}} \dfrac{1}{2E_{\vec{p}}} \ket{\vec{p}\,} \bra{\vec{p}\,}
\end{align}
This implies
\begin{align}
    \ket{\vec{k}\,} = \int \dfrac{d^{3}\vec{p}}{(2\pi)^{3}} \dfrac{1}{2E_{\vec{p}}} \ket{\vec{p}\,} \braket{\vec{p}\,}{\vec{k}\,}
\end{align}
which is achieved so long as we choose our normalization such that
\begin{align}
    \braket{\vec{p}\,}{\vec{p}^{\,\prime}} = (2\pi)^{3} \, (2E_{\vec{p}}) \, \delta^{3}(\vec{p}-\vec{p}^{\,\prime}) \label{3momentumnormalization}
\end{align}
and so we do. A simultaneous eigenstate of $P^{2}$ and $H$ is also an eigenstate of $\vec{P}^{\,2}$, so we could use $|\vec{p}\,|$ instead of $E$ as a quantum number.

This section will focus on single-particle states of the form $\ket{\vec{p}\,}$. However, other single-particles kets exist which are useful in different contexts. For example, we might define an alternate collection of $3$-momentum kets $\ket{|\vec{p}\,|,\theta,\phi}$, which are normalized like Eq. \eqref{3momentumnormalization} but expressed in spherical coordinates. Note that, because $d^{3}\vec{p} = |\vec{p}\,|^{2}\, d|\vec{p}\,|\, d\Omega$ (where $d\Omega = d(\cos\theta)\, d\phi$),
\begin{align}
    \delta^{3}(\vec{p}-\vec{p}^{\,\prime}) &= \dfrac{1}{|\vec{p}\,|^{2}}\, \delta(|\vec{p}\,|-|\vec{p}^{\,\prime}|) \, \delta^{2}( \Omega - \Omega^{\prime} )
\end{align}
where
\begin{align}
    \delta^{2}( \Omega - \Omega^{\prime} ) \equiv \delta( \phi - \phi^{\prime} ) \, \delta(\cos\theta - \cos\theta^{\prime} )
\end{align}
and $\theta,\theta^{\prime}\in [0,\pi]$ and $\phi,\phi^{\prime} \in [0,2\pi)$. Thus, the kets $\ket{|\vec{p}\,|,\theta,\phi}$ are normalized analogous to Eq. \eqref{3momentumnormalization} if we define,
\begin{align}
    \braket{|\vec{p}\,|,\theta,\phi}{|\vec{p}^{\,\prime}|,\theta^{\prime},\phi^{\prime}} &= (2\pi)^{3} \, (2E_{\vec{p}}) \,  \dfrac{1}{|\vec{p}\,|^{2}} \, \delta(|\vec{p}\,|-|\vec{p}^{\,\prime}|) \, \delta^{2}(\Omega - \Omega^{\prime}) \\
    &= (2\pi) \, \dfrac{8 \pi^{2} E_{\vec{p}}}{|\vec{p}\,|^{2}} \, \delta(|\vec{p}\,|-|\vec{p}^{\,\prime}|) \, \delta^{2}(\Omega - \Omega^{\prime}) \label{3momentumnormalizationB}
\end{align}
such that
\begin{align}
    \mathbbm{1} &= \int |\vec{p}\,|^{2}\, d|\vec{p}\,| \, d\Omega \hspace{5 pt} \dfrac{1}{(2\pi)^{3}\, 2E_{\vec{p}}} \,  \ket{|\vec{p}\,|,\theta,\phi} \bra{|\vec{p}\,|,\theta,\phi}\nonumber\\
    &= \int \dfrac{d|\vec{p}\,|}{2\pi} \, d\Omega \, \hspace{5 pt} \dfrac{|\vec{p}\,|^{2}}{8\pi^{2} E_{\vec{p}}} \,  \ket{|\vec{p}\,|,\theta,\phi} \bra{|\vec{p}\,|,\theta,\phi}
\end{align}
on this space. We can go a step further and define kets $\ket{E,\theta,\phi}$, where the $3$-momentum quantum number $|\vec{p}\,|$ has been replaced with energy $E_{\vec{p}} = \sqrt{m^{2} + |\vec{p}\,|^{2}}$. Angular momentum kets defined in analogy to these kets will be useful for deriving the partial wave unitarity constraints later in this chapter. Note that, dropping the $\vec{p}$ subscript on $E_{\vec{p}}$,
\begin{align}
    d|\vec{p}\,| = \bigg|\dfrac{\partial \vec{p}}{\partial E} \bigg| \, dE = \dfrac{E}{|\vec{p}\,|} \, dE \hspace{15 pt}\text{ and }\hspace{15 pt}
    \delta(|\vec{p}\,|-|\vec{p}^{\,\prime}|) = \dfrac{|\vec{p}\,|}{E} \, \delta(E- E^{\prime})
\end{align}
such that we can define a collection of kets $\ket{E, \theta, \phi}$ which maintain the normalization defined in Eqs. \eqref{3momentumnormalization} and \eqref{3momentumnormalizationB} as long as
\begin{align}
    \braket{E,\theta,\phi}{E^{\prime},\theta^{\prime},\phi^{\prime}} = (2\pi)^{3} \, \dfrac{2}{|\vec{p}\,|}\,\delta(E-E^{\prime}) \, \delta^{2}(\Omega - \Omega^{\prime}) \label{3momentumnormalizationC1}
\end{align}
with
\begin{align}
    \mathbbm{1} = \int \dfrac{dE}{2\pi}\, d\Omega\hspace{5 pt} \dfrac{|\vec{p}\,|}{8 \pi^{2}}\, \ket{E,\theta,\phi}\bra{E,\theta,\phi} \label{3momentumnormalizationC2}
\end{align}
where $|\vec{p}\,| = \sqrt{E^{2} - m^{2}}$. For succinctness, we sometimes write Eqs. \eqref{3momentumnormalizationC1} and \eqref{3momentumnormalizationC2} using $w_{E} \equiv |\vec{p}|/8\pi^{2} = \sqrt{E^{2}-m^{2}}/8\pi^{2}$. Having derived these, let us return to considering the kets $\ket{\vec{p}\,}$.

Eq. \eqref{3momentumnormalization} expresses a lot of information about the space of $3$-momentum kets, but we can add further structure to this space by using our knowledge of Lorentz transformations: we know from our considerations of the Lorentz group in Section \ref{Classical - Minkowski} that any two $4$-momenta on the same mass hyperboloid can be related via a Lorentz transformation. Consequently, given a Lorentz transformation $\Lambda$ that maps a $4$-momentum $p$ to a $4$-momentum $p^{\prime}$, there exists a unitary operator $\mathcal{U}[\Lambda]$ that maps $\ket{\vec{p}\,}$ to $\ket{\vec{p}^{\,\prime}}$ up to a phase:
\begin{align}
    \ket{\vec{p}^{\,\prime}} \propto \mathcal{U}[\Lambda]\, \ket{\vec{p}\,}
\end{align}
While it may be tempting to set this to an equality, such an equality would not be well-defined because there are {\it many} distinct Lorentz transformations that take $p$ to $p^{\prime}$. Therefore, to uniquely identify individual kets we follow Wigner's lead \cite{Wigner:1939cj} and choose a standard $4$-momentum $k$ on each Lorentz invariant $4$-momentum hypersurface. Then, for every non-standard $4$-momentum $p$ on a given hypersurface, we choose a standard Lorentz transformation that maps the corresponding standard $4$-momentum $k$ to $p$. By choosing these standard $4$-momenta and transformations, we eliminate the ambiguity of the above proportionality and can establish a well-defined equality.

Our particular choice of standard 4-momentum depends on the mass of the single-particle state in question:
\begin{itemize}
    \item[$\bullet$] {\bf Massive:} For a single-particle state with mass $m>0$, we choose the rest frame $4$-momentum $k^{\mu}=(m,\vec{0}\,)$. To obtain any other $4$-momentum $p$ having equal mass, we first boost along $z$ until it has $3$-momentum $|\vec{p}\,|\hat{z}$ and then rotate via $R(\theta,\phi)$ to attain a $3$-momentum $\vec{p}$. This allows us to define, unambiguously,
    \begin{align}
        \ket{\vec{p}\,} = \mathcal{U}[R(\phi,\theta)]\,\mathcal{U}[B_{z}(\beta_{k\rightarrow p})]\,\ket{\vec{0}}
    \end{align}
    where $\beta_{k\rightarrow p} = \text{arccosh}(E_{\vec{p}}/m)$.
    \item[$\bullet$] {\bf Massless:} There is no rest frame for a single-particle state with vanishing mass $m=0$, so we instead choose a standard light-like $4$-momentum $(E_{\vec{k}},E_{\vec{k}}\hat{z})$ for some choice of energy $E_{\vec{k}}$. From here the procedure mimics the massive case: to obtain any other $4$-momentum $p$ on the light cone, we first boost along $z$ until it has $3$-momentum $|\vec{p}\,|\hat{z}$ and then rotate via $R(\theta,\phi)$ to attain a $3$-momentum $\vec{p}$. This allows us to define, unambiguously,
    \begin{align}
        \ket{\vec{p}\,} = \mathcal{U}[R(\phi,\theta)]\,\mathcal{U}[B_{z}(\beta_{k\rightarrow p})]\,\ket{\vec{k}\,}
    \end{align}
    where now $\beta_{k\rightarrow p} = \ln(E_{\vec{p}}/E_{\vec{k}})$.
\end{itemize}
We will revisit these procedures when constructing helicity eigenstates in Section \ref{S - Helicity}.

The above discussion glosses over an important (but ultimately inconsequential) technicality. The only physical states are those which have finite normalizations. Because the $3$-momentum kets are normalized to a Dirac delta, they are unphysical, and thus in principll cannot serve as external states in physical scattering processes. This reflects the fact that we cannot in practice construct a system with definite $3$-momentum. Even in the most ideal of experimental conditions, the existence of such a state is forbidden by the Heisenberg uncertainty principle. Therefore, we should actually perform calculations in quantum field theory using wavepacket superpositions of states. For example, rather than using a ket $\ket{\vec{p}\,}$, we might instead use the wavepacket
\begin{align}
    \ket{\psi_{\vec{p}}\,} = \int \dfrac{d^{3}\vec{q}}{(2\pi)^{3}}\dfrac{1}{2E_{\vec{q}}}\hspace{5 pt} \psi_{\vec{p}}(\vec{q}\,) \, \ket{\vec{q}\,}
\end{align}
where $\psi_{\vec{p}}(\vec{q}\,)$ is a three-dimensional Gaussian sharply peaked as $\vec{q}=\vec{p}$. The smoothing this wavepacket provides is sufficient to yield a finite normalization:
\begin{align}
    \braket{\psi_{\vec{p}}\,}{\psi_{\vec{p}}\,} &= \int \dfrac{d^{3}\vec{k}}{(2\pi)^{3}}\dfrac{d^{3}\vec{q}}{(2\pi)^{3}}\dfrac{1}{2E_{\vec{k}}}\dfrac{1}{2E_{\vec{q}}}\hspace{5 pt} \psi^{*}_{\vec{p}}(\vec{k}\,) \, \psi_{\vec{p}}(\vec{q}\,) \, \braket{\vec{k}\,}{\vec{q}\,}\\
    &= \int \dfrac{d^{3}\vec{q}}{(2\pi)^{3}} \dfrac{1}{2E_{\vec{q}}} \hspace{5 pt} |\psi_{\vec{p}}(\vec{q}\,)|^{2}
\end{align}
This is important when deriving results like the LSZ reduction formula (which relates external states to quantum fields), but as far as matrix elements are concerned we can always take the limit as the wavepacket becomes a Dirac delta and thereby use the $3$-momentum kets as external states (even if technically we should not). Because this dissertation does not derive results sensitive to this technicality, it will be ignored.

\subsection{Multi-Particle States: Definite 4-Momentum} \label{SS - Multiparticle}
A basis of multi-particle states can be formed by combining single-particle states that each have a well-defined mass and total spin. For our single-particle kets $\ket{\vec{p}\,}$ with well-defined 4-momenta, such an $n$-particle basis state would be labeled $\ket{\vec{p}_{1},\,\vec{p}_{2}\,,\cdots\,,\vec{p}_{n}}$ where each 3-momentum $\vec{p}_{i}$ labels a particle with definite mass $m_{i}$. Mathematically, such a basis state is related to the single-particle kets up to a phase like so:
\begin{align}
    \ket{\vec{p}_{1},\,\vec{p}_{2}\,,\cdots\,,\vec{p}_{n}} \propto \ket{\vec{p}_{1}}\otimes \ket{\vec{p}_{2}}\otimes\cdots\otimes\ket{\vec{p}_{n}} \label{multiparticlestatedefinition}
\end{align}
These single-particle states are assumedly distinguishable (more on identical particles soon) and arranged in some canonical ordering based on their distinguishability, e.g. electrons are listed left of muons and so-on, and electron kets vanish when contracted with muon bras. We choose the free phase in Eq. \eqref{multiparticlestatedefinition} to be $+1$ so that equality replaces the proportionality. However, regardless of the particular phase selected, the multi-particle normalization is implied by the single-particle normalization Eq. \eqref{3momentumnormalization}. By complex squaring both sides of Eq. \ref{multiparticlestatedefinition}, the multi-particle normalization is found to equal
\begin{align}
    \braket{\vec{p}_{1}\,,\vec{p}_{2}\,,\cdots\,,\vec{p}_{n}}{\vec{k}_{1}\,,\vec{k}_{2}\,,\cdots\,,\vec{k}_{n}} = \prod_{i=1}^{n} (2\pi)^{3}\, (2 E_{\vec{p}_{i}})\, \delta^{3}(\vec{p}_{i} - \vec{k}_{i}) \label{distinguishedmultiparticlenormalization}
\end{align}
from which the $n$-particle resolution of identity (without identical particles) equals
\begin{align}
    \mathbbm{1} = \int \prod_{i=1}^{n} \bigg[\dfrac{d^{3}p_{i}}{(2\pi)^{3}} \dfrac{1}{2E_{p_{i}}}\bigg] \hspace{5 pt} \ket{\vec{p}_{1}\,,\vec{p}_{2}\,,\cdots\,,\vec{p}_{n}}\bra{\vec{p}_{1}\,,\vec{p}_{2}\,,\cdots\,,\vec{p}_{n}} \label{distinguishableresolutionofidentity}
\end{align}
Note the square brackets contain the $n$-particle Lorentz-invariant phase space measure $d\Pi_{n}$ modulo the Dirac delta to conserve total 4-momentum. The above construction is sufficient if all particles are distinguishable. In that case, we can imagine an additional indicator being added to each $3$-momentum label in the ket that gives a unique name to each particle beyond its $3$-momentum content. Then, when we perform the inner product described in Eq. \eqref{distinguishedmultiparticlenormalization}, we could pair up particles in the bra and ket based on matching their names to obtain the correct Dirac deltas (and if we cannot find such a collection of pairs then we know the inner product vanishes). However, if any number of the particles involved are instead identical, then we must be more careful in our construction of the ket space.

Two particles are identical if they share all of the same intrinsic quantum numbers---such as mass, total spin, and gauge transformation properties---and a particular set of such properties defines a particle species. For example, as listed in Figure \ref{Figure - SM}, the particle species known as ``top quark" is characterized by a mass of $173\text{ GeV}$, total spin $\tfrac{1}{2}$, electric charge $+\tfrac{2}{3}$, and triplet transformation behavior under the color gauge group $\mathbf{SU(3)_{C}}$. Because they are spin-$\tfrac{1}{2}$ particles, each top quark can be measured as either spin up ($m = +\tfrac{1}{2}$) or spin down ($m = -\tfrac{1}{2}$) with respect to a given projection axis; however, the need for a projection axis indicates that although projected spin is an {\it internal} quantum number, it is not an {\it intrinsic} quantum number. Thus, spin up and spin down top quarks are still regarded as identical in the technical sense. Frame-dependence similarly indicates that 4-momentum (a possible choice for an extrinsic quantum number) cannot be an intrinsic quantum number, and particles with different 4-momenta can still be identical. These considerations apply to color charge as well: the status of a top quark as red, green, or blue (or a specific superposition of those colors) is a gauge-dependent quality, and so color charge is not an intrinsic quantum number. (This contrasts with electric charge, which {\it does} possess a gauge-independent value.) Meanwhile, despite the charm quark sharing many of the same intrinsic quantum numbers as the top quark, the two quarks differ in mass and thus every charm quark is distinguishable from every top quark regardless of further details.

To demonstrate that the existing machinery is insufficient for the construction of multi-particle states involving identical particles, suppose we try to use the previous construction to describe a $2$-particle state consisting of identical particles with distinct 3-momenta $\vec{p}_{1}$ and $\vec{p}_{2}$. If the previous construction truly is sufficient, then (because the particles are identical) the kets $\ket{\vec{p}_{1}\,,\vec{p}_{2}}$ or $\ket{\vec{p}_{1}\,,\vec{p}_{2}}$ describe indistinguishable physical realities and thus must be equal up to a phase $\chi$:
\begin{align}
    \ket{\vec{p}_{1}\,,\vec{p}_{2}} \stackrel{?}{=} \chi \ket{\vec{p}_{2}\,,\vec{p}_{1}} \label{chiintroduction}
\end{align}
If we swap the order of the labels in the RHS ket once more (and assume $\chi$ is agnostic to the details of the $3$-momenta encoded by the ket\footnote{This is a nontrivial assumption. Thankfully, even when this assumption is dropped one can still recover the same end result, although doing so requires a good amount of homotopy theory to demonstrate that the $3$-momentum-dependent phase is always removable via ket redefinitions.}), then we return to the original ordering and gain another factor of $\chi$
\begin{align}
    \ket{\vec{p}_{1}\,,\vec{p}_{2}} \stackrel{?}{=} \chi^{2} \ket{\vec{p}_{1}\,,\vec{p}_{2}}
\end{align}
where equality only holds true if $\chi^{2} = 1$. Note that $\chi^{2}$ is a regular square (i.e. not a complex square), so this restricts $\chi$ to equaling $+1$ or $-1$. The exact choice of one sign over the other is an intrinsic property of the particle being considered and is ultimately tied to the spin of the given particle. Unfortunately, Eq. \eqref{chiintroduction} is inconsistent with the normalization defined in Eq. \eqref{distinguishedmultiparticlenormalization}: specifically,
\begin{align}
    0 = \braket{\vec{p}_{1}\,,\vec{p}_{2}}{\vec{p}_{2}\,,\vec{p}_{1}} \stackrel{?}{=} \chi \braket{\vec{p}_{1}\,,\vec{p}_{2}}{\vec{p}_{1}\,,\vec{p}_{2}} = \chi \prod_{i=1}^{2} (2\pi)^{3}\, (2 E_{\vec{p}_{i}})\, \delta^{3}(0)
\end{align}
which is zero on the LHS, but infinite on the RHS. The origin of this obstruction lies in Eq. \eqref{multiparticlestatedefinition}, where we expressed an $n$-particle ket as a direct product of single-particle kets. The ordering in the direct product $\ket{\vec{p}_{1}}\otimes\ket{\vec{p}_{2}}$ is absolute and lacks the exchange symmetry we desire, e.g.
\begin{align}
    \ket{\vec{p}_{1}}\otimes\ket{\vec{p}_{2}} \neq \chi \ket{\vec{p}_{2}} \otimes \ket{\vec{p}_{1}}
\end{align}
To remedy this, we define the following symmetric and antisymmetric kets:
\begin{align}
    \cket{\vec{p}_{1}\,,\vec{p}_{2}\,,\cdots\,,\vec{p}_{n}} &=\dfrac{1}{\sqrt{n!}} \Bigg[\prod_{i=1}^{N} \dfrac{1}{\sqrt{n_{i}!}}\Bigg] \sum_{\pi \in \pi_{n}} \ket{\pi(\vec{p}_{1},\vec{p}_{2},\cdots,\vec{p}_{n})}\label{bosonicstate}\\
    \sket{\vec{p}_{1}\,,\vec{p}_{2}\,,\cdots\,,\vec{p}_{n}} &=\dfrac{1}{\sqrt{n!}} \sum_{\pi \in \pi_{n}} \text{sign}(\pi)\, \ket{\pi(\vec{p}_{1},\vec{p}_{2},\cdots,\vec{p}_{n})}\label{fermionicstate}
\end{align}
where $\pi_{n}$ denotes the set of all $n$-element permutations and $\text{sign}(\pi)$ refers to the parity of a permutation $\pi$ ($+1$ for even permutations, $-1$ for odd permutations). The exact prefactors in front of each permutation sum are chosen to guarantee upcoming normalization formulas (Eqs. \eqref{bosonmultiparticlenormalization} and \eqref{fermionmultiparticlenormalization}). Within the symmetrized case in particular, care must be taken to account for potential repeats of particle information, e.g. (because we continue to neglect other quantum numbers) when two identical particles have identical 3-momentum $\vec{p}_{1} = \vec{p}_{2}$. To be explicit, suppose among the $n$ particle labels there is only $N$ unique labels present. The $n_{i}$ present in Eq. \eqref{bosonicstate} takes into account possible label repeats and equals how many times a given unique label occurs in the list $(\vec{p}_{1}\,,\,\vec{p}_{2}\,,\,\cdots\,,\,\vec{p}_{n})$. Thus, $n = n_{1} + \cdots + n_{N}$. For future use, it is useful to define a symbol $\mathcal{S}(\vec{p}_{1},\vec{p}_{2},\cdots,\vec{p}_{n})$ for this repeated label information:
\begin{align}
    \mathcal{S}(\vec{p}_{1},\vec{p}_{2},\cdots,\vec{p}_{n}) \equiv \prod_{i=1}^{N} n_{i}! \label{symmetryfactorsortof}
\end{align}
where $n_{i}$ and $N$ are defined for the list $(\vec{p}_{1},\vec{p}_{2},\cdots,\vec{p}_{n})$ in the same way as they are defined in the preceding paragraph. The identical particle kets defined in Eqs. \eqref{bosonicstate} and \eqref{fermionicstate} are fully symmetric and antisymmetric in their particle labeling respectively: that is, given a permutation $\pi$, they satisfy
\begin{align}
    \cket{\pi(\vec{p}_{1}\,,\vec{p}_{2}\,,\cdots\,,\vec{p}_{n})} &= \cket{\vec{p}_{1}\,,\vec{p}_{2}\,,\cdots\,,\vec{p}_{n}}\\
    \sket{\pi(\vec{p}_{1}\,,\vec{p}_{2}\,,\cdots\,,\vec{p}_{n})} &= \text{sign}(\pi)\, \sket{\vec{p}_{1}\,,\vec{p}_{2}\,,\cdots\,,\vec{p}_{n}}
\end{align}
Particles described by the multi-particle symmetrized kets ($\chi = +1$) are bosons and particles described by the multi-particle antisymmetrized kets ($\chi = -1$) are fermions \cite{Fierz:1939ix,PhysRev.58.716}. The antisymmetry of the latter kets is why we need not worry about repeated labels when normalizing that case; if any labels are repeated (e.g. two particles have identical quantum numbers, which at present means identical 3-momenta), then the ket will automatically vanish:
\begin{align}
    \sket{\vec{p}\,,\vec{p}\,,\vec{p}_{3}\,,\cdots\,,\vec{p}_{n}} = -\sket{\vec{p}\,,\vec{p}\,,\vec{p}_{3}\,,\cdots\,,\vec{p}_{n}}\hspace{15 pt}\implies\hspace{15 pt} \sket{\vec{p}\,,\vec{p}\,,\vec{p}_{3}\,,\cdots\,,\vec{p}_{n}} = 0
\end{align}
This is an expression of the Pauli exclusion principle \cite{PhysRev.58.716}, which states that identical fermions are forbidden from having fully identical quantum numbers.

We now address the normalizations of these identical particle states. For multi-particle states composed of a bosonic species,
\begin{align}
    \cbracket{\vec{p}_{1}\,,\vec{p}_{2}\,,\cdots\,,\vec{p}_{n}}{\vec{p}_{1}^{\,\prime}\,,\vec{p}_{2}^{\,\prime}\,,\cdots\,,\vec{p}_{n}^{\,\prime}} &= \dfrac{1}{n!} \Bigg[\prod_{i=1}^{N} \dfrac{1}{\sqrt{n_{i}!}}\Bigg] \Bigg[\prod_{j=1}^{N^{\prime}} \dfrac{1}{\sqrt{n^{\prime}_{j}!}}\Bigg]\nonumber\\
    &\hspace{35 pt}\times \sum_{\pi \in \pi_{n}} \sum_{\pi^{\prime} \in \pi_{n}} \braket{\pi(\vec{p}_{1},\vec{p}_{2},\cdots,\vec{p}_{n})}{\pi^{\prime}(\vec{p}_{1}^{\,\prime},\vec{p}_{2}^{\,\prime},\cdots,\vec{p}_{n}^{\,\prime})}
\end{align}
\begin{align}
    &= \dfrac{1}{n!} \Bigg[\prod_{i=1}^{N} \dfrac{1}{\sqrt{n_{i}!}}\Bigg] \Bigg[\prod_{j=1}^{N^{\prime}} \dfrac{1}{\sqrt{n^{\prime}_{j}!}}\Bigg]\, n! \sum_{\pi \in \pi_{n}} \braket{\pi(\vec{p}_{1},\vec{p}_{2},\cdots,\vec{p}_{n})}{\vec{p}_{1}^{\,\prime}\,,\vec{p}_{2}^{\,\prime}\,,\cdots\,,\vec{p}_{n}^{\,\prime}}\\
    &= \dfrac{1}{n!} \Bigg[\prod_{i=1}^{N} \dfrac{1}{\sqrt{n_{i}!}}\Bigg] \Bigg[\prod_{j=1}^{N^{\prime}} \dfrac{1}{\sqrt{n^{\prime}_{j}!}}\Bigg]\, n! \Bigg[\prod_{i=1}^{N} n_{i}!\Bigg] \sum_{\text{unique }\pi \in \pi_{n}} \braket{\pi(\vec{p}_{1},\vec{p}_{2},\cdots,\vec{p}_{n})}{\vec{p}_{1}^{\,\prime}\,,\vec{p}_{2}^{\,\prime}\,,\cdots\,,\vec{p}_{n}^{\,\prime}}\\
    &
    =\Bigg[\prod_{i=1}^{N} \sqrt{n_{i}!}\Bigg]\,\Bigg[\prod_{j=1}^{N^{\prime}}\dfrac{1}{\sqrt{n_{j}^{\prime}!}}\Bigg] \sum_{\text{unique }\pi \in \pi_{n}} \braket{\pi(\vec{p}_{1},\vec{p}_{2},\cdots,\vec{p}_{n})}{\vec{p}_{1}^{\,\prime}\,,\vec{p}_{2}^{\,\prime}\,,\cdots\,,\vec{p}_{n}^{\,\prime}}
\end{align}
where ``unique $\pi\in\pi_{n}$" means only summing over a subset of permutations $\pi$ that yield unique lists $\pi(\vec{p}_{1},\vec{p}_{2},\cdots,\vec{p}_{n})$. Consequently, if there is no permutation $\pi$ such that $\pi(\vec{p}_{1},\vec{p}_{2},\cdots,\vec{p}_{n}) = (\vec{p}_{1}^{\,\prime},\vec{p}_{2}^{\,\prime},\cdots,\vec{p}_{n}^{\,\prime})$, then the RHS vanishes. However, if such a permutation $\pi$ does exist, then $N=N^{\prime}$, $\{n_{i}\} = \{n_{j}^{\prime}\}$, and
\begin{align}
    \cbracket{\vec{p}_{1}\,,\vec{p}_{2}\,,\cdots\,,\vec{p}_{n}}{\vec{p}_{1}^{\,\prime}\,,\vec{p}_{2}^{\,\prime}\,,\cdots\,,\vec{p}_{n}^{\,\prime}} &= \prod_{i=1}^{n} (2\pi)^{3}\, (2 E_{\vec{p}_{i}})\, \delta^{3}(0)
\end{align}
Therefore, returning to the general case,
\begin{align}
    \cbracket{\vec{p}_{1}\,,\vec{p}_{2}\,,\cdots\,,\vec{p}_{n}}{\vec{p}_{1}^{\,\prime}\,,\vec{p}_{2}^{\,\prime}\,,\cdots\,,\vec{p}_{n}^{\,\prime}} &= \sum_{\text{unique }\pi \in \pi_{n}} \braket{\pi(\vec{p}_{1},\vec{p}_{2},\cdots,\vec{p}_{n})}{\vec{p}_{1}^{\,\prime}\,,\vec{p}_{2}^{\,\prime}\,,\cdots\,,\vec{p}_{n}^{\,\prime}} \label{bosonmultiparticlenormalization}
\end{align}
which is the normalization we would have obtained from distinguishable particles. For multi-particle states composed of a fermionic species, the procedure is similar, except that no labels in the bra nor ket may be repeated (otherwise they will vanish by antisymmetry, as remarked previously) such that all permutations automatically yield a unique ordering of labels. We must also be cautious of the parity of the permutations involved. After taking these facets into account, we ultimately find
\begin{align}
    \sbrasket{\vec{p}_{1}\,,\vec{p}_{2}\,,\cdots\,,\vec{p}_{n}}{\vec{p}_{1}^{\,\prime}\,,\vec{p}_{2}^{\,\prime}\,,\cdots\,,\vec{p}_{n}^{\,\prime}} &= \sum_{\pi \in \pi_{n}} \text{sign}(\pi)\, \braket{\pi(\vec{p}_{1},\vec{p}_{2},\cdots,\vec{p}_{n})}{\vec{p}_{1}^{\,\prime}\,,\vec{p}_{2}^{\,\prime}\,,\cdots\,,\vec{p}_{n}^{\,\prime}} \label{fermionmultiparticlenormalization}
\end{align}
which is again consistent with the normalization we would have obtained from an analogous assortment of distinguishable particles, aside from an overall phase factor (a potential multiplicative $-1$).

These normalizations imply corresponding resolutions of identity. Let us first consider the bosonic case. To avoid overcounting states, we use the symmetrization of the bosonic kets to arrange the $3$-momentum labels in some canonical ordering. The specific canonical ordering is unimportant at present, but one such choice is to rewrite all kets $\cket{\vec{p}_{1}\,,\cdots\,,\vec{p}_{n}}$ so that the 3-momentum are organized from smallest-to-largest in magnitude (with some additional criteria for breaking ties). Whatever the specific choice of canonical ordering, the resulting resolution of identity equals
\begin{align}
    \mathbbm{1} = \int_{\text{unique}} \prod_{i=1}^{n} \bigg[\dfrac{d^{3}p_{i}}{(2\pi)^{3}} \dfrac{1}{2E_{p_{i}}}\bigg] \hspace{5 pt} \cket{\vec{p}_{1}\,,\vec{p}_{2}\,,\cdots\,,\vec{p}_{n}}\cbra{\vec{p}_{1}\,,\vec{p}_{2}\,,\cdots\,,\vec{p}_{n}}
\end{align}
where the ``unique" label on the integral indicates that, for instance, if $(\vec{p}_{1},\vec{p}_{2},\cdots,\vec{p}_{n}) = (\vec{p}_{1}^{\,\prime},\vec{p}_{2}^{\,\prime},\cdots,\vec{p}_{n}^{\,\prime})$ is included in the integral, then no distinct permutation of $(\vec{p}_{1}^{\,\prime},\vec{p}_{2}^{\,\prime},\cdots,\vec{p}_{n}^{\,\prime})$ is also included in the integral. Although in principle this uniquely identifies the bosonic resolution of identity, we would like to rewrite it in a way that does not depend on a specific canonical ordering. To do so, suppose we lift the ``unique" label from the RHS of the previous equation so that we integrate over all $3$-momentum combinations (regardless if any are related via permutation) and act the resulting operator on a ket $\cket{\vec{k}_{1}\,,\vec{k}_{2}\,,\cdots\,,\vec{k}_{n}}$ where all $3$-momentum $\vec{k}_{i}$ are unique. Because $\cket{\vec{k}_{1}\,,\vec{k}_{2}\,,\cdots\,,\vec{k}_{n}}$ is symmetric in its labels, it will yield a nonzero result when projected onto any of the bras $\cbra{\vec{p}_{1}\,,\vec{p}_{2}\,,\cdots\,,\vec{p}_{n}}$ wherein $\pi(\vec{p}_{1}\,,\vec{p}_{2}\,,\cdots\,,\vec{p}_{n}) = (\vec{k}_{1}\,,\vec{k}_{2}\,,\cdots\,,\vec{k}_{n})$ for some permutation $\pi$. Because there are $n!$ such permutations,
\begin{align}
    &\left[\int \prod_{i=1}^{n} \bigg[\dfrac{d^{3}p_{i}}{(2\pi)^{3}} \dfrac{1}{2E_{p_{i}}}\bigg] \hspace{5 pt} \cket{\vec{p}_{1}\,,\cdots\,,\vec{p}_{n}}\cbra{\vec{p}_{1}\,,\cdots\,,\vec{p}_{n}}\right] \cket{\vec{k}_{1}\,,\cdots\,,\vec{k}_{n}}= n! \cket{\vec{k}_{1}\,,\cdots\,,\vec{k}_{n}}
\end{align}
Therefore, when acting on a ket wherein no set of quantum numbers is repeated,
\begin{align}
    \mathbbm{1} &= \dfrac{1}{n!} \int \prod_{i=1}^{n} \bigg[\dfrac{d^{3}p_{i}}{(2\pi)^{3}} \dfrac{1}{2E_{p_{i}}}\bigg] \hspace{5 pt} \cket{\vec{p}_{1}\,,\vec{p}_{2}\,,\cdots\,,\vec{p}_{n}}\cbra{\vec{p}_{1}\,,\vec{p}_{2}\,,\cdots\,,\vec{p}_{n}}
\end{align}
The above resolution of identity will not work on a state where there are repeated sets of quantum numbers, because the coincidence of those sets is not overcounted as much by the integral. For instance, if $\vec{p}_{1}\neq \vec{p}_{2}$ then the integral over all momentum would catch both $(\vec{p}_{1}\,,\vec{p}_{2})$ and $(\vec{p}_{2}\,,\vec{p}_{1})$ despite their equivalence as far as the corresponding symmetrized ket is concerned, whereas if $\vec{p}_{1} = \vec{p}_{2} = \vec{p}$ then only the single phase space point $(\vec{p}\,,\vec{p}\,)$ will contribute. Thus, repeated labels yield fewer than $n!$ contributing instances in the integral. When these considerations are generally applied, we obtain a resolution of identity on the whole space of symmetrized kets that does not rely on a specific canonical ordering:
\begin{align}
    \mathbbm{1} = \int \prod_{i=1}^{n} \bigg[\dfrac{d^{3}p_{i}}{(2\pi)^{3}} \dfrac{1}{2E_{p_{i}}}\bigg]\hspace{5 pt}\bigg[\dfrac{1}{n!} \mathcal{S}(\vec{p}_{1}\,,\cdots\,,\vec{p}_{n})\bigg]\, \cket{\vec{p}_{1}\,,\cdots\,,\vec{p}_{n}}\cbra{\vec{p}_{1}\,,\cdots\,,\vec{p}_{n}} \label{bosonicresolutionofidentityproto}
\end{align}
where $\mathcal{S}$ is defined as in Eq. \eqref{symmetryfactorsortof}. Furthermore, because we will always be acting the bosonic $n$-particle identity on bosonic $n$-particle states and
\begin{align}
    \cbracket{\vec{p}_{1}\,,\vec{p}_{2}\,,\cdots\,,\vec{p}_{n}}{\vec{k}_{1}\,,\vec{k}_{2}\,,\cdots\,,\vec{k}_{n}} = \bracket{\vec{p}_{1}\,,\vec{p}_{2}\,,\cdots\,,\vec{p}_{n}}{\vec{k}_{1}\,,\vec{k}_{2}\,,\cdots\,,\vec{k}_{n}}
\end{align}
(note the bra on the RHS is {\it not} symmetrized) we can replace the symmetrized states in Eq. \ref{bosonicresolutionofidentityproto} with distinguishable states. In doing so, we obtain our final result:
\begin{align}
    \mathbbm{1} = \int \prod_{i=1}^{n} \bigg[\dfrac{d^{3}p_{i}}{(2\pi)^{3}} \dfrac{1}{2E_{p_{i}}}\bigg]\hspace{5 pt}\bigg[\dfrac{1}{n!} \mathcal{S}(\vec{p}_{1}\,,\cdots\,,\vec{p}_{n})\bigg]\, \ket{\vec{p}_{1}\,,\cdots\,,\vec{p}_{n}}\bra{\vec{p}_{1}\,,\cdots\,,\vec{p}_{n}}
\end{align}
When expressed in this form, the bosonic resolution of identity only differs from the distinguishable resolution of identity Eq. \eqref{distinguishableresolutionofidentity} in its multiplicative $\mathcal{S}/n!$ factor. As a result, it is common practice to perform derivations in quantum field theory as if all the particles involved are distinguishable (e.g. without the factor of $\mathcal{S}/n!$) and then reintroduce the $\mathcal{S}/n!$ factor as necessary in closing. This occurs frequently when considering 2-to-2 scattering in the center-of-momentum frame. Because the particles in such a process have equal-and-opposite $3$-momentum (which must be nonzero in order to describe nontrivial scattering: $\vec{p}_{1} \neq \vec{p}_{2}$), each identical incoming or outgoing pair contributes a factor of $\mathcal{S}(\vec{p}_{1},\vec{p}_{2})/2! = 1/2$ relative to the equivalent integral involving distinguishable particles. Formulas throughout textbooks and the literature will often come with a caveat that an additional $1/2$ must be tacked on for each initial or final pair of identical bosons. This will be the case when we derive the elastic/inelastic unitarity constraints in Subsection \ref{SS - Elastic, Inelastic Unitarity Constraints}.

Although we will not need it in this dissertation, for completeness let us next consider the fermionic resolution of identity. Because a coincidence of particle labels causes antisymmetrized kets to vanish, the concerns regarding the repetition factor $\mathcal{S}$ do not carry over to the fermionic case. Thus, the fermionic resolution of identity expressed in terms of canonical momentum ordering is
\begin{align}
    \mathbbm{1} = \int_{\text{unique}} \prod_{i=1}^{n} \bigg[\dfrac{d^{3}p_{i}}{(2\pi)^{3}} \dfrac{1}{2E_{p_{i}}}\bigg] \hspace{5 pt} \sket{\vec{p}_{1}\,,\cdots\,,\vec{p}_{n}}\sbra{\vec{p}_{1}\,,\cdots\,,\vec{p}_{n}}
\end{align}
and generalizes to
\begin{align}
    \mathbbm{1} = \int \prod_{i=1}^{n} \bigg[\dfrac{d^{3}p_{i}}{(2\pi)^{3}} \dfrac{1}{2E_{p_{i}}}\bigg]\hspace{5 pt}\dfrac{1}{n!}\, \ket{\vec{p}_{1}\,,\cdots\,,\vec{p}_{n}}\bra{\vec{p}_{1}\,,\cdots\,,\vec{p}_{n}}
\end{align}
As mentioned following the derivation of the bosonic resolution of identity, derivations in quantum field theory are often performed while assuming all particles are distinguishable and any necessary factors due to identical particles are appended after the fact. In the fermionic case, that factor is $1/n!$, which again simplifies to $1/2$ for each identical fermion pair in 2-to-2 scattering processes.

\subsection{External States: General Quantum Numbers} \label{SS - General Quantum Numbers}
While the previous results were derived and motivated by considering $4$-momentum eigenstates, they readily generalize to kets labeled by other sets of quantum numbers. Suppose we have a complete set of single-particle kets $\ket{\alpha}$ that resolve the single-particle identity according to
\begin{align}
    \mathbbm{1} = \int d\Pi(\alpha)\hspace{5 pt}\ket{\alpha}\bra{\alpha}
\end{align}
where $\int d\Pi(\alpha)$ is in principle some combination of sums (for discrete quantum numbers), integrals (for continuous quantum numbers), and multiplicative weights, and with normalization
\begin{align}
    \braket{\alpha}{\alpha^{\prime}} = w(\alpha)\,\delta_{\alpha,\alpha^{\prime}}
\end{align}
where $\delta_{\alpha,\alpha^{\prime}}$ is a product of Kronecker deltas (for discrete quantum numbers) and Dirac deltas (for continuous quantum numbers). Together, these imply
\begin{align}
    \ket{\alpha^{\prime}} = \int d\Pi(\alpha)\hspace{5 pt}w(\alpha)\,\delta_{\alpha,\alpha^{\prime}}\, \ket{\alpha^{\prime}} \hspace{35 pt}\implies\hspace{35 pt} d\Pi(\alpha) = \dfrac{1}{w(\alpha)} d\alpha
\end{align}
where $d\alpha$ is the differential integration element of the continuous quantum numbers specified by $\ket{\alpha}$. For example, in the previous subsection, $\alpha = \vec{p}$, such that $w(\vec{p}\,) = (2\pi)^{3}(2E_{\vec{p}})$ and $d\alpha = d^{3}\vec{p}$. Because kets labeled by continuous quantum numbers have Dirac delta normalizations, wavepackets corresponding to those continuous quantum numbers must be utilized in practice (refer to the discussion at the end of Subsection \ref{External States - Single-Particle} for more details on this use of wavepackets). The construction of multi-particle states goes through without significant modification (e.g. two labels $\alpha$ and $\alpha^{\prime}$ are now considered repeated if all of the quantum numbers between them are equal), such that we define the distinguishable $n$-particle state as
\begin{align}
    \ket{\alpha_{1}\,,\cdots \,,\alpha_{n}} = \ket{\alpha_{1}}\otimes \cdots \otimes \ket{\alpha_{n}}
\end{align}
and the identical $n$-particle states as
\begin{align}
    \cket{\alpha_{1}\,,\cdots\,,\alpha_{n}} &= \dfrac{1}{\sqrt{n!\, \mathcal{S}(\alpha_{1}\,,\cdots\,,\alpha_{n})}} \sum_{\pi \in \pi_{n}} \ket{\pi(\alpha_{1}\,,\alpha_{2}\,,\cdots\,,\alpha_{n})}\\
    \sket{\alpha_{1}\,,\cdots\,,\alpha_{n}} &= \dfrac{1}{\sqrt{n!}}\, \sum_{\pi \in \pi_{n}}\text{sign}(\pi)\, \ket{\pi(\alpha_{1}\,,\alpha_{2}\,,\cdots\,,\alpha_{n})}
\end{align}
for bosons and fermions respectively. In that same order, the resolutions of identity for each of these spaces equal
\begin{align}
    \mathbbm{1} &= \int \prod_{i=1}^{n} d\Pi(\alpha_{i})\hspace{5 pt}\, \ket{\alpha_{1}\,,\cdots \,,\alpha_{n}}\bra{\alpha_{1}\,,\cdots\,, \alpha_{n}}\\
    \mathbbm{1} &= \int \prod_{i=1}^{n} d\Pi(\alpha_{i})\hspace{5 pt}\dfrac{1}{n!}\, \mathcal{S}(\alpha_{1}\,,\cdots\,,\alpha_{n})\, \ket{\alpha_{1}\,,\cdots \,,\alpha_{n}}\bra{\alpha_{1}\,,\cdots\,, \alpha_{n}}\\
    \mathbbm{1} &= \int \prod_{i=1}^{n} d\Pi(\alpha_{i})\hspace{5 pt}\dfrac{1}{n!} \, \ket{\alpha_{1}\,,\cdots\,, \alpha_{n}}\bra{\alpha_{1}\,,\cdots\,, \alpha_{n}}
\end{align}
and the kets have normalizations
\begin{align}
    \braket{\alpha_{1}\,,\cdots\,,\alpha_{n}}{\alpha_{1}^{\prime}\,,\cdots\,,\alpha_{n}^{\prime}} &= \prod_{i=1}^{n} w(\alpha_{i})\, \delta_{\alpha_{i},\alpha_{i}^{\prime}}\\
    \cbracket{\alpha_{1}\,,\cdots\,,\alpha_{n}}{\alpha_{1}^{\prime}\,,\cdots\,,\alpha_{n}^{\prime}} &= \sum_{\text{unique }\pi \in \pi_{n}} \braket{\pi(\alpha_{1},\cdots,\alpha_{n})}{\alpha_{1}^{\prime}\,,\cdots\,,\alpha_{n}^{\prime}}\\
    \sbrasket{\alpha_{1}\,,\cdots\,,\alpha_{n}}{\alpha_{1}^{\prime}\,,\cdots\,,\alpha_{n}^{\prime}} &= \sum_{\pi \in \pi_{n}} \text{sign}(\pi)\, \braket{\pi(\alpha_{1}\,,\cdots\,,\alpha_{n})}{\alpha_{1}^{\prime}\,,\cdots\,,\alpha_{n}^{\prime}}
\end{align}
where $\mathcal{S}$ is defined as in Eq. \eqref{symmetryfactorsortof}. These general results will become relevant as we consider maximally-commuting sets of observables and thereby introduce more quantum numbers to our state labels. Note the fermionic states still obey the Pauli exclusion principle ($\sbra{\alpha\,,\alpha\,,\cdots} = 0$). Also note the rule of thumb that an extra factor of $1/2$ should be included per identical particle pair in a 2-to-2 COM scattering calculation that was otherwise performed with distinguishable particles carries over to these more general descriptions as well.

\subsection{S-Matrix, Matrix Element}
\label{SS - S-Matrix}
We can use the multi-particle states defined in the previous subsection as our initial and final states in scattering processes. The collection of all states regardless of differing particle numbers and particle species content yields a Fock space, which equals the direct sum of the zero-particle, single-particle, two-particle, etc. Hilbert spaces. Scattering processes are modeled as beginning in the infinite past (at time $t = -\infty$) and ending in the infinite future (at time $t = +\infty$) with the interesting dynamics occurring near $t = 0$. A Fock space state set up in the infinite past is called an ``in state", whereas a Fock space state set up in the infinite future is called an ``out state." We can evolve an in state to an analogous out state via a generalization of the time-evolution operator $\hat{S}$ called the $S$-matrix:
\begin{align}
    \hat{S}\ket{i}_{\text{in}} &= \ket{i}_{\text{out}}
\end{align}
from which the probability that an initial particle scattering state $\ket{i}_{\text{in}}$ becomes a final particle scattering state $\ket{f}_{\text{out}}$ can be calculated via
\begin{align}
     {}_{\text{out}}\braket{f}{i}_{\text{out}} &=  {}_{\text{out}}\bra{f}\hat{S}\ket{i}_{\text{in}}
\end{align}
The $S$-matrix $\hat{S}$ by construction commutes with $P^{\mu}$ and $\vec{J}$ because of its relation to the time-evolution operator. Because our in and out states will always have definite total $4$-momentum, this means we will always generate a total $4$-momentum conserving Dirac delta function when calculating ${}_{\text{out}}\braket{f}{i}_{\text{out}}$, which we can preemptively factor out:
\begin{align}
     {}_{\text{out}}\braket{f}{i}_{\text{out}} &=  {}_{\text{out}}\bra{f}\hat{S}\ket{i}_{\text{in}} = {}_{\text{out}}\braket{f}{i}_{\text{in}} + i (2\pi)^{4}\delta^{4}(p_{f}-p_{i})\, {}_{\text{out}}\bra{f}\hat{T}(p_{i}=p_{f})\ket{i}_{\text{in}} \label{71220x}
\end{align}
where the argument ``$p_{i} = p_{f}$" reminds us that the newly-defined $T$-matrix $\hat{T}$ has already had a total momentum conserving Dirac delta removed. Relative to the $T$-matrix, the (Lorentz-invariant) matrix element equals
\begin{align}
    \mathcal{M}_{i\rightarrow f} \equiv \,{}_{\text{out}}\bra{f}\hat{T}(p_{i}=p_{f})\ket{i}_{\text{in}} \label{MEdef}
\end{align}
The square of a matrix element $\mathcal{M}_{i\rightarrow f}$ is related to the probability that a given scattering process $i\rightarrow f$ will occur, and is a central topic of this dissertation. A matrix element is also sometimes called a scattering amplitude.

Before moving on to a general discussion of 2-to-2 scattering in the next section, let us consider the energy units of a matrix element (keeping in mind that we use natural units, i.e. $c=\hbar = 1$). Its units will depend on the units of our out states, which are in turn determined through the out state normalization. For example, we previously described single-particle kets $\ket{\vec{p}\,}$ normalized via $\braket{\vec{p}\,}{\vec{p}^{\,\prime}} = (2\pi)^{3}\, (2E_{\vec{p}})\, \delta^{3}(\vec{p}-\vec{p}^{\,\prime})$, such that the inner product $\braket{\vec{p}\,}{\vec{p}^{\,\prime}}$ has units of $(\text{Energy})^{-2}$ and the single-particle ket $\ket{\vec{p}\,}$ has units of $(\text{Energy})^{-1}$. This means the units of an $n$-particle ket $\ket{\vec{p}_{1},\,\vec{p}_{2}\,,\cdots\,,\vec{p}_{n}} \equiv \ket{\vec{p}_{1}}\otimes \ket{\vec{p}_{2}}\otimes\cdots\otimes\ket{\vec{p}_{n}}$ are $(\text{Energy})^{-n}$ and depend on the number of particles considered. Thus, by using such kets to describe our out states, the inner product ${}_{\text{out}}\braket{f}{i}_{\text{out}}$ corresponding to an $n_{i}$-to-$n_{f}$ scattering process will have units $(\text{Energy})^{-(n_{i} + n_{f})}$. Therefore, in order to be consistent with Eq. \eqref{71220x}, the matrix element $\mathcal{M}_{i\rightarrow f}$ must have units $(\text{Energy})^{4-(n_{i} + n_{f})}$. Although our external state kets will ultimately have a helicity quantum number in addition to well-defined 4-momentum, this will not change the units of our out states, and so this unit argument also carries through there. Note that, in particular, a 2-to-2 scattering matrix element ($n_{i} = n_{f} = 2$) is dimensionless.

\section{2-to-2 Scattering} \label{S - 2to2}
This dissertation is largely concerned with 2-to-2 scattering processes, so it is important that we establish a consistent choice of conventions relating to those processes. Subsection \ref{SS - 2to2 - Mandelstam} describes our parameterization of 2-to-2 scattering processes in terms of the Mandelstam variables $s$, $t$, and $u$. Subsection \ref{SS - 2to2 - COM Frame} defines the center-of-momentum (COM) frame and (in this frame) rewrites the aforementioned $t$ and $u$ in terms of $s$ and the outgoing scattering angles $\theta$, $\phi$. Subsection \ref{SS - 2to2 - Integral} describes how to reduce a generic Lorentz-invariant integral over the final state particle pair degrees of freedom into a standard angular integral in the COM frame.

\subsection{Mandelstam Variables} \label{SS - 2to2 - Mandelstam}
A 2-to-2 scattering process refers to the evolution of a two-particle state in the infinite past into a two-particle state in the infinite future. For the time being, we will label the particles in the incoming pair as $1$ and $2$, and the particles in the outgoing pair as $3$ and $4$. The initial and final two-particle states can be defined by various quantum numbers. For the duration of this dissertation, we will choose each external single-particle state to have definite 4-momentum $p_{i}$ and helicity $\lambda_{i}$. The discussion of helicity is delayed until Section \ref{S - Helicity}. By definition, an external particle with $4$-momentum $p_{i}$ has mass $m_{i} = \sqrt{p_{i}^{2}}$.

Diagrammatically, we express the aforementioned generic 2-to-2 scattering process by:
\begin{center}
    \begin{tikzpicture}
    \begin{feynman}[medium]
        \vertex[blob] (c) at (0, 0) {};
        \vertex (n1) at (-1.5,1.5) {$1$};
        \vertex (a1) at (-1.7,1.2) {};
        \vertex (b1) at (-0.75,0.25) {};
        
        \vertex (n2) at (-1.5,-1.5) {$2$};
        \vertex (a2) at (-1.7,-1.2) {};
        \vertex (b2) at (-0.75,-0.25) {};
        
        \vertex (n3) at (1.5,1.5) {$3$};
        \vertex (a3) at (0.75,0.25) {};
        \vertex (b3) at (1.7,1.2) {};
        
        \vertex (n4) at (1.5,-1.5) {$4$};
        \vertex (a4) at (0.75,-0.25) {};
        \vertex (b4) at (1.7,-1.2) {};
        
        \diagram* {
            (n1) -- [photon] (c) -- [photon] (n3),
            (n2) -- [photon] (c) -- [photon] (n4),
            (n1) -- [] (c) -- [] (n3),
            (n2) -- [] (c) -- [] (n4),
            (a1) -- [arrow, edge label' = $p_{1}$] (b1),
            (a2) -- [arrow, edge label = $p_{2}$] (b2),
            (a3) -- [arrow, edge label' = $p_{3}$] (b3),
            (a4) -- [arrow, edge label = $p_{4}$] (b4)
        };
    \end{feynman}
    \end{tikzpicture}
\end{center}
which is intended to be read from left to right, where 4-momentum conservation guarantees
\begin{align}
    p_{1} + p_{2} = p_{3} + p_{4}
\end{align}
and arrows indicate the flow of 4-momentum through the diagram. A 2-to-2 scattering process can often occur in a variety of ways via a variety of interactions. For example, depending on the details of the field theory describing this scattering process, the $(1,2)$ pair might be able to directly become a $(3,4)$ pair through a local quartic interaction. We call a diagram corresponding to this specific subprocess a contact diagram:
\begin{center}
    \begin{samepage}
    \begin{tikzpicture}
    \begin{feynman}[medium]
        \vertex[dot] (c) at (0, 0) {};
        \vertex (n1) at (-1.5,1.5) {$1$};
        \vertex (a1) at (-1.7,1.2) {};
        \vertex (b1) at (-0.75,0.25) {};
        
        \vertex (n2) at (-1.5,-1.5) {$2$};
        \vertex (a2) at (-1.7,-1.2) {};
        \vertex (b2) at (-0.75,-0.25) {};
        
        \vertex (n3) at (1.5,1.5) {$3$};
        \vertex (a3) at (0.75,0.25) {};
        \vertex (b3) at (1.7,1.2) {};
        
        \vertex (n4) at (1.5,-1.5) {$4$};
        \vertex (a4) at (0.75,-0.25) {};
        \vertex (b4) at (1.7,-1.2) {};
        
        \diagram* {
            (n1) -- [photon] (n4),
            (n2) -- [photon] (n3),
            (a1) -- [arrow, edge label' = $p_{1}$] (b1),
            (a2) -- [arrow, edge label = $p_{2}$] (b2),
            (a3) -- [arrow, edge label' = $p_{3}$] (b3),
            (a4) -- [arrow, edge label = $p_{4}$] (b4)
        };
    \end{feynman}
    \end{tikzpicture}\\*
    contact
    \end{samepage}
\end{center}
Furthermore, if the appropriate cubic interactions are present, then this 2-to-2 scattering process is also facilitated by various channels of virtual particle exchange, i.e.
\begin{center}
\begin{tabular}{c @{\hskip 35 pt} c @{\hskip 35 pt} c}
    \begin{tikzpicture}
    \begin{feynman}[medium]
        \vertex[dot] (c1) at (-0.5, 0) {};
        \vertex[dot] (c2) at (0.5, 0) {};
        \vertex (ac1) at (-0.5, 0.25) {};
        \vertex (ac2) at (0.5, 0.25) {};
        
        \vertex (n1) at (-1.5,1.5) {$1$};
        \vertex (a1) at (-1.7,1.2) {};
        \vertex (b1) at (-1.05,0.25) {};
        
        \vertex (n2) at (-1.5,-1.5) {$2$};
        \vertex (a2) at (-1.7,-1.2) {};
        \vertex (b2) at (-1.05,-0.25) {};
        
        \vertex (n3) at (1.5,1.5) {$3$};
        \vertex (a3) at (1.05,0.25) {};
        \vertex (b3) at (1.7,1.2) {};
        
        \vertex (n4) at (1.5,-1.5) {$4$};
        \vertex (a4) at (1.05,-0.25) {};
        \vertex (b4) at (1.7,-1.2) {};
        
        \diagram* {
            (n1) -- [photon] (c1),
            (n2) -- [photon] (c1),
            (n3) -- [photon] (c2),
            (n4) -- [photon] (c2),
            (c1) -- [photon, edge label' = $5$] (c2),
            (a1) -- [arrow, edge label' = $p_{1}$] (b1),
            (a2) -- [arrow, edge label = $p_{2}$] (b2),
            (a3) -- [arrow, edge label' = $p_{3}$] (b3),
            (a4) -- [arrow, edge label = $p_{4}$] (b4),
            (ac1) -- [arrow, edge label = $p_{s}$] (ac2)
        };
    \end{feynman}
    \end{tikzpicture}
    &
    \begin{tikzpicture}
    \begin{feynman}[medium]
        \vertex[dot] (c1) at (0, 0.5) {};
        \vertex[dot] (c2) at (0, -0.5) {};
        \vertex (ac1) at (-0.25, 0.5) {};
        \vertex (ac2) at (-0.25, -0.5) {};
        
        \vertex (n1) at (-1.5,1.5) {$1$};
        \vertex (a1) at (-1.7,1.2) {};
        \vertex (b1) at (-0.75,0.55) {};
        
        \vertex (n2) at (-1.5,-1.5) {$2$};
        \vertex (a2) at (-1.7,-1.2) {};
        \vertex (b2) at (-0.75,-0.55) {};
        
        \vertex (n3) at (1.5,1.5) {$3$};
        \vertex (a3) at (0.75,0.55) {};
        \vertex (b3) at (1.7,1.2) {};
        
        \vertex (n4) at (1.5,-1.5) {$4$};
        \vertex (a4) at (0.75,-0.55) {};
        \vertex (b4) at (1.7,-1.2) {};
        
        \diagram* {
            (n1) -- [photon] (c1),
            (n2) -- [photon] (c2),
            (n3) -- [photon] (c1),
            (n4) -- [photon] (c2),
            (c1) -- [photon, edge label = $5$] (c2),
            (a1) -- [arrow, edge label' = $p_{1}$] (b1),
            (a2) -- [arrow, edge label = $p_{2}$] (b2),
            (a3) -- [arrow, edge label' = $p_{3}$] (b3),
            (a4) -- [arrow, edge label = $p_{4}$] (b4),
            (ac1) -- [arrow, edge label' = $p_{t}$] (ac2)
        };
    \end{feynman}
    \end{tikzpicture}
    &
    \begin{tikzpicture}
    \begin{feynman}[medium]
        \vertex[dot] (c1) at (0, 0.5) {};
        \vertex[dot] (c2) at (0, -0.5) {};
        \vertex (ac1) at (-0.5, 0.5) {};
        \vertex (ac2) at (-0.5, -0.5) {};
        
        \vertex (n1) at (-1.5,1.5) {$1$};
        \vertex (a1) at (-1.7,1.2) {};
        \vertex (b1) at (-0.75,0.55) {};
        
        \vertex (n2) at (-1.5,-1.5) {$2$};
        \vertex (a2) at (-1.7,-1.2) {};
        \vertex (b2) at (-0.75,-0.55) {};
        
        \vertex (n3) at (1.5,1.5) {$3$};
        \vertex (a3) at (1.05,0.25) {};
        \vertex (b3) at (1.7,1.2) {};
        
        \vertex (n4) at (1.5,-1.5) {$4$};
        \vertex (a4) at (1.05,-0.25) {};
        \vertex (b4) at (1.7,-1.2) {};
        
        \diagram* {
            (n1) -- [photon] (c1),
            (n2) -- [photon] (c2),
            (n3) -- [photon] (c2),
            (n4) -- [photon] (c1),
            (c1) -- [photon, edge label' = $5$] (c2),
            (a1) -- [arrow, edge label' = $p_{1}$] (b1),
            (a2) -- [arrow, edge label = $p_{2}$] (b2),
            (a3) -- [arrow, edge label' = $p_{3}$] (b3),
            (a4) -- [arrow, edge label = $p_{4}$] (b4),
            (ac1) -- [arrow, edge label' = $p_{u}$] (ac2)
        };
    \end{feynman}
    \end{tikzpicture}\\
    s-channel & t-channel & u-channel
\end{tabular}
\end{center}
where $5$ denotes the virtual particle being exchanged in each diagram. $4$-momentum is conserved at each vertex, such that $p_{s} = p_{1} + p_{2}$, and $p_{1} = p_{t} + p_{3}$, and so-on. These diagrams are the motivation for the Mandelstam variables \cite{PhysRev.112.1344}, which are defined as follows:
\begin{align}
    s &\equiv p_{s}^{2} = (p_{1} + p_{2})^{2} = (p_{3} + p_{4})^{2}\\
    t &\equiv p_{t}^{2} = (p_{1} - p_{3})^{2} = (p_{4} - p_{2})^{2}\\
    u &\equiv p_{u}^{2} = (p_{1} - p_{4})^{2} = (p_{3} - p_{2})^{2}
\end{align}
Note that $s$ ($t$; $u$) is the invariant momentum-squared that flows through the virtual particle in an $s$-channel ($t$-channel; $u$-channel) exchange diagram. Although the Mandelstam variables are motivated by these exchange diagrams, we may express any $2$-to-$2$ scattering process in terms of $s$, $t$, and $u$. Indeed, we will be using $s$ as a convenient variable to track energy growth for all kinds of diagrams.

Mandelstam $s$, $t$, and $u$ are not independent variables. For example, their sum is constrained: through direct evaluation, we find
\begin{align}
    s + t + u &= (p_{1} + p_{2})^{2} + (p_{1} - p_{3})^{2} + (p_{1} - p_{4})^{2}\\
    &= p_{1}^{2} + p_{2}^{2} + p_{3}^{2} + p_{4}^{2} + \underbrace{2p_{1}\cdot (p_{1} + p_{2} - p_{3} - p_{4})}_{=0\text{ by 4-momentum conservation}}
\end{align}
such that
\begin{align}
    s + t + u &= \sum_{i=1}^{4} m_{i}^{2}
\end{align}
Furthermore, the Mandelstam variables are real-valued with restricted range when describing experimentally-allowed processes. Mandelstam $s$, for example, is never smaller than
\begin{align}
    s_{\text{min}} \equiv \text{max}\left[(m_{1} + m_{2})^{2},(m_{3} + m_{4})^{2} \right]
\end{align}
which corresponds to both particles of either the initial or final particle pair being at rest, depending on which pair is more massive overall (because of $4$-momentum conservation, heavier particles at rest can become lighter particles in motion, but lighter particles at rest cannot become heavier particles).  Consequently, Mandelstam $s$ only vanishes when all external particles are massless and the $3$-momenta between the particles in each pair are parallel. Because parallel massless wavepackets will never collide, $s$ will never vanish for nontrivial scattering processes.

Until now, our discussion has been frame independent. Let us now consider a special frame that is particularly useful for simplifying scattering calculations: the center-of-momentum frame.

\subsection{Center-Of-Momentum Frame} \label{SS - 2to2 - COM Frame}
As remarked in the previous subsection, $s = (p_{1} + p_{2})^{2}$ is nonzero for any nontrivial 2-to-2 scattering process. Like a massive single-particle state with positive squared 4-momentum, such a process possesses a rest frame, wherein the particle pair's total 3-momentum vanishes: $\vec{p}_{1}+\vec{p}_{2} = \vec{0}$. This property (in addition to some coordinate choices we detail shortly) defines the center-of-momentum (COM) frame. So long as $s>0$, we may always use some combinations of boosts and rotations to enter the COM frame. For example, we only need an appropriately-chosen boost to ensure the total 3-momentum of the system vanishes, or in other words that the incoming particles have equal-and-opposite $3$-momenta:
\begin{align}
    \vec{p}_{1} + \vec{p}_{2} = \vec{0}
\end{align}
which (via 4-momentum conservation) implies the outgoing particles have equal-and-opposite 3-momenta as well:
\begin{align}
    \vec{p}_{3} + \vec{p}_{4} = \vec{0}
\end{align}
Geometrically, this means that in the COM frame the 3-momentum of the incoming particle pair lie on a common line through the origin and the 3-momentum of the final particle pair lie on another. Furthermore, this boost uniquely determines the 3-momentum magnitudes of the external particles: namely,
\begin{align}
    |\vec{p}_{1}| &= |\vec{p}_{2}| = \mathbbm{P}(1,2)\hspace{35 pt}|\vec{p}_{3}| = |\vec{p}_{4}| = \mathbbm{P}(3,4) \label{COMp1p2p3p4}
\end{align}
where
\begin{align}
    \mathbbm{P}(i,j) &= \sqrt{\dfrac{1}{4s}\bigg[s-(m_{i}-m_{j})^{2}\bigg]\,\bigg[s-(m_{i}+m_{j})^{2}\bigg]} \label{Pij}
\end{align}
Next, we can use a rotation to orient the $3$-momentum of particle $1$ in the $\hat{z}$ direction (or, equivalently, we can define the $\hat{z}$ direction of our coordinate system such that it follows $\vec{p}_{1}$ so long as $|\vec{p}_{1}|$ is nonzero), such that
\begin{align}
    p_{1} &= E_{1}\,\hat{t} + |\vec{p}_{1}|\, \hat{z}\\
    p_{2} &= E_{2}\,\hat{t} - |\vec{p}_{1}|\, \hat{z}
\end{align}
and
\begin{align}
    p_{3} &= E_{3}\,\hat{t} + |\vec{p}_{3}|\, \hat{p}_{3}\\
    p_{4} &= E_{4}\,\hat{t} - |\vec{p}_{3}|\, \hat{p}_{3}
\end{align}
where the basis $4$-vectors were defined at the end of Section \eqref{Classical - Minkowski}. This completes our definition of the COM frame. We choose to express $\hat{p}_{3}$ in spherical coordinates with respect to $\hat{z}$ in the usual way, such that $[\hat{p}_{3}^{\,\mu}] = (0,c_{\theta}s_{\phi},s_{\theta}s_{\phi},c_{\phi})$. We remind the reader that all of the external energies are restricted by the on-shell condition $m_{i}^{2} = p_{i}^{2} = E_{i}^{2} - |\vec{p}_{i}|^{2}$, such that (via Eq. \eqref{COMp1p2p3p4}) all external 4-momenta can be expressed in terms of the $s$, $\theta$, $\phi$, and the particle masses.

Because the 3-momenta of the incoming particles $1$ and $2$ are equal-and-opposite in the COM frame, Mandelstam $s$ reduces to the square of the total incoming energy, which we denote $E_{\text{COM}}$:
\begin{align}
    s = (p_{1} + p_{2})^{2} = (E_{1} + E_{2})^{2} \equiv E_{\text{COM}}^{2}
\end{align}
When context makes ambiguity unlikely (i.e. it is apparent that we are not referring to a single-particle energy), we will drop the label from $E_{\text{COM}}$ and simply write $s = E^{2}$.

Like the external 4-momenta, we can express the Mandelstam variables $t$ and $u$ in terms of $s$, $\theta$, and $\phi$. To do so succinctly, it is useful to define
\begin{align}
    \mathbbm{P}(i,j,k,l) &= \sqrt{\dfrac{1}{4s}\bigg[s^{2} - (m_{k}^{2} + m_{l}^{2} + m_{m}^{2} + m_{n}^{2})s + (m_{k}^{2} - m_{l}^{2})(m_{m}^{2}-m_{n}^{2})\bigg]}
\end{align}
where the previously-defined $\mathbbm{P}(i,j)$ equals $\mathbbm{P}(i,j,i,j)$. Then the Mandelstam variables equal
\begin{align}
    t(s,\theta) &= 2\bigg[-\mathbbm{P}(1,2,3,4)^{2} + \cos(\theta)\, \mathbbm{P}(1,2)\cdot\mathbbm{P}(3,4)\bigg]\\
    u(s,\theta) &= 2\bigg[-\mathbbm{P}(1,2,4,3)^{2} - \cos(\theta)\, \mathbbm{P}(1,2)\cdot\mathbbm{P}(3,4)\bigg]
\end{align}
Note these are all independent of $\phi$, which cancels out despite its presence in $p_{3}$ and $p_{4}$.

For future use in elastic processes, it is useful to define one last simplification of $\mathbbm{P}(i,j,k,l)$:
\begin{align}
    \mathbbm{P}(i) &= \mathbbm{P}(i,i,i,i) = \dfrac{1}{2}\sqrt{s- 4m_{i}^{2}}
\end{align}
For example, in elastic scattering (where all external particles are of identical particle species, say, $1$),
\begin{align}
    \left.t(s,\theta)\right|_{\text{elastic}} = 2\mathbbm{P}(1)^{2} \bigg[-1 + \cos(\theta)\bigg] &= -\dfrac{1}{2}(s-4m_{1}^{2}) [1 - \cos(\theta)]\\
    \left.u(s,\theta)\right|_{\text{elastic}} = 2\mathbbm{P}(1)^{2} \bigg[-1 - \cos(\theta)\bigg] &= -\dfrac{1}{2}(s-4m_{1}^{2}) [1 + \cos(\theta)]
\end{align}
Before discussing the quantum theory of 2-to-2 scattering, there is one more result we require. This subsection demonstrated that once an incoming energy $E_{\text{COM}} = \sqrt{s}$ is set, the only remaining degrees of freedom (ignoring internal degrees of freedom like helicity) correspond to the outgoing angles $\theta$ and $\phi$. To derive the optical theorem (in Subsection \ref{SS - Generalized Optical Theorem}) in a form that then allows us to derive the partial wave elastic/inelastic unitarity constraints (in Subsection \ref{SS - Elastic, Inelastic Unitarity Constraints}), we would like to rewrite a 2-particle Lorentz invariant integral in terms of the remaining variables $\theta$ and $\phi$. This is the subject of the next subsection.

\subsection{2-Particle Lorentz Invariant Integrals in the COM Frame} \label{SS - 2to2 - Integral}
There are several occasions when an integral over a final state particle pair is necessary. For example, such an integral is required when we calculate the total cross-section for a given 2-to-2 scattering process and are uninterested in the specific outgoing angle of the final pair. This kind of integral also occurs when deriving the partial wave elastic/inelastic unitarity constraints, which are important for this dissertation.

For the 2-to-2 scattering process $(1 , 2) \rightarrow (3 , 4)$, an outgoing particle pair integral is typically written as
\begin{align}
    \mathfrak{F} \equiv \int \underbrace{\bigg[\dfrac{d^{3} p_{3}}{(2\pi)^{3}} \dfrac{1}{2E_{3}}\bigg]\, \bigg[\dfrac{d^{3} p_{4}}{(2\pi)^{3}} \dfrac{1}{2E_{4}}\bigg]\, \bigg[(2\pi)^{4}\delta^{4}(p_{1} + p_{2} - p_{3} - p_{4})\bigg]}_{\text{2-Particle Lorentz-Invariant Phase Space}}\, F(p_{3},p_{4})
\end{align}
independent of frame, where $F$ is a generic function of the final particle 4-momenta. We aim to use the four Dirac deltas present to eliminate four of the six integration parameters and thereby rewrite $\mathfrak{F}$ as a two-dimensional integral. In particular, we perform this integral in the COM frame, and so the goal is to have those final two integration parameters be $\theta$ and $\phi$, which describe the direction of $\hat{p}_{3}$ relative to $\hat{p}_{1} = \hat{z}$.

In the COM frame, $p_{1} = (E_{1},\vec{p}_{1})$ and $p_{2} = (E_{2},-\vec{p}_{1})$, and the Dirac delta becomes
\begin{align}
    \delta^{4}(p_{1}+p_{2} - p_{3} - p_{4}) = \delta(E_{\text{COM}} - E_{3} - E_{4}) \, \delta^{3}(\vec{p}_{3} + \vec{p}_{4})
\end{align}
where $E_{\text{COM}} = E_{1} + E_{2}$. The $3$-vector Dirac delta $\delta^{3}(\vec{p}_{3}+\vec{p}_{4})$ allows us to immediately eliminate the $d^{3}p_{4}$ integral by constraining $\vec{p}_{4} = -\vec{p}_{3}$, such that we may write
\begin{align}
    \mathfrak{F} = \dfrac{1}{16\pi^{2}} \int \dfrac{d^{3}p_{3}}{E_{3}E_{4}} \, \delta(E_{\text{COM}} - E_{3} - E_{4}) \, F(p_{3},p_{4})\bigg|_{\vec{p}_{4} = -\vec{p}_{3}}
\end{align}
Meanwhile, the integration measure $d^{3}p_{3}$ is expressible in spherical coordinates like so
\begin{align}
    d^{3}p_{3} = |\vec{p}_{3}|^{2} \, d|\vec{p}_{3}|\, d\Omega = \dfrac{1}{2}\,|\vec{p}_{3}|\, d|\vec{p}_{3}|^{2} \, d\Omega
\end{align}
where $d\Omega = d\cos\theta \, d\phi$ contains the integration variables we wish to retain. Therefore, we want to use the final Dirac delta $\delta(E_{\text{COM}} - E_{3} - E_{4})$ remaining in $\mathfrak{F}$ to eliminate the $d|\vec{p}_{3}|^{2}$ integral. To do so, we must reparameterize the Dirac delta using the following property:
\begin{align}
    \delta(f(x)) = \sum_{x_{*}\text{ s.t. }f(x_{*}) = 0} \dfrac{\delta(x-x_{*})}{|f^{\prime}(x_{*})|}
\end{align}
which sums over zeroes of $f(x)$. As mentioned in the previous section, 4-momentum conservation is satisfied (and thus $E_{\text{COM}} = E_{3} + E_{4}$) precisely when $|\vec{p}_{3}| = \mathbbm{P}(3,4)$. Furthermore, using the existing $\vec{p}_{4} = -\vec{p}_{3}$ constraint,
\begin{align}
    \dfrac{\partial}{\partial |\vec{p}_{3}|^{2}} \bigg[E_{\text{COM}} - E_{3} - E_{4}\bigg] &= \dfrac{\partial}{\partial |\vec{p}_{3}|^{2}} \bigg[E_{\text{COM}} - \sqrt{m_{3}^{2} + |\vec{p}_{3}|^{2}} - \sqrt{m_{4}^{2} + |\vec{p}_{3}|^{2}}\bigg]\\
    &= -\dfrac{1}{2}\left[\dfrac{1}{\sqrt{m_{3}^{2} + |\vec{p}_{3}|^{2}}} + \dfrac{1}{\sqrt{m_{4}^{2} + |\vec{p}_{3}|^{2}}} \right]\\
    &= -\dfrac{1}{2}\dfrac{E_{3} + E_{4}}{E_{3} E_{4}}
\end{align}
Hence, utilizing the fact that the Dirac delta vanishes whenever $E_{\text{COM}} \neq E_{3} + E_{4}$,
\begin{align}
    \delta(E_{\text{COM}} - E_{3} - E_{4}) = \dfrac{2E_{3}E_{4}}{E_{\text{COM}}} \, \delta\bigg(|\vec{p}_{3}|^{2} - \mathbbm{P}(3,4)^{2}\bigg)
\end{align}
and, thus,
\begin{align}
    \mathfrak{F} = \dfrac{\mathbbm{P}(3,4)}{16\pi^{2} E_{\text{COM}}} \int d\Omega\hspace{10 pt}F(p_{3},p_{4})\bigg|_{\vec{p}_{3} = \mathbbm{P}(3,4) \hat{p}_{3} = -\vec{p}_{4}}
\end{align}
where $\mathbbm{P}(3,4)$ is defined in Eq. \ref{Pij}. This is the desired result.

\subsection{The Optical Theorem} \label{SS - Generalized Optical Theorem}
The $S$-matrix (defined in Subsection \ref{SS - S-Matrix}) is a unitary operator on Fock space that encodes how initial particle configurations evolve into final state particle configurations. Because the $S$-matrix is unitary, $S$-matrix elements must satisfy
\begin{align}
    {}_{\text{in}}\braket{\overline{i}}{i}_{\text{in}} = {}_{\text{in}}\bra{\overline{i}}\hat{S}^{\dagger} \hat{S}\ket{i}_{\text{in}} &= \sum_{f} \int d\Pi(f)\hspace{10 pt} {}_{\text{in}}\bra{\overline{i}}\hat{S}^{\dagger}\ket{f}_{\text{out}}\,\,\, {}_{\text{out}}\bra{f}\hat{S}\ket{i}_{\text{in}}\\
    &= \sum_{f} \int d\Pi(f)\hspace{10 pt} {}_{\text{out}}\bra{f}\hat{S}\ket{\overline{i}}^{*}_{\text{in}}\,\,\, {}_{\text{out}}\bra{f}\hat{S}\ket{i}_{\text{in}}\label{UnitarityOfSMatrix}
\end{align}
where we have inserted the Fock space resolution of identity and embedded the necessary state normalization weights into $d\Pi(f)$. We would like to recast this constraint in terms of the corresponding matrix elements $\mathcal{M}_{i\rightarrow f}$ and $\mathcal{M}_{\overline{i}\rightarrow f}$. To do so, suppose $p_{i} = p_{\overline{i}}$, and note
\begin{align}
    {}_{\text{out}}\bra{f}\hat{S}\ket{\overline{i}}^{*}_{\text{in}}\,\,\, {}_{\text{out}}\bra{f}\hat{S}\ket{i}_{\text{in}} &= \bigg[{}_{\text{out}}\braket{f}{\overline{i}}_{\text{in}}^{*} - i (2\pi)^{4}\delta^{4}(p_{\overline{i}}-p_{f})\, \mathcal{M}_{\overline{i}\rightarrow f}^{*}\bigg]\nonumber\\
    &\hspace{35 pt}\cdot\bigg[{}_{\text{out}}\braket{f}{i}_{\text{in}} + i (2\pi)^{4}\delta^{4}(p_{f}-p_{i})\, \mathcal{M}_{i\rightarrow f}\bigg]\\
    &\hspace{-70 pt}= {}_{\text{in}}\braket{\overline{i}}{f}_{\text{out}}\,\, {}_{\text{out}}\braket{f}{i}_{\text{in}} + i(2\pi)^{4} \delta^{4}(p_{f} - p_{i}) \bigg[\mathcal{M}_{i\rightarrow f}\,\, {}_{\text{out}}\braket{f}{\overline{i}}_{\text{in}} - \mathcal{M}^{*}_{\overline{i}\rightarrow f}\,\,{}_{\text{in}} \braket{i}{f}_{\text{out}}\bigg]\nonumber\\
    &\hspace{35 pt} + \bigg[(2\pi)^{4}\delta^{4}(p_{i}-p_{f})\bigg]^{2} \mathcal{M}^{*}_{\overline{i}\rightarrow f} \mathcal{M}_{i\rightarrow f}
\end{align}
The squared Dirac delta in the final term can be understood by considering a finite volume universe wherein the Dirac delta is replaced with a Kronecker delta; however, we simply use this expression as written in the RHS of Eq. \eqref{UnitarityOfSMatrix}, and eliminate one Dirac delta from the pair via $\sum_{f} \int d\Pi(f)$. (If we had not assumed $p_{i} = p_{\overline{i}}$ before now, the Dirac delta pair would have enforced their equality for this term.) In entirety, this substitution yields
\begin{align}
    -i \bigg[\mathcal{M}_{i\rightarrow \overline{i}} - \mathcal{M}^{*}_{\overline{i}\rightarrow i}\bigg] = \sum_{f} \int d\Pi(f)\hspace{10 pt}(2\pi)^{4}\delta^{4}(p_{i}-p_{f}) \mathcal{M}^{*}_{\overline{i}\rightarrow f} \mathcal{M}_{i\rightarrow f}
\end{align}
In particular, if $\overline{i} = i$ (and not just $p_{i} = p_{\overline{i}}$ as previously assumed), then
\begin{align}
    2\mathfrak{I}[\mathcal{M}_{i\rightarrow i}] = \sum_{f} \int d\Pi(f)\hspace{10 pt}(2\pi )^{4} \delta^{4}(p_{i}-p_{f}) |\mathcal{M}_{i\rightarrow f}|^{2} \label{GeneralizedOpticalTheorem}
\end{align}
where $\mathfrak{I}$ denotes the imaginary part of its argument ($\mathfrak{R}$ similarly denotes a real part). Eq. \eqref{GeneralizedOpticalTheorem} is the optical theorem, which says twice the imaginary part of the forward scattering amplitude $\mathcal{M}_{i\rightarrow i}$ strictly equals $\mathcal{M}_{i\rightarrow f}$ squared and summed over all possible final states. In other words, the contribution of any individual channel $|\mathcal{M}_{i\rightarrow f}|^{2}$ is bounded above by $\mathfrak{I}[\mathcal{M}_{i\rightarrow i}]$.

We are interested in applying the optical theorem to 2-to-2 scattering processes in the COM frame. To facilitate this application, first divide the sum over processes on the RHS of Eq. \eqref{GeneralizedOpticalTheorem} into two groups: $n$-to-2 scattering ($f = f_{2}$) processes, and the rest. This yields two sums
\begin{align}
    \sum_{f_{2}} \int d\Pi(f_{2})\hspace{10 pt}(2\pi)^{4}\delta^{4}(p_{f_{2}}-p_{i})|\mathcal{M}_{i\rightarrow f_{2}}|^{2} + \underbrace{\sum_{f \neq f_{2}} \int d\Pi(f)\hspace{10 pt}(2\pi )^{4} \delta^{4}(p_{i}-p_{f}) |\mathcal{M}_{i\rightarrow f}|^{2}}_{\equiv C_{f\neq f_{2}} \geq 0}
\end{align}
If we assume our external states have well-defined $4$-momentum quantum numbers, then (in addition to any sums and integrals over other quantum numbers) the first term contains an integral precisely of the form we simplified in the previous subsection. Therefore, we can rewrite it as
\begin{align}
    \int \Pi(f_{2})\hspace{10 pt}(2\pi)^{4}\delta^{4}(p_{i} - p_{3} - p_{4}) f(\theta,\phi) = \dfrac{\mathbbm{P}(3,4)}{16 \pi^{2} E_{i}} \int d\Pi(f^{*}_{2})\, \int d\Omega\hspace{10 pt}f(\theta,\phi)
\end{align}
where $d\Pi(f^{*}_{2})$ includes any sums or integrals quantum numbers besides $4$-momenta. Substituting this into Eq. \eqref{GeneralizedOpticalTheorem}, the optical theorem now equals
\begin{align}
    2\mathfrak{I}[\mathcal{M}_{i\rightarrow i}] = \sum_{f_{2}} \dfrac{\mathbbm{P}(3,4)}{16\pi^{2} E_{i}} \int d\Omega\hspace{10 pt}|\mathcal{M}_{i\rightarrow f_{2}}|^{2} + C_{f\neq f_{2}} \label{OpticalTheoremOnPause}
\end{align}
We will further reduce this in Section \ref{S - Helicity} with the help of the partial wave amplitude decomposition. However, before we define the partial wave decomposition of a matrix element, we first recount the rotational machinery, notation, and conventions of quantum mechanics which the decomposition relies on.

\section{Angular Momentum} \label{Angular Momentum}
As remarked in Subsection \ref{Quantum - Generators}, angular momentum operators $\vec{J}$ generate representations of the Lie group $\mathbf{SU(2)}$ despite being associated with representations of $\mathbf{SO(3)}$ before their quantum promotion. This section reviews the derivation of all irreducible finite-dimensional unitary representations of $\mathbf{SU(2)}$, how $\mathbf{SU(2)}$ representations are combined using Clebsh-Gordan coefficients, and the Wigner D-matrix. Because these topics are standard in quantum mechanics texts, we outline results for the sake of reference (and establishing convention) rather than pedagogy. For those readers interested in further details, \cite{osti_4389568} is a particularly complete resource on these topics (especially for proving certain Wigner D-matrix properties which we recount without proof).

\subsection{Finite-Dimensional Angular Momentum Representations} \label{Finite Dim Angular Momentum Reps}
The angular momentum operators satisfy the $\mathbf{SU}(2)$ commutation relations
\begin{align}
    [J_{i},J_{j}] &= i\epsilon_{ijk} J_{k}\hspace{35 pt}\implies\hspace{35 pt}\vec{J}\times\vec{J} = i\vec{J}
\end{align}
for $i,j\in \{x,y,z\}$, which we obtain from the $4$-vector equivalent Eq. \eqref{Jcommutators} by replacing $J_{i}\mapsto -iJ_{i}$ according to the quantum promotion procedure described in Subsection \ref{SS - QM}. Because we desire unitary representations of $\mathbf{SU(2)}$, we assume each angular momentum operator $J_{i}$ is Hermitian.

As before, each angular momentum operator commutes with the total angular momentum operator $\vec{J}^{\,2}$, which is the only Casimir operator of $\mathbf{SU(2)}$: for example,
\begin{align}
    [J_{z},\vec{J}^{\,2}] &= \sum_{j=1}^{3} [J_{z},J_{j}] J_{j} + J_{j}[J_{z},J_{j}]\\
    &= [J_{z},J_{x}] J_{x} + [J_{z},J_{y}] J_{y} + J_{x} [J_{z},J_{x}] + J_{y} [J_{z},J_{y}]\\
    &= i J_{y} J_{x} - i J_{x} J_{y} + i J_{x} J_{y} - i J_{y} J_{x}\\
    &= 0
\end{align}
which, by cyclic symmetry, means
\begin{align}
    [\vec{J},\vec{J}^{\,2}] = \vec{0}
\end{align}
As is standard, we choose our maximally-commuting set of observables in $\mathbf{SU(2)}$ to be $\{J^{\,2},J_{z}\}$, such that our kets satisfy
\begin{align}
    \vec{J}^{\,2} \ket{j,m} = c_{j} \ket{j,m}\hspace{35 pt}J_{z} \ket{j,m} = m \ket{j,m}
\end{align}
for a soon-to-be-determined real number $c_{j}$. We also choose to normalize these states such that
\begin{align}
    \braket{j,m}{j^{\prime},m^{\prime}} = \delta_{j,j^{\prime}}\, \delta_{m,m^{\prime}}
\end{align}
At this stage, $j$ is simply a label associated with the eigenvalue $c_{j}$, and has not been defined as any particular number (yet). It is in this basis that we begin the process of deriving all irreducible finite-dimensional representations.

Just as we were able to relate kets with different $4$-momentum on the same mass hyperboloid using Lorentz transformations, we can relate different eigenstates of $J_{z}$ having the same eigenvalue of $\vec{J}^{\,2}$ via the ladder operators
\begin{align}
    J_{\pm} = J_{x} \pm iJ_{y}
\end{align}
The ladder operators cannot change the eigenvalue of $\vec{J}^{\,2}$ because $\vec{J}^{\,2}$ commutes with every angular momentum operator and thus $J_{\pm}$ as well. Note that $J_{\pm}^{\dagger} = J_{\mp}$, where $\dagger$ denotes the Hermitian conjugate. Also note that
\begin{align}
    [J_{z},J_{\pm}] &= \pm J_{\pm}\hspace{35 pt}[J_{+},J_{-}] = 2J_{z}
\end{align}
and
\begin{align}
    J_{\pm} J_{\mp} = (J_{x} \pm i J_{y})(J_{x} \mp iJ_{y}) = J_{x}^{2} + J_{y}^{2} \mp i [J_{x},J_{y}] = \vec{J}^{\,2} - J_{z}^{2} \pm J_{z}
\end{align}
such that
\begin{align}
    \vec{J}^{\, 2} = J_{\pm}J_{\mp} + J_{z}^{2} \mp J_{z}
\end{align}
The $[J_{z},J_{\pm}]$ commutator allows us to confirm that the ladder operators do in fact change the eigenvalue of $J_{z}$ in a well-defined way:
\begin{align}
    J_{z} J_{\pm} \ket{jm} &= \bigg[J_{\pm}J_{z} + [J_{z},J_{\pm}] \bigg] \ket{j,m}\\
    &= \bigg[ J_{\pm} J_{z} \pm J_{\pm} \bigg] \ket{j,m}\\
&= (m\pm 1) J_{\pm} \ket{j,m} \label{2223}
\end{align}
or, in other words,
\begin{align}
    J_{\pm}\ket{j,m} \propto \ket{j,m\pm 1}
\end{align}
up to some overall phase and normalization. Therefore, by repeatedly applying instances of $J_{+}$ and $J_{-}$ to a ket $\ket{j,m}$, we can seemingly construct a ket $\ket{j,m+n}$ with $J_{z}$ eigenvalue $m+n$ for any integer $n$. However, we desire a finite-dimensional representation, for which there must exist some real number $m_{\text{max}} \equiv m + n$ such that its eigenvalue cannot be raised any further, e.g. $J_{+}\ket{j,m_{\text{max}}} = 0$. For this state,
\begin{align}
    \vec{J}^{\,2} \ket{j,m_{\text{max}}} = \bigg[ J_{-} J_{+} + J_{z}^{2} + J_{z} \bigg] \ket{j,m_{\text{max}}} = m_{\text{max}}(m_{\text{max}}+1)\ket{j,m_{\text{max}}}
\end{align}
Thus, for this maximal $J_{z}$ state with $J_{z}$ eigenvalue $m_{\text{max}}$, it has definite $\vec{J}^{\,2}$ eigenvalue $m_{\text{max}}(m_{\text{max}}+1)$. Because $[J_{z},\vec{J}^{\,2}] = 0$, all $J_{z}$ eigenkets that are related to each other by ladder operators have the same $\vec{J}^{\,2}$ eigenvalue. With this information, we now imbue $j$ with a definite meaning by defining $j \equiv m_{\text{max}}$. Hence, $\ket{j,m}\equiv\ket{m_{\text{max}},m}$ and the earlier $c_{j}$ equals $j(j+1)$, such that
\begin{align}
    J_{z} \ket{j,m} &= m\ket{j,m}\hspace{35 pt}
    \vec{J}^{\,2} \ket{j,m} = j(j+1)\ket{j,m}
\end{align}
By combining $J_{z} J_{\pm} \ket{j,m} = (m\pm 1) J_{\pm} \ket{j,m}$  from Eq. \eqref{2223} and
\begin{align}
    \bra{j,m}J_{\pm}^{\dagger}J_{\pm}\ket{j,m} &= \bra{j,m}J_{\mp}J_{\pm}\ket{j,m}\\
    &= \bra{j,m}\bigg[\vec{J}^{\,2} - J_{z}^{2} \mp J_{z} \bigg]\ket{j,m} \\
    &= \left[j(j+1) - m^{2} \mp m\right]\delta_{j,j^{\prime}} \,\delta_{m,m^{\prime}}
\end{align}
we find (noting $j(j+1) - m^{2} \mp m = (j\mp m)(j\pm m+1)$ as to rewrite the denominator factor into a standard form),
\begin{align}
    \ket{j,m\pm 1} = \dfrac{J_{\pm}}{\sqrt{(j\mp m)(j\pm m+1)}} \ket{j,m} \label{2230}
\end{align}
where an undetermined phase has been set to $1$; this is called the Condon-Shortley phase convention.

Note that the demand for a finite-dimensional representation works on both extremes of the $J_{z}$ eigenvalue spectrum: instead of demanding $J_{+}\ket{j,m}$ vanish for some value of $m = j \equiv m_{\text{max}}$ (i.e. the $J_{z}$ eigenvalue can be raised no further), we can seek the value $m = m_{\text{min}}$ such that $J_{-}\ket{j,m}$ vanishes (i.e. the $J_{z}$ eigenvalue can be lowered no further). For this value, we find
\begin{align}
    j(j+1) \ket{j,m_{\text{min}}} &= \vec{J}^{\,2} \ket{j,m_{\text{min}}} = \bigg[ J_{+} J_{-} + J_{z}^{2} - J_{z} \bigg]\ket{j,m_{\text{min}}} = m_{\text{min}}(m_{\text{min}}-1)\ket{j,m_{\text{min}}}
\end{align}
which implies $m_{\text{min}}$ must equal either $-j$ or $j+1$. Because $m_{\text{min}}$ cannot exceed $m_{\text{max}}$ by definition, it must be the case that $m_{\text{min}} = -j$. Finally, because the ladder operators only change $J_{z}$ eigenvalues by integer amounts, the range of the spectrum $j-(-j)=2j$ must be an integer as well, and thus $j$ must be either a nonnegative integer or a positive half-integer. With this, our construction of the representation is complete.

To summarize: there exists a $(2j+1)$-dimensional unitary representation of $\mathbf{SU(2)}$ for every $j\in\{0,\tfrac{1}{2},1,\tfrac{3}{2},\dots\}$, each of which is composed of kets $\ket{jm}$ that satisfy $\vec{J}^{\,2}\ket{jm} = j(j+1)\ket{jm}$ and $J_{z} \ket{jm} = m\ket{jm}$ for $m \in \{-j,-j+1,\dots,j\}$. We choose our normalizations and phases for these states as follows:
\begin{align}
    \braket{jm}{j^{\prime}m^{\prime}} = \delta_{j,j^{\prime}}\, \delta_{m,m^{\prime}}
\end{align}
such that
\begin{align}
    \mathbbm{1} = \sum_{j=0}^{+\infty} \sum_{m=-j}^{+j} \ket{j,m}\bra{j,m}
\end{align}
and
\begin{align}
    \ket{j,m\pm 1} = \dfrac{J_{\pm}}{\sqrt{(j\mp m)(j\pm m+1)}} \ket{j,m}
\end{align}
where $J_{\pm} = J_{x} \pm i J_{y}$. For each rotation $R(\vec{\alpha}\,)$ (an element of $\mathbf{SO(3)}$) describing a rotation by an angle $|\vec{\alpha}|$ about a rotation axis $\hat{\alpha}$, we can write a unitary rotation operator $\mathcal{U}[R(\vec{\alpha}\,)]$ (an element of a representation of $\mathbf{SU(2)}$) via exponentiation of the generators in the usual way:
\begin{align}
    \mathcal{U}[R(\vec{\alpha})]  = \text{Exp}\bigg[ -i\vec{\alpha}\cdot\vec{J}\,\bigg]
\end{align}
A $(2j+1)$-dimensional unitary representation is useful when, for example, describing the physics of a spin-$j$ massive particle.

These representations are also useful for describing the helicity eigenstates of a spin-$j$ massless particle, for which two helicity values are possible: $\lambda = \pm j$ (unless $j=0$, in which case only $\lambda = 0$ is available). Because massless particles lack longitudinal helicity modes, we typically cannot automatically relate the $\lambda = +j$ and $\lambda = -j$ helicity states via the ladder operators. Instead, we relate them via the reflection operator
\begin{align}
    Y \equiv \mathcal{U}[R_{y}(\pi)]\, \mathcal{U}[P] \label{reflection operator1}
\end{align}
where $\mathcal{U}[P]$ is a unitary quantum equivalent of the $4$-vector parity operator $P$ \cite{Haber:1994pe}. Because the angular momentum generators commute with the parity operator ($[J^{i},P]=0$), the angular momentum eigenstates are at most changed by a phase
\begin{align}
    \mathcal{U}[P] \ket{j,m} \propto \ket{j,m}
\end{align}
whereas, as remarked (for instance) in \cite{Haber:1994pe},
\begin{align}
    \mathcal{U}[R_{y}(\pi)] \ket{j,m} = e^{-i\pi J_{y}} \ket{j,m} = (-1)^{j-m} \ket{j,-m} \label{FlipTheMagneticQuantumNumber}
\end{align}
In all, we choose these phases such that
\begin{align}
    Y\,\ket{j,m} = \eta\, \ket{j,-m} \label{reflection operator2}
\end{align}
where the phase $\eta = \pm 1$ is called the parity factor and its precise value depends on the species of particle considered. Note that when acted on a $4$-momentum $p$, the equivalent $4$-vector representation of $Y$ yields ${Y^{\mu}}_{\nu} p^{\nu} = {{R_{y}(\pi)}^{\mu}}_{\nu} \, (E,-\vec{p}\,)^{\nu} = (E,p_{x},-p_{y},p_{z})$, such that $Y$ leaves (for example) $p_{z}$ invariant.

\subsection{Adding Angular Momentum Representations} \label{SS - Clebsch Gordan}
Angular momentum eigenstates can be combined via a direct product in the usual way to form a state $\ket{j_{1},m_{1},j_{2},m_{2}}$ defined as
\begin{align}
    \ket{j_{1},m_{1},j_{2},m_{2}} \equiv \ket{j_{1},m_{1}}\otimes \ket{j_{2},m_{2}}
\end{align}
with eigenvalue content
\begin{align}
    \vec{J}_{1}^{\,2} \, \ket{j_{1},m_{1},j_{2},m_{2}} &= j_{1}(j_{1}+1)\, \ket{j_{1},m_{1},j_{2},m_{2}}\\
    (J_{1})_{z} \, \ket{j_{1},m_{1},j_{2},m_{2}} &= m_{1}\, \ket{j_{1},m_{1},j_{2},m_{2}}\\
    \vec{J}_{2}^{\,2} \, \ket{j_{1},m_{1},j_{2},m_{2}} &= j_{2}(j_{2}+1)\, \ket{j_{1},m_{1},j_{2},m_{2}}\\
    (J_{2})_{z} \, \ket{j_{1},m_{1},j_{2},m_{2}} &= m_{2}\, \ket{j_{1},m_{1},j_{2},m_{2}}
\end{align}
However, there is another basis for these two-particle states which is sometimes more useful. Define the two-particle total angular momentum operator as
\begin{align}
    \vec{J} = \vec{J}_{1}\otimes \mathbbm{1}_{2} + \mathbbm{1}_{1} \otimes \vec{J}_{2}
\end{align}
wherein $\mathbbm{1}_{1}$ and $\mathbbm{1}_{2}$ are the identity operators on the first and second particle Hilbert spaces respectively. Usually the identity operators are understood from context, and we simply write $\vec{J} = \vec{J}_{1} + \vec{J}_{2}$. Because $[(\vec{J}_{1})_{i},(\vec{J}_{2})_{j}] = 0$ for all $i,j\in\{x,y,z\}$,
\begin{align}
    [J_{i},J_{j}] &= [(J_{1})_{i},(J_{1})_{j}] + [(J_{2})_{i},(J_{2})_{j}] = \epsilon_{ijk} \left[(J_{1})_{k} + (J_{2})_{k}\right] = \epsilon_{ijk} J_{k}
\end{align}
such that $\vec{J}$ acts like the usual total angular momentum operator. Furthermore, $[\vec{J}^{\, 2},\vec{J}_{1}^{\, 2}] = [\vec{J}^{\, 2},\vec{J}_{2}^{\, 2}] = 0$, and so we can choose $\{\vec{J}^{\,2}_{1},\vec{J}^{\,2}_{2},\vec{J}^{\,2},J_{z}\}$ as a maximally-commuting set of observables for a basis of states $\ket{j_{1},j_{2},J,M}$, with eigenvalue content
\begin{align}
    \vec{J}_{1}^{\,2} \, \ket{j_{1},j_{2},J,M} &= j_{1}(j_{1}+1)\, \ket{j_{1},j_{2},J,M}\\
    \vec{J}_{2}^{\,2} \, \ket{j_{1},j_{2},J,M} &= j_{2}(j_{2}+1)\, \ket{j_{1},j_{2},J,M}\\
    \vec{J}^{\,2} \, \ket{j_{1},j_{2},J,M} &= J(J+1)\, \ket{j_{1},j_{2},J,M}\\
    \vec{J}_{z}^{\,2} \, \ket{j_{1},j_{2},J,M} &= M\, \ket{j_{1},j_{2},J,M}
\end{align}
Given eigenvalues $j_{1}$ and $j_{2}$, the $\vec{J}^{\,2}$ eigenvalue only exists for $J\in \{|j_{1} - j_{2}|, \dots, j_{1} + j_{2}\}$.

We can convert between the $\ket{j_{1},m_{1},j_{2},m_{2}}$ and $\ket{j_{1},j_{2},J,M}$ representations using completeness, e.g.
\begin{align}
    \ket{j_{1},j_{2},J,M} &= \sum_{m_{1} = -j_{1}}^{+j_{1}} \sum_{m_{2} = -j_{2}}^{+j_{2}} \ket{j_{1},m_{1},j_{2},m_{2}} \braket{j_{1},m_{1},j_{2},m_{2}}{j_{1},j_{2},J,M}
\end{align}
where each $\braket{j_{1},m_{1},j_{2},m_{2}}{j_{1},j_{2},J,M}$ is called a Clebsch-Gordan (CG) coefficient. Physicists typically use existing resources (such as \cite{PhysRevD.98.030001}) rather than calculating CG coefficients themselves. The particular CG coefficients we require in this chapter are those used to combine two $j_{1}=j_{2}=1$ representations into a $J=2$ representation. Explicitly, this $J=2$ representation equals
\begin{align}
    \ket{2,\pm2} &= \ket{1,\pm1}\otimes\ket{1,\pm1}\nonumber\\
    \ket{2,\pm1} &= \dfrac{1}{\sqrt{2}}\bigg[\ket{1,\pm1}\otimes\ket{1,0} + \ket{1,0}\otimes\ket{1,\pm1}\bigg]\label{1plus1is2}\\
    \ket{2,0} &= \dfrac{1}{\sqrt{6}}\bigg[\ket{1,\pm1}\otimes\ket{1,\mp1} + \ket{1,\mp1}\otimes\ket{1,\pm1} + 2\,\ket{1,0}\otimes\ket{1,0}\bigg]\nonumber
\end{align}
where we suppress the $j_{1}=j_{2}=1$ labels of the $\ket{j_{1},j_{2},J,M}$ kets on the LHS.

\subsection{Wigner D-Matrix}
In quantum mechanics, each rotation is replaced with a corresponding unitary operator. Thus, the generic rotation $R(\phi,\theta,\psi)$ expressed in terms of Euler angles $(\phi,\theta,\psi)$ becomes
\begin{align}
    \mathcal{U}[R(\phi,\theta,\psi)] \equiv \mathcal{U}[R_{z}(\phi)]\, \mathcal{U}[R_{y}(\theta)] \, \mathcal{U}[R_{z}(\psi)] \label{URphithetagamma}
\end{align}
where
\begin{align}
    \mathcal{U}[R_{i}(\alpha)] &= \text{Exp}[-i\alpha J_{i}]
\end{align}
for $i\in\{x,y,z\}$, and $\vec{J} = (J_{x},J_{y},J_{z})$ are the angular momentum operators. We previously defined a restricted version of the Euler angle decomposition $\mathcal{U}[R(\hat{p})]=\mathcal{U}[R(\phi,\theta)]=\mathcal{U}[R(\phi,\theta,-\phi)]$ which is sufficient for mapping a $3$-momentum $|\vec{p}\,|\hat{z}$ to a $3$-momentum $\vec{p}$. The inverse of $\mathcal{U}[R(\phi,\theta,\psi)]$ is $\tilde{\mathcal{U}}[R(\phi,\theta,\psi)] = \mathcal{U}[R(-\psi,-\theta,-\phi)]$.

Keeping in mind that the rotation operator (being a function of the angular momentum operators alone) cannot influence the eigenvalue of $\vec{J}^{\,2}$, the Wigner D-matrix $\mathcal{D}^{j}_{m_{f},m_{i}}$ is defined as follows:
\begin{align}
    \mathcal{D}^{j_{i}}_{m_{f},m_{i}}(\phi,\theta,\psi)\, \delta_{j_{f},j_{i}} \equiv \bra{j_{f},m_{f}}\mathcal{U}[R(\phi,\theta,\psi)]\ket{j_{i},m_{i}}
\end{align}
Note the Kronecker delta $\delta_{j_{f},j_{i}}$ on the LHS. The Wigner D-matrix is sometimes referred to as the wavefunction of a symmetric top due to the Euler angles providing a natural coordinate system for a symmetric top. Because $J_{z}\ket{j,m} = m\ket{j,m}$, we can simplify the $z$-axis rotations from Eq. \eqref{URphithetagamma} when evaluating the Wigner D-matrix. Doing so defines the Wigner (small) d-matrix $d^{j}_{m_{f},m_{i}}$ relative to the Wigner D-matrix:
\begin{align}
    \mathcal{D}^{j_{i}}_{m_{f},m_{i}}(\phi,\theta,\psi)\, \delta_{j_{f},j_{i}} &= e^{-i(m_{f}\phi + m_{i}\psi)} \bra{j_{f},m_{f}}\mathcal{U}[R_{y}(\theta)]\ket{j_{i},m_{i}}\\
    &\equiv e^{-i(m_{f}\phi + m_{i}\psi)}\,d^{j_{i}}_{m_{f},m_{i}}(\theta)\, \delta_{j_{f},j_{i}}
\end{align}
In particular, when using the restricted Euler angle decomposition, we find
\begin{align}
    \mathcal{D}^{j}_{m_{f},m_{i}}(\phi,\theta) = \bra{j,m_{f}}\mathcal{U}[R(\phi,\theta,-\phi)]\ket{j,m_{i}} = e^{i(m_{f} - m_{i})\phi} d^{j}_{m_{f},m_{i}}(\theta) \label{WignerDDefinition}
\end{align}
having set $j=j_{i}=j_{f}$.

The Wigner D-matrix satisfies several convenient properties. For example, if $\theta = 0$, then $\mathcal{U}[R(\phi,\theta)] = \mathcal{U}[R(\phi,0)] = \mathbbm{1}$, such that
\begin{align}
    \mathcal{D}^{j}_{m_{f} m_{i}}(\phi,0) = \braket{j,m_{f}}{j,m_{i}} = \delta_{m_{f},m_{i}}
\end{align}
Other convenient properties include a relation describing orthogonality among instances of the restricted Wigner D-matrix:
\begin{align}
    \int d\Omega\hspace{5 pt}\mathcal{D}^{j_{1}*}_{m_{1} \lambda }(\hat{p}) \, \mathcal{D}^{j_{2}}_{m_{2} \lambda }(\hat{p}) = \dfrac{4\pi}{2j_{1}+1} \, \delta_{j_{1},j_{2}} \, \delta_{m_{1},m_{2}} \label{WignerDOrthonormality}
\end{align}
and the ability to construct a $\theta$-dependent Dirac delta function from the Wigner small d-matrices:
\begin{align}
    \delta(\cos\overline{\theta} - \cos\theta) &= \sum_{j} \left(\dfrac{2j+1}{2}\right)\, d^{j*}_{m,\lambda}(\overline{\theta}) \, d^{j}_{m,\lambda}(\theta) \label{2311yy}
\end{align}
These are proved in, for example, \cite{osti_4389568}.

The Wigner D-matrix is an important element of relativistic scattering calculations involving helicity eigenstates, which we are now prepared to address.

\section{Helicity} \label{S - Helicity}
\subsection{Single-Particle States} \label{SS - Helicity - Operator}
In Subsection \ref{SS - QM - Helicity Operator}, we refined our focus to eigenstates of the Hamiltonian $H$ with definite mass $M$ and spin $s$, and thereby defined the helicity operator $\Lambda$ as
\begin{align}
    \Lambda \equiv \dfrac{\vec{J}\cdot \vec{P}}{\sqrt{E^{2} - M^{2}}}
\end{align}
on those states. As demonstrated then, $\Lambda$ commutes with $P^{i}$, $\vec{P}^{\,2}$, $J^{i}$, $\vec{J}^{\,2}$, and $P^{2}$. This yields (among others) two maximally-commuting sets of observable operators, both of which involve the helicity operator:
\begin{itemize}
    \item[$\bullet$] {\bf Option 1:} $P^{\mu}$, $\Lambda$
    \item[$\bullet$] {\bf Option 2:} $H$, $\vec{J}^{\,2}$, $J_{z}$, $\Lambda$
\end{itemize}
in addition to the Poincar\'{e} Casimir operators, the mass operator $P^{2}$ and the Pauli-Lubanski pseudovector (which determines internal spin/helicity). The first option will describe our external one-particle states. However, the second option allows us utilize symmetries of the $S$-matrix and thereby derive the partial wave unitarity constraints. This section investigates the relationship between these two options.

Suppose we utilize Option 1, so that our one-particle states $\ket{p,\lambda}$ satisfy
\begin{align}
    H \, \ket{p,\lambda} = E \, \ket{p,\lambda} \hspace{35 pt}\vec{P} \, \ket{p,\lambda} = \vec{p} \, \ket{p,\lambda} \hspace{35 pt}\Lambda \, \ket{p,\lambda} = \lambda \, \ket{p,\lambda}
\end{align}
and are normalized according to
\begin{align}
    \braket{p,\lambda}{p^{\prime},\lambda^{\prime}} = (2\pi)^{3} \, (2E_{\vec{p}}) \,  \delta^{3}(\vec{p}-\vec{p}^{\,\prime}) \, \delta_{\lambda,\lambda^{\prime}}
\end{align}
The collection of helicity eigenstates having $3$-momentum $\vec{p}$ in the $+\hat{z}$ direction, i.e. $4$-momentum $p^{\mu} = (E,0,0,\sqrt{E^{2}-M^{2}})$, are automatically also $J_{z}$ eigenstates:
\begin{align}
     J_{z} \ket{p^{\prime},\lambda} = \Lambda \ket{p^{\prime},\lambda} = \lambda \ket{p^{\prime},\lambda}
\end{align}
This feature allows us to derive helicity eigenstates from $J_{z}$ eigenstates (and is a large part of why Section \ref{Angular Momentum} is included in this dissertation). In doing so, we also require several other features of the helicity operator:
\begin{itemize}
    \item[$\bullet$] {\bf Rotations Preserve Helicity:} Because $[\Lambda,\vec{J}]=0$, the helicity eigenvalue of a 4-momentum eigenstate is unchanged by rotations.
    
    Explicitly, given a generic rotation $R(\alpha)$, the $4$-momentum eigenvalue will transform in the usual way, but we might expect mixing of helicity eigenvalues:
    \begin{align}
        \mathcal{U}[R(\alpha)] \ket{p,\lambda} = e^{-i\vec{\alpha}\cdot\vec{J}} \ket{p,\lambda} = \sum_{\overline{\lambda}} c_{\overline{\lambda}} \ket{R(\alpha)p,\overline{\lambda}}
    \end{align}
    where $c_{\overline{\lambda}}$ are complex coefficients. However,
    \begin{align}
        \Lambda \, \mathcal{U}[R(\alpha)] \ket{p,\lambda} = \Lambda e^{-i\vec{\alpha}\cdot\vec{J}} \ket{p,\lambda} = e^{-i\vec{\alpha}\cdot\vec{J}} \Lambda \ket{p,\lambda} =  \mathcal{U}[R(\alpha)] \Lambda \ket{p,\lambda} = \lambda \, \mathcal{U}[R(\alpha)] \ket{p,\lambda}
    \end{align}
    Therefore,
    \begin{align}
        \mathcal{U}[R(\alpha)] \, \ket{p,\lambda} \propto \ket{R(\alpha)p,\lambda} \label{1259}
    \end{align}
    up to a phase, as desired.
    
    \item[$\bullet$] {\bf Certain Boosts Preserve Helicity:} Because $[J^{i},K^{i}]=0$, the helicity eigenvalue of a 4-momentum eigenstate is unchanged by any boost along the direction of motion that preserves the $3$-momentum direction.
    
    Consider a ket $\ket{p,\lambda}$ for which $p = (E,0,0,\sqrt{E^{2}-M^{2}})$. Under a generic boost $B_{z}(\beta)$ along the $z$-axis, the $4$-momentum eigenvalue will be changed in the usual way, but the helicity eigenvalue might be changed:
    \begin{align}
        \mathcal{U}[B_{z}(\beta)] \ket{p,\lambda} = e^{-i\beta K_{z}} \ket{p,\lambda} = \sum_{\overline{\lambda}} c_{\overline{\lambda}} \ket{B_{z}(\beta)p,\overline{\lambda}}
    \end{align}
    where $c_{\overline{\lambda}}$ are complex coefficients. Additionally suppose the boost $B_{z}(\beta)$ preserves the $3$-momentum direction of $p$ (so if $p^{\prime} = B_{z}(\beta)\,p$, then $\hat{p}^{\,\prime} = \hat{p} = \hat{z}$), such that
    \begin{align}
        J_{z} \ket{p,\lambda} = \Lambda \ket{p,\lambda} = \lambda \ket{p,\lambda} \hspace{15 pt}\text{ and }\hspace{15 pt} J_{z} \ket{p^{\prime},\overline{\lambda}} = \Lambda \ket{p^{\prime},\overline{\lambda}} = \overline{\lambda} \ket{p^{\prime},\overline{\lambda}}
    \end{align}
    Consequently, for this restricted set of kets and boosts,
    \begin{align}
        \lambda \, \mathcal{U}[B_{z}(\beta)] \, \ket{p,\lambda} = \mathcal{U}[B_{z}(\beta)] \, \Lambda \, \ket{p,\lambda} = \mathcal{U}[B_{z}(\beta)] \, J_{z} \, \ket{p,\lambda}  = J_{z} \, \mathcal{U}[B_{z}(\beta)] \, \ket{p,\lambda}
    \end{align}
    and
    \begin{align}
        J_{z} \, \mathcal{U}[B_{z}(\beta)] \, \ket{p,\lambda} =  \sum_{\overline{\lambda}} c_{\overline{\lambda}} \, J_{z} \, \ket{B_{z}(\beta)p,\overline{\lambda}} =  \sum_{\overline{\lambda}} c_{\overline{\lambda}} \, \Lambda \, \ket{B_{z}(\beta)p,\overline{\lambda}} = \Lambda \, \mathcal{U}[B_{z}(\beta)] \, \ket{p,\lambda}
    \end{align}
    such that
    \begin{align}
        \Lambda \, \mathcal{U}[B_{z}(\beta)] \, \ket{p,\lambda} = \lambda \, \mathcal{U}[B_{z}(\beta)] \, \ket{p,\lambda}\textbf{}
    \end{align}
    Therefore, so long as $B_{z}(\beta)$ preserves the $3$-direction of $p$,
    \begin{align}
        \mathcal{U}[B_{z}(\beta)] \, \ket{p,\lambda} \propto \ket{B_{z}(\beta) p,\lambda} \label{1265}
    \end{align}
    up to a phase, as desired. Note that if $\ket{p,\lambda}$ describes a massless state, then {\it all} boosts along the direction of motion preserve helicity.
\end{itemize}
The process of using phase conventions to eliminate proportionalities like the ones in Eqs. \eqref{1259} and \eqref{1265} has been handled on several occasions throughout this chapter. Specifically, Subsection \ref{External States - Single-Particle} described the process of relating single-particle $4$-momentum eigenstates on the same Lorentz-invariant hypersurface (i.e. the same mass hyperboloid or light cone). There we chose a standard $4$-momentum $k^{\mu}$ per hypersurface with $3$-momentum $\vec{k}$ pointing along the $+\hat{z}$ direction (or $\vec{k} = \vec{0}$, in the massive case). To obtain any another $4$-momentum $p^{\mu}$ on the same Lorentz-invariant hypersurface, we boosted $k^{\mu}$ along the $z$-direction to obtain the desired $3$-momentum magnitude $|\vec{p}\,|$ (without flipping the $3$-momentum direction) and then rotated the resultant $4$-momentum until its $3$-momentum aimed in the desired direction as well. We now modify the massive and massless versions of this procedure to include the helicity eigenvalue.

For the massive case, the standard $4$-momentum is $k^{\mu} = (M,0,0,0) = M\,\hat{t}^{\mu}$. To obtain a $4$-momentum $p^{\mu} = E\,\hat{t}^{\mu} + \sqrt{E^{2}-M^{2}}\, \hat{p}^{\mu}$ where $\hat{p}^{\mu} = (0, c_{\phi}s_{\theta}, s_{\phi}s_{\theta},c_{\theta})$, we can apply a boost and then a rotation like so:
\begin{align}
    p = R(\phi,\theta) B_{z}(\beta_{k\rightarrow p}) k\hspace{15 pt}\text{ where }\hspace{15 pt}\beta_{k\rightarrow p} = \text{arccosh}(E_{\vec{p}}/m) \label{1266}
\end{align}
There are other Lorentz transformations that map $k^{\mu}$ to $p^{\mu}$ (the Lorentz group is six-dimensional whereas the mass hyperboloid is only three-dimensional), but Eq. \eqref{1266} will be our canonical Lorentz transformation for taking $k^{\mu}$ to $p^{\mu}$. In the quantum equivalent, we will use $\ket{k,\lambda}$ as our standard eigenket. However, we encounter an obstacle. Because $\vec{k} = \vec{0}$, the application of the helicity operator $\Lambda$ to $\ket{k,\lambda}$ is not automatically well-defined: $\Lambda \ket{k,\lambda} = (\vec{J}\cdot \vec{k}\,)\ket{k,\lambda}/\sqrt{M^{2}-M^{2}} = (0/0)\ket{k,\lambda}$. To patch this, we modify our definition of $\ket{k,\lambda}$ and assert that $k^{\mu}$ should be interpreted as having an infinitesimal $3$-momentum in the $+\hat{z}$ direction, such that $\Lambda \ket{k,\lambda} = J_{z} \ket{k,\lambda}$, thereby avoiding any reference to $3$-momentum at $\vec{k}=\vec{0}$. With this solved, the quantum equivalent of the RHS of Eq. \eqref{1266} is
\begin{align}
    \mathcal{U}[R(\phi,\theta)]\,\mathcal{U}[B_{z}(\beta_{k\rightarrow p})]\,\ket{k,\lambda}
\end{align}
We would like to use this to define single-particle states having definite $4$-momentum and helicity, and thankfully we can: as previously established, the choices of $\mathcal{U}[B_{z}(\beta_{k\rightarrow p})]$ and $\mathcal{U}[R(\phi,\theta)]$ above preserve the helicity eigenvalue, and thus we can choose our phases such that
\begin{align}
    \ket{p,\lambda} \equiv \mathcal{U}[R(\phi,\theta)]\,\mathcal{U}[B_{z}(\beta_{k\rightarrow p})]\,\ket{k,\lambda} \label{1267}
\end{align}
for any massive single-particle state $\ket{p,\lambda}$. For later convenience, we define the symbol
\begin{align}
    \ket{p_{z},\lambda} \equiv \mathcal{U}[B_{z}(\beta_{k\rightarrow p})]\,\ket{k,\lambda}\hspace{35 pt}\text{ such that }\hspace{35 pt}\ket{p,\lambda} = \mathcal{U}[R(\phi,\theta)]\ket{p_{z},\lambda}
\end{align}
There remains one ambiguity in this definition, which occurs when applying Eq. \eqref{1267} to a state with $4$-momentum $-p_{z}\equiv(E_{\vec{p}},-|\vec{p}\,|\hat{z})$. In this case, $\phi$ is not uniquely defined and typically does not cancel from the final result, leading to an ambiguous phase $C_{\pi}$ that we will parameterize like so:
\begin{align}
    \ket{-p_{z},\lambda} \equiv C_{\pi} \, \mathcal{U}[R(0,\pi)] \, \ket{p_{z},\lambda}
\end{align}
As per usual, setting this phase is a matter of convention. We will use the Jacob-Wick (2nd particle) convention \cite{Jacob:1959at,Haber:1994pe}, which is motivated as follows: in the limit that the particle's 3-momentum vanishes, $-p_{z}$ and $+p_{z}$ both go to the rest frame $4$-momentum $(m,\vec{0}\,)$. In this same limit, the helicity operator acting on a state with $4$-momentum $\pm p_{z}$ will go to $\pm J_{z}$. Therefore, up to a phase, $\lim_{|\vec{p}\,|\rightarrow 0} \ket{\pm p_{z},\lambda} \propto \ket{(m,\vec{0}\,),\pm\lambda}$. Eq. \eqref{1267} already establishes an equality in the $+p_{z}$ case; the Jacob-Wick convention chooses $C_{\pi}$ so that equality will also hold in the $-p_{z}$ case. Because the total angular momentum and helicity operators equal the total spin and $J_{z}$ operators respectively in the rest frame, we can use Eq. \eqref{FlipTheMagneticQuantumNumber} to find
\begin{align}
    \lim_{|\vec{p}\,|\rightarrow 0} \ket{-p_{z},\lambda} &= C_{\pi} \,  \lim_{|\vec{p}\,|\rightarrow 0}  \mathcal{U}[R(0,\pi)] \, \ket{p_{z},\lambda}\\
    &=  C_{\pi} \, \mathcal{U}[R(0,\pi)] \, \ket{(m,\vec{0}\,),\lambda}\\
    &= C_{\pi} \, (-1)^{s-\lambda} \, \ket{(m,\vec{0}\,),-\lambda}
\end{align}
and therefore $C_{\pi} = (-1)^{s-\lambda}$, such that
\begin{align}
    \ket{-p_{z},\lambda} = (-1)^{s-\lambda} \, \mathcal{U}[R(0,\pi)] \, \ket{p_{z},\lambda} \label{JWPhaseConvention}
\end{align}
and this completes the construction of massive single-particle helicity eigenstates. Before moving to the massless case, we note that it is useful to define a conversion factor $\xi_{\lambda}(\phi)$ from the convention established in Eq. \eqref{1267} to the Jacob-Wick convention in Eq. \eqref{JWPhaseConvention}:
\begin{align}
    \ket{-p_{z},\lambda} = \xi_{\lambda}(\phi)\,\mathcal{U}[R(\phi,\pi)] \, \ket{p_{z},\lambda} \label{JWconversiondefinition}
\end{align}
or, by combining the last two equations,
\begin{align}
    \bigg[(-1)^{s-\lambda}\,\mathcal{U}[R(\phi,-\pi)] \, \mathcal{U}[R(0,\pi)]\bigg]\, \ket{p_{z},\lambda} = \xi_{\lambda}(\phi)\, \ket{p_{z},\lambda}
\end{align}
which will depend on the specific representation of the helicity eigenstates.

For the massless case, the same procedure applies in essence, but we no longer have access to a rest frame, so $\vec{k}$ cannot be made to vanish. Instead, we choose $k^{\mu} = E_{k}(\hat{t}^{\mu}+\hat{z}^{\mu})$ for some value of energy $E_{k}$ (the specific choice will not matter). Any other light-like $4$-momentum $p^{\mu} = E(\hat{t}^{\mu} + \hat{p}^{\mu})$ on the same lightcone can then be attained via a boost and rotation just like in Eq. \eqref{1266}, although now $\beta_{k\rightarrow p} = \ln(E_{\vec{p}}/E_{\vec{k}})$. Finally, by going over to the quantum equivalent, we can choose our phases such that Eq. \eqref{1267} also holds for any massless eigenstate $\ket{p,\lambda}$. Recall that for non-scalar massless particles the two available helicity states are related via the reflection operator (Eq. \ref{reflection operator1}), and thus massless expressions may include an additional parity factor $\eta$ relative to the massive case.

Next consider Option 2, wherein our single-particle states $\ket{E,j,m,\lambda}$ satisfy
\begin{align*}
    H \, \ket{E,j,m,\lambda} = E \, \ket{E,j,m,\lambda} \hspace{35 pt} \vec{J}^{\,2} \, \ket{E,j,m,\lambda} = j(j+1)\, \ket{E,j,m,\lambda}
\end{align*}
\begin{align}
    J_{z} \, \ket{E,j,m,\lambda} = m \, \ket{E,j,m,\lambda} \hspace{35 pt}  \Lambda \, \ket{E,j,m,\lambda} = \lambda \, \ket{E,j,m,\lambda}
\end{align}
and are normalized such that
\begin{align}
    \braket{\overline{E},\overline{j},\overline{m},\overline{\lambda}}{E,j,m,\lambda} = (2\pi)^{3} \, \dfrac{2}{|\vec{p}\,|}\,\delta(\overline{E}-E) \, \delta_{\overline{j},j} \, \delta_{\overline{m},m} \, \delta_{\overline{\lambda},\lambda}
\end{align}
with
\begin{align}
    \mathbbm{1} = \sum_{j,m,\lambda} \int \dfrac{dE}{2\pi}\hspace{5 pt} \dfrac{|\vec{p}\,|}{8 \pi^{2}}\, \ket{E,j,m,\lambda}\bra{E,j,m,\lambda}
\end{align}
where $|\vec{p}\,| = \sqrt{E^{2} - m^{2}}$, as motivated by Eqs. \eqref{3momentumnormalizationC1} and \eqref{3momentumnormalizationC2}. As remarked previously, because $P^{2} = H^{2} - \vec{P}^{\,2}$, each state $\ket{E,j,m,\lambda}$ is also an eigenstate of $\vec{P}^{\,2}$ with eigenvalue $E^{2} - M^{2}$. As a result, these states are sometimes labeled by $|\vec{p}\,| = \sqrt{E^{2} - M^{2}}$ in place of $E$ in the literature. Normalizations vary between resources as well.

Using properties of the Wigner D-matrix and the above definitions, we now derive the following expression:
\begin{align}
    \ket{p,\lambda} = \sum_{j,m} \sqrt{\dfrac{2j+1}{4\pi}} \mathcal{D}^{j}_{m,\lambda}(\phi,\theta)\, \ket{E,j,m,\lambda} \label{plambdafromEjmlambda}
\end{align}
This defines the single-particle state $\ket{p,\lambda}$ in terms of the angular momentum eigenstates $\ket{E,j,m,\lambda}$ \cite{Jacob:1959at,Haber:1994pe}. For completeness, we note that inverting Eq. \eqref{plambdafromEjmlambda} (via orthonormality of the Wigner D-matrix, Eq. \eqref{WignerDOrthonormality}) yields,
\begin{align}
    \ket{E,j,m,\lambda} = \sqrt{\dfrac{2j+1}{4\pi}} \int d\Omega\hspace{5 pt} \mathcal{D}^{j*}_{m,\lambda}(\phi,\theta)\,\ket{p,\lambda}
\end{align}
To derive Eq. \eqref{plambdafromEjmlambda}, we first insert the $\ket{E,j,m,\lambda}$ identity twice on the RHS of $\ket{p,\lambda} = \mathcal{U}[R(\phi,\theta)]\ket{p_{z},\lambda}$:
\begin{align}
    \ket{p,\lambda} &= \sum_{\overline{\lambda}}\sum_{\overline{j},\overline{m}} \int \dfrac{d\overline{E}}{2\pi} \sum_{\overline{\overline{\lambda}}}\sum_{\overline{\overline{j}},\overline{\overline{m}}} \int \dfrac{d\overline{\overline{E}}}{2\pi} \hspace{5 pt} w_{\overline{E}}\, w_{\overline{\overline{E}}}\, \ket{\overline{\overline{E}},\overline{\overline{j}},\overline{\overline{m}},\overline{\overline{\lambda}}}\nonumber\\
    &\hspace{35 pt} \cdot \bra{\overline{\overline{E}},\overline{\overline{j}},\overline{\overline{m}},\overline{\overline{\lambda}}} \mathcal{U}[R(\phi,\theta)] \ket{\overline{E},\overline{j},\overline{m},\overline{\lambda}}\, \braket{\overline{E},\overline{j},\overline{m},\overline{\lambda}}{p_{z},\lambda} \label{2297yy}
\end{align}
where $w_{E} \equiv |\vec{p}\,|/8\pi^{2} = \sqrt{E^{2} - m^{2}}/8\pi^{2}$. The quantity containing $\mathcal{U}[R(\phi,\theta)]$ is proportional to the Wigner D-matrix (originally defined in Eq. \eqref{WignerDDefinition}):
\begin{align}
    \bra{\overline{\overline{E}},\overline{\overline{j}},\overline{\overline{m}},\overline{\overline{\lambda}}} \mathcal{U}[R(\phi,\theta)] \ket{\overline{E},\overline{j},\overline{m},\overline{\lambda}} = \mathcal{D}^{\overline{j}}_{\overline{\overline{m}},\overline{m}}(\phi,\theta) \,\, \dfrac{2\pi}{w_{\overline{E}}}\,\,\delta(\overline{\overline{E}}-\overline{E})\, \delta_{\overline{\overline{j}},\overline{j}} \, \delta_{\overline{\overline{\lambda}},\overline{\lambda}}
\end{align}
The energy-dependent multiplicative coefficient is determined by setting $\phi = \theta = 0$ (and thus $\mathcal{U}[R(\phi,\theta)]=\mathcal{U}[R(0,0)]=\mathbbm{1}$) and recalling that rotations do not change helicity nor 3-momentum magnitude. Meanwhile, because $J_{z} = \Lambda$ on states with $z$-directional momenta, we may write
\begin{align}
    \braket{\overline{E},\overline{j},\overline{m},\overline{\lambda}}{p_{z},\lambda} \equiv c_{\overline{E},\overline{j},\overline{\lambda}}  \,\,\dfrac{2\pi}{w_{\overline{E}}}\,\, \delta(\overline{E} - E_{\vec{p}})\, \delta_{\overline{m},\lambda}\, \delta_{\overline{\lambda},\lambda}
\end{align}
where $c_{E,j,\lambda}$ is a soon-to-be-determined quantity. Therefore, returning to Eq. \eqref{2297yy}, we find (after relabeling indices)
\begin{align}
    \ket{p,\lambda} = \sum_{j,m} c_{E,j,\lambda} \, \mathcal{D}^{j}_{m,\lambda}(\phi,\theta) \, \ket{E,j,m,\lambda} \label{2300yy}
\end{align}
where $E \equiv \sqrt{m^{2} + |\vec{p}\,|^{2}}$. To determine $c_{E,j,\lambda}$, consider squaring both sides of the above expression:
\begin{align}
    \braket{\overline{p},\overline{\lambda}}{p,\lambda} = \bigg[\sum_{\overline{j},\overline{m}} c^{*}_{\overline{E},\overline{j},\overline{\lambda}} \, \mathcal{D}^{\overline{j}*}_{\overline{m},\overline{\lambda}}(\overline{\phi},\overline{\theta}) \, \bra{\overline{E},\overline{j},\overline{m},\overline{\lambda}}\bigg]\,\bigg[\sum_{j,m} c_{E,j,\lambda} \, \mathcal{D}^{j}_{m,\lambda}(\phi,\theta) \, \ket{E,j,m,\lambda}\bigg]
\end{align}
The LHS is the usual normalization equation, which we cast in energy-spherical coordinates (Eq. \eqref{3momentumnormalizationC1}) for upcoming convenience:
\begin{align}
    \braket{\overline{p},\overline{\lambda}}{p,\lambda} &= \dfrac{2\pi}{w_{E}}\,\delta(\overline{E}-E) \, \delta^{2}(\overline{\Omega} - \Omega) \, \delta_{\overline{\lambda},\lambda} \label{2302yy}
\end{align}
Meanwhile, the RHS equals
\begin{align}
    &\sum_{\overline{j},\overline{m}}\sum_{j,m} c^{*}_{\overline{E},\overline{j},\overline{\lambda}}\, c_{E,j,\lambda} \, \mathcal{D}^{\overline{j}*}_{\overline{m},\overline{\lambda}}(\overline{\phi},\overline{\theta}) \, \mathcal{D}^{j}_{m,\lambda}(\phi,\theta) \, \braket{\overline{E},\overline{j},\overline{m},\overline{\lambda}}{E,j,m,\lambda}\\
    &=  \sum_{\overline{j},\overline{m}}\sum_{j,m} c^{*}_{\overline{E},\overline{j},\overline{\lambda}}\, c_{E,j,\lambda} \, \mathcal{D}^{\overline{j}*}_{\overline{m},\overline{\lambda}}(\overline{\phi},\overline{\theta}) \, \mathcal{D}^{j}_{m,\lambda}(\phi,\theta) \,\dfrac{2\pi}{w_{E}}\,\delta(\overline{E}-E) \, \delta_{\overline{j},j} \, \delta_{\overline{m},m} \, \delta_{\overline{\lambda},\lambda}\\
    &= \dfrac{2\pi}{w_{E}}\,\delta(\overline{E}-E) \,\bigg[\sum_{j,m} |c_{E,j,\lambda}|^{2} \, \mathcal{D}^{j*}_{m,\lambda}(\overline{\phi},\overline{\theta}) \, \mathcal{D}^{j}_{m,\lambda}(\phi,\theta)\bigg] \, \delta_{\overline{\lambda},\lambda} \label{2305yy}
\end{align}
Eqs. \eqref{2302yy} and \eqref{2305yy} are equal if and only if
\begin{align}
    \delta^{2}(\overline{\Omega}-\Omega) &= \sum_{j,m} |c_{E,j,\lambda}|^{2} \, \mathcal{D}^{j*}_{m,\lambda}(\overline{\phi},\overline{\theta}) \, \mathcal{D}^{j}_{m,\lambda}(\phi,\theta)\\
    &= \sum_{j,m} |c_{E,j,\lambda}|^{2} \, e^{i(m-\lambda)(\phi-\overline{\phi})} \, d^{j*}_{m,\lambda}(\overline{\theta})\, d^{j}_{m,\lambda}(\theta) \label{2308yy}
\end{align}
where we have used the definition of the Wigner small d-matrix relative to the Wigner D-matrix, Eq. \eqref{WignerDDefinition}. Consider integrating both sides of this equation with respect to $\phi$. Because $m-\lambda$ must be an integer, the RHS will vanish unless $m=\lambda$:
\begin{align}
    \int_{0}^{2\pi} d\phi\hspace{5 pt} e^{i(m-\lambda)(\phi-\overline{\phi})} &= 2\pi \, \delta_{m,\lambda}
\end{align}
Meanwhile, the LHS can be integrated by recalling that $\delta^{2}(\overline{\Omega}-\Omega) =\delta( \overline{\phi} - \phi ) \, \delta(\cos\overline{\theta} - \cos\theta)$. Thus, after integration over $\phi$, Eq. \eqref{2308yy} becomes
\begin{align}
    \delta(\cos\overline{\theta} - \cos\theta) &= \sum_{j} (2\pi) \, |c_{E,j,\lambda}|^{2}\, d^{j*}_{m,\lambda}(\overline{\theta}) \, d^{j}_{m,\lambda}(\theta)
\end{align}
Compare this to Eq. \eqref{2311yy}, one of the relations we introduced (without proof) when we first defined the Wigner D-matrix:
\begin{align}
    \delta(\cos\overline{\theta} - \cos\theta) &= \sum_{j} \left(\dfrac{2j+1}{2}\right)\, d^{j*}_{m,\lambda}(\overline{\theta}) \, d^{j}_{m,\lambda}(\theta)
\end{align}
These become equal when $|c_{E,j,\lambda}|^{2} = (2j+1)/4\pi$, or (by choosing an otherwise arbitrary phase) $c_{E,j,\lambda} = \sqrt{(2j+1)/4\pi}$. Substituting this solution into Eq. \eqref{2300yy} yields Eq. \eqref{plambdafromEjmlambda}, as desired.

As in Subsections \ref{SS - Multiparticle} and \ref{SS - General Quantum Numbers}, we can combine single-particle states to form multi-particle states. If we follow that procedure, we would define a (distinguishable) two-particle state as
\begin{align}
    \ket{p_{1},\lambda_{1}} \otimes \ket{p_{2},\lambda_{2}}
\end{align}
where each single-particle state is defined according to Eqs. \eqref{1267} and (when $\vec{p} = -|\vec{p}\,|\hat{z}$) \eqref{JWPhaseConvention}. However when considering two-particle states in the center-of-momentum frame, this is not the convention typically adopted.

Instead, it is conventional to define the two-particle COM states as
\begin{align}
    \ket{\vec{p},\lambda_{1},\lambda_{2}} &\equiv \bigg(\mathcal{U}[R(\phi,\theta)]\, \ket{(E_{1},+|\vec{p}\,|\hat{z}),\lambda_{1}}\bigg) \otimes \bigg(\mathcal{U}[R(\phi,\theta)]\, \ket{(E_{2},-|\vec{p}\,|\hat{z}),\lambda_{2}}\bigg) \label{TwoParticleHelicityState}
\end{align}
This is why the phase convention for $\ket{-p_{z},\lambda}$ chosen in Eq. \eqref{JWPhaseConvention} for single-particle states is typically called the Jacob-Wick {\it 2nd particle} convention. We also define the two-particle total and relative helicity operators as $\Lambda_{\text{total}} = \Lambda_{1} + \Lambda_{2}$ and $\Lambda = \Lambda_{1} - \Lambda_{2}$ respectively, where
\begin{align}
    \Lambda_{1}\pm\Lambda_{2} = \dfrac{\vec{J}_{1}\cdot \vec{P}_{1}}{\sqrt{E_{1}^{2} - m_{1}^{2}}} \pm \dfrac{\vec{J}_{2}\cdot \vec{P}_{2}}{\sqrt{E_{2}^{2} - m_{2}^{2}}} \hspace{5 pt}\mathrel{\mathop{=}^{\text{COM}}_{\text{frame}}}\hspace{5 pt} (\vec{J}_{1}\mp\vec{J}_{2})\cdot \hat{p}
\end{align}
and the last equality in each line assumes it acts on a state with definite $3$-momentum $\vec{p}$. Note that the relative helicity $\Lambda$ is related to the two-particle angular momentum operator $\vec{J} = \vec{J}_{1} + \vec{J}_{2}$.

The single-particle argument that allowed $\ket{p,\lambda}$ to be rewritten as a superposition of $\ket{E,j,m,\lambda}$ carries through essentially unchanged for $\ket{\vec{p},\lambda_{1},\lambda_{2}}$ in terms of the relative helicity $\lambda = \lambda_{1}-\lambda_{2}$, such that we may write the state $\ket{\vec{p},\lambda_{1},\lambda_{2}}$ in terms of two-particle angular momentum eigenstates as
\begin{align}
    \ket{\vec{p},\lambda_{1},\lambda_{2}} = \sum_{J,M} \sqrt{\dfrac{2J+1}{4\pi}} \, \mathcal{D}^{J}_{M,\lambda_{1}-\lambda_{2}}(\phi,\theta) \,  \ket{\sqrt{s},J,M,\lambda_{1},\lambda_{2}} \label{twoparticlehelicitydecomposition}
\end{align}
Because they occur regularly in 2-to-2 scattering calculations, the relative helicities of the initial and final particle pairs are given special symbols: $\lambda_{i} \equiv \lambda_{1} - \lambda_{2}$ and $\lambda_{f} \equiv \lambda_{3} - \lambda_{4}$.

\subsection{Partial Wave Amplitudes}
Because the S-matrix commutes with $\vec{J}$, it can be put into a block-diagonal form wherein each block has a definite total angular momentum $\vec{J}^{\,2}$ eigenvalue. This implies a similar decomposition of the $T$-matrix, so that we may write
\begin{align}
    \bra{\sqrt{s},J,M,\lambda_{3},\lambda_{4}} \hat{T}(p_{i}=p_{f}) \ket{\sqrt{s},J^{\prime},M^{\prime},\lambda_{1},\lambda_{2}} \equiv \delta_{J^{\prime},J}\, \delta_{M^{\prime},M}\, \bra{\lambda_{3},\lambda_{4}}\,\hat{T}^{J}(s)\,\ket{\lambda_{1},\lambda_{2}}
\end{align}
when considering 2-to-2 scattering. Using the definition of the matrix element Eq. \eqref{MEdef}, the decomposition of helicity eigenstates in terms of angular momentum eigenstates Eq. \eqref{twoparticlehelicitydecomposition}, and the fact that $\mathcal{D}^{j}_{m_{1},m_{2}}(\phi,0) = \delta_{m_{1},m_{2}}$,
\begin{align}
    \mathcal{M}_{i\rightarrow f} &= \bra{\vec{p}_{f},\lambda_{3},\lambda_{4}} \, \hat{T}(p_{i} = p_{f}) \, \ket{\vec{p}_{i},\lambda_{1},\lambda_{2}}\\
    &= \sum_{J,M,J^{\prime},M^{\prime}} \sqrt{\dfrac{2J+1}{4\pi}} \sqrt{\dfrac{2J^{\prime}+1}{4\pi}} \,  \mathcal{D}^{J*}_{M,\lambda_{f}}(\phi,\theta) \, \mathcal{D}^{J^{\prime}}_{M^{\prime},\lambda_{i}}(0,0) \nonumber\\
    &\hspace{105 pt}\times \bra{\sqrt{s},J,M,\lambda_{3},\lambda_{4}}\, \hat{T}(p_{i} = p_{f})\, \ket{\sqrt{s},J^{\prime},M^{\prime},\lambda_{1},\lambda_{2}}\\
    &= \sum_{J} \bigg(\dfrac{2J+1}{4\pi}\bigg) \, \mathcal{D}^{J*}_{\lambda_{i},\lambda_{f}}(\phi,\theta) \, \bra{\lambda_{3},\lambda_{4}}\, \hat{T}^{J}(s)\, \ket{\lambda_{1},\lambda_{2}} \label{NearlyToPWA}
\end{align}
where $\lambda_{i} =\lambda_{1}-\lambda_{2}$ and $\lambda_{f} = \lambda_{3}-\lambda_{4}$. We define the partial wave amplitude (PWA) $a^{J}(s)$ as\footnote{Definitions of the partial wave amplitude vary throughout the literature, with (for example) some authors choosing a $1/32\pi^{2}$ factor in place of $1/64\pi^{2}$. The particular choice of convention impacts other expressions, including the form of the partial wave unitarity constraints.}
\begin{align}
    a^{J}(s) \equiv \dfrac{1}{64\pi^{2}} \bra{\lambda_{3},\lambda_{4}}\, \hat{T}^{J}(s)\, \ket{\lambda_{1},\lambda_{2}}
\end{align}
such that the matrix element may be written as, via Eq. \eqref{NearlyToPWA},
\begin{align}
    \mathcal{M}(s,\theta,\phi) &= \sum_{J} 16\pi\, (2J+1)\, a^{J}(s)\, \mathcal{D}^{J*}_{\lambda_{i},\lambda_{f}}(\phi,\theta) \label{MEtoPWA}
\end{align}
or, equivalently, via Eq. \eqref{WignerDOrthonormality},
\begin{align}
    a^{J}(s) = \dfrac{1}{64\pi^{2}} \int d\Omega\hspace{10 pt} {\mathcal{D}}_{\lambda_i,\lambda_f}^{J}(\phi,\theta)\, \mathcal{M}(s,\phi,\theta)
\end{align}
In the next subsection, we use the partial wave decomposition of 2-to-2 scattering matrix elements in order to derive the elastic and inelastic partial wave unitarity constraints from the optical theorem.

\subsection{Elastic, Inelastic Unitarity Constraints} \label{SS - Elastic, Inelastic Unitarity Constraints}
Recall Eq. \eqref{OpticalTheoremOnPause}, wherein we reduced the optical theorem to
\begin{align}
    2\mathfrak{I}[\mathcal{M}_{i\rightarrow i}] = \sum_{f_{2}} \dfrac{\mathbbm{P}(3,4)}{16\pi^{2} E_{i}} \int d\Omega\hspace{10 pt}|\mathcal{M}_{i\rightarrow f_{2}}|^{2} + C_{f\neq f_{2}} \label{2263}
\end{align}
Decompose the matrix element on the RHS of Eq. \eqref{2263} via Eq. \eqref{MEtoPWA}, such that
\begin{align}
    \int d\Omega\hspace{10 pt}|\mathcal{M}_{i\rightarrow f_{2}}|^{2} &= \mathcal{M}_{i\rightarrow f_{2}}^{*} \mathcal{M}_{i\rightarrow f_{2}}\\
    &= \int d\Omega\hspace{10 pt}\bigg[\sum_{J^{\prime}} 16\pi (2J^{\prime}+1)\, a_{i\rightarrow f_{2}}^{J^{\prime}*}(s)\, \mathcal{D}^{J^{\prime}}_{\lambda_{i},\lambda_{f}}(\phi,\theta)\bigg]\nonumber\\
    &\hspace{35 pt}\cdot\,\bigg[\sum_{J} 16\pi (2J+1)\, a_{i\rightarrow f_{2}}^{J}(s)\, \mathcal{D}^{J*}_{\lambda_{i},\lambda_{f}}(\phi,\theta)\bigg]\\
    &= \sum_{J} \sum_{J^{\prime}} 256\pi^{2}\,(2J+1)\,(2J^{\prime}+1)\, a_{i\rightarrow f_{2}}^{J^{\prime}*}(s)\,a_{i\rightarrow f_{2}}^{J}(s)\nonumber\\
    &\hspace{35 pt}\cdot \int d\Omega\hspace{10 pt}  \mathcal{D}^{J^{\prime}}_{\lambda_{i},\lambda_{f}}(\phi,\theta) \, \mathcal{D}^{J*}_{\lambda_{i},\lambda_{f}}(\phi,\theta)\\
    &= \sum_{J} 1024 \pi^{3} (2J+1) |a_{i\rightarrow f_{2}}^{J}(s)|^{2}
\end{align}
where we used orthonormality of the Wigner D-matrices to evaluate the angular integral, Eq. \eqref{WignerDOrthonormality}. Thus, overall the RHS of Eq. \eqref{2263} becomes
\begin{align}
    64 \pi \sum_{f_{2}} \dfrac{\mathbbm{P}(3,4)}{E_{i}} \sum_{J} (2J+1) |a_{i\rightarrow f_{2}}^{J}(s)|^{2} + C_{f\neq f_{2}}
\end{align}
In this same reference frame, the matrix element on the LHS of Eq. \eqref{2263} equals
\begin{align}
    \mathcal{M}_{i\rightarrow i} = 16\pi \sum_{J} (2J+1)\, a_{i\rightarrow i}^{J}(s)\, \mathcal{D}^{J*}_{\lambda_{i},\lambda_{i}}(0,0) = 16\pi \sum_{J} (2J+1) \, a_{i\rightarrow i}^{J}(s)
\end{align}
such that the LHS equals, overall,
\begin{align}
    2\mathfrak{I}[\mathcal{M}_{i\rightarrow i}] = 32\pi \sum_{J} (2J+1)\,\mathfrak{I}[a_{i\rightarrow i}^{J}(s)]
\end{align}
and all together Eq. \eqref{2263} implies 
\begin{align}
    32\pi \sum_{J} (2J+1) \, \mathfrak{I}[a_{i\rightarrow i}^{J}(s)] =  64 \pi \sum_{f_{2}} \dfrac{\mathbbm{P}(3,4)}{E_{i}} \sum_{J} (2J+1)\, |a_{i\rightarrow f_{2}}^{J}(s)|^{2} + C_{f\neq f_{2}}
\end{align}
or, focusing on the 2-to-2 scattering and dropping the nonnegative constant $C_{f\neq f_{2}}$,
\begin{align}
    \sum_{J} (2J+1)\, \mathfrak{I}[a_{i\rightarrow i}^{J}(s)] \geq \sum_{f_{2}} 2\dfrac{\mathbbm{P}(3,4)}{E_{i}} \sum_{J} (2J+1)\,  |a_{i\rightarrow f_{2}}^{J}(s)|^{2}
\end{align}
We can isolate individual angular momentum components by employing superpositions of helicity eigenstates that reconstruct the angular momentum eigenstates, and thereby demonstrate this inequality holds component-by-component:
\begin{align}
     \mathfrak{I}[a_{i\rightarrow i}^{J}(s)] \geq \sum_{f_{2}} 2\dfrac{\mathbbm{P}(3,4)}{E_{i}} |a_{i\rightarrow f_{2}}^{J}(s)|^{2}
\end{align}
The RHS of this inequality can be further reduced by dividing the expression into elastic ($i = f_{2}$, aside from the values of $(\theta,\phi)$ describing the pair, which each PWA does not depend on) and inelastic ($i\neq f_{2}$) pieces,
\begin{align}
    \mathfrak{I}[a_{i\rightarrow i}^{J}(s)] \geq 2\dfrac{\mathbbm{P}(1,2)}{E_{i}}\,|a_{i\rightarrow i}^{J}(s)|^{2} + \sum_{f_{2}\neq i} 2\dfrac{\mathbbm{P}(3,4)}{E_{i}}\, |a_{i\rightarrow f_{2}}^{J}(s)|^{2} 
\end{align}
or, equivalently,
\begin{align}
    \mathfrak{I}[a_{i\rightarrow i}^{J}(s)] \geq \beta_{12}\, |a_{i\rightarrow i}^{J}(s)|^{2} + \sum_{f_{2}\neq i} \beta_{34}\,|a_{i\rightarrow f_{2}}^{J}(s)|^{2} 
\end{align}
where
\begin{align}
    \beta_{jk} \equiv 2\dfrac{\mathbbm{P}(j,k)}{E_{i}} = \dfrac{1}{s} \sqrt{\bigg[s-(m_{j}-m_{k})^{2}\bigg]\,\bigg[s-(m_{j}+m_{k})^{2}\bigg]}
\end{align}
because $E_{i} = E_{1} + E_{2} = \sqrt{s}$.
By definition, $|a^{J}_{i\rightarrow i}(s)|^{2} = \mathfrak{R}[a^{J}_{i\rightarrow i}(s)]^{2} + \mathfrak{I}[a^{J}_{i\rightarrow i}(s)]^{2}$, so the previous inequality can also be expressed as, after multiplying both sides by $\beta_{12}$, adding $(1/2)^{2}$ to both sides, and rearranging,
\begin{align}
    \bigg[\dfrac{1}{2}\bigg]^{2} - \sum_{f_{2}\neq i} \beta_{12} \, \beta_{34} \, |a_{i\rightarrow f_{2}}^{J}(s)|^{2} \geq \bigg[\beta_{12}\, \mathfrak{R}[a^{J}_{i\rightarrow i}(s)]\bigg]^{2} +  \bigg[\beta_{12}\,\mathfrak{I}[a^{J}_{i\rightarrow i}(s)] - \dfrac{1}{2} \bigg]^{2} \label{2274}
\end{align}
 Thus, the values of $\beta_{12}\, a_{i\rightarrow i}^{J}(s)$ are bounded by a circle in the complex plane centered at $i/2$ and with radius at most equal to $1/2$, where the radius shrinks as inelastic contributions grow in magnitude. Therefore, the real and imaginary parts of the elastic amplitudes must satisfy
\begin{align}
    \bigg|\beta_{12}\, \mathfrak{R}[a^{J}_{i\rightarrow i}]\bigg| \leq \dfrac{1}{2} \hspace{35 pt} 0 \leq \beta_{12}\, \mathfrak{I}[a^{J}_{i\rightarrow i}] \leq 1 \label{elasticPWAconstraints}
\end{align}
Meanwhile, the RHS of Eq. \eqref{2274} must be nonnegative (the radius of the circle cannot be imaginary), so the net sum of squares of inelastic amplitudes are bounded from above
\begin{align}
    \sum_{f_{2}\neq i} \beta_{12}\,\beta_{34}\, |a^{J}_{i\rightarrow f_{2}}(s)|^{2} < \dfrac{1}{4} \label{inelasticPWAconstraints}
\end{align}
These are the inequalities we sought to derive: the elastic and inelastic partial wave unitarity constraints \cite{Endo:2014mja}. For most of the processes in which we are interested, $\mathcal{M}$ grows like $\mathcal{O}(s^{k})$ at large $s$ for $k\geq 1$, such that (via Eq. \eqref{MEtoPWA}) $a^{J}(s) \sim \mathcal{O}(s^{k})$ as well. If these inequalities happen to be satisfied for such a partial wave amplitude at some energy scale, then there necessarily exists a higher energy scale $\Lambda_{\text{strong}}$ for which $a^{J}(s)$ contradicts these inequalities for all $s\geq \Lambda_{\text{strong}}^{2}$, and thus contradicts the optical theorem, and thus contradicts unitarity of the $S$-matrix.

Lastly, we note that an additional factor of $1/2$ should be included in $\beta_{jk}$ if the particles associated with it are identical, per the discussion at the end of Subsection \ref{SS - General Quantum Numbers}. Thus, when a process describes elastic scattering of identical particles ($i=f=(1,1)$), we set
\begin{align}
    \beta_{11} = \dfrac{1}{2}\sqrt{1- \dfrac{4m_{1}^{2}}{s}}
\end{align}
such that the relevant partial wave unitarity constraints equal
\begin{align}
     \sqrt{1- \dfrac{s_{\text{min}}}{s}} \, \left|\mathfrak{R}[a^{J}_{i\rightarrow i}]\right| \leq 1 \hspace{35 pt} 0 \leq \sqrt{1- \dfrac{s_{\text{min}}}{s}} \, \left|\mathfrak{R}[a^{J}_{i\rightarrow i}]\right| \leq 2
\end{align}
where $s_{\text{min}} = 4m_{1}^{2}$.\footnote{For the remainder of this dissertation, we will refer to elastic scattering of identical particles simply as ``elastic scattering" even when elastic scattering of distinguishable particles is also allowed. Where ambiguity is likely, we indicate the relevant process.}

\section{Polarization Tensors and Lagrangians} \label{S - Fields}
\subsection{Derivation of the Spin-1 and Spin-2 Polarizations} \label{SS - Polarizations Derived}

In Subsection \ref{FiniteDimLorentzReps}, we demonstrated that the $4$-vector representation is the $(\tfrac{1}{2},\tfrac{1}{2})$ Lorentz representation (an irreducible finite-dimensional non-unitary representation of $\mathbf{SL(2,\mathbb{C})}$) and remarked that it contains both spin-$0$ and spin-$1$ representations with respect to the $\mathbf{SU(2)}$ rotation subgroup. This subsection now isolates that spin-1 representation, and then uses Clebsch-Gordan coefficients (defined in Subsection \ref{SS - Clebsch Gordan}) to build a spin-2 representation from two copies of the spin-1 representation. The end products of this procedure are the spin-1 and spin-2 polarization structures, which accompany external states when calculating certain matrix elements.

To derive these structures, we revisit the Lorentz generators $J^{i} = i(J^{i})_{\text{4-vector}}$ and $K^{i} = i(K^{i})_{\text{4-vector}}$ of the $4$-vector representation defined in Subsection \ref{FiniteDimLorentzReps}, which equal
\begin{align}
    J^{1} = \left(\hspace{-5 pt}\begin{tabular}{ c c c c}
    ${0}$ & ${0}$ & ${0}$ & ${0}$\\
    ${0}$ & ${0}$ & ${0}$ & ${0}$\\
    ${0}$ & ${0}$ & ${0}$ & ${-i}$\\
    ${0}$ & ${0}$ & ${+i}$ & ${0}$
    \end{tabular}\hspace{-5 pt}\right)\hspace{35 pt}J^{2} = \left(\hspace{-5 pt}\begin{tabular}{ c c c c}
    ${0}$ & ${0}$ & ${0}$ & ${0}$\\
    ${0}$ & ${0}$ & ${0}$ & ${+i}$\\
    ${0}$ & ${0}$ & ${0}$ & ${0}$\\
    ${0}$ & ${-i}$ & ${0}$ & ${0}$
    \end{tabular}\hspace{-5 pt}\right)\hspace{35 pt}J^{3} = \left(\hspace{-5 pt}\begin{tabular}{ c c c c}
    ${0}$ & ${0}$ & ${0}$ & ${0}$\\
    ${0}$ & ${0}$ & ${-i}$ & ${0}$\\
    ${0}$ & ${+i}$ & ${0}$ & ${0}$\\
    ${0}$ & ${0}$ & ${0}$ & ${0}$
    \end{tabular}\hspace{-5 pt}\right)
\end{align}
and
\begin{align}
    K^{1} = \left(\hspace{-5 pt}\begin{tabular}{ c c c c}
    ${0}$ & ${+i}$ & ${0}$ & ${0}$\\
    ${+i}$ & ${0}$ & ${0}$ & ${0}$\\
    ${0}$ & ${0}$ & ${0}$ & ${0}$\\
    ${0}$ & ${0}$ & ${0}$ & ${0}$
    \end{tabular}\hspace{-5 pt}\right)\hspace{35 pt}K^{2} = \left(\hspace{-5 pt}\begin{tabular}{ c c c c}
    ${0}$ & ${0}$ & ${+i}$ & ${0}$\\
    ${0}$ & ${0}$ & ${0}$ & ${0}$\\
    ${+i}$ & ${0}$ & ${0}$ & ${0}$\\
    ${0}$ & ${0}$ & ${0}$ & ${0}$
    \end{tabular}\hspace{-5 pt}\right)\hspace{35 pt}K^{3} = \left(\hspace{-5 pt}\begin{tabular}{ c c c c}
    ${0}$ & ${0}$ & ${0}$ & ${+i}$\\
    ${0}$ & ${0}$ & ${0}$ & ${0}$\\
    ${0}$ & ${0}$ & ${0}$ & ${0}$\\
    ${+i}$ & ${0}$ & ${0}$ & ${0}$
    \end{tabular}\hspace{-5 pt}\right)
\end{align}
such that a generic rotation and boost equal $R(\vec{\alpha}) = \text{Exp}[-i\vec{\alpha}\cdot\vec{J}]$ and $B(\vec{\beta}) = \text{Exp}[-i\vec{\beta}\cdot\vec{K}]$ respectively. Note that the boost generators $\{K^{i}\}$ are anti-Hermitian, cementing the fact that this representation is non-unitary. Using these generators, we can define a generic 4-vector Lorentz transformation $\Lambda$, which we will act on {\it complex} $4$-vectors $\epsilon^{\mu}$ in the usual way ($\epsilon^{\mu}\mapsto {\Lambda^{\mu}}_{\nu} \epsilon^{\nu}$); we utilize complex $4$-vectors to ensure we can eventually solve for all eigenvectors of the helicity operator (and note that $\{J^{i},K^{i}\}$ still only span the real Lie algebra $\mathfrak{so}(1,3)\cong \mathfrak{sl}(2,\mathbb{C})$).

In particular, suppose the complex $4$-vectors $\epsilon^{\mu}$ encode single-particle states with definite $4$-momentum $p$, helicity $\lambda$, internal spin $s$, and mass $m$, i.e. there exists a $4$-vector single-particle basis $\epsilon^{\mu}_{s,\lambda}(p)$ (where $p$ is restricted by the on-shell condition $p^{2} = m^{2}$). We can construct these states explicitly by using the techniques explained in Subsection \ref{SS - Helicity - Operator}, wherein a standard $4$-momentum $k^{\mu}$ per Lorentz-invariant hypersurface is used to define any other state having $4$-momentum $p^{\mu}$ on that same hypersurface.

For a single-particle state with nonzero mass $m$, consider $\epsilon^{\mu}_{s,\lambda}(p)$ in the rest frame, when its $4$-momentum $p^{\mu}$ equals the standard $4$-momentum $k^{\mu} = (m,\vec{0}\,)$. In this frame, the helicity operator $\Lambda = (\vec{J}\cdot\vec{p}\,)/\sqrt{E^{2}-m^{2}}$ reduces to $J_{z} = J^{3}$, so finding helicity eigenstates $\epsilon^{\mu}_{\lambda}(k)$ amounts to finding $J_{z}$ eigenstates. To do this, note that the total angular momentum operator in this representation equals
\begin{align}
    \vec{J}^{\, 2} \equiv \vec{J}\cdot \vec{J} = \left(\hspace{-5 pt}\begin{tabular}{ c c c c}
    ${0}$ & ${0}$ & ${0}$ & ${0}$\\
    ${0}$ & ${+2}$ & ${0}$ & ${0}$\\
    ${0}$ & ${0}$ & ${+2}$ & ${0}$\\
    ${0}$ & ${0}$ & ${0}$ & ${+2}$
    \end{tabular}\hspace{-5 pt}\right)
\end{align}
The total angular momentum operator in general has eigenvalues of the form $j(j+1)$ and thus we immediately recognize that the time-time and space-space blocks of the $4$-vector representation of $\vec{J}^{\,2}$ encode the anticipated $j=0$ and $j=1$ representations respectively. Because $\vec{J}^{\,2}$ is block-diagonal in this way, we can directly construct projection operators $\mathfrak{P}_{0}(k)$ and $\mathfrak{P}_{1}(k)$ that (when acted on a generic complex $4$-vector in the rest frame) will isolate the $j=0$ and $j=1$ representations therein:
\begin{align}
    [{\mathfrak{P}_{0}(k)^{\mu}}_{\nu}] = \left(\hspace{-5 pt}\begin{tabular}{ c c c c}
    ${1}$ & ${0}$ & ${0}$ & ${0}$\\
    ${0}$ & ${0}$ & ${0}$ & ${0}$\\
    ${0}$ & ${0}$ & ${0}$ & ${0}$\\
    ${0}$ & ${0}$ & ${0}$ & ${0}$
    \end{tabular}\hspace{-5 pt}\right) \hspace{35 pt} [{\mathfrak{P}_{1}(k)^{\mu}}_{\nu}] = \left(\hspace{-5 pt}\begin{tabular}{ c c c c}
    ${0}$ & ${0}$ & ${0}$ & ${0}$\\
    ${0}$ & ${1}$ & ${0}$ & ${0}$\\
    ${0}$ & ${0}$ & ${1}$ & ${0}$\\
    ${0}$ & ${0}$ & ${0}$ & ${1}$
    \end{tabular}\hspace{-5 pt}\right)
\end{align}
It is useful to cast these into a Lorentz covariant form via the standard $4$-momentum $k^{\mu}$ and the Minkowski metric $\eta_{\mu\nu}$. Doing so yields
\begin{align}
    [{\mathfrak{P}_{0}(k)^{\mu}}_{\nu}] = \dfrac{k^{\mu}k_{\nu}}{m^{2}} \hspace{35 pt} [{\mathfrak{P}_{1}(k)^{\mu}}_{\nu}] = {\eta^{\mu}}_{\nu} - \dfrac{k^{\mu} k_{\nu}}{m^{2}}
\end{align}
In the rest frame of a massive particle, the total angular momentum and total internal spin operators are identical, and the $j=0$ and $j=1$ $4$-vector representations are equivalent to the spin-$0$ and spin-$1$ $4$-vector representations respectively. After applying a boost, we leave the rest frame, and the $j=0$ and $j=1$ representations mix; however, because internal spin is a Casimir operator of the Poincar\'{e} group, the above projection operators will (once transformed according to their Lorentz index structures) still project onto the spin-$0$ and spin-$1$ representations. We will consider the helicity eigenstates in a generic frame after we solve for them in the rest frame.

To find a spin-$j$ helicity eigenstate in the rest frame, we act the spin-$j$ projection operator $\mathfrak{P}_{j}(k)$ on a generic complex $4$-vector $\epsilon^{\mu}(k)$ and then solve for eigenstates of $\Lambda = J_{z}$. Specifically, we solve $J_{z} \, [\mathfrak{P}_{j}(k) \, \epsilon(k)] = \lambda \, [\mathfrak{P}_{j}(k) \, \epsilon(k)]$ for the helicities available to the specific choice of $j$. For example: when $j=0$, the only helicity available is $\lambda = 0$, so we aim to solve $J_{z} \, [\mathfrak{P}_{0}(k) \, \epsilon(k)] = 0$ for $[\mathfrak{P}_{0}(k) \, \epsilon(k)]$. Because
\begin{align}
    [\mathfrak{P}_{0}(k) \, \epsilon(k)] = \left(\hspace{-5 pt}\begin{tabular}{ c c c c}
    ${1}$ & ${0}$ & ${0}$ & ${0}$\\
    ${0}$ & ${0}$ & ${0}$ & ${0}$\\
    ${0}$ & ${0}$ & ${0}$ & ${0}$\\
    ${0}$ & ${0}$ & ${0}$ & ${0}$
    \end{tabular}\hspace{-5 pt}\right) \left(\hspace{-5 pt}\begin{tabular}{ c }
    ${\epsilon^{0}(k)}$\\
    ${\epsilon^{1}(k)}$\\
    ${\epsilon^{2}(k)}$\\
    ${\epsilon^{3}(k)}$
    \end{tabular}\hspace{-5 pt}\right) = \left(\hspace{-5 pt}\begin{tabular}{ c }
    ${\epsilon^{0}(k)}$\\
    $0$\\
    $0$\\
    $0$
    \end{tabular}\hspace{-5 pt}\right)
\end{align}
and
\begin{align}
    J_{z} \, [\mathfrak{P}_{0}(k) \, \epsilon(k)] =  \left(\hspace{-5 pt}\begin{tabular}{ c c c c}
    ${0}$ & ${0}$ & ${0}$ & ${0}$\\
    ${0}$ & ${0}$ & ${-i}$ & ${0}$\\
    ${0}$ & ${+i}$ & ${0}$ & ${0}$\\
    ${0}$ & ${0}$ & ${0}$ & ${0}$
    \end{tabular}\hspace{-5 pt}\right) \left(\hspace{-5 pt}\begin{tabular}{ c c c c}
    ${1}$ & ${0}$ & ${0}$ & ${0}$\\
    ${0}$ & ${0}$ & ${0}$ & ${0}$\\
    ${0}$ & ${0}$ & ${0}$ & ${0}$\\
    ${0}$ & ${0}$ & ${0}$ & ${0}$
    \end{tabular}\hspace{-5 pt}\right) \left(\hspace{-5 pt}\begin{tabular}{ c }
    ${\epsilon^{0}(k)}$\\
    ${\epsilon^{1}(k)}$\\
    ${\epsilon^{2}(k)}$\\
    ${\epsilon^{3}(k)}$
    \end{tabular}\hspace{-5 pt}\right) = \left(\hspace{-5 pt}\begin{tabular}{ c }
    $0$\\
    $0$\\
    $0$\\
    $0$
    \end{tabular}\hspace{-5 pt}\right)
\end{align}
we find $J_{z} \, [\mathfrak{P}_{0}(k) \, \epsilon_{0,0}(k)] = \lambda \, [\mathfrak{P}_{0}(k) \, \epsilon_{0,0}(k)] = 0$ for any 4-vector $\epsilon_{0,0}^{\mu}(k) = (\epsilon^{0}_{0,0}(k),\vec{0}\,) \propto k^{\mu}/m$, which determines the spin-$0$ $4$-vector representation up to its normalization and choice of phase. However, this representation has little use in actual quantum field theory calculations because there exists a more succinct Lorentz covariant spin-$0$ representation: the Lorentz scalar $\epsilon_{0,0}(k) = 1$. Thus, we consider the spin-$0$ part of this representation no further, and simply write $\epsilon^{\mu}_{\lambda}(p)$ instead of $\epsilon^{\mu}_{1,\lambda}(p)$ for the spin-$1$ representation, as is conventional.

The process of finding the spin-$1$ helicity eigenstates in the rest frame proceeds similarly: we aim to solve $J_{z} \, [\mathfrak{P}_{1}(k) \, \epsilon(k)] = \lambda \, [\mathfrak{P}_{1}(k) \, \epsilon(k)]$ for helicities $\lambda \in \{ -1, 0, +1 \}$. Note that
\begin{align}
    [\mathfrak{P}_{1}(k) \, \epsilon(k)] = \left(\hspace{-5 pt}\begin{tabular}{ c c c c}
    ${0}$ & ${0}$ & ${0}$ & ${0}$\\
    ${0}$ & ${1}$ & ${0}$ & ${0}$\\
    ${0}$ & ${0}$ & ${1}$ & ${0}$\\
    ${0}$ & ${0}$ & ${0}$ & ${1}$
    \end{tabular}\hspace{-5 pt}\right) \left(\hspace{-5 pt}\begin{tabular}{ c }
    ${\epsilon^{0}(k)}$\\
    ${\epsilon^{1}(k)}$\\
    ${\epsilon^{2}(k)}$\\
    ${\epsilon^{3}(k)}$
    \end{tabular}\hspace{-5 pt}\right) = \left(\hspace{-5 pt}\begin{tabular}{ c }
    $0$\\
    ${\epsilon^{1}(k)}$\\
    ${\epsilon^{2}(k)}$\\
    ${\epsilon^{3}(k)}$
    \end{tabular}\hspace{-5 pt}\right)
\end{align}
and
\begin{align}
    J_{z} \, [\mathfrak{P}_{1}(k) \, \epsilon(k)] =  \left(\hspace{-5 pt}\begin{tabular}{ c c c c}
    ${0}$ & ${0}$ & ${0}$ & ${0}$\\
    ${0}$ & ${0}$ & ${-i}$ & ${0}$\\
    ${0}$ & ${+i}$ & ${0}$ & ${0}$\\
    ${0}$ & ${0}$ & ${0}$ & ${0}$
    \end{tabular}\hspace{-5 pt}\right) \left(\hspace{-5 pt}\begin{tabular}{ c c c c}
    ${0}$ & ${0}$ & ${0}$ & ${0}$\\
    ${0}$ & ${1}$ & ${0}$ & ${0}$\\
    ${0}$ & ${0}$ & ${1}$ & ${0}$\\
    ${0}$ & ${0}$ & ${0}$ & ${1}$
    \end{tabular}\hspace{-5 pt}\right) \left(\hspace{-5 pt}\begin{tabular}{ c }
    ${\epsilon^{0}(k)}$\\
    ${\epsilon^{1}(k)}$\\
    ${\epsilon^{2}(k)}$\\
    ${\epsilon^{3}(k)}$
    \end{tabular}\hspace{-5 pt}\right) = \left(\hspace{-5 pt}\begin{tabular}{ c }
    $0$\\
    $-i\epsilon^{2}(k)$\\
    $+i\epsilon^{1}(k)$\\
    $0$
    \end{tabular}\hspace{-5 pt}\right)
\end{align}
Thus, when $\lambda = 0$, we require $(0,-i\epsilon^{2},+i\epsilon^{1},0) = (0,0,0,0)$, such that $\epsilon_{0}^{\mu}(k) \propto (0,0,0,\epsilon^{3}(k))$. It is conventional to set the magnitude of $\epsilon_{0}(k)$, which equals
\begin{align}
    \epsilon_{0}(k) \cdot \epsilon_{0}(k) = -\epsilon^{3}(k)^{2}
\end{align}
to $-1$, and to set the remaining phase such that $\epsilon_{0}^{\mu}(k) = (0,0,0,1)$.

Although we could find the $\lambda = \pm1$ solutions by solving a similar eigenvalue problem, it is quicker to use the ladder operators
\begin{align}
    J_{\pm} = J_{x}\pm iJ_{y} = \left(\hspace{-5 pt}\begin{tabular}{ c c c c}
    ${0}$ & ${0}$ & ${0}$ & ${0}$\\
    ${0}$ & ${0}$ & ${0}$ & ${\mp 1}$\\
    ${0}$ & ${0}$ & ${0}$ & ${-i}$\\
    ${0}$ & ${\pm 1}$ & ${+i}$ & ${0}$
    \end{tabular}\hspace{-5 pt}\right)
\end{align}
This has the added benefit of automatically setting phases and normalizations in a way consistent with our existing assumptions. Using the raising/lowering formula Eq. \eqref{2230} (and noting $\sqrt{(j\mp m)(j\pm m +1)} \rightarrow \sqrt{j(j+1)} = \sqrt{2}$ in this case), we calculate
\begin{align}
    [\epsilon_{\pm1}(k)^{\mu}] = \dfrac{1}{\sqrt{2}} \left(\hspace{-5 pt}\begin{tabular}{ c c c c}
    ${0}$ & ${0}$ & ${0}$ & ${0}$\\
    ${0}$ & ${0}$ & ${0}$ & ${\mp 1}$\\
    ${0}$ & ${0}$ & ${0}$ & ${-i}$\\
    ${0}$ & ${\pm 1}$ & ${+i}$ & ${0}$
    \end{tabular}\hspace{-5 pt}\right) \matrixda{0}{0}{0}{1} =  \dfrac{1}{\sqrt{2}} \matrixda{0}{\mp 1}{-i}{0}
\end{align}
and in doing so have found the final rest frame helicity eigenstates.

All together, the polarization vectors $\{\epsilon^{\mu}_{-1}(k),\epsilon^{\mu}_{0}(k),\epsilon^{\mu}_{+1}(k)\}$ form the desired spin-$1$ representation in the rest frame. Explicit calculation reveals they are orthonormal and transverse,
\begin{align}
    \epsilon_{\lambda}(k)^{*} \cdot \epsilon_{\lambda^{\prime} }(k) = -\delta_{\lambda,\lambda^{\prime}} \hspace{35 pt} k \cdot \epsilon_{\lambda}(k) = 0
\end{align}
and as a basis for the spin-$1$ representation they naturally resolve the projection operator $\mathfrak{P}_{1}(k)$, which is the identity on the spin-$1$ subspace:
\begin{align}
    [\mathfrak{P}_{1}(k)^{\mu\nu}] = -\sum_{\lambda = -1}^{+1} \epsilon^{\mu}_{\lambda}(k)^{*} \epsilon^{\nu}_{\lambda}(k)
\end{align}
This completes the derivation of the massive spin-$1$ representation in the rest frame.

To obtain this representation in all other frames, we apply the standard Lorentz transformation $\Lambda_{k\rightarrow p} = R(\phi,\theta)\, B_{z}(\beta_{k\rightarrow p})$ (defined in Eq. \eqref{1266} of Section \ref{S - Helicity}) to each polarization vector $\epsilon^{\mu}_{\lambda}(k)$, and define
\begin{align}
    \epsilon^{\mu}_{\lambda}(p) \equiv {(\Lambda_{k\rightarrow p})^{\mu}}_{\nu} \, \epsilon^{\mu}_{\lambda}(k)
\end{align}
As mentioned previously, the internal spin of a particle corresponds to a Casimir operator of the Lorentz group and thus is invariant under the Lorentz transformation $\Lambda_{k\rightarrow p}$. In the $4$-vector representation, the standard (massive) Lorentz transformation equals
\begin{align}
    [{(\Lambda_{k\rightarrow p})^{\mu}}_{\nu}] = \left(\hspace{-5 pt}\begin{tabular}{ c c c c}
    ${1}$ & ${0}$ & ${0}$ & ${0}$\\
    ${0}$ & ${c_{\phi}^{2}c_{\theta} + s_{\phi}^{2}}$ & ${c_{\phi}s_{\phi}(c_{\theta}-1)}$ & ${c_{\phi}s_{\theta}}$\\
    ${0}$ & ${c_{\phi}s_{\phi}(c_{\theta}-1)}$ & ${c_{\phi}^{2} + c_{\theta} s_{\phi}^{2}}$ & ${s_{\phi}s_{\theta}}$\\
    ${0}$ & ${-c_{\phi}s_{\theta}}$ & ${-s_{\phi}s_{\theta}}$ & ${c_{\theta}}$
    \end{tabular}\hspace{-5 pt}\right) \, \dfrac{1}{m} \left(\hspace{-5 pt}\begin{tabular}{ c c c c}
    ${E}$ & ${0}$ & ${0}$ & ${\vec{p}}$\\
    ${0}$ & ${1}$ & ${0}$ & ${0}$\\
    ${0}$ & ${0}$ & ${1}$ & ${0}$\\
    ${\vec{p}}$ & ${0}$ & ${0}$ & ${E}$
    \end{tabular}\hspace{-5 pt}\right) \, 
\end{align}
such that the spin-$1$ polarization tensors equal, in a generic frame,
\begin{align}
    [\epsilon^{\mu}_{\pm 1}(p)] = \pm \dfrac{e^{\pm i \phi}}{\sqrt{2}} \matrixda{0}{-c_{\theta}c_{\phi} \pm is_{\phi}}{-c_{\theta}s_{\phi}\mp i c_{\phi}}{s_{\theta}} \hspace{35 pt} [\epsilon^{\mu}_{0}(p)] = \dfrac{1}{m} \matrixda{|\vec{p}\,|}{E_{\vec{p}}\, c_{\phi} s_{\theta}}{E_{\vec{p}}\, s_{\phi} s_{\theta}}{E_{\vec{p}}\, c_{\theta}} = \dfrac{1}{m} \matrixba{|\vec{p}\,|}{E_{\vec{p}}\,\hat{p}} \label{C1Spin1Polarizations}
\end{align}
where $\hat{p} = \hat{z}$ when $\vec{p} = \vec{0}$ per the helicity eigenstate convention established in Subsection \ref{SS - Helicity - Operator}. Note that the helicity-zero polarization tensor grows like $\mathcal{O}(E)$ whereas the others do not depend on energy at all. Because of Lorentz covariance, the spin-$1$ polarization vectors $\epsilon^{\mu}_{\lambda}(p)$ retain their rest frame properties (orthogonal, transverse),
\begin{align}
    \epsilon_{\lambda}(p)^{*} \cdot \epsilon_{\lambda^{\prime} }(p) = -\delta_{\lambda,\lambda^{\prime}} \hspace{35 pt} p \cdot \epsilon_{\lambda}(p) = 0
\end{align}
and the spin-$1$ projection operator becomes
\begin{align}
    [\mathfrak{P}_{1}(k)^{\mu\nu}] = -\sum_{\lambda = -1}^{+1} \epsilon^{\mu}_{\lambda}(k)^{*} \epsilon^{\nu}_{\lambda}(k) = \eta^{\mu\nu} - \dfrac{p^{\mu} p^{\nu}}{m^{2}}
\end{align}
This completes the derivation of the massive spin-1 polarization vectors.

From these expressions, we can directly calculate the Jacob-Wick 2nd particle conversion factor $\xi_{\lambda}(\phi)$ from Eq. \eqref{JWconversiondefinition}. Because
\begin{align}
    [{R(\phi,-\pi)^{\mu}}_{\nu} {R(0,\pi)^{\nu}}_{\rho}] &= \left(\hspace{-5 pt}\begin{tabular}{ c c c c}
    ${1}$ & ${0}$ & ${0}$ & ${0}$\\
    ${0}$ & ${+c_{2\phi}}$ & ${-s_{2\phi}}$ & ${0}$\\
    ${0}$ & ${+s_{2\phi}}$ & ${+c_{2\phi}}$ & ${0}$\\
    ${0}$ & ${0}$ & ${0}$ & ${1}$
    \end{tabular}\hspace{-5 pt}\right)
\end{align}
we find
\begin{align}
    \bigg[(-1)^{1-\lambda} \, {R(\phi,-\pi)^{\mu}}_{\nu} \, {R(0,\pi)^{\nu}}_{\rho} \bigg]\, \epsilon^{\rho}_{\lambda}(p_{z}) = (-1)^{1-\lambda} \, e^{-2\lambda i\phi} \, \epsilon^{\rho}_{\lambda}(p_{z})
\end{align}
such that $\xi_{\lambda}(\phi) = (-1)^{1-\lambda}e^{-2\lambda i\phi}$.

The derivation of the massless spin-1 polarization vectors follows the same trajectory as the massive case, but now there is no rest frame and their helicities are restricted to $\lambda = \pm1$. However, we already have helicity eigenstates corresponding to $\lambda = \pm1$ which work in any frame, and sure enough the existing polarization vectors $\epsilon^{\mu}_{\pm 1}(p)$ are admissible helicity eigenstates for massless spin-1 particles. It is possible a relative parity factor $\eta$ may occur between the two massless helicity states because they are not directly related via ladder operators, depending on the particle species in question. In this representation, the reflection operator equals $Y = \text{Diag}(1,1,-1,1)$, such that
\begin{align}
    \begin{cases}
    {Y^{\mu}}_{\nu}\,\epsilon^{\nu}_{0}(p_{z}) = \epsilon^{\mu}_{0}(p_{z})\\
    {Y^{\mu}}_{\nu}\,\epsilon^{\nu}_{\pm 1}(p_{z}) = - \epsilon^{\mu}_{\mp 1}(p_{z})
    \end{cases} \hspace{15 pt}\implies\hspace{15 pt} {Y^{\mu}}_{\nu} \, \epsilon^{\nu}_{\lambda}(p_{z}) = -(-1)^{1-\lambda} \epsilon^{\mu}_{-\lambda}(p_{z})
\end{align}
Having completed our derivation of the spin-$1$ polarization vectors, let us now derive the spin-2 polarization tensors.

As described in Subsection \ref{SS - Clebsch Gordan}, any two angular momentum representations can be combined to form a new angular momentum representation via the Clebsch-Gordan coefficients. Thus, we can combine two copies of our (massive or massless) spin-1 polarization vectors $\epsilon^{\mu}_{\lambda}(p)$ and thereby obtain a Lorentz-covariant representation of spin-2 particles in the form of polarization tensors $\epsilon^{\mu\nu}_{\lambda}(p)$. Explicitly, these spin-2 polarization tensors equal, using Eq. \eqref{1plus1is2},
\begin{align}
    \epsilon_{\pm 2}^{\mu\nu}(p) &= \epsilon_{\pm1}^{\mu}(p)\, \epsilon_{\pm 1}^{\nu}(p)~,\\
    \epsilon_{\pm 1}^{\mu\nu}(p) &= \dfrac{1}{\sqrt{2}}\left[\epsilon_{\pm 1}^{\mu}(p)\,\epsilon_{0}^{\nu}(p) + \epsilon_{0}^{\mu}(p)\,\epsilon_{\pm 1}^{\nu}(p)\right]~\label{ep21}\\
    \epsilon_{0}^{\mu\nu} &= \dfrac{1}{\sqrt{6}}\bigg[\epsilon^{\mu}_{+1}(p)\, \epsilon^{\nu}_{-1}(p) + \epsilon^{\mu}_{-1}(p)\,\epsilon^{\nu}_{+1}(p) + 2\epsilon^{\mu}_{0}(p)\,\epsilon^{\nu}_{0}(p)\bigg]~, \label{C1Spin2Polarizations}
\end{align}
where the massive case has access to all five helicity states ($\lambda = \pm2, \pm1, 0$) and the massless case only has access to two ($\lambda = \pm 2$). Via the properties of the polarization vectors that compose them, each polarization tensor is traceless, symmetric, and transverse: 
\begin{align}
    \eta_{\mu\nu}\epsilon^{\mu\nu}_{\lambda}(p) = 0\hspace{35 pt} \epsilon^{\mu\nu}_{\lambda}(p) = \epsilon^{\nu\mu}_{\lambda}(p) \hspace{35 pt} p_{\mu} \epsilon^{\mu\nu}_{\lambda}(p) = 0
\end{align}
By applying the appropriate generalization of the helicity reflection operator ${Y^{\mu\nu}}_{\rho\sigma} = {Y^{\mu}}_{\rho}{Y^{\nu}}_{\sigma}$, we find
\begin{align}
    {Y^{\mu\nu}}_{\rho\sigma} \, \epsilon^{\rho\sigma}_{\lambda}(p_{z}) = (-1)^{2-\lambda}\epsilon^{\mu\nu}_{-\lambda}(p_{z})
\end{align}
Finally, the spin-2 Jacob-Wick 2nd particle conversion factor can be determined by applying the spin-1 conversion factor to each spin-1 polarization vector in the definitions of the spin-2 polarization tensor, thereby yielding $\xi_{\lambda}(\phi) = (-1)^{2-\lambda}e^{-2\lambda i\phi}$.

\subsection{Quadratic Lagrangians and Propagators}
This chapter has largely focused on the construction of external particle states as eigenstates of $4$-momentum and helicity. In order to calculate matrix elements describing scattering processes between these external states, we must encode those external states into quantum fields which then compose carefully-chosen Lagrangians. The quadratic terms of a Lagrangian will determine the masses and spins of the particles encoded within the fields, whereas higher-order terms determine interactions between various particles.

Perhaps the simplest field (and Lagrangian) corresponds to a spin-0 massless particle. A scalar field $\hat{r}(x)$ will encode (real) massless spin-0 particles if our overall Lagrangian possesses the following quadratic terms:
\begin{align}
    \mathcal{L}^{(s=0)}_{\text{massless}} &\equiv \dfrac{1}{2} (\partial_{\mu} \hat{r})^{2} \label{Ls0massless1}
\end{align}
To derive the propagator associated with this Lagrangian,\footnote{This is a trick for getting the right answer; typically the propagator should be more carefully derived.} we
\begin{itemize}
    \item[$\bullet$] Rewrite the Lagrangian into the form $\hat{r}\,(D\,\hat{r})$.
    \item[$\bullet$] Fourier transform to $4$-momentum space, effectively replacing $\partial_{\mu}$ with $-iP_{\mu}$, where $P_{\mu}$ is the $4$-momentum carried through the propagator (such that $(\partial_{\mu}^{2} \hat{r}) \mapsto - P^{2}\tilde{\hat{r}}$),
    \item[$\bullet$] take the functional derivative with respect to the Fourier-transformed field $\tilde{\hat{r}}$ twice to isolate $\tilde{D}(P)$,
    \item[$\bullet$] invert $\tilde{D}(P)$ and then multiply by $i$, and in doing so effectively solve the momentum space equation $\tilde{D}(P)\,\Delta(P) = i$ for the momentum space propagator $\Delta(P)$.
\end{itemize}
Applying this procedure to Eq. \eqref{Ls0massless1} yields
\begin{align}
    \mathcal{L}^{(s=0)}_{\text{massless}} \hspace{15 pt}\rightarrow\hspace{15 pt} \hat{r}\bigg[-\dfrac{1}{2}\partial^{2}\bigg]\hat{r}\hspace{15 pt}\rightarrow\hspace{15 pt} \dfrac{1}{2} P^{2} \hat{r}^{\,2} \hspace{15 pt}\rightarrow\hspace{15 pt} P^{2} \hspace{15 pt}\rightarrow\hspace{15 pt} \dfrac{i}{P^{2}} \label{masslessspin0propagator1}
\end{align}
and, thus, we find the (momentum space) massless spin-0 propagator equals
\begin{center}
    \begin{tikzpicture}
    \begin{feynman}[medium]
        \vertex (n1) at (-1,0) {};
        \vertex (a1) at (-0.7,0.35) {};
        \vertex (b1) at (0.7,0.35) {};
        \vertex (n2) at (1,0) {};
        
        \diagram* {
            (n1) -- [] (n2),
            (a1) -- [arrow, edge label = $P$] (b1),
        };
    \end{feynman}
    \end{tikzpicture} $= \dfrac{i}{P^{2}}$
\end{center}

If we instead desire a (real) {\it massive} spin-0 field $\hat{r}(x)$, we can add a mass term $-(1/2)M^{2} \hat{r}^{\,2}$ to the massless spin-0 Lagrangian:
\begin{align}
    \mathcal{L}^{(s=0)}_{\text{massive}} &\equiv \dfrac{1}{2} (\partial_{\mu} \hat{r})^{2} - \dfrac{1}{2}M^{2} \hat{r}^{2} \label{Ls0massive1}
\end{align}
in which case the same procedure instead yields
\begin{center}
    \begin{tikzpicture}
    \begin{feynman}[medium]
        \vertex (n1) at (-1,0) {};
        \vertex (a1) at (-0.7,0.35) {};
        \vertex (b1) at (0.7,0.35) {};
        \vertex (n2) at (1,0) {};
        
        \diagram* {
            (n1) -- [] (n2),
            (a1) -- [arrow, edge label = $P$] (b1),
        };
    \end{feynman}
    \end{tikzpicture} $= \dfrac{i}{P^{2} - M^{2}}$
\end{center}

As derived by Fierz and Pauli \cite{Fierz:1939ix,fierz2017relativistic}, massless and massive spin-2 particles can be embedded in a symmetric rank-2 Lorentz tensor field $\hat{h}_{\mu\nu}(x)$ which is transverse and traceless for on-shell excitations. Using a tensor field is convenient because it possesses manifest Lorentz covariance with which we can directly construct Lorentz scalar Lagrangians. Because index symmetry, transversality, and tracelessness reduce its otherwise $4^{2}=16$ available degrees of freedom by $6$, $4$, and $1$ respectively, the spin-2 field $\hat{h}_{\mu\nu}$ only propagates five degrees of freedom, precisely the correct number to describe a massive spin-2 particle. Unfortunately, these constraints are still insufficient for the description of a massless spin-2 particle, which requires only two propagating degrees of freedom. As a result, $\hat{h}_{\mu\nu}$ possesses gauge freedoms when utilized for a massless spin-2 particle. Thus, although we may write the canonical massless spin-2 quadratic Lagrangian as
\begin{align}
    \mathcal{L}^{(s=2)}_{\text{massless}} &\equiv (\partial \hat{h})_{\mu} (\partial^{\mu} \hat{h}) - (\partial\hat{h})_{\mu}^{2} +\dfrac{1}{2} (\partial_{\mu} \hat{h}_{\nu\rho})^2 - \dfrac{1}{2} (\partial_{\mu} \hat{h})^2 \label{Ls2massless1}
\end{align}
we cannot directly apply the previous procedure to obtain the massless spin-2 propagator: the differential operator defined in Eq. \eqref{Ls2massless1} is not invertible due to gauge redundancies. Specifically, this manifests as invariance of the massless spin-2 Lagrangian under the following gauge transformation:
\begin{align}
    \hat{h}_{\mu\nu} \hspace{15 pt}\longrightarrow\hspace{15 pt} \hat{h}_{\mu\nu} + (\partial_{\mu}\epsilon_{\nu}) + (\partial_{\nu}\epsilon_{\mu}) \label{hmunugaugefreedom}
\end{align}
for a generic 4-vector field $\epsilon_{\mu}(x)$. In fact, Eq. \eqref{Ls2massless1} is the only (properly normalized) combination of quadratic-level kinetic terms for $\hat{h}_{\mu\nu}$ that is invariant under this gauge transformation, such that we could have started by demanding invariance under transformations of the form Eq. \eqref{hmunugaugefreedom} and thereby derived $\mathcal{L}^{(s=2)}_{\text{massless}}$.

In order to invert Eq. \eqref{Ls2massless1} and obtain a massless spin-2 propagator, we must somehow break this aforementioned gauge invariance. This can be done in a multitude of ways, whether it be by employing a specific gauge condition or adding a gauge-fixing term to the Lagrangian. A popular gauge choice is the harmonic gauge, which is defined by setting
\begin{align}
    \partial^{\mu}\hat{h}^{(0)}_{\mu\nu} = \tfrac{1}{2} \partial_{\nu} \ltr \hat{h}^{(0)} \rtr
\end{align}
This isolates a specific gauge orbit, thereby breaking the gauge invariance of the quadratic Lagrangian Eq. \eqref{Ls2massless1} and allowing it to be inverted into a propagator. However, this dissertation does not use harmonic gauge (or any other gauge condition), instead opting to add a gauge-fixing term $\mathcal{L}_{\text{gf}}$ to the massless spin-2 Lagrangian. Specifically, we employ the de Donder gauge, which has a gauge-fixing term
\begin{align}
    \mathcal{L}_{\text{gf}} &\equiv -\bigg(\partial^{\mu}\hat{h}_{\mu\nu} - \dfrac{1}{2} \partial_{\nu} \hat{h} \bigg)^{2}
\end{align}
Rather than isolate any single gauge orbit, de Donder gauge averages over a continuum of gauge orbits. This averaging is weighted in favor of the harmonic gauge condition, the bias of which successfully breaks the troublesome gauge invariance of Eq. \eqref{Ls2massless1}. The resulting de Donder gauge massless spin-2 propagator equals
\begin{center}
    \begin{tikzpicture}
    \begin{feynman}[medium]
        \vertex (n1) at (-1,0) {$\mu\nu$};
        \vertex (a1) at (-0.7,0.35) {};
        \vertex (b1) at (0.7,0.35) {};
        \vertex (n2) at (1,0) {$\rho\sigma$};
        
        \diagram* {
            (n1) -- [photon] (n2),
            (n1) -- [] (n2),
            (a1) -- [arrow, edge label = $P$] (b1),
        };
    \end{feynman}
    \end{tikzpicture} $= \dfrac{i B_{0}^{\mu\nu,\rho\sigma}}{P^{2}}$
\end{center}
where
\begin{align}
    B_{0}^{\mu\nu,\rho\sigma} &\equiv \dfrac{1}{2}\bigg[\eta^{\mu\rho}\eta^{\nu\sigma}+\eta^{\mu\sigma}\eta^{\nu\rho} - \eta^{\mu\nu}\eta^{\rho\sigma}\bigg]
\end{align}

In the same way that we went from massless to massive spin-0 Lagrangian, the massive spin-2 Lagrangian is obtained from the massless spin-2 Lagrangian (without the gauge-fixing term) by adding a mass term. As it turns out, there is only one non-kinetic quadratic combination of the field $\hat{h}_{\mu\nu}$ which simultaneously yields a propagator pole at $P^{2} = M^{2}$ and does not introduce ghosts \cite{Fierz:1939ix}. This combination defines the Fierz-Pauli mass terms,
\begin{align}
    \mathcal{L}_{\text{FP}}(m,\hat{h}) \equiv m^{2}\bigg[\dfrac{1}{2} \hat{h}^{2} - \dfrac{1}{2} \ltr \hat{h}\hat{h} \rtr \bigg]
\end{align}
which when added to the massless spin-2 Lagrangian yields the canonical massive spin-2 quadratic Lagrangian:
\begin{align}
    \mathcal{L}^{(s=2)}_{\text{massive}} &\equiv \mathcal{L}^{(s=2)}_{\text{massless}} + m^{2}\bigg[\dfrac{1}{2} \hat{h}^{2} - \dfrac{1}{2} \ltr \hat{h}\hat{h} \rtr \bigg] \label{Ls2massive1}
\end{align}
Because the Fierz-Pauli mass term breaks the gauge invariance of the massless Lagrangian, all five degrees of freedom in the symmetric traceless field $\hat{h}_{\mu\nu}$ can propagate, which is in agreement with the five helicity states we expect from a massive spin-2 particle. This also allows us to invert $\mathcal{L}^{(s=2)}_{\text{massive}}$ and obtain the massive spin-2 propagator:
\begin{center}
    \begin{tikzpicture}
    \begin{feynman}[medium]
        \vertex (n1) at (-1,0) {$\mu\nu$};
        \vertex (a1) at (-0.7,0.35) {};
        \vertex (b1) at (0.7,0.35) {};
        \vertex (n2) at (1,0) {$\rho\sigma$};
        
        \diagram* {
            (n1) -- [photon] (n2),
            (n1) -- [] (n2),
            (a1) -- [arrow, edge label = $P$] (b1),
        };
    \end{feynman}
    \end{tikzpicture} $= \dfrac{i B^{\mu\nu,\rho\sigma}}{P^{2} - M^{2}}$
\end{center}
where
\begin{align}
    B^{\mu\nu,\rho\sigma} &= \dfrac{1}{2}\bigg[\overline{B}^{\mu\rho}\overline{B}^{\nu\sigma}+\overline{B}^{\mu\sigma}\overline{B}^{\nu\rho} - \dfrac{2}{3}\overline{B}^{\mu\nu}\overline{B}^{\rho\sigma}\bigg] \label{Bmunu}
\end{align}
This is the last piece of four-dimensional quantum field theory information that we require for calculating the desired scattering amplitudes. In the next chapter, we introduce the necessary information about five-dimensional field theories, including the machinery of general relativity machinery and the definition of the Randall Sundrum 1 model.
\chapter{The 5D RS1 Model} \label{C - 5D RS1}

\section{Chapter Summary}
The previous chapter introduced important definitions and conventions regarding 4D quantum field theory, including discussions of 2-to-2 scattering, helicity eigenstates, and partial wave unitarity constraints. It also defined the twice-squared bracket notation which is used often throughout the remainder of this dissertation: given a collection of spin-2 fields $\{\hat{h}^{(1)},\hat{h}^{(2)},\dots,\hat{h}^{(n)}\}$, we define the $\ltr\cdots\rtr_{\alpha\beta}$ and $\ltr\cdots\rtr$ symbols according to
\begin{align}
    \ltr \hat{h}^{(1)} \hat{h}^{(2)} \cdots \hat{h}^{(n)} \rtr_{\alpha\beta} &\equiv \hat{h}^{(1)}_{\alpha\mu_{1}}\, \eta^{\mu_{1}\mu_{2}}\, \hat{h}^{(2)}_{\mu_{1}\mu_{2}}\, \eta^{\mu_{2}\mu_{3}} \,\cdots \, \hat{h}^{(n)}_{\mu_{n}\beta}\\
    \ltr \hat{h}^{(1)} \hat{h}^{(2)} \cdots \hat{h}^{(n)} \rtr &\equiv \eta^{\alpha\beta} \, \ltr \hat{h}^{(1)} \hat{h}^{(2)} \cdots \hat{h}^{(n)} \rtr_{\alpha\beta}
\end{align}
such that, for example, $\ltr 1 \rtr_{\alpha\beta} = \eta_{\alpha\beta}$ and $\ltr 1\rtr = 4$. The previous chapter also established the use of tildes to denote inverse quantities, e.g. $\tilde{A} \equiv A^{-1}$ for an invertible matrix $A$.

This chapter introduces important definitions and conventions regarding general relativity, as well as introducing the Randall-Sundrum 1 (RS1) model which is the primary theory considered in this dissertation. It also introduces several original results, including an updated 5D weak field expanded (WFE) RS1 Lagrangian, which we originally published in Appendix A of \cite{Chivukula:2020hvi} using a different form of the Einstein-Hilbert Lagrangian. We also demonstrate for the first time that all terms in the 5D WFE RS1 Lagrangian which are proportional to $(\partial_{\varphi}^{2}|\varphi|)$ and $(\partial_{\varphi}|\varphi|)$ can be repackaged into a physically-irrelevant total derivative.
\begin{itemize}
    \item[$\bullet$] Section \ref{S - GR Intro} establishes our tensor conventions, including the covariant derivative, Riemann curvature, Ricci scalar, and Einstein-Hilbert Lagrangian; rewrites the Einstein-Hilbert Lagrangian into a more convenient form; and isolates the extra-dimensional graviton resulting from the Einstein-Hilbert Lagrangian.
    \item[$\bullet$] Section \ref{S - Intro RS1} motivates the Randall-Sundrum 1 background metric and Lagrangian by considering what modifications are required in order to accommodate a nonzero extrinsic curvature at its branes. The background metric is then perturbed to generate the full 5D RS1 model, with a metric that depends on 5D fields $\hat{h}_{\mu\nu}(x,y)$ and $\hat{r}(x)$. The final subsection demonstrates that terms proportional to $(\partial_{\varphi}^{2}|\varphi|)$ and $(\partial_{\varphi}|\varphi|)$ combine to form physically-irrelevant total derivatives, and then introduces a new term $\Delta\mathcal{L}$ to the 5D RS1 model Lagrangian to automate the removal of such terms.
    \item[$\bullet$] Section \ref{AppendixWFE} weak field expands the 5D RS1 model Lagrangian as a series in the 5D fields $\hat{h}_{\mu\nu}(x,y)$ and $\hat{r}(x)$ to second order in the 5D coupling, $\mathcal{O}(\kappa_{\text{5D}}^{2})$. Each term in the expansion can be classified as an A-type or B-type term, depending on if it contains two four-dimensional or two extra-dimensional derivatives respectively. This 5D weak field expanded (WFE) RS1 Lagrangian is the principal result of this chapter, and updates the expressions we originally published in Appendix A of \cite{Chivukula:2020hvi}.
    \item[$\bullet$] Section \ref{GRAppendices - WFE} is an appendix which details certain formulas used in the weak field expansion procedure.
\end{itemize}

\section{Motivations, Definitions, and Conventions} \label{S - GR Intro}

\subsection{Revisiting the Metric}
The previous chapter explored the consequences of demanding that the speed of light be globally conserved between inertial reference frames in flat 4D spacetime, i.e. that every finite spacetime interval that is light-like according to one observer is also light-like to all other observers. This led us to the Poincar\'{e} group and eventually the characterization of external particles on that spacetime. This chapter generalizes those assumptions.

Instead of a 4D spacetime with coordinates $[x^{\mu}] = (x^{0},x^{1},x^{2},x^{3})$, we consider an $X$-dimensional spacetime with coordinates $[x^{M}] = (x^{0},x^{1},x^{2},x^{3},x^{5},\dots,x^{X})$ (for example, when $X=5$, $[x^{M}] = (x^{0},x^{1},x^{2},x^{3},x^{5})$). In the previous chapter, the 4D Minkowski metric $[\eta_{\mu\nu}]=\text{Diag}(+1,-1,-1,-1)$ defined 4-vector inner products, including the invariant spacetime interval $ds^{2} = \eta_{\mu\nu}\, dX^{\mu}\, dX^{\nu}$. Given a specific choice of coordinates, the $X$-dimensional metric $G$ also defines an invariant spacetime interval, this time defined as
\begin{align}
    ds^{2} \equiv G_{MN}\, dX^{M} \, dX^{N}
\end{align}
for any infinitesimal displacement $dX$. By assumption, the tensor $G_{MN}$ is symmetric and nondegenerate. Consequently, the matrix $[G_{MN}]$ is invertible, with its inverse $[\tilde{G}^{MN}]$ defined such that the components satisfy $\tilde{G}^{MN} G_{NP} = \delta_{M,P}$. We still use the ``mostly-minus" convention, which in this framework means that $G_{MN}$ has a single positive eigenvalue among otherwise negative eigenvalues regardless of the specific coordinates we use.

Different choices of coordinates correspond to different observers, and a key feature of general relativity is that observer-independent quantities should be invariant under coordinate transformations, which are also known as diffeomorphisms. In this sense, the group of $X$-dimensional diffeomorphisms comprise a symmetry group on $X$-dimensional spacetime. In particular, the speed of light remains an invariant between all inertial reference frames in this framework, although only locally: if an invariant interval $ds^{2}$ vanishes for one observer, then it must vanish for all other observers as well, such that $ds^{2} = 0$ is diffeomorphism invariant and we may meaningfully declare the corresponding infinitesimal displacement $dX$ to be light-like. Finite light-like displacements typically do not exhibit the same frame invariance.

Just as we did in the last chapter, we can generalize beyond infinitesimal spacetime displacements, and declare a generic spacetime vector $v$ with components $[v^{M}] = (v^{0},\dots,v^{X})$ as light-like, time-like, or space-like based on the value of its magnitude with respect to the metric $G$, i.e. whether the inner product $G_{MN} v^{M} v^{N}$ vanishes, is positive, or is negative respectively. Given a pair of spacetime vectors $v$ and $w$, we also define the inner product $G_{MN} v^{M} w^{N}$.

In the last chapter, the metric $G_{MN}$ was assumed to equal $\eta_{\mu\nu}$ and we only considered linear transformations that mapped $\eta_{\mu\nu}$ to itself. We now relax those requirements: the metric $G_{MN}$ can be a nontrivial function of the coordinates $x^{M}$, and we consider (possibly nonlinear) coordinate transformations that map $x^{M}$ to new coordinates $\overline{x}^{\overline{M}}$ which thereby map $G_{MN}$ to a new form $\overline{G}_{\overline{M}\overline{N}}$. This is the topic of the next subsection.

\subsection{Diffeomorphisms, Tensors}
A diffeomorphism is a transformation that maps the coordinates of one reference frame to the coordinates of another reference frame. In order to locally preserve the speed of light between any two reference frames, we demand $ds^{2}$ be invariant under diffeomorphisms. This implies how the metric $G$ must be transformed. Specifically, if $G_{MN}$ describes spacetime in coordinates $x^{M}$ and $\overline{G}_{\overline{M}\overline{N}}$ describes spacetime in coordinates $\overline{x}^{\overline{M}}$, then the infinitesimal displacements at an equivalent point in either description are related according to
\begin{align}
    d\overline{x}^{\overline{M}} =  \overline{\mathfrak{D}}{{}^{\overline{M}}}_{M} dx^{M}\hspace{20 pt}\text{ where }\hspace{20 pt}\overline{\mathfrak{D}}{{}^{\overline{M}}}_{M} \equiv \left(\dfrac{\partial \overline{x}^{\overline{M}}}{\partial x^{M}}\right)
\end{align}
We can similarly convert the $dx^{M}$ on the RHS of this expression to $d\overline{x}^{\overline{M}}$, and thereby we obtain
\begin{align}
    d\overline{x}^{\overline{M}} = \overline{\mathfrak{D}}{{}^{\overline{M}}}_{N} \mathfrak{D}{{}^{M}}_{\overline{N}} dx^{\overline{N}}\hspace{20 pt}\text{ where }\hspace{20 pt}\mathfrak{D}{{}^{M}}_{\overline{M}} \equiv \left(\dfrac{\partial x^{M}}{\partial \overline{x}^{\overline{M}}}\right)
\end{align}
which implies, recalling that we use tildes to denote inverses,
\begin{align}
    \overline{\mathfrak{D}}{{}^{\overline{M}}}_{M} \mathfrak{D}{{}^{M}}_{\overline{N}} = \delta_{\overline{M},\overline{N}}\hspace{20 pt}\text{ such that }\hspace{20 pt}\tilde{\mathfrak{D}}{{}^{\overline{M}}}_{M} = \overline{\mathfrak{D}}{{}^{\overline{M}}}_{M}
\end{align}
The requirement that a coordinate transformation leaves the invariant spacetime interval unchanged, i.e.
\begin{align}
    G_{MN} \, dx^{M} dx^{N} = ds^{2} = \overline{G}_{\overline{M}\overline{N}} \, d\overline{x}^{\overline{M}} d\overline{x}^{\overline{N}}
\end{align}
implies that the metric transforms according to
\begin{align}
    \overline{G}_{\overline{M}\overline{N}} = \mathfrak{D}{{}^{M}}_{\overline{M}} \, \mathfrak{D}{{}^{N}}_{\overline{N}} \, G_{MN} \label{GMNtransformationrule}
\end{align}
The transformation properties of other spacetime tensors can be derived via Eq. \eqref{GMNtransformationrule}, which we do now.

By definition, any object that transforms like $dx^{M}$ under a diffeomorphism is called a vector, i.e. $v$ is a vector if
\begin{align}
    \overline{v}^{\overline{M}} = \overline{\mathfrak{D}}{{}^{\overline{M}}}_{M} v^{M}
\end{align}
and is said to have a contravariant index. The vector transformation rule in combination with Eq. \eqref{GMNtransformationrule} implies that the covector $(Gv)_{M}\equiv (G_{MN} v^{N})$ corresponding to the vector $v^{M}$ must transform under diffeomorphisms according to
\begin{align}
    \overline{(Gv)}_{\overline{M}} = (\overline{G}_{\overline{M}\overline{N}}\overline{v}^{\overline{N}}) = \mathfrak{D}{{}^{M}}_{\overline{M}} \, \mathfrak{D}{{}^{N}}_{\overline{N}} \, G_{MN} \, \overline{\mathfrak{D}}{{}^{\overline{M}}}_{N} \, v^{N} = \mathfrak{D}{{}^{M}}_{\overline{M}} \, (Gv)_{M}
\end{align}
and $(Gv)_{M}$ is said to have a covariant index. More generally, any index that transforms via $\overline{\mathfrak{D}}$ ($\mathfrak{D}$) is termed contravariant (covariant), and an object having $m$ contravariant and $n$ covariant indices is called a rank-$(m,n)$ tensor. A tensor is said to transform covariantly under diffeomorphisms. By contracting all contravariant indices with covariant indices and evaluating all fields at equivalent spacetime points, we guarantee the construction of a diffeomorphism-invariant quantity. For example, the inner product $G_{MN} v^{M} w^{N}$ of any tangent space vector fields $v$ and $w$ at a spacetime point $x$ is diffeomorphism invariant.

In the gravity literature, the symbol $v_{M}$ is commonly used to denote the covector $(Gv)_{M}$. This is a specific instance of a more general rule wherein indices are lowered via the metric $G$ and raised via its inverse $\tilde{G}$. This rule is quite convenient because allows us to immediately know how an index transforms based on whether it is written as a superscript or a subscript. Unfortunately, this convention is not particularly useful for the goals of this dissertation. As demonstrated in this chapter and the next, the metric (when perturbed relative to a background solution) contains particle content, and allowing the metric to be buried in raising and lowering indices will obscure where instances of various fields occur. Therefore, we avoid absorbing the metric into tensors by instead raising or lowering indices via a flat metric $[\eta_{MN}] \equiv \text{Diag}(+1,-1,\cdots,-1)$, which is a popular convention in the weak field expansion literature. Therefore, given a vector $v$, we define $v_{M} \equiv (\eta v)_{M} = \eta_{MN} v^{N}$. This means that, although the index $M$ in $(G v)_{M}$ is covariant, the index $M$ in $ v_{M} = (\eta v)_{M}$ is still contravariant:
\begin{align}
    \overline{(Gv)}_{\overline{M}} = \mathfrak{D}{{}^{M}}_{\overline{M}} (Gv)_{M}\hspace{20 pt}\text{ versus }\hspace{20 pt} \overline{v}_{\overline{M}} = \eta_{\overline{M}\overline{N}}\, \overline{v}^{\overline{N}} = \overline{\mathfrak{D}}{}_{\overline{M}N} v^{N}
\end{align}
where we treat $\eta$ as a coordinate-independent quantity: $[\overline{\eta}_{\overline{M}\overline{N}}] = [\eta_{MN}]$.

When constructing a Lagrangian theory of gravity, a diffeomorphism-invariant integration element is vital for defining spacetime integrals. To begin, consider the typical volume element $d^{X} x$. This is not invariant under the coordinate transformation $x\rightarrow \overline{x} = \overline{\mathfrak{D}} x$, yielding instead
\begin{align}
    d^{X}\overline{x} = \left|\det \overline{\mathfrak{D}}\right|\, d^{X}x \label{310}
\end{align}
where $\det \overline{\mathfrak{D}} \equiv \det [\overline{\mathfrak{D}}{{}^{\overline{M}}}_{M}]$. Our goal is to combine this with other objects as to create a diffeomorphism-invariant measure. Thankfully, we immediately have access to another object that transforms proportional to factors of $|\det\overline{\mathfrak{D}}|$: by taking the determinant of the transformation rule of the metric Eq. \eqref{GMNtransformationrule}, we find that $|\det G|$ and $|\det \overline{G}|$ are related according to
\begin{align}
    |\det \overline{G}| = \left|\det \mathfrak{D} \right|^{2} |\det G| = \dfrac{|\det G|}{\left|\det \overline{\mathfrak{D}} \right|^{2}}
\end{align}
where we have used that $\overline{\mathfrak{D}} = \tilde{\mathfrak{D}}$, such that
\begin{align}
    \sqrt{|\det \overline{G}|} = \dfrac{\sqrt{|\det G|}}{\left|\det \overline{\mathfrak{D}} \right|} \label{39}
\end{align}
Combining Eqs. \eqref{310} and \eqref{39}, we find that $\sqrt{|\det G|}\, d^{X}x$ is diffeomorphism invariant:
\begin{align}
    \sqrt{|\det \overline{G}|}\, d^{X}\overline{x} = \dfrac{\sqrt{|\det G|}}{\left|\det \overline{\mathfrak{D}} \right|} \, \left|\det \overline{\mathfrak{D}}\right|\, d^{X}x  = \sqrt{|\det G|}\, d^{X}x 
\end{align}
This is the invariant (spacetime) volume element we desired. Because we use the mostly-minus convention, $\text{sign}(\det G) = (-1)^{X-1}$, such that $\sqrt{|\det G|} = \sqrt{\mp \det G}$ if $X$ is even or odd respectively. For succinctness, we define $\sqrt{G} \equiv \sqrt{|\det G|}$. If $\phi(x)$ is a diffeomorphism invariant scalar field, then $\int d^{X}x\, \sqrt{G} \, \phi(x)$ is diffeomorphism invariant as well, such that we can construct a coordinate-independent action.

On occasion, it is useful to purposefully symmetrize (antisymmetrize) some collection of indices, which we denote with parentheses (brackets). For example,
\begin{align}
    T_{(a_{1}\cdots a_{\ell})} \equiv \dfrac{1}{\ell!}\sum_{\pi} T_{a_{\pi(1)}\cdot a_{\pi(\ell)}}\hspace{35 pt}T_{[a_{1}\cdots a_{\ell}]} \equiv \dfrac{1}{\ell!}\sum_{\pi} \text{sign}(\pi)\, T_{a_{\pi(1)}\cdot a_{\pi(\ell)}}
\end{align}
where $\text{sign}(\pi) = \pm 1$ if the permutation $\pi$ is even (odd). Sometimes symmetrization (antisymmetrization) will occur for indices across multiple tensors; in any case, the indices contained between the parentheses (brackets) are included in the procedure.

\subsection{Covariant Derivative, Christoffel Symbol, Lie Derivative}
Beyond any specific coordinate-dependent effects, the metric encodes curvature inherent to spacetime. This curvature implies that the usual coordinate derivative $\partial_{M}\equiv (\partial/\partial x^{M})$ is not necessarily a natural derivative on spacetime, e.g. although $\partial_{M}$ dictates translations in the coordinate $x^{M}$, information about vectors or covectors is not necessarily translated in a coordinate-covariant way. Furthermore, although the index $M$ of $\partial_{M}\phi$ (where $\phi$ is a generic spacetime scalar field) is covariant under diffeomorphisms,
\begin{align}
    \partial_{M}\phi \equiv \dfrac{\partial \phi}{\partial x^{M}} \hspace{20 pt}\mapsto\hspace{20 pt} \overline{\partial}_{\overline{M}}\overline{\phi} \equiv \dfrac{\partial \overline{\phi}}{\partial \overline{x}^{\overline{M}}} = {\mathfrak{D}^{M}}_{\overline{M}} (\partial_{M} \phi)
\end{align}
the equivalent index on the derivative of a more complicated tensor such as $\partial_{M}v^{N}$ (where $v$ is a generic spacetime vector field) is not diffeomorphism covariant,
\begin{align}
    \partial_{M}v^{N} \equiv \dfrac{\partial v^{N}}{\partial x^{M}} \hspace{20 pt}\mapsto\hspace{20 pt} \overline{\partial}_{\overline{M}} v^{\overline{N}} = {\mathfrak{D}^{M}}_{\overline{M}}\, \overline{\partial}_{M}[ \overline{\mathfrak{D}}{{}^{\overline{N}}}_{N} v^{N} ] \neq {\mathfrak{D}^{M}}_{\overline{M}}\, \overline{\mathfrak{D}}{{}^{\overline{N}}}_{N}\, \overline{\partial}_{M}v^{N}
\end{align}
This presents an obstacle when constructing a diffeomorphism-invariant action. To address these problems, we require a derivative that incorporates the structure of spacetime.

Two derivatives of this sort commonly occur in general relativity calculations: the covariant derivative and the Lie derivative. Both are derivatives in the traditional sense---i.e. they are linear maps which obey the Leibniz rule $f(xy) = f(x)y + xf(y)$\,---although they differ in their details and applications. The covariant derivative is particularly useful when constructing Lagrangians on curved spacetimes, depends on the metric $G$, and transforms a rank-$(m,n)$ tensor into a rank-$(m,n+1)$ tensor. In contrast, the Lie derivative generalizes the directional derivative of flat spacetime, is independent of the metric $G$, and transforms a rank-$(m,n)$ tensor into another rank-$(m,n)$ tensor.

For the covariant derivative, we utilize what is called the Levi-Civita connection $\nabla_{A}$, which is the unique affine connection that is simultaneously compatible with the metric ($\nabla_{A} G_{MN} = 0$) and torsion-free. Its action on a given tensor depends on the rank of that tensor, e.g. for a scalar field $\phi(x)$ the covariant derivative reduces to the usual derivative,
\begin{align}
    \nabla_{A} \phi = \partial_{A} \phi
\end{align}
whereas for a vector field $v^{M}(x)$, the covariant derivative contains an additional term,
\begin{align}
    \nabla_{A} v^{M} = \partial_{A} v^{M} + \Gamma^{M}_{AN} v^{N}
\end{align}
where $\Gamma^{P}_{MN}$ is the Christoffel symbol,
\begin{align}
    \Gamma^{P}_{MN} \equiv \dfrac{1}{2} \tilde{G}^{PQ} (\partial_{M} G_{NQ} + \partial_{N} G_{MQ} - \partial_{Q} G_{MN})
\end{align}
Note that the Christoffel symbol is symmetric in its lower indices, i.e. $\Gamma_{MN}^{P} = \Gamma_{NM}^{P}$. Despite its suggestive index structure, the Christoffel symbol does not transform like a spacetime tensor (because, for example, $(\partial_{A} v^{M})$ is not a spacetime tensor but $\nabla_{A} v^{M}$ is). Taking the covariant derivative of a tensor possessing multiple contravariant indices proceeds similarly, with as many additional terms as there are indices and where each term contains a Christoffel symbol contracted with a different index. When covariant indices are present, the Christoffel symbol terms are instead subtracted, e.g. the covariant derivative of a covector field $v_{M}(x)$ equals
\begin{align}
    \nabla_{A} v_{M} = \partial_{A} v_{M} - \Gamma^{N}_{AM} v_{N}
\end{align}
Multiple covariant indices generalize accordingly via the additional subtraction of a Christoffel symbol-containing term per covariant index. Combining the contravariant and covariant behaviours yields the formula for a generic rank-$(m,n)$ tensor. Because of its compatibility with the metric, any function that depends on the metric alone has vanishing covariant derivative.

The Lie derivative is a coordinate-invariant measure of the change in a spacetime tensor with respect to a vector field. It is the generalization of the standard directional derivative in flat spacetimes. Like the covariant derivative, its exact operation depends on the rank of the tensor it operates on. For example, given a vector field $v^{M}(x)$, the Lie derivative of a scalar field $\phi(x)$ with respect to $v^{M}(x)$ is
\begin{align}
    \LD_{v} \phi \equiv (v \cdot \partial) \phi
\end{align}
whereas the same Lie derivative of a vector field $w^{M}(x)$ is
\begin{align}
    \LD_{v} w^{M} \equiv (v\cdot \partial)w^{M} - (\partial_{N}v^{M})w^{N}
\end{align}
and of a covector field $w_{M}$ is
\begin{align}
    \LD_{v} w_{M} \equiv (v\cdot \partial)w_{M} + (\partial_{M}v^{N})w_{N}
\end{align}
where $v\cdot \partial \equiv v^{M} \partial_{M}$. A rank-$(m,n)$ tensor will have $m$ subtracted terms and $n$ added terms, each involving the contraction of an index from $(\partial_{M} v^{N})$ with a different contravariant or covariant index respectively. These equations for the Lie derivative also hold true if the usual derivatives $\partial_{A}$ are replaced with covariant derivatives $\nabla_{A}$. We will utilize the Lie derivative when we calculate extrinsic curvature in the RS1 model.

\subsection{Curvature}
The metric expressed in a given coordinate system enables a quantitative measure of the curvature of spacetime. For example, the Riemann curvature (tensor) ${R_{ABC}}^{D}$ measures spacetime curvature via the failure of covariant derivatives to commute when acting on a generic covector field:
\begin{align}
    {R_{ABC}}^{D} w_{D} \equiv (\nabla_{A}\nabla_{B} - \nabla_{B} \nabla_{A}) w_{C}
\end{align}
By replacing the covariant derivatives with their expression in terms of Christoffel symbols, we attain a formula for the Riemann curvature that will prove more useful for our computations:
\begin{align}
    {R_{ABC}}^{D} &\equiv (\partial_{B} \Gamma^{D}_{AC}) - (\partial_{A} \Gamma^{D}_{BC}) + \Gamma^{E}_{AC}\Gamma^{D}_{BE} - \Gamma^{E}_{BC} \Gamma^{D}_{AE}\\
    &= (\partial_{[B} \Gamma^{D}_{A]C}) + \Gamma^{E}_{C[A} \Gamma^{D}_{B]E}
\end{align}
Whether or not an additional minus sign is included in the above definition amounts to a convention; across the literature, both choices are used with nearly equal frequency and without much consistency across in any given subfield. Consequently, ambiguity in this convention can be a source of many headaches. For this dissertation, we use the Riemann curvature as written above (which contrasts the convention we used in \cite{Chivukula:2020hvi}).

The Riemann curvature is frequently self-contracted to form the Ricci tensor,
\begin{align}
    R_{AC} \equiv {R_{ABC}}^{B}
\end{align}
from which a subsequent contraction with the inverse metric yields the Ricci scalar (or scalar curvature),
\begin{align}
    R \equiv \tilde{G}^{AC} R_{AC}
\end{align}
The Ricci scalar is an important constituent of the Einstein-Hilbert Lagrangian, the foundation on which all gravitational Lagrangians are built.

\subsection{Einstein-Hilbert Lagrangian, Cosmological Constant, Einstein Field Equations}
The Einstein-Hilbert action $S_{\text{EH}}$ and the Einstein-Hilbert Lagrangian $\mathcal{L}_{\text{EH}}$ are defined according to
\begin{align}
    S_{\text{EH}} \equiv -\dfrac{2}{\kappa_{\text{XD}}^{2}} \int d^{X}x\hspace{5 pt} \sqrt{G}\, R \equiv \int d^{X}x\hspace{5 pt}\mathcal{L}_{\text{EH}}
\end{align}
where $\sqrt{G} \equiv \sqrt{|\det G|}$. The negative prefactor $(-2/\kappa_{\text{XD}}^{2})$ is directly tied to the sign of the Riemann curvature which we chose in the previous section, and ensures properly normalized (positive energy) graviton modes. To derive the equations of motion for the metric, consider varying the Einstein-Hilbert action with respect to the inverse metric $\tilde{G}^{AB}$. Because
\begin{align}
    \dfrac{\delta}{\delta \tilde{G}^{AB}}\left[\sqrt{G}\right] = -\dfrac{1}{2}\sqrt{G}\, G_{AB}\hspace{20 pt}\text{ and }\hspace{20 pt}\dfrac{\delta}{\delta \tilde{G}^{AB}}\left[R \right] = R_{AB}
\end{align}
the first variation of $S_{\text{EH}}$ yields, assuming vanishing surface terms,
\begin{align}
    \delta S_{\text{EH}} = -\dfrac{2}{\kappa_{\text{XD}}^{2}} \int d^{X}x\hspace{5 pt}\sqrt{G}\,\bigg[R_{AB} - \dfrac{1}{2} G_{AB} R\bigg]\delta\tilde{G}^{AB}
\end{align}
such that, without additional modifications, the equations of motion equal
\begin{align}
    \mathcal{G}_{AB} \equiv R_{AB} - \dfrac{1}{2} G_{AB} R = 0
\end{align}
where $\mathcal{G}_{AB}$ is the Einstein tensor.

There are two other Lagrangians commonly added to $\mathcal{L}_{\text{EH}}$. The first we consider is the cosmological constant Lagrangian,
\begin{align}
    \mathcal{L}_{\text{CC}} \equiv -\dfrac{4}{\kappa_{\text{XD}}^{2}} \sqrt{G}\, \Lambda
\end{align}
where $\Lambda$ is a real number. The variation of $\mathcal{L}_{\text{CC}}$ yields
\begin{align}
    \dfrac{\delta}{\delta \tilde{G}^{AB}}\left[\mathcal{L}_{\text{CC}}\right] = -\dfrac{2}{\kappa_{\text{XD}}^{2}}\sqrt{G}\, (-\Lambda \, G_{AB})
\end{align}
The second is the matter Lagrangian, the form of which is left mostly ambiguous unless applied to a specific choice of matter fields. Its contribution is typically written with a factor of the invariant volume element already accounted for but (in contrast to the previous two Lagrangian contributions considered) without any factors of $\kappa_{\text{XD}}$, as $\sqrt{G}\,\mathcal{L}_{\text{M}}$. Its variation with respect to the inverse metric equals
\begin{align}
    \dfrac{\delta}{\delta \tilde{G}^{AB}} \bigg[ \sqrt{G}\, \mathcal{L}_{\text{M}} \bigg] &= \dfrac{1}{2}\sqrt{G}\, \mathcal{T}_{AB}
\end{align}
where $\mathcal{T}_{AB}$ is the stress-energy tensor
\begin{align}
     \mathcal{T}_{AB} \equiv 2 \dfrac{\delta \mathcal{L}_{\text{M}}}{\delta \tilde{G}^{AB}} - G_{AB} \mathcal{L}_{\text{M}}
\end{align}
which expresses the stress-energy content generated by the matter fields.

Therefore, for the Lagrangian,
\begin{align}
    \mathcal{L}_{\text{EH}} + \mathcal{L}_{\text{CC}} + \sqrt{G}\, \mathcal{L}_{\text{M}}
\end{align}
the equations of motion equal
\begin{align}
    \mathcal{G}_{AB} - \Lambda G_{AB} = \dfrac{\kappa_{\text{XD}}^{2}}{4} \mathcal{T}_{AB}
\end{align}
These gravitational equations of motion (and extensions thereof) are the Einstein field equations, and imply that matter or a cosmological constant can influence the Einstein tensor $\mathcal{G}_{AB}$ and thereby curve spacetime. The curvature of spacetime is closely tied to the presence of fields on that spacetime, not unlike the close ties between electric fields and electric charges.

The aforementioned Lagrangians describe bulk gravitational physics; when it becomes necessary, we will extend these ideas to incorporate spacetime matter and/or energy localized to submanifolds, such as branes. 

To conclude this section, we note that the Einstein-Hilbert Lagrangian can be rewritten using integration-by-parts into a form wherein any given instance of the metric is never differentiated more than once \cite{Dyer:2008hb}:
\begin{align}
    \mathcal{L}_{\text{EH}} \cong -\dfrac{2}{\kappa_{\text{XD}}^{2}} \sqrt{G} \tilde{G}^{MN}\bigg[\Gamma^{Q}_{NP} \Gamma^{P}_{MQ} - \Gamma^{P}_{QP} \Gamma^{Q}_{MN}\bigg]
\end{align}
The symbol $\cong$ denotes equality as an action integrand via integration by parts. This alternate form is derived in next subsection.

\subsection{Rewriting the Einstein-Hilbert Lagrangian} \label{SS - Rewriting EH}
The Einstein-Hilbert Lagrangian is defined, traditionally, in terms of the scalar curvature as
\begin{align}
    \mathcal{L}_{\text{EH}} &= -\dfrac{2}{\kappa^{2}_{\text{XD}}} \, \sqrt{G}\, R\\
    &= -\dfrac{2}{\kappa_{\text{XD}}^{2}} \, \sqrt{G}\,\tilde{G}^{MN}\,\bigg[ (\partial_{P} \Gamma^{P}_{MN}) - (\partial_{M} \Gamma^{P}_{PN}) + \Gamma^{Q}_{MN} \Gamma^{P}_{PQ} - \Gamma^{Q}_{MP} \Gamma^{P}_{NQ} \bigg] \label{LagEHChristoffels}
\end{align}
However, we find it more useful to work with an alternate form of $\mathcal{L}_{\text{EH}}$ which is attained through integration by parts. Integration by parts will move the derivatives acting on Christoffel symbols in the first two terms of Eq. \eqref{LagEHChristoffels} onto $\sqrt{G}\,\tilde{G}^{MN}$, such that all Christoffel symbols are no longer differentiated. This will eliminate all twice-differentiated quantities from the Einstein-Hilbert Lagrangian, and yield a form of the Lagrangian utilized by Einstein \cite{PhysRevD.79.024028}.

In order to eventually simplify the expressions we obtain from this procedure, recall that any function which only depends on the metric has vanishing covariant derivative. Therefore,
\begin{align}
    0 = \nabla_{C} \tilde{G}^{MN} = (\partial_{C} \tilde{G}^{MN}) + \Gamma^{M}_{AC} \tilde{G}^{AN} + \Gamma^{N}_{AC} \tilde{G}^{MA}
\end{align}
such that
\begin{align}
    (\partial_{C} \tilde{G}^{MN}) = - \tilde{G}^{AN} \Gamma^{M}_{AC} - \tilde{G}^{MA} \Gamma^{N}_{AC}
\end{align}
and\footnote{That $\sqrt{G} = \sqrt{\det G}$ possesses a nontrivial covariant derivative arises from the fact that $\det G$ transforms nontrivially under diffeomorphisms, as originally mentioned in Eq. \eqref{39}. In particular, $\sqrt{G}$ is a scalar density with unit weight, where weight refers to the constant multiplying $-\sqrt{G}\, \Gamma^{A}_{AC}$ in Eq. \eqref{339}. For example, $\det G$ has weight $+2$, and thus its covariant derivative contains instead the term $-2\sqrt{G}\,\Gamma^{A}_{AC}$.}
\begin{align}
    0 = \nabla_{C} \sqrt{G} = (\partial_{C} \sqrt{G}) - \sqrt{G}\, \Gamma^{A}_{AC} \label{339}
\end{align}
such that
\begin{align}
    (\partial_{C}\sqrt{G}) = \sqrt{G}\, \Gamma^{A}_{AC} 
\end{align}
Together these results imply that
\begin{align}
    \partial_{C} \left(\sqrt{G}\,\tilde{G}^{MN}\right) &= (\partial_{C} \sqrt{G})\, \tilde{G}^{MN} + \sqrt{G}\,(\partial_{C} \tilde{G}^{MN})\\
    &= \sqrt{G}\,\bigg[\tilde{G}^{MN}\,\Gamma^{A}_{AC}  - \tilde{G}^{AN} \Gamma^{M}_{AC} - \tilde{G}^{MA} \Gamma^{N}_{AC} \bigg]
\end{align}
and we are now ready to begin rewriting the Einstein-Hilbert Lagrangian.

Consider the first term of Eq. \eqref{LagEHChristoffels}. It is proportional to
\begin{align}
    +\sqrt{G} \, \tilde{G}^{MN} \, (\partial_{P} \Gamma^{P}_{MN}) &= - \partial_{P} \left[ \sqrt{G}\, \tilde{G}^{MN} \right] \, \Gamma^{P}_{MN}+\partial_{P}\left[ \sqrt{G}\,\tilde{G}^{MN}\, \Gamma^{P}_{MN} \right]\label{Eq3350}\\
    &\cong \sqrt{G}\,\bigg[ -\tilde{G}^{MN}\,\Gamma^{A}_{AP} \Gamma^{P}_{MN} + \tilde{G}^{AN} \Gamma^{M}_{AP} \Gamma^{P}_{MN} + \tilde{G}^{MA} \Gamma^{N}_{AP} \Gamma^{P}_{MN} \bigg]\\
    &= \sqrt{G}\,\tilde{G}^{MN}\,\bigg[ - \Gamma^{Q}_{MN}\,\Gamma^{P}_{PQ} + 2 \, \Gamma^{Q}_{MP} \Gamma^{P}_{NQ} \bigg]
\end{align}
where integration by parts was used in the first line, and the last line utilizes both index relabeling and the index symmetries of $\tilde{G}^{MN}$ and $\Gamma^{P}_{MN}$. Similarly, the second term of Eq. \eqref{LagEHChristoffels} is proportional to
\begin{align}
    -\sqrt{G} \, \tilde{G}^{MN} \, (\partial_{M} \Gamma^{P}_{PN}) &= +\partial_{M} \left[ \sqrt{G}\, \tilde{G}^{MN} \right] \, \Gamma^{P}_{PN}-\partial_{M} \left[ \sqrt{G}\, \tilde{G}^{MN}  \, \Gamma^{P}_{PN}\right]\label{Eq3353}\\
    &\cong \sqrt{G}\,\bigg[ +\tilde{G}^{MN}\,\Gamma^{A}_{AM} \Gamma^{P}_{PN}  - \tilde{G}^{AN} \Gamma^{M}_{AM} \Gamma^{P}_{PN} - \tilde{G}^{MA} \Gamma^{N}_{AM} \Gamma^{P}_{PN} \bigg]\\
    &= \sqrt{G}\,\tilde{G}^{MN}\,\bigg[ - \Gamma^{Q}_{MN} \Gamma^{P}_{PQ} \bigg]
\end{align}
Substituting these results into Eq. $\eqref{LagEHChristoffels}$ yields the desired alternate form of the Einstein-Hilbert Lagrangian, sometimes called the gamma-gamma Lagrangian:
\begin{align}
    \mathcal{L}_{\text{EH}} &\cong \mathcal{L}_{\Gamma\Gamma} \equiv -\dfrac{2}{\kappa_{\text{XD}}^{2}}\,\sqrt{G}\,\tilde{G}^{MN}\,\bigg[ \Gamma^{Q}_{MP} \Gamma^{P}_{NQ} - \Gamma^{Q}_{MN} \Gamma^{P}_{PQ} \bigg] \label{LagEHDef}
\end{align}
By keeping the total derivatives from Eqs. \eqref{Eq3350} and \eqref{Eq3350}, we have the exact relation
\begin{align}
    \mathcal{L}_{\text{EH}} = \mathcal{L}_{\Gamma\Gamma} - \dfrac{2}{\kappa_{\text{XD}}^{2}}\, \partial_{P} \bigg[ \sqrt{G} \, \tilde{G}^{M[N}\,\Gamma^{P]}_{MN} \bigg]
\end{align}
Because each Christoffel symbol contains exactly one derivative per term by definition, $\mathcal{L}_{\text{EH}}$ contains exactly two derivatives per term. One advantage of this alternate form (which lacks the $\partial \Gamma \supset \partial\partial G$ terms of the traditional form) is that it ensures those two derivatives are never applied to the same object in any given term. Despite the total derivative relating $\mathcal{L}_{\text{EH}}$ and $\mathcal{L}_{\Gamma\Gamma}$ typically yielding a nonzero contribution to the action, we may drop it without altering the physics (although doing so typically alters how physical information is stored in the fields).

\subsection{Deriving the Graviton} \label{SS - Deriving Graviton}
Consider the aforementioned $X$-dimensional gravitational Lagrangian in the absence of a cosmological constant and matter, so that the relevant Lagrangian is exclusively the Einstein-Hilbert Lagrangian, Eq. \eqref{LagEHDef}. The corresponding Einstein field equations are then $\mathcal{G}_{AB} = 0$, which can be trivially satisfied by the flat metric $\eta_{MN} = \text{Diag}(+1,-1,\cdots,-1)$ (for which the Riemann curvature vanishes). Choose this solution as a background metric, and consider the metric $G_{MN}$ present in the Einstein-Hilbert Lagrangian as only slightly perturbed away from $\eta_{MN}$, e.g. $G_{MN} \equiv \eta_{MN} + \kappa_{\text{XD}} \hat{H}_{MN}$ for some spacetime-dependent perturbation $\hat{H}_{MN}(x)$. This enables us to calculate $\mathcal{L}_{\text{EH}}$ as a perturbative series in $\hat{H}$. In general, the process of expanding a metric about a background metric that solves the Einstein field equations is called weak field expansion (WFE).  At present, we will weak field expand the Einstein-Hilbert Lagrangian through $\mathcal{O}(\hat{H}^{2})$.

First, note that weak field expansion of the Christoffel symbol corresponding to the $G_{MN}$ described above yields
\begin{align}
    \Gamma^{P}_{MN} &\equiv \dfrac{1}{2}\tilde{G}^{PQ}(\partial_{M}G_{NQ} + \partial_{N} G_{MQ} - \partial_{Q}G_{MN})\\
    &=\dfrac{\kappa_{\text{XD}}}{2}\bigg[\sum_{n=0}^{+\infty}(-1)^{n}\ltr (\kappa_{\text{XD}}\hat{H})^{n}\rtr^{PQ}\bigg]\,\bigg[(\partial_{M} \hat{H}_{NQ}) + (\partial_{N} \hat{H}_{MQ}) - (\partial_{Q} \hat{H}_{MN})\bigg]\\
    &= \dfrac{\kappa_{\text{XD}}}{2}\bigg[(\partial_{M} \hat{H}^{P}_{N}) + (\partial_{N} \hat{H}^{P}_{M}) - (\partial^{P}\hat{H}_{MN})\bigg] +\mathcal{O}(\hat{H}^{2})
\end{align}
where we utilize the twice-squared bracket notation introduced in Chapter \ref{Classical - Minkowski}. We need only expand the Christoffel symbols to first order in the field $\hat{H}$ to obtain an overall $\mathcal{O}(\hat{H}^{2})$ result because they begin at that order and $\mathcal{L}_{\text{EH}}$ is composed of products of pairs of Christoffel symbols.

When these expansions are substituted into the Einstein-Hilbert Lagrangian, we find
\begin{align}
    \mathcal{L}_{\text{EH}} &\cong -\dfrac{2}{\kappa_{\text{XD}}^{2}} \eta^{MN} \bigg[\Gamma^{Q}_{NP} \Gamma^{P}_{MQ} - \Gamma^{P}_{QP} \Gamma^{Q}_{MN}\bigg] + \mathcal{O}(H^{3})\\
    &=(\partial^{A} \hat{H}_{AB})(\partial^{B} \hat{H}) - (\partial_{A} \hat{H}_{BC}) (\partial^{C} \hat{H}^{AB}) +\dfrac{1}{2} (\partial_{A} \hat{H}_{BC})^2 - \dfrac{1}{2} (\partial_{B} \hat{H})^2+ \mathcal{O}(\hat{H}^{3}) \label{354}
\end{align}
where the $\sqrt{G}\, \tilde{G}^{MN}$ prefactor has already been expanded in the first line (more information about the weak field expansion of $\sqrt{G}$ and $\tilde{G}^{MN}$ can be found in Section \ref{GRAppendices - WFE}). When $X=4$, Eq. \eqref{354} is precisely the massless spin-2 Lagrangian from Section \ref{S - Fields}. When $X\neq 4$, the equations of motion still go through as-is and constrain the propagation of $\hat{H}_{MN}$ such that the field must be transverse and traceless when on shell: $(\partial^{M} \hat{H}_{MN}) = \hat{H}^{M}_{M} = 0$. In general, after applying the equations of motion, an $X$-dimensional graviton has $(X+1)X/2 - 2X = (X-3)X/2$ degrees of freedom. Therefore, a 4D graviton has $2$ degrees of freedom, whereas a 5D graviton has $5$ degrees of freedom.

Consider the effect of a coordinate transformation $x\rightarrow \overline{x} = \overline{\mathfrak{D}} x$ on the field $\hat{H}_{MN}(x)$, as transmitted through the known transformation properties of $G_{MN}(x)$. In particular, suppose the diffeomorphism is of the form of a coordinate-dependent spacetime translation $\overline{x}^{M} = x^{M} + \epsilon^{M}(x)$ for some vector field $\epsilon^{M}(x)$, and that the vector components $\epsilon^{M}$ are at most comparable in magnitude to the field components $\hat{H}_{MN}$ so that we may simultaneously expand in $\epsilon$, e.g. $\mathcal{O}(\epsilon)\sim \mathcal{O}(\hat{H})$. We now demonstrate that this spacetime translation exactly reproduces the gauge freedom of the massless spin-2 Lagrangian when $X=4$.

The aforementioned diffeomorphism implies $\overline{\mathfrak{D}}{{}^{\overline{M}}}_{M} = (\partial \overline{x}^{\overline{M}}/\partial x^{M}) = \eta^{\overline{M}}_{M} + (\partial_{M} \epsilon^{\overline{M}})$, so that diffeomorphism invariance demands
\begin{align}
    G_{MN} &= \overline{\mathfrak{D}}{{}^{\overline{M}}}_{M}\, \overline{\mathfrak{D}}{{}^{\overline{N}}}_{N} \,\overline{G}_{\overline{M}\overline{N}}\\
    & =\bigg[\eta^{\overline{M}}_{M} + (\partial_{M}\epsilon^{\overline{M}})\bigg] \,  \bigg[\eta^{\overline{N}}_{N} + (\partial_{N}\epsilon^{\overline{N}})\bigg] \, \overline{G}_{\overline{M}\overline{N}}\\
    &= \overline{G}_{MN} + (\partial_{M}\epsilon^{\overline{M}}) \overline{G}_{\overline{M}N} + (\partial_{N}\epsilon^{\overline{N}}) \overline{G}_{M\overline{N}}  + (\partial_{M}\epsilon^{\overline{M}}) (\partial_{N}\epsilon^{\overline{N}}) \overline{G}_{\overline{M}\overline{N}}  
\end{align}
which is an exact result. To proceed further, series expand the quantity $\overline{G}(\overline{x}) = \overline{G}(x+\epsilon)$ in $\epsilon$ through $\mathcal{O}(\epsilon)$:
\begin{align}
    \overline{G}_{MN}(\overline{x}) = \overline{G}_{MN}(x) +  \epsilon^{\overline{M}} \partial_{\overline{M}} \overline{G}_{MN}(x) + \mathcal{O}(\epsilon^{2})
\end{align}
such that,
\begin{align}
    G_{MN} = \overline{G}_{MN} + (\epsilon\cdot \partial)\overline{G}_{MN} + (\partial_{M} \epsilon^{P})\overline{G}_{PN} + (\partial_{N} \epsilon^{P})\overline{G}_{MP} + \mathcal{O}(\epsilon^{2}) \label{359}
\end{align}
where all fields are expressed as functions of the coordinates $x$. This completes the expansion in $\epsilon$. Note that this can be succinctly expressed in terms of the Lie derivative
\begin{align}
     \LD_{\epsilon} \overline{G}_{MN} = (\overline{G}_{MN} - G_{MN}) + \mathcal{O}(\epsilon^{2})
\end{align}
which---given that we performed an infinitesimal coordinate translation---confirms the Lie derivative's role as a direction derivative. Next, expand each term in powers of $\hat{H}$, and remember that $\hat{H}$ and $\epsilon$ are componentwise comparable in magnitude: per term of Eq. \eqref{359}, we find
\begin{align}
    G_{MN} &= \eta_{MN} + \kappa_{\text{XD}} \hat{H}_{MN}\\
    \overline{G}_{MN} &= \eta_{MN} + \kappa_{\text{XD}} \hat{\overline{H}}_{MN}\\
    (\epsilon\cdot \partial)\overline{G}_{MN} &= (\epsilon \cdot \partial)\hat{\overline{H}}_{MN} \hspace{85 pt}= \mathcal{O}(\epsilon^{2},\epsilon\hat{\overline{H}},\hat{\overline{H}}{}^{2})\\
    (\partial_{M}\epsilon^{P}) \overline{G}_{PN} &= (\partial_{M} \epsilon^{P}) (\eta_{PN} + \hat{\overline{H}}_{PN}) \hspace{35 pt}= (\partial_{M} \epsilon_{N}) + \mathcal{O}(\epsilon^{2},\epsilon\hat{\overline{H}},\hat{\overline{H}}{}^{2})\\ 
    (\partial_{N}\epsilon^{P}) \overline{G}_{MP} &= (\partial_{N} \epsilon^{P}) (\eta_{MP} + \hat{\overline{H}}_{MP}) \hspace{35 pt}= (\partial_{N} \epsilon_{M}) + \mathcal{O}(\epsilon^{2},\epsilon\hat{\overline{H}},\hat{\overline{H}}{}^{2})
\end{align}
such that
\begin{align}
    \kappa_{\text{XD}} \hat{H}_{MN} &= \kappa_{\text{XD}} \hat{\overline{H}}_{MN} + (\partial_{M} \epsilon_{N}) + (\partial_{N}\epsilon_{M}) + \LD_{\epsilon} \hat{\overline{H}}_{MN} + \mathcal{O}(\epsilon^{2}) \\
    &= \kappa_{\text{XD}} \hat{\overline{H}}_{MN} + (\partial_{M} \epsilon_{N}) + (\partial_{N}\epsilon_{M}) +  \mathcal{O}(\epsilon^{2},\epsilon\hat{\overline{H}},\hat{\overline{H}}{}^{2})
\end{align}
This mean that (dropping the distinction between the new and old field labels from here), as far as the field $\hat{H}_{MN}$ is concerned, an infinitesimal coordinate translation corresponds to the field transformation
\begin{align}
    \kappa_{\text{XD}} \hat{H}_{MN}  \rightarrow \kappa_{\text{XD}} \hat{H}_{MN} + (\partial_{M} \epsilon_{N}) + (\partial_{N}\epsilon_{M}) + \mathcal{O}(\epsilon^{2}) 
\end{align}
which is precisely the gauge invariance exhibited by the massless spin-2 Lagrangian when $X=4$.

\section{The Randall-Sundrum 1 Model} \label{S - Intro RS1}
\subsection{Deriving the Background Metric}
In this subsection, the Randall-Sundrum 1 (RS1) model background metric is motivated and derived. In the next subsection, we perturb this background metric and thereby obtain the full RS1 theory.

As mentioned in this dissertation's introduction, the RS1 model is a five-dimensional model of gravity with nonfactorizable geometry that was introduced in 1999 in order to solve the hierarchy problem. Relative to the usual four-dimensional spacetime, the RS1 model adds a finite extra dimension of space parameterized by a coordinate $y$ ranging from $y=0$ to $y = \pi r_{c}$, where $r_{c}$ is called the compactification radius. The size $\pi r_{c}$ of the extra-dimension is assumed small so that the five-dimensional nature of spacetime remains hidden at low energies. The four-dimensional hypersurfaces defined by $y=0$ and $y=\pi r_{c}$ are called branes, and the five-dimensional region between those branes is called the bulk.

The RS1 construction possesses two additional features not mentioned in the previous paragraph: warping of the 4D spacetime relative to the extra dimension and orbifold invariance. Because we will discuss the former property at length later in this section, let us first focus on orbifold invariance. In order that spacetime truly be truncated at the branes, any physically-relevant 5D fields cannot be allowed to oscillate beyond the branes, and thereby their derivatives with respect to $y$ must vanish at the branes. This can be ensured by extending the extra dimension so that $y$ covers $[-\pi r_{c},+\pi r_{c}]$ and then demanding that the so-called orbifold reflection $y \rightarrow -y$ is a symmetry of the invariant spacetime interval $ds^{2}$. Having done this, we can extend $y$ to the entire real line by also declaring the discrete translation $y \rightarrow y + 2\pi r_{c}$ as another symmetry of $ds^{2}$. This discrete translational symmetry suggests we can just as well think of the extra dimension as a circle of radius $r_{c}$ parameterized by an angle $\varphi \equiv y/r_{c}$, with the discrete translation corresponding to rotating the entire circle about its center by $2\pi$ radians. (Despite this extension, we will limit any integrals over the extra dimension to the finite domain $y \in [-\pi r_{c},+\pi r_{c}]$, or equivalently $\varphi \in [-\pi, +\pi]$.) If we imagine this circle to be drawn on a piece of paper, then the identification of points via the orbifold reflection corresponds to folding the paper in half along the line between the points at $\phi = 0$ and $\phi = \pm \pi$ and declaring any points which overlap afterwards to be equivalent. From this perspective, if we once again unfold the paper, then the orbifold reflection transformation swaps points across the folding line, such that the only points unchanged by the transformation are the branes at $\varphi =0$ and $\varphi=\pi$. In other words, the two branes are uniquely determined as the orbifold fixed points of the RS1 spacetime.

With descriptions of the RS1 coordinates and spacetime symmetries out of the way, we now aim to find a Lagrangian description of the RS1 background metric, although to do so we must include types of terms we have not yet discussed in this chapter. We begin by searching for a background metric of the form
\begin{align}
    [G_{MN}] = \matrixbb{a(y)\, \eta_{\mu\nu}}{0}{0}{-1} \label{374}
\end{align}
that is consistent with the Einstein field equations, where $a(y)$ is a nontrivial positive real function of the extra-dimensional coordinate $y$. The function $a(y)$ provides the aforementioned warping of 4D spacetime relative to the extra dimension. Eq. \eqref{374} is intentionally written so that the $x^{\mu}$ coordinates are all treated on equal footing, as would be expected from a 4D Poincar\'{e}-invariant geometry. If $a(y) = 1$, we recover the flat 5D metric $\eta_{MN}$; otherwise, this metric (combined with the orbifold condition) necessarily implies a discontinuity in the curvature at the interval endpoints. As will be detailed in a moment, this introduces Dirac delta terms to the Einstein tensor which provide an obstacle to solving the Einstein field equations. Overcoming this obstacle requires extending the techniques utilized thus far to include brane-localized stress-energy content.

First, note that $G_{MN}$ as written above only depends on $y$, so $\partial_{\alpha} G_{MN} = 0$, whereas $\partial_{y} G_{MN} = (\partial_{y}a)\, \delta_{M}^{\mu}\delta_{N}^{\nu} \, \eta_{\mu\nu}$. Consequently, the only independent non-zero Christoffel symbols equal
\begin{align}
    \Gamma^{5}_{\mu\nu} &= -\dfrac{1}{2}\tilde{G}^{55}(\partial_{y} G_{\mu\nu}) = \dfrac{1}{2}\, (\partial_{y} a)\, \eta_{\mu\nu} \label{375}\\
    \Gamma^{\rho}_{\mu 5} &= +\dfrac{1}{2} \tilde{G}^{\rho\sigma} (\partial_{y} G_{\mu\sigma}) = \dfrac{1}{2}\, a^{-1}(\partial_{y}a)\, \eta^{\rho}_{\mu} \label{376}
\end{align}
which once again only depend on $y$. Thus, we may calculate
\begin{align}
    \partial_{P} \Gamma^{P}_{MN} &= \dfrac{1}{2}\,(\partial_{y}^{2} a)\,\delta^{\mu}_{M}\delta^{\nu}_{N}\,\eta_{\mu\nu}\\
    \partial_{N}\Gamma^{P}_{MP} &=  \left[2\,a^{-1}(\partial_{y}^{2} a)-2\,a^{-2}(\partial_{y} a)^2\right]\,\delta^{5}_{M}\delta^{5}_{N}\\
    \Gamma^{P}_{PQ}\Gamma^{Q}_{MN} &=  a^{-1}(\partial_{y} a)^{2}\,\delta^{\mu}_{M}\delta^{\nu}_{N}\,\eta_{\mu\nu}\\
    \Gamma^{P}_{NQ}\Gamma^{Q}_{MP} &= \dfrac{1}{2}\,a^{-1}(\partial_{y} a)^{2}\,\delta^{\mu}_{M}\delta^{\nu}_{N}\,\eta_{\mu\nu} + a^{-2}(\partial_{y}a)^{2}\,\delta^{5}_{M}\delta^{5}_{N}
\end{align}
which collectively yield the Ricci tensor 
\begin{align}
    [R_{MN}] &= \matrixbb{\tfrac{1}{2}\left[(\partial^{2}_{y} a) + a^{-1}(\partial_{y}a)^{2}\right]\eta_{\mu\nu}}{0}{0}{-a^{-1}\left[2(\partial_{y}^{2}a) - a^{-1}(\partial_{y} a)^{2}\right]}
\end{align}
and the scalar curvature
\begin{align}
    R = 4 a^{-1}(\partial_{y}^{2} a) + a^{-2}(\partial_{y} a)^{2}
\end{align}
This allows us to calculate the Einstein tensor, which equals
\begin{align}
    [\mathcal{G}_{AB}] = \left[R_{AB} - \tfrac{1}{2} G_{AB} R\right] = \matrixbb{-\tfrac{3}{2}(\partial_{y}^{2} a) \eta_{\alpha\beta}}{0}{0}{\tfrac{3}{2} a^{-2}(\partial_{y}a)^{2}}
\end{align}
Without an additional cosmological constant or matter content, the Einstein field equations demand that the Einstein tensor vanish ($\mathcal{G}_{AB} = 0$). This implies $(\partial^{2}_{y} a) = (\partial_{y} a) = 0$, which is only achievable by setting $a$ to a constant; however, a constant $a$ just describes the flat 5D metric up to a coordinate rescaling. We desire a more interesting geometry.

By adding a cosmological constant throughout 5D spacetime (a ``bulk" cosmological constant), we instead obtain $\mathcal{G}_{AB} - \Lambda G_{AB} = 0$ as our Einstein field equations, wherein the precise value of $\Lambda$ can be tuned as necessary. Now our constraints read
\begin{align}
    -\dfrac{3}{2}(\partial_{y}^{2} a) - \Lambda a &= 0 \label{384}\\
    \dfrac{3}{2}(\partial_{y}a)^2 + \Lambda a^{2} &= 0 \label{385}
\end{align}
We focus on this second equation first. Immediately, we note a solution cannot exist if $\Lambda > 0$ because $(\partial_{y} a)^{2}/a^{2}$ is necessarily nonnegative. This plus the fact that we already ruled out the $\Lambda = 0$ case as being uninteresting leaves us to consider $\Lambda < 0$, which allows us to solve for $(\partial_{y} a)$ up a sign: $(\partial_{y} a) = \pm (\sqrt{-2\Lambda/3}) a$, corresponding to $a(y) \propto e^{\pm (\sqrt{-2\Lambda/3}) y}$. Define the so-called warping parameter $k\equiv \sqrt{-\Lambda/6}$ for ease of writing, and remove the proportionality so that $a(y) = e^{\pm 2 k y}$ via coordinate rescaling.\footnote{The metric corresponding to $a(y) = e^{- 2 k y}$ describes 5D anti-de Sitter space (AdS${}_{5}$). More specifically, because the RS1 model has branes at $y = 0$ and $y = \pi r_{c}$, the RS1 model is a finite interval of AdS${}_{5}$, wherein the brane at $y = \pi r_{c}$ explicitly breaks the conformal invariance of the infinite AdS${}_{5}$.} Because this solution also satisfies the first constraint, all may seem well. However, this solution does not respect the orbifold reflection symmetry: neither solution is individually invariant under the replacement $y\rightarrow -y$. To fix this, we can patch together separate solutions in the regions $y<0$ and $y>0$ to form the continuous \& orbifold-even solution $a(y) = e^{\pm 2k|y|}$. Differentiating this new solution yields, keeping in mind the orbifold symmetry and periodic nature of the extra dimension,
\begin{align}
    (\partial_{y} a)^{2} &= \left[\pm 2k\,\text{sign}(y)\, a\right]^{2} = 4k^{2}\, a^{2}\\
    (\partial_{y}^{2} a) &= \left[4k^{2} \pm 4k\left(\delta_{0} - \delta_{\pi r_{c}}\right)\right] a
\end{align}
where we used
\begin{align}
    (\partial_{y}|y|) = \text{sign}(y)\hspace{35 pt} (\partial_{y}^{2}|y|) = 2\left(\delta_{0} - \delta_{\pi r_{c}}\right)
\end{align}
and define $\delta_{\overline{y}}\equiv \delta(y-\overline{y})$. Although this orbifold-even solution solves Eq. \eqref{385}, it does not solve Eq. \eqref{384}. In fact, any attempt to modify the action (and therein the Einstein field equations) that treats all of 5D spacetime on equal footing is doomed to fail. We will need to further extend the types of terms we include in the action in order to overcome this difficulty

To better understand why we are running into trouble, let us divide the 5D RS1 spacetime (which has coordinates $(x,y)$) into a collection of constant $y$ slices, e.g. hypersurfaces consisting of points $(x,\overline{y})$ for some $\overline{y} \in [0,\pi r_{c}]$. This defines what is called a ``foliation" of 5D RS1 spacetime into time-like 4D hypersurfaces\footnote{A time-like 4D hypersurface is a hypersurface where the normal vector at every point is space-like, such that the hypersurface itself resembles a 4D spacetime.}, where the hypersurfaces at $\overline{y}=0$ and $\overline{y}=\pi r_{c}$ are the RS1 branes. Choose one such hypersurface in this foliation. Because this hypersurface is itself a submanifold of spacetime, we can calculate its curvature. Furthermore, because it exists within a larger spacetime, it has two kinds of curvature: intrinsic (curvature tangent to the hypersurface) and extrinsic (curvature normal to the hypersurface). The extrinsic curvature of such a hypersurface is given by
\begin{align}
    K_{MN} = -\dfrac{1}{2}  \overline{G}_{MP} \, \overline{G}_{NQ} \, \tilde{G}^{PR} \, \tilde{G}^{QS} \, \LD_{n} \overline{G}_{RS}
\end{align}
where $n^{M}$ is a vector field of unit $5$-vectors normal to our hypersurface, $\LD_{n}$ denotes the Lie derivative along $n^{M}$, and $\overline{G}$ is the projection of the metric $G$ onto the hypersurface at $y=\overline{y}$. By choosing $n^{M}\equiv (0,0,0,0,1)$ as our hypersurface normals, the projected metric equals
\begin{align}
    [\overline{G}_{MN}(\overline{y})] = [G_{MN}(\overline{y}) + n_{M}n_{N}] = \matrixbb{a(\overline{y})\, \eta_{\mu\nu}}{0}{0}{0}
\end{align}
Thus $\overline{G}_{MN} \tilde{G}^{NR} = (\delta^{\mu}_{M} \delta^{R}_{\rho})\delta^{\rho}_{\mu}$ and the extrinsic curvature simplifies to
\begin{align}
    K_{MN} = -\dfrac{1}{2} \LD_{n} \overline{G}_{MN}
\end{align}
When acting on a rank-2 covariant tensor such as $\overline{G}_{MN}$, the Lie derivative $\LD_{n}$ equals
\begin{align}
    \LD_{n} \overline{G}_{MN} = n^{A} (\partial_{A} \overline{G}_{MN}) + (\partial_{M} n^{A}) \overline{G}_{AN} + (\partial_{N} n^{A}) \overline{G}_{AM}
\end{align}
Because $\overline{G}_{MN}$ is only nonzero in its upper $4\times4$ block and $n^{A}$ is only nonzero in its fifth component, only the first term of the Lie derivative contributes, and
\begin{align}
    [\LD_{n} \overline{G}_{MN}] = [n^{A} (\partial_{A} \overline{G}_{MN})] = [(\partial_{y} \overline{G}_{MN})] = \matrixbb{(\partial_{y} a)\eta_{\mu\nu}}{0}{0}{0}
\end{align}
such that the extrinsic curvature of a constant $y$ hypersurface in a spacetime with metric Eq. \eqref{374} equals
\begin{align}
    K_{MN}(\overline{y}) = -\dfrac{1}{2} (\partial_{y} a)\, \delta_{M}^{\mu}\delta_{N}^{\nu}\, \eta_{\mu\nu}
\end{align}
This extrinsic curvature poses a problem when trying to solve the Einstein field equations in the presence of an orbifold-even function like $a(y) = e^{\pm 2k|y|}$. In this case, $K_{MN}$ is nonzero, and thus necessarily implies additional warping in the spacetime geometry not accounted for solely by the standard Einstein-Hilbert Lagrangian nor an additional bulk cosmological constant. In particular, the orbifold symmetry demands that $K_{MN}(0^{+}) = - K_{MN}(0^{-})$ across the orbifold fixed point at $y = 0$ and $K_{MN}(r_{c}^{-}) = - K_{MN}((-r_{c})^{+})$ across the orbifold fixed point at $y = r_{c}$, which subsequently imply jumps in the extrinsic curvature at the branes, i.e.
\begin{align}
    \left.[K_{MN}]\right|_{y=\overline{y}} &\equiv K^{MN}(\overline{y}^{+}) - K^{MN}(\overline{y}^{-})\\
    &= 2 K^{MN}(\overline{y}^{+})\\
    &= -\left.(\partial_{y}a)\right|_{\overline{y}^{+}\rightarrow \overline{y}}\,\delta^{\mu}_{M}\delta^{\nu}_{N}\eta_{\mu\nu}
\end{align}
To accomplish a jump in the extrinsic curvature like this, we need a surface source of stress-energy (not unlike using a surface charge density to cause a jump in the electric field in classical E\&M). In analogy with our previous (bulk-based) situation, we have two immediate options for trying to achieve this: either embedding matter into the branes, or introducing a surface cosmological constant on each brane. We opt for the latter to keep things purely gravitational, and call each of these new surface cosmological constants a brane tension.

As far as the Einstein field equations are concerned, this means introducing new terms into the action. For terms evaluated on the brane, we use the appropriate brane-projected metric $\overline{G}$, but otherwise the new brane tension terms closely resemble our bulk cosmological constant term: we include them in our existing cosmological constant Lagrangian like so,
\begin{align}
    S_{\text{CC}} = -\dfrac{4}{\kappa_{\text{5D}}^{2}} \int d^{5}x \bigg[\Lambda \sqrt{G} + \lambda_{0}\,\sqrt{\overline{G}(0)}\, \delta(y) + \lambda_{\pi r_{c}}\,\sqrt{\overline{G}(\pi r_{c})}\, \delta(y-\pi r_{c}) \bigg] \label{398}
\end{align}
The constants $\Lambda$, $\lambda_{0}$, and $\lambda_{\pi r_{c}}$ will be determined soon using the Einstein field equations. The variation of the new terms with respect to $\tilde{G}^{MN}$ proceeds similarly to the bulk cosmological constant term so long as we are careful to continue projecting onto each respective brane: using the Lagrangian implied by Eq. \eqref{398}, we find
\begin{align}
    \dfrac{\delta}{\delta\tilde{G}^{AB}} \left[\mathcal{L}_{\text{CC}}\right] = -\dfrac{2}{\kappa_{\text{5D}}^{2}} \bigg[-\Lambda \sqrt{G} \, G_{AB} - \sum_{\overline{y}\in\{0,\pi r_{c}\}} \lambda_{\overline{y}} \sqrt{\overline{G}(\overline{y})}\, \overline{G}_{AB}(\overline{y}) \, \delta_{\overline{y}}  \bigg]
\end{align}
where $\delta_{\overline{y}}\equiv \delta(y-\overline{y})$. Therefore, the Einstein fields equations derived from combining Eq. \eqref{398} and the usual Einstein-Hilbert Lagrangian (in the absence of matter) are
\begin{align}
    \sqrt{G}\left[\mathcal{G}_{AB} - \Lambda G_{AB}\right] -\sum_{\overline{y}\in\{0,\pi r_{c}\}} \lambda_{\overline{y}} \sqrt{\overline{G}(\overline{y})}\, \overline{G}_{AB}(\overline{y}) \, \delta_{\overline{y}} = 0 \label{3100}
\end{align}
After substituting explicit values into the Einstein field equations Eq. \eqref{3100}, including
\begin{align}
    \sqrt{G} = a(y)^{2} \hspace{35 pt}\sqrt{\overline{G}(\overline{y})} = a(\overline{y})^{2}
\end{align}
we obtain
\begin{align}
    -\tfrac{3}{2}a^{2}(\partial^{2}_{y}a)-\Lambda\, a^{3} - \sum_{\overline{y}\in\{0,\pi r_{c}\}} \lambda_{\overline{y}}\, a(\overline{y})^{3} \delta_{\overline{y}} &= 0\\
    \dfrac{3}{2} (\partial_{y}a)^{2} + \Lambda a^{2} &= 0
\end{align}
The second equation was solved previously and led us (after a coordinate rescaling) to the orbifold-even function $a(y) = e^{\pm 2 k|y|}$ and bulk cosmological constant $\Lambda = - 6k^{2}$. When this solution is substituted into the first equation, all terms lacking Dirac deltas are automatically cancelled, and the residual Dirac deltas only cancel if
\begin{align}
    \mp 6k - \lambda_{0} &= 0\hspace{35 pt} \pm 6k - \lambda_{\pi r_{c}} = 0
\end{align}
Hence, each brane requires a different-signed tension, where the sign of the exponential in $a(y)$ determines which brane gets which sign. It is conventional to choose the sign such that the $y=0$ brane (sometimes called the hidden or Planck brane) has positive tension and the $y = \pi r_{c}$ brane (sometimes called the visible or TeV brane) has negative tension. Thus, we choose the lower sign option and find the Einstein field equations are solved by taking
\begin{align}
    a(y) = e^{-2k|y|}\hspace{35 pt}\Lambda = -k\lambda_{0} = k \lambda_{\pi r_{c}} = -6k^{2}
\end{align}
This completes the construction of the RS1 background metric.

We now summarize the results of the above derivation, but add the label ``(bkgd)" while doing so as to emphasize that these results are specific to the RS1 background metric. The background metric 5D RS1 Lagrangian equals
\begin{align}
    \mathcal{L}_{\text{5D}}^{(\text{bkgd})} = \mathcal{L}_{\text{EH}}^{(\text{bkgd})} + \mathcal{L}_{\text{CC}}^{(\text{bkgd})} =  -\dfrac{2}{\kappa_{\text{5D}}^{2}} \bigg[\sqrt{G^{(\text{bkgd})}}\, R - 12k^{2} \sqrt{G^{(\text{bkgd})}}  + 6k\sqrt{\overline{G}{}^{(\text{bkgd})}}\, (\partial_{y}^{2}|y|) \bigg] \label{L5Dvacuum}
\end{align}
wherein the Einstein-Hilbert and cosmological constant Lagrangians equal
\begin{align}
    \mathcal{L}_{\text{EH}}^{(\text{bkgd})} &= -\dfrac{2}{\kappa_{\text{5D}}^{2}} \sqrt{G^{(\text{bkgd})}} \, R^{(\text{bkgd})}\\
    \mathcal{L}_{\text{CC}}^{(\text{bkgd})} &= \dfrac{12 k}{\kappa_{\text{5D}}^{2}} \bigg[ 6k \sqrt{G^{(\text{bkgd})}} + \sqrt{\overline{G}{}^{(\text{bkgd})}}(\partial_{y}^{2}|y|)\bigg] \label{LEHandLCCvacuum}
\end{align}
with corresponding background metric and 4D projection
\begin{align}
    [G^{(\text{bkgd})}_{MN}] = \matrixbb{e^{-2k|y|}\eta_{\mu\nu}}{0}{0}{-1}\hspace{20 pt}\text{ and }\hspace{25 pt} [\overline{G}{}^{(\text{bkgd})}_{MN}] = \matrixbb{e^{-2k|y|}\eta_{\mu\nu}}{0}{0}{0} \label{GandGbar}
\end{align}
In order to obtain a particle theory of RS1 gravity, we must perturb the background solution summarized in Eqs. \eqref{L5Dvacuum}-\eqref{GandGbar} by field-dependent amounts. This is the topic of the next subsection.

\subsection{Perturbing the Background Metric} \label{SS - Perturbing The Vacuum}
The last subsection constructed the RS1 background metric, which is ultimately described by Eqs. \eqref{L5Dvacuum}-\eqref{GandGbar}. The particle theory is subsequently obtained by perturbing this background metric, but we must take care to correctly distinguish physical and unphysical degrees of freedom when doing so. For example, one way to parameterize a generic perturbed metric $G$ relative to the background metric $G^{(\text{bkgd})}$ is
\begin{align}
    [G_{MN}] = \matrixbb{e^{-2\big[k|y|+\hat{u}(x,y)\big]}\left(\eta_{\mu\nu} + \kappa_{\text{5D}} \hat{h}_{\mu\nu}(x,y)\right)}{\kappa_{\text{5D}}\hat{\rho}_{\mu}(x,y)}{\kappa_{\text{5D}}\hat{\rho}_{\nu}(x,y)}{-\big[1 + 2\hat{u}(x,y)\big]^{2}} \label{3110}
\end{align}
in coordinates $x^{M} = (x^{\mu},y)$, where $x^{\mu}$ are the usual 4D coordinates and $y \in [0,\pi r_{c}]$ is the extra-dimensional spatial coordinate (which is extended to $y \in [-\pi r_{c},+\pi r_{c}]$ by imposing orbifold invariance). Note that Eq. \eqref{3110} recovers $G^{(\text{bkgd})}$ when $\hat{h} = \hat{\rho} = \hat{u} = 0$. Via coordinate transformations, Eq. \eqref{3110} can always be brought into the form
\begin{align}
    [G_{MN}] = \matrixbb{e^{-2\big[k|y|+\hat{u}(x,y)\big]}\left(\eta_{\mu\nu} + \kappa_{\text{5D}} \hat{h}_{\mu\nu}(x,y)\right)}{0}{0}{-\big[1 + 2\hat{u}(x,y)\big]^{2}}
\end{align}
where $\rho_{\mu}$ is made to vanish via orbifold symmetry, and $\hat{u}(x,y)$ equals
\begin{align}
    \hat{u}(x,y) \equiv \dfrac{\kappa_{\text{5D}} \hat{r}(x)}{2\sqrt{6}} e^{+k(2|y|-\pi r_{c})} \label{udefinition}
\end{align}
in terms of a $y$-independent field $\hat{r}(x)$ \cite{Charmousis:1999rg}. The 5D fields $\hat{h}(x,y)$ and $\hat{r}(x)$ contain all dynamical degrees of freedom of the RS1 model \cite{Charmousis:1999rg}, and will be the source of our 4D particle content in the next chapter. By demanding that $ds^{2}$ be invariant under the orbifold symmetry, $\hat{h}_{\mu\nu}(x,y)$ and $\hat{r}(x)$ are necessarily even functions of $y$; in other words, these fields are ``orbifold even." Furthermore, because $G_{MN}$ is symmetric in its indices, $\hat{h}_{\mu\nu}(x,y)$ is symmetric as well.

For convenience, we will often parameterize the perturbed metric $G$ (and its projection onto a constant $y$ hypersurface, $\overline{G}$) as
\begin{align}
    [G_{MN}] = \matrixbb{w(x,y)\, g_{\mu\nu}}{0}{0}{-v(x,y)^{2}}\hspace{35 pt}[\overline{G}_{MN}] = \matrixbb{w(x,y)\, g_{\mu\nu}}{0}{0}{0} \label{GMNwvform}
\end{align}
where
\begin{align}
    g_{\mu\nu}(x,y) &\equiv \eta_{\mu\nu} + \kappa_{\text{5D}} h_{\mu\nu}(x,y)\\
    w(x,y) &\equiv \vep^{-2} e^{-2\hat{u}(x)}\\
    v(x,y) &\equiv 1 + 2\hat{u}(x) \label{endoffielddefinitions}
\end{align}
and $\vep \equiv e^{+k|y|}$. Replacing $G^{(\text{bkgd})}$ with $G$ (and $\overline{G}^{(\text{bkgd})}$ with $\overline{G}$) in Eqs. \eqref{L5Dvacuum}-\eqref{GandGbar} yields the 5D RS1 theory:
\begin{align}
    \mathcal{L}_{\text{5D}} = \mathcal{L}_{\text{EH}} + \mathcal{L}_{\text{CC}}
\end{align}
where
\begin{align}
    \mathcal{L}_{\text{EH}} &\equiv -\dfrac{2}{\kappa_{\text{5D}}^{2}} \sqrt{G}\, R\hspace{10 pt} \cong\hspace{10 pt} \mathcal{L}_{\Gamma\Gamma} \equiv -\dfrac{2}{\kappa_{\text{5D}}^{2}} \sqrt{G} \, \tilde{G}^{MN}\bigg[\Gamma^{Q}_{MP} \Gamma^{P}_{NQ} - \Gamma^{Q}_{MN}\Gamma^{P}_{PQ}\bigg] \label{Lag5DEH}\\
    \mathcal{L}_{\text{CC}} &= -\dfrac{2}{\kappa_{\text{5D}}^{2}} \bigg[- 12k^{2} \sqrt{G}  + 6k\sqrt{\overline{G}}\, (\partial_{y}^{2}|y|) \bigg] \label{Lag5DCC}
\end{align}
The alternate form of $\mathcal{L}_{\text{EH}}$ included on the RHS of Eq. \eqref{Lag5DEH} was derived in Subsection \ref{SS - Rewriting EH}.

In this parameterization, the invariant spacetime interval equals
\begin{align}
    ds^{2} = (G_{MN})\, dx^{M} \, dx^{N} = (w\, g_{\mu\nu})\, dx^{\mu}\, dx^{\nu} - (v^{2}) \, dy^{2}
\end{align}
Furthermore, the inverse metric $\tilde{G}^{MN}$ equals
\begin{align}
    [\tilde{G}^{MN}] = \matrixbb{w(x,y)^{-1}\,\tilde{g}^{\mu\nu}}{0}{0}{-v(x,y)^{-2}}
\end{align}
where $\tilde{g}^{\mu\nu}$ is the inverse of $g_{\mu\nu} = \eta_{\mu\nu} + \kappa_{\text{5D}} \hat{h}_{\mu\nu}$ such that $\tilde{g}^{\mu\nu} g_{\nu\rho} = \eta^{\mu}_{\rho}$, and the invariant volume element nicely decomposes into four-dimensional and extra-dimensional weights:
\begin{align}
    \sqrt{\det G}\, d^{4}x\, dy = \bigg[w^{2}\,\sqrt{-\det g}\, d^{4}x\bigg]\cdot (v\, dy) = (\sqrt{\det \overline{G}}\, d^{4}x)\cdot(v\, dy)
\end{align}
For use in the next subsection, note that $(\partial_{y}u) = +2k(\partial_{y}|y|) u$, such that
\begin{align}
    (\partial_{y} w) &= -2 w \left[k(\partial_{y}|y|) +(\partial_{y}u) \right]\\
    &= -2k(\partial_{y}|y|)\, (1+2u)\, w\\
    &= -2k(\partial_{y}|y|)\, v\, w \label{Dyw}
\end{align}
The extrinsic curvature $K_{MN}$ is now
\begin{align}
    K_{MN} = -\dfrac{1}{2} \LD_{n}\overline{G}_{MN} = -\dfrac{1}{2} n^{A}(\partial_{A}\overline{G}_{MN}) = -\dfrac{1}{2v}\, \delta^{\mu}_{M}\delta^{\nu}_{N}\,\partial_{y}(wg_{\mu\nu})
\end{align}
where the normal vector field equals $[n^{A}] = (0,0,0,0,1/v)$, such that
\begin{align}
    K \equiv \tilde{G}^{MN} K_{MN} &= \bigg[\delta_{\mu}^{M}\delta_{\nu}^{N}\,\dfrac{\tilde{g}^{\mu\nu}}{w}\bigg]\, \bigg[-\dfrac{1}{2v}\, \delta^{\mu}_{M}\delta^{\nu}_{N}\,\partial_{y}(wg_{\mu\nu})\bigg]\\
    &= - \dfrac{1}{2wv} \tilde{g}^{\mu\nu} \,\partial_{y}(wg_{\mu\nu})\\
    &= - \dfrac{1}{2v}\bigg[4\dfrac{(\partial_{y}w)}{w} + \ltr \tilde{g}g^{\prime}\rtr \bigg]
\end{align}
and
\begin{align}
    \sqrt{\overline{G}} \, K = - \dfrac{w^{2}}{2v}\,\sqrt{-g}\,\bigg[4\dfrac{(\partial_{y}w)}{w} + \ltr \tilde{g}g^{\prime}\rtr \bigg] \label{sqrtGbarK}
\end{align}

In order to eventually obtain the 4D effective RS1 model, its particle content, and its interactions (which are necessary to analyze the processes in which we are interested), we must weak field expand (WFE) the 5D RS1 Lagrangian. That is, we must series expand the 5D RS1 Lagrangian in powers of the 5D fields $\hat{h}_{\mu\nu}$ and $\hat{r}$. In principle, we could begin the weak field expansion now, but it is worthwhile to first modify $\mathcal{L}_{\text{5D}}$ by the addition of a total derivative $\Delta \mathcal{L}$ which will eliminate any terms proportional to $(\partial_{y}|y|)$ and $(\partial_{y}^{2}|y|)$ from the Lagrangian. This is achieved in the next subsection.

\subsection{Eliminating ``Cosmological Constant"-Like Terms} \label{SS- Cancel CC-Like}
The cosmological constant Lagrangian Eq. \eqref{Lag5DCC} contains terms that potentially complicate our analysis. For example, the terms proportional to $(\partial_{y}^{2}|y|)$ introduce Dirac deltas. When going from the 5D theory to the 4D effective theory, we must integrate the Lagrangian over the extra dimension, and the presence of Dirac deltas would replace what would otherwise become coupling integrals with evaluations of extra-dimensional wavefunctions at the branes. Thankfully, such terms in the cosmological constant Lagrangian combine with similar terms in the Einstein-Hilbert Lagrangian Eq. \eqref{Lag5DEH} to form physically-irrelevant total derivatives, and in this way all terms proportional to $(\partial_{y}|y|)$ or $(\partial_{y}^{2}|y|)$ are eliminated. The present subsection will explicitly demonstrate the elimination of these terms in the $\mathcal{L}_{\Gamma\Gamma}$ form of $\mathcal{L}_{\text{EH}}$ to all orders in the 5D fields as well as introducing a new term $\Delta \mathcal{L}$ to the RS1 Lagrangian which automates this elimination.

The terms in $\mathcal{L}_{\text{EH}}$ which cancel $\mathcal{L}_{\text{CC}}$ arise when an extra-dimensional derivative $\partial_{y}$ acts on a $y$-dependent multiplicative factor such as $\vep$ or $(\partial_{y}|y|)$ instead of the 5D field $\hat{h}_{\mu\nu}$ (recall that $\hat{r}$ is $y$-independent by construction). Hence, for the purposes of this subsection, we seek to isolate all such terms in $\mathcal{L}_{\Gamma\Gamma}$. To begin, we recalculate the Christoffel symbols (originally calculated in Eqs. \eqref{375}-\eqref{376} for the RS1 background solution) for the perturbed theory: recall
\begin{align}
    \Gamma^{P}_{MN} \equiv \dfrac{1}{2} \tilde{G}^{PQ} (\partial_{M} G_{NQ} + \partial_{N} G_{MQ} - \partial_{Q} G_{MN})
\end{align}
such that, using the fact that $G_{MN}$ and its inverse $\tilde{G}^{MN}$ are block-diagonal,
\begin{align}
    \Gamma^{5}_{\mu\nu} &= -\dfrac{1}{2}\tilde{G}^{55} (\partial_{5} G_{\mu\nu})\\
    \Gamma^{\rho}_{5\nu} &= +\dfrac{1}{2}\tilde{G}^{\rho\sigma} (\partial_{5} G_{\nu\sigma}) \hspace{35 pt}\implies\hspace{35 pt} \Gamma^{\rho}_{5\rho} =  +\dfrac{1}{2} \ltr \tilde{G} G^{\prime} \rtr\\
    \Gamma^{5}_{5\nu} &= +\dfrac{1}{2} \tilde{G}^{55} (\partial_{\nu} G_{55})\\
    \Gamma^{\rho}_{55} &= -\dfrac{1}{2} \tilde{G}^{\rho\sigma} (\partial_{\sigma} G_{55})\\
    \Gamma^{5}_{55} &= +\dfrac{1}{2} \tilde{G}^{55} (\partial_{5} G_{55})
\end{align}
where $\partial_{5} \equiv \partial_{y}$. Because $\tilde{G}^{MN}$ is block-diagonal, the index summations on the RHS of Eq. \eqref{Lag5DEH} only yield nonzero contributions when $(M,N)=(\mu,\nu)$ and $(M,N)=(5,5)$. Consider when $(M,N)=(\mu,\nu)$. The first product of Christoffel symbols in the $(M,N)=(\mu,\nu)$ case equals
\begin{align}
    \Gamma^{Q}_{\mu P} \Gamma^{P}_{\nu Q} = \Gamma^{\sigma}_{\mu\rho} \Gamma^{\rho}_{\nu\sigma} + \Gamma^{5}_{\mu\rho} \Gamma^{\rho}_{\nu5} + \Gamma^{\sigma}_{\mu 5} \Gamma^{5}_{\nu\sigma} + \Gamma^{5}_{\mu 5} \Gamma^{5}_{\nu5}
\end{align}
of which the second and third terms contain $y$-derivatives. Their contributions are identical and yield, when combined,
\begin{align}
    \Gamma^{Q}_{\mu P} \Gamma^{P}_{\nu Q} \mathrel{\mathop{\supset}^{\text{$\partial_{y}$ not on}}_{\text{a field}}} -\dfrac{1}{2} \tilde{G}^{55}\ltr G^{\prime}\tilde{G} G^{\prime}\rtr_{\mu\nu}
\end{align}
The second product of Christoffel symbols in the $(M,N)=(\mu,\nu)$ case equals
\begin{align}
    \Gamma^{Q}_{\mu \nu} \Gamma^{P}_{PQ} = \Gamma^{\sigma}_{\mu \nu} \Gamma^{\rho}_{\rho\sigma} + \Gamma^{5}_{\mu \nu} \Gamma^{\rho}_{\rho5} + \Gamma^{\sigma}_{\mu \nu} \Gamma^{5}_{5\sigma} + \Gamma^{5}_{\mu \nu} \Gamma^{5}_{55}
\end{align}
of which the second and fourth terms contain $y$-derivatives, such that
\begin{align}
    \Gamma^{Q}_{\mu \nu} \Gamma^{P}_{PQ} \mathrel{\mathop{\supset}^{\text{$\partial_{y}$ not on}}_{\text{a field}}} -\dfrac{1}{4} \tilde{G}^{55}  \ltr \tilde{G} G^{\prime} \rtr (\partial_{5}G_{\mu\nu}) - \dfrac{1}{4} \tilde{G}^{55} \tilde{G}^{55} (\partial_{5} G_{55}) (\partial_{5} G_{\mu\nu})
\end{align}
Hence, when contracted with $\tilde{G}^{\mu\nu}$, the net contributions coming from the $(M,N)=(\mu,\nu)$ case equal
\begin{align}
    \tilde{G}^{\mu\nu} \bigg[\Gamma^{Q}_{\mu P} \Gamma^{P}_{\nu Q} - \Gamma^{Q}_{\mu \nu} \Gamma^{P}_{PQ}\bigg] &\mathrel{\mathop{\supset}^{\text{$\partial_{y}$ not on}}_{\text{a field}}} -\dfrac{1}{2}\tilde{G}^{55}\ltr \tilde{G}G^{\prime}\tilde{G}G^{\prime}\rtr + \dfrac{1}{4} \tilde{G}^{55} \ltr \tilde{G} G^{\prime} \rtr^{2}\nonumber\\
    &\hspace{70 pt}+ \dfrac{1}{4} \tilde{G}^{55} \tilde{G}^{55} (\partial_{5} G_{55}) \ltr \tilde{G} G^{\prime}\rtr \label{badtermsEHmunu}
\end{align}
Meanwhile, the equivalent expression in the $(5,5)$ case equals, thanks to cancellations,
\begin{align}
    \tilde{G}^{55}\bigg[\Gamma^{Q}_{5P} \Gamma^{P}_{5Q} - \Gamma^{Q}_{55} \Gamma^{P}_{PQ}\bigg] &= \tilde{G}^{55}\bigg[\Gamma^{\sigma}_{5\rho} \Gamma^{\rho}_{5\sigma} + \Gamma^{5}_{5\rho} \Gamma^{\rho}_{55} - \Gamma^{\sigma}_{55} \Gamma^{\rho}_{\rho\sigma} - \Gamma^{5}_{55} \Gamma^{\rho}_{\rho5} \bigg]
\end{align}
of which the first and third terms contain $y$-derivatives, contributing overall
\begin{align}
    \tilde{G}^{55}\bigg[\Gamma^{Q}_{5P} \Gamma^{P}_{5Q} - \Gamma^{Q}_{55} \Gamma^{P}_{PQ}\bigg] &\mathrel{\mathop{\supset}^{\text{$\partial_{y}$ not on}}_{\text{a field}}} +\dfrac{1}{4}\tilde{G}^{55}\ltr\tilde{G}G^{\prime}\tilde{G}G^{\prime}\rtr - \dfrac{1}{4} \tilde{G}^{55}\tilde{G}^{55}(\partial_{5} G_{55}) \ltr \tilde{G} G^{\prime} \rtr \label{badtermsEH55}
\end{align}
Combining Eqs. \eqref{badtermsEHmunu} and \eqref{badtermsEH55} yields, at the level of the Einstein-Hilbert Lagrangian,
\begin{align}
    \mathcal{L}_{\text{EH}} \cong \mathcal{L}_{\Gamma\Gamma} \mathrel{\mathop{\supset}^{\text{$\partial_{y}$ not on}}_{\text{a field}}} -\dfrac{2}{\kappa_{\text{5D}}^{2}}\sqrt{G} \, \tilde{G}^{55}\bigg[\dfrac{1}{4} \ltr \tilde{G} G^{\prime} \rtr^{2} - \dfrac{1}{4} \ltr \tilde{G} G^{\prime} \tilde{G} G^{\prime} \rtr \bigg] \label{3142}
\end{align}
However, this expression contains more than just the terms we desire: some of the $y$-derivatives in this expression will end up acting on fields and, thus, not help eliminate $\mathcal{L}_{\text{CC}}$. To refine this expression further, we utilize the explicit form of $G$ in terms of $w$ and $v$ from Eq. \eqref{GMNwvform}. For example, with this parameterization the prefactor $\sqrt{G}\,\tilde{G}^{55}$ becomes $(v\, w^{2}\,\sqrt{-g}) (-1/v^{2}) = -(w^{2}/v)\,\sqrt{-g}$. This decomposition also allows $\ltr\tilde{G} G^{\prime}\rtr$ to be rewritten as
\begin{align}
    \ltr \tilde{G}G^{\prime} \rtr &= \ltr (\tilde{g}/w)\, \partial_{y}(wg)\rtr \\
    &= \dfrac{(\partial_{y} w)}{w} \ltr \tilde{g} g\rtr + \ltr \tilde{g} g^{\prime}\rtr
\end{align}
where we utilized the fact that $\ltr \tilde{g} g\rtr = \ltr \eta \rtr = 4$. Squaring this, we then obtain
\begin{align}
    \ltr \tilde{G}G^{\prime} \rtr^{2} = 16\,\dfrac{(\partial_{y} w)^{2}}{w^{2}} + 8 \, \dfrac{(\partial_{y} w)}{w}\, \ltr \tilde{g} g^{\prime}\rtr + \ltr \tilde{g} g^{\prime}\rtr^{2} \label{3137}
\end{align}
The final term in Eq. \eqref{3137} only contains $y$-derivatives acting on fields and thus can be ignored from here on. Similarly, the second term in Eq. \eqref{3142} is proportional to
\begin{align}
    \ltr \tilde{G} G^{\prime} \tilde{G} G^{\prime} \rtr = 4\,\dfrac{(\partial_{y} w)^{2}}{w^{2}} + 2\,\dfrac{(\partial_{y} w)}{w}\, \ltr \tilde{g} g^{\prime}\rtr + \ltr\tilde{g}g^{\prime}\tilde{g}g^{\prime}\rtr
\end{align}
wherein the first two terms involve $(\partial_{y}w)\propto (\partial_{y}|y|)$ via Eq. \eqref{Dyw} and the final term can be ignored. By keeping these distinctions in mind, the only terms in $\mathcal{L}_{\text{EH}}$ where $y$-derivatives do not act on fields are
\begin{align}
    \mathcal{L}_{\text{EH}} \cong \mathcal{L}_{\Gamma\Gamma} \mathrel{\mathop{\supset}^{\text{$\partial_{y}$ not on}}_{\text{a field}}} -\dfrac{2}{\kappa^{2}_{\text{5D}}} \,\left(-\dfrac{w^{2}}{v} \,\sqrt{-g}\right)\bigg[ 3 \,\dfrac{(\partial_{y} w)^{2}}{w^{2}} + \dfrac{3}{2}\, \,\dfrac{(\partial_{y} w)}{w} \, \ltr \tilde{g} g^{\prime} \rtr \bigg]
\end{align}
But $(\partial_{y} w)/w = -2k(\partial_{y}|y|)\, v$ via Eq. \eqref{Dyw}, such that
\begin{align}
    \mathcal{L}_{\text{EH}} \cong \mathcal{L}_{\Gamma\Gamma} \mathrel{\mathop{\supset}^{\text{$\partial_{y}$ not on}}_{\text{a field}}} -\dfrac{2}{\kappa^{2}_{\text{5D}}} w^{2}\,\sqrt{-g} \, \bigg[ -12k^{2}v + 3k(\partial_{y}|y|)\ltr\tilde{g} g^{\prime}\rtr  \bigg] \label{3140}
\end{align}
This completes our manipulations of the Einstein-Hilbert Lagrangian. We can apply a similar decomposition to $\mathcal{L}_{\text{CC}}$ in Eq. \eqref{Lag5DCC}:
\begin{align}
    \mathcal{L}_{\text{CC}} = -\dfrac{2}{\kappa^{2}_{\text{5D}}} w^{2} \, \sqrt{-g} \, \bigg[ -12 k^{2} v + 6k(\partial_{y}^{2}|y|) \bigg] \label{3141}
\end{align}
where $\sqrt{\overline{G}} = w^{2}\,\sqrt{-g}$ because $\overline{G}$ only includes the $4$-by-$4$ part of the metric $G$. Combining Eq. \eqref{3141} in its entirety with the terms we isolated from $\mathcal{L}_{\Gamma\Gamma}$ in Eq. \eqref{3140} yields, in total,
\begin{align}
    \mathcal{L}_{\text{5D}} \cong \mathcal{L}_{\Gamma\Gamma} + \mathcal{L}_{\text{CC}} \mathrel{\mathop{\supset}^{\text{$\partial_{y}$ not on}}_{\text{a field}}} -\dfrac{6 k}{\kappa^{2}_{\text{5D}}} w^{2} \, \sqrt{-g} \, \bigg[ -8 k v + (\partial_{y}|y|)\ltr\tilde{g} g^{\prime}\rtr + 2(\partial_{y}^{2}|y|) \bigg]
\end{align}
Thankfully, this collection of terms actually forms the total derivative $\partial_{y}[w^{2}\, \sqrt{-g}\, (\partial_{y}|y|)]$ up to multiplicative constants:
\begin{align}
    \partial_{y}\left[w^{2}\, \sqrt{-g}\,(\partial_{y}|y|)\right] &= \sqrt{-g}\bigg[ 2w(\partial_{y}w) (\partial_{y}|y|) + \dfrac{1}{2}w^{2}\ltr \tilde{g} g^{\prime}\rtr (\partial_{y}|y|) + w^{2}\,(\partial_{y}^{2}|y|)  \bigg]\\
    &= \dfrac{1}{2} w^{2}\, \sqrt{-g} \, \bigg[ 4\,\dfrac{(\partial_{y} w)}{w} \, (\partial_{y}|y|) + (\partial_{y}|y|) \ltr \tilde{g} g^{\prime} \rtr + 2(\partial_{y}^{2}|y|) \bigg]\\
    &= \dfrac{1}{2} w^{2}\, \sqrt{-g} \, \bigg[ -8kv + (\partial_{y}|y|) \ltr \tilde{g} g^{\prime} \rtr + 2(\partial_{y}^{2}|y|) \bigg]
\end{align}
Therefore, all terms in $\mathcal{L}_{\text{5D}}$ that resemble contributions from the cosmological constant Lagrangian combine to form a total derivative,
\begin{align}
    \mathcal{L}_{\text{5D}} \cong \mathcal{L}_{\Gamma\Gamma} + \mathcal{L}_{\text{CC}} \mathrel{\mathop{\supset}^{\text{$\partial_{y}$ not on}}_{\text{a field}}} -\dfrac{12 k}{\kappa_{\text{5D}}^{2}} \partial_{y}\left[w^{2}\, \sqrt{-g}\,(\partial_{y}|y|)\right] \cong 0 \label{allCCtermscancel}
\end{align}
and only terms where derivatives are applied to fields contribute to the physics.

To avoid performing the integration by parts implied by Eq. \eqref{allCCtermscancel} in the future, we can manually subtract the total derivative we eliminated from the 5D Lagrangian and use, in practice,
\begin{align}
    \mathcal{L}_{\text{5D}}^{(\text{RS})} = \mathcal{L}_{\Gamma\Gamma} + \mathcal{L}_{\text{CC}} + \Delta \mathcal{L}_{\Gamma\Gamma} \label{L5DwithDeltaL}
\end{align}
where
\begin{align}
    \Delta \mathcal{L}_{\Gamma\Gamma} \equiv \dfrac{12 k}{\kappa_{\text{5D}}^{2}}\, \partial_{y}\left[w^{2}\, \sqrt{-g}\,(\partial_{y}|y|)\right]\hspace{10 pt}\bigg(= -\dfrac{6}{\kappa_{\text{5D}}^{2}}\, \partial_{y}\left[\dfrac{w^{2}}{v}\,\sqrt{-g}\,\dfrac{(\partial_{y}w)}{w}\right] \bigg) \label{DeltaLValue}
\end{align}
This $\Delta\mathcal{L}_{\Gamma\Gamma}$ is different than the $\Delta\mathcal{L}_{\text{EH}}$ used in \cite{Chivukula:2020hvi} because the present dissertation uses $\mathcal{L}_{\Gamma\Gamma}$ derived in Subsection \ref{SS - Rewriting EH} instead of $\mathcal{L}_{\text{EH}}$. Specifically, \cite{Chivukula:2020hvi} uses
\begin{align}
    \Delta \mathcal{L}_{\text{EH}} = \dfrac{2}{\kappa_{\text{5D}}^{2}} \partial_{y}\bigg[\dfrac{w^{2}}{v}\,\sqrt{-g}\, \bigg(\ltr \tilde{g}g^{\prime}\rtr + \dfrac{(\partial_{y}w)}{w}\bigg) \bigg]
\end{align}
Note their difference equals
\begin{align}
    \Delta\mathcal{L}_{\text{EH}} - \Delta\mathcal{L}_{\Gamma\Gamma} = \dfrac{2}{\kappa_{\text{5D}}^{2}} \partial_{y}\bigg[\dfrac{w^{2}}{v}\,\sqrt{-g}\, \bigg(\ltr \tilde{g}g^{\prime}\rtr + 4\,\dfrac{(\partial_{y}w)}{w}\bigg) \bigg]
\end{align}
which equals, via Eq. \eqref{sqrtGbarK},
\begin{align}
    \Delta\mathcal{L}_{\text{EH}} - \Delta\mathcal{L}_{\Gamma\Gamma} = -\dfrac{2}{\kappa_{\text{5D}}^{2}} \,\partial_{y}\left[2\,\sqrt{\overline{G}}\,K\right] = \mathcal{L}_{\text{GHY}}
\end{align}
where the final term yields the Gibbons-Hawking-York boundary term from general relativity \cite{PhysRevD.79.024028}.

Because of the structure of $\mathcal{L}_{\Gamma\Gamma}$ in Eq. \eqref{Lag5DEH}, there are two derivatives in every term of $\mathcal{L}_{\text{5D}}^{(\text{RS})}$ and those derivatives never act on the same field instance. This fact is useful in the next chapter, when we analyze the coupling structures present in the 4D effective RS1 theory. Having obtained Eq. \eqref{DeltaLValue}, we now weak field expand the 5D RS1 Lagrangian.

\section[5D Weak Field Expanded RS1 Lagrangian]{5D Weak Field Expanded RS1 Lagrangian\footnote{This section was originally published as Appendix A of \cite{Chivukula:2020hvi}. The content has been updated to reflect the new form of the Einstein-Hilbert Lagrangian, and material has been added to connect this section to the rest of this dissertation.}}\label{AppendixWFE}

This section details the weak field expansion of the RS1 model Lagrangian, Eq. \eqref{L5DwithDeltaL}, including explicit expressions for all terms in the Lagrangian having four or fewer instances of the 5D fields $\hat{h}_{\mu\nu}(x,y)$ and $\hat{r}(x)$.

\subsection{General Considerations}
The matter-free RS1 model Lagrangian $\mathcal{L}^{(\text{RS})}_{\text{5D}}$ is defined by Eq. \eqref{L5DwithDeltaL} and is perturbed relative to a background metric according to Eqs. \eqref{udefinition}-\eqref{endoffielddefinitions}. By expanding in field content, we obtain a dual power series in the 5D fields $\hat{h}_{\mu\nu}$ and $\hat{r}$:
\begin{align}
    \mathcal{L}_{\text{5D}}^{(\text{RS})}(\hat{h},\hat{r}) = \sum_{H,R=0}^{+\infty} \mathcal{L}^{(\text{RS})}_{h^{H} r^{R}}\hspace{20 pt}\text{ where }\hspace{20 pt}\mathcal{L}^{(\text{RS})}_{h^{H} r^{R}} \propto \kappa^{H+R-2} \, (\hat{h}_{\mu_{1}\nu_{1}}\cdots \hat{h}_{\mu_{H}\nu_{H}}) \, \hat{r}^{\,R}
\end{align}
and $\kappa\equiv \kappa_{\text{5D}}$. A power series of this sort is called a weak field expansion, and the Lagrangian that results is the 5D weak field expanded (WFE) RS1 Lagrangian.

As remarked at the end of the last section, each term in $\mathcal{L}_{\text{5D}}$ contains exactly two derivatives, which by construction must act on (different) 5D fields. In order to contract all Lorentz indices, that pair of derivatives is necessarily either a pair of 4D derivatives or a pair of extra-dimensional derivatives (i.e. there are no terms containing a mixture of both). We call a term wherein both derivatives are four-dimensional an A-type term whereas we call a term wherein both derivatives are extra-dimensional a B-type term. By partitioning all terms containing $H$ $\hat{h}_{\mu\nu}$ fields and $R$ $\hat{r}$ fields into A-type and B-type terms, we obtain the following decomposition:
\begin{align}
    \mathcal{L}^{(\text{RS})}_{h^{H}r^{R}} &= \kappa^{H+R-2} \left[e^{-\pi kr_{c}}\varepsilon^{+2}\right]^{R}\bigg[\varepsilon^{-2}\overline{\mathcal{L}}_{A:h^{H}r^{R}}+\varepsilon^{-4}\overline{\mathcal{L}}_{B:h^{H}r^{R}}\bigg]~, \label{LRShHrR}
\end{align}
where $\varepsilon \equiv e^{-kr_{c}|\varphi|}$. By definition, the quantities $\overline{\mathcal{L}}_{A:h^{H}r^{R}}$ and $\overline{\mathcal{L}}_{B:h^{H}r^{R}}$ contain exclusively A-type and B-type terms respectively, and all warp factors $\vep$ have been organized in Eq. \eqref{LRShHrR} such that $\overline{\mathcal{L}}_{A:h^{H}r^{R}}$ and $\overline{\mathcal{L}}_{B:h^{H}r^{R}}$ only depend on the extra-dimensional coordinate $y$ through the $y$-dependence of the 5D field $\hat{h}_{\mu\nu}(x,y)$.

Having established these notations and organization, we must next answer a practical question: to what order in each field should we expand the 5D RS1 Lagrangian? For the processes relevant to this dissertation (tree-level $2$-to-$2$ KK mode scattering), we require the cubic $\hat{h}\hat{h}\hat{h}$ and $\hat{h}\hat{h}\hat{r}$ interactions and the quartic $\hat{h}\hat{h}\hat{h}\hat{h}$ interaction. This latter interaction occurs at $\mathcal{O}(\kappa^{2})$ in the Lagrangian. Because we have already calculated them anyway, we actually provide {\it all} terms of the 5D WFE RS1 Lagrangian at $\mathcal{O}(\kappa^{2})$ and lower, which we organize based on field content:
\begin{align}
    \mathcal{L}^{(\text{RS})}_{\text{5D}} &= \mathcal{L}^{(\text{RS})}_{hh} + \mathcal{L}^{(\text{RS})}_{rr} + \mathcal{L}^{(\text{RS})}_{hhh} + \cdots +  \mathcal{L}^{(\text{RS})}_{rrr}\\
    &\hspace{15 pt}+ \mathcal{L}^{(\text{RS})}_{hhhh} + \cdots +  \mathcal{L}^{(\text{RS})}_{rrrr} + \mathcal{O}(\kappa^{3}) ~.\nonumber
\end{align}
In principle, these interaction Lagrangians enable the calculation of every 2-to-2 tree-level scattering matrix element in the matter-free RS1 model.

The next subsection reviews several notations and formula that are useful for the weak field expansion of $\mathcal{L}^{(\text{RS})}_{\text{5D}}$. The remaining subsections then summarize the 5D WFE RS1 Lagrangian through $\mathcal{O}(\kappa^{2})$, which is the principal result of this chapter. Afterwards, an appendix derives various weak field expansion formulas there were used to obtain that result.

\subsection{Notations and Useful Formulas}
The 5D RS1 Lagrangian, Eq. \eqref{L5DwithDeltaL}, is composed of various functions of the metric $G_{MN}(x,y)$ and, thus, the $4\times4$ quantity $g_{\mu\nu}(x,y) \equiv \eta_{\mu\nu} + \kappa\, \hat{h}_{\mu\nu}(x,y)$. This includes the inverse quantity $\tilde{g}^{\mu\nu}$ and the determinant $\sqrt{g}\equiv\sqrt{-\det [g_{\mu\nu}]}$, both of which must be expanded in powers of $\hat{h}_{\mu\nu}$. It is in these expansions that the twice-squared bracket and tilde-as-inverse notations prove particularly useful.

Recall that the twice-squared bracket notation is used to indicate sequential Lorentz index contractions of rank-2 tensors, e.g.
\begin{align}
    \ltr\hat{h}^{\prime}\rtr \equiv (\partial_{y}\hat{h}^{\alpha}_{\alpha})\hspace{35 pt}\ltr \hat{h}\hat{h}\rtr_{\alpha\beta} = \hat{h}_{\alpha\gamma}\hat{h}^{\gamma}_{\beta}\hspace{35 pt} \ltr \hat{h}\hat{h}\hat{h}\rtr = \hat{h}^{\alpha}_{\beta} \hat{h}^{\beta}_{\gamma}\hat{h}^{\gamma}_{\alpha}
\end{align}
where a prime indicates differentiation with respect to $y$. When writing the 5D WFE RS1 Lagrangian, we also utilize the following abbreviations
\begin{align}
    \hat{h} \equiv \hat{h}^{\alpha}_{\alpha} \hspace{35 pt}(\partial_{\alpha} \hat{h}) \equiv (\partial_{\alpha}\hat{h}_{\beta}^{\beta}) \hspace{35 pt}(\partial\hat{h})_{\alpha} \equiv (\partial^{\beta}\hat{h}_{\beta\alpha})
\end{align}

As mentioned above, the $4\times4$ quantity $g_{\mu\nu}$ exactly satisfies
\begin{align}
    g_{\alpha\beta} = \eta_{\alpha\beta} +\kappa \hat{h}_{\alpha\beta}~.
\end{align}
From this, the inverse quantity $\tilde{g}^{\mu\nu}$ may be solved for order-by-order by imposing its defining condition, $g_{\alpha\beta}\tilde{g}^{\beta\gamma} = \eta_{\alpha}^{\gamma}$. This process yields
\begin{align}
\tilde{g}^{\alpha\beta} = \eta^{\alpha\beta} + \sum_{n=1}^{+\infty} (- \kappa)^{n} \ltr\hat{h}^{n}\rtr^{\alpha\beta}~.
\end{align}
Meanwhile, weak field expansion of the determinant $\sqrt{g} \equiv \sqrt{-\det[g_{\mu\nu}]}$ yields
\begin{align}
    \sqrt{g} &= \prod_{n=1}^{+\infty} \exp\left[\dfrac{(-1)^{n-1}}{2n}\kappa^n \hspace{2 pt}\ltr \hat{h}^{n}\rtr \right]~.
\end{align}
which equals, to fourth order in the fields,
\begin{align}
\sqrt{g}&=1 + \dfrac{\kappa}{2}\hspace{2 pt}\hat{h} + \dfrac{\kappa^2}{8}\left(\hat{h}^{2} - 2 \ltr \hat{h}\hat{h}\rtr\right)+ \dfrac{\kappa^3}{48}\left(\hat{h}^3 - 6 \hat{h}\ltr\hat{h}\hat{h}\rtr + 8\ltr\hat{h}\hat{h}\hat{h}\rtr\right)\nonumber\\
&\hspace{10 pt}+\dfrac{1}{384}\left(\hat{h}^{4} - 12 \hat{h}^{2} \ltr \hat{h}\hat{h}\rtr + 12 \ltr \hat{h}\hat{h}\rtr^{2} + 32 \hat{h} \ltr \hat{h}\hat{h}\hat{h}\rtr - 48 \ltr \hat{h}\hat{h}\hat{h}\hat{h}\rtr \right)+\mathcal{O}(\kappa^{5})~.\nonumber
\end{align}
Derivations of these weak field expansion formulas are included in the appendix of this chapter.

The remainder of this section summarizes the 5D WFE RS1 Lagrangian at quadratic, cubic, and quartic order.

\subsection{\label{sec:levelAa}Quadratic-Level Results}

\begin{align}
    \overline{\mathcal{L}}_{A:hh} &= (\partial \hat{h})_{\mu} (\partial^{\mu} \hat{h})  - (\partial\hat{h})_{\mu}^{2} + \dfrac{1}{2} (\partial_{\mu}\hat{h}_{\nu\rho})^{2}- \dfrac{1}{2} (\partial_{\mu}\hat{h})^{2} \label{LAhh}\\
  \overline{\mathcal{L}}_{B:hh} &= \dfrac{1}{2} \ltr \hat{h}^\prime \rtr^2 - \dfrac{1}{2} \ltr \hat{h}^\prime \hat{h}^\prime \rtr \label{LBhh}\\
    &\text{ }\nonumber\\
    \overline{\mathcal{L}}_{A:rr} &= \frac{1}{2} (\partial_\mu \hat{r})^{2} \label{LArr}\\
    \overline{\mathcal{L}}_{B:rr} &= 0 \label{LBrr}
\end{align}

\subsection{\label{sec:levelAb}Cubic-Level Results}

\begin{align}
    \overline{\mathcal{L}}_{A:hhh} &= \dfrac{1}{2} \hat{h} (\partial \hat{h})_{\mu} (\partial^{\mu}\hat{h}) - \hat{h}_{\mu\nu} (\partial\hat{h})^{\mu}(\partial^{\nu}\hat{h}) - \dfrac{1}{4}\hat{h}(\partial_{\mu}\hat{h})^{2} - \hat{h}_{\nu\rho} (\partial\hat{h})_{\mu} (\partial^{\mu} \hat{h}^{\nu\rho})\nonumber\\
    &\hspace{10 pt}+ \hat{h}_{\nu\rho}(\partial_{\mu}\hat{h})(\partial^{\mu}\hat{h}^{\nu\rho}) + \dfrac{1}{4}\hat{h}(\partial_{\mu}\hat{h}_{\nu\rho})^{2} - \hat{h}^{\rho}_{\sigma} (\partial_{\mu}\hat{h}_{\nu\rho})(\partial^{\mu}\hat{h}^{\nu\sigma})\nonumber\\
    &\hspace{10 pt}+ \dfrac{1}{2}\hat{h}_{\mu\nu}(\partial^{\mu}\hat{h})(\partial^{\nu}\hat{h}) - \hat{h}_{\mu\rho}(\partial^{\mu}\hat{h}^{\nu\rho})(\partial_{\nu}\hat{h}) - \dfrac{1}{2}\hat{h}(\partial_{\mu}\hat{h}_{\nu\rho})(\partial^{\nu}\hat{h}^{\mu\rho})\nonumber\\
    &\hspace{10 pt}+\hat{h}_{\rho}^{\sigma}(\partial_{\mu}\hat{h}_{\nu\sigma})(\partial^{\nu}\hat{h}^{\mu\rho}) + 2\hat{h}^{\sigma}_{\rho} (\partial_{\mu}\hat{h}_{\nu\sigma})(\partial^{\rho}\hat{h}^{\mu\nu}) - \dfrac{1}{2}\hat{h}^{\sigma}_{\rho} (\partial_{\sigma}\hat{h}_{\mu\nu})(\partial^{\rho}\hat{h}^{\mu\nu})\\
    \overline{\mathcal{L}}_{B:hhh} &= \dfrac{1}{4}\hat{h}\ltr\hat{h}^{\prime}\rtr^{2} - \ltr\hat{h}^{\prime}\rtr\ltr\hat{h}\hat{h}^{\prime}\rtr - \dfrac{1}{4} \hat{h} \ltr\hat{h}^{\prime}\hat{h}^{\prime}\rtr + \ltr\hat{h}\hat{h}^{\prime}\hat{h}^{\prime}\rtr\\
    &\text{ }\nonumber\\
    \overline{\mathcal{L}}_{A:hhr} &= 0\\
    \overline{\mathcal{L}}_{B:hhr} &= \dfrac{1}{2}\sqrt{\dfrac{3}{2}} \bigg[ \ltr\hat{h}^{\prime}\hat{h}^{\prime}\rtr -\ltr\hat{h}^{\prime}\rtr^{2} \bigg] \hat{r} \\
    &\text{ }\nonumber\\
    \overline{\mathcal{L}}_{A:hrr} &= -\dfrac{1}{3}(\partial\hat{h})_{\mu}\hat{r}(\partial^{\mu}\hat{r}) + \dfrac{1}{3}(\partial_{\mu}\hat{h})\hat{r}(\partial^{\mu}\hat{r}) + \dfrac{1}{4}\hat{h}(\partial_{\mu}\hat{r})^{2} - \dfrac{1}{2}\hat{h}_{\mu\nu} (\partial^{\mu}\hat{r})(\partial^{\nu}\hat{r})\\
    \overline{\mathcal{L}}_{B:hrr} &= 0\\
    &\text{ }\nonumber\\
    \overline{\mathcal{L}}_{A:rrr} &= -\dfrac{1}{\sqrt{6}} \hat{r}(\partial_{\mu}\hat{r})^{2}\\
    \overline{\mathcal{L}}_{B:rrr} &= 0
\end{align}

\subsection{\label{sec:levelAc}Quartic-Level Results}

\begin{align}
    \overline{\mathcal{L}}_{A:hhhh} &= \dfrac{1}{8}\hat{h}^{2}(\partial\hat{h})_{\mu} (\partial^{\mu}\hat{h}) - \dfrac{1}{4} \ltr\hat{h}\hat{h}\rtr (\partial\hat{h})_{\mu}(\partial^{\mu}\hat{h}) - \dfrac{1}{2}\hat{h} \hat{h}_{\mu\nu} (\partial\hat{h})^{\mu}(\partial^{\nu}\hat{h}) + \ltr \hat{h}\hat{h}\rtr_{\mu\nu}(\partial\hat{h})^{\mu}(\partial^{\nu}\hat{h})\nonumber\\
    &\hspace{10 pt}-\dfrac{1}{16}\hat{h}^{2}(\partial_{\mu}\hat{h})^{2} +\dfrac{1}{8} \ltr\hat{h}\hat{h}\rtr (\partial_{\mu}\hat{h})^{2} - \dfrac{1}{2}\hat{h} \hat{h}_{\mu\nu} (\partial\hat{h})_{\rho} (\partial^{\rho}\hat{h}^{\mu\nu}) + \ltr \hat{h}\hat{h}\rtr_{\mu\nu} (\partial\hat{h})_{\rho} (\partial^{\rho}\hat{h}^{\mu\nu})\nonumber\\
    &\hspace{10 pt}+\hat{h}_{\mu\nu}\hat{h}_{\rho\sigma}(\partial\hat{h})^{\mu} (\partial^{\nu}\hat{h}^{\rho\sigma}) + \dfrac{1}{2}\hat{h}\hat{h}_{\mu\nu}(\partial_{\rho}\hat{h})(\partial^{\rho}\hat{h}^{\mu\nu}) - \ltr\hat{h}\hat{h}\rtr_{\mu\nu}(\partial_{\rho}\hat{h})(\partial^{\rho}\hat{h}^{\mu\nu})\nonumber\\
    &\hspace{10 pt} + \dfrac{1}{16}\hat{h}^{2}(\partial_{\rho}\hat{h}_{\mu\nu})^{2} - \dfrac{1}{8} \ltr \hat{h}\hat{h} \rtr (\partial_{\rho}\hat{h}_{\mu\nu})^{2} - \dfrac{1}{2}\hat{h}\hat{h}^{\sigma}_{\rho} (\partial_{\mu}\hat{h}_{\nu\sigma})(\partial^{\mu}\hat{h}^{\nu\rho})\nonumber\\
    &\hspace{10 pt} + \ltr\hat{h}\hat{h}\rtr^{\sigma}_{\rho} (\partial_{\mu}\hat{h}_{\nu\sigma})(\partial^{\mu}\hat{h}^{\nu\rho}) - \dfrac{1}{2} \hat{h}_{\mu\nu} \hat{h}_{\rho\sigma} (\partial_{\tau}\hat{h}^{\mu\nu})(\partial^{\tau}\hat{h}^{\rho\sigma}) + \dfrac{1}{2} \hat{h}_{\mu\sigma} \hat{h}_{\rho\nu} (\partial_{\tau}\hat{h}^{\mu\nu})(\partial^{\tau}\hat{h}^{\rho\sigma})\nonumber\\
    &\hspace{10 pt} +\dfrac{1}{4}\hat{h}\hat{h}_{\mu\nu}(\partial^{\mu}\hat{h})(\partial^{\nu}\hat{h}) - \dfrac{1}{2} \ltr \hat{h}\hat{h} \rtr_{\mu\nu} (\partial^{\mu}\hat{h})(\partial^{\nu}\hat{h}) - \dfrac{1}{2} \hat{h} \hat{h}_{\mu\nu} (\partial_{\rho}\hat{h}) (\partial^{\mu} \hat{h}^{\nu\rho})\nonumber\\
    &\hspace{10 pt} +\ltr \hat{h}\hat{h}\rtr_{\mu\nu} (\partial_{\rho}\hat{h}) (\partial^{\mu}\hat{h}^{\nu\rho}) + \hat{h}_{\mu\rho}\hat{h}_{\nu\sigma} (\partial^{\mu}\hat{h})(\partial^{\nu}\hat{h}^{\rho\sigma}) - \hat{h}_{\mu\nu}\hat{h}_{\rho\sigma} (\partial^{\mu}\hat{h})(\partial^{\nu}\hat{h}^{\rho\sigma})\nonumber\\
    &\hspace{10 pt}-\dfrac{1}{8}\hat{h}^{2} (\partial_{\mu}\hat{h}_{\nu\rho})(\partial^{\nu}\hat{h}^{\mu\rho}) + \dfrac{1}{4} \ltr \hat{h}\hat{h} \rtr (\partial_{\mu}\hat{h}_{\nu\rho})(\partial^{\nu}\hat{h}^{\mu\rho}) + \dfrac{1}{2}\hat{h}\hat{h}^{\sigma}_{\rho} (\partial_{\mu}\hat{h}_{\nu\sigma})(\partial^{\nu}\hat{h}^{\mu\rho})\nonumber\\
    &\hspace{10 pt} - \ltr\hat{h}\hat{h}\rtr^{\sigma}_{\rho} (\partial_{\mu}\hat{h}_{\nu\sigma})(\partial^{\nu}\hat{h}^{\mu\rho}) + \hat{h}\hat{h}^{\sigma}_{\rho} (\partial_{\mu}\hat{h}_{\nu\sigma})(\partial^{\rho}\hat{h}^{\mu\nu}) - 2\ltr\hat{h}\hat{h}\rtr^{\sigma}_{\rho} (\partial_{\mu}\hat{h}_{\nu\sigma})(\partial^{\rho}\hat{h}^{\mu\nu})\nonumber\\
    &\hspace{10 pt} - 2\hat{h}_{\mu\nu} \hat{h}_{\rho\sigma} (\partial_{\tau} \hat{h}^{\nu\rho}) (\partial^{\sigma} \hat{h}^{\tau\mu}) + \hat{h}_{\mu\sigma}\hat{h}_{\nu\rho} (\partial_{\tau} \hat{h}^{\nu\rho}) (\partial^{\sigma} \hat{h}^{\tau\mu}) - \dfrac{1}{4} \hat{h}\hat{h}^{\sigma}_{\rho} (\partial_{\sigma}\hat{h}_{\mu\nu})(\partial^{\rho}\hat{h}^{\mu\nu})\nonumber\\
    &\hspace{10 pt}+\dfrac{1}{2} \ltr\hat{h}\hat{h}\rtr^{\sigma}_{\rho} (\partial_{\sigma}\hat{h}_{\mu\nu})(\partial^{\rho}\hat{h}^{\mu\nu}) - \hat{h}^{\mu\nu}\hat{h}_{\rho\sigma} (\partial_{\mu}\hat{h}^{\sigma\tau})(\partial^{\rho}\hat{h}_{\nu\tau}) + \hat{h}^{\mu}_{\rho} \hat{h}^{\nu}_{\sigma} (\partial_{\mu}\hat{h}^{\sigma\tau})(\partial^{\rho}\hat{h}_{\nu\tau})\\
    \overline{\mathcal{L}}_{B:hhhh} &= \dfrac{1}{16}\hat{h}^{2} \ltr\hat{h}^{\prime}\rtr^{2} - \dfrac{1}{8} \ltr\hat{h}\hat{h} \rtr \ltr \hat{h}^{\prime} \rtr^{2} - \dfrac{1}{2}\hat{h} \ltr\hat{h}^{\prime}\rtr \ltr \hat{h} \hat{h}^{\prime} \rtr + \ltr\hat{h}^{\prime}\rtr \ltr \hat{h}\hat{h} \hat{h}^{\prime}\rtr - \dfrac{1}{16} \hat{h}^{2} \ltr \hat{h}^{\prime}\hat{h}^{\prime} \rtr\nonumber\\
    &\hspace{10 pt}+\dfrac{1}{8} \ltr\hat{h}\hat{h}\rtr \ltr \hat{h}^{\prime}\hat{h}^{\prime}\rtr + \dfrac{1}{2} \hat{h} \ltr \hat{h} \hat{h}^{\prime}\hat{h}^{\prime} \rtr - \ltr\hat{h}\hat{h}\hat{h}^{\prime}\hat{h}^{\prime}\rtr + \dfrac{1}{2} \ltr \hat{h} \hat{h}^{\prime} \rtr^{2} - \dfrac{1}{2} \ltr \hat{h} \hat{h}^{\prime}\hat{h} \hat{h}^{\prime}\rtr\\
    &\text{ }\nonumber\\
    \overline{\mathcal{L}}_{A:hhhr} &= 0\\
    \overline{\mathcal{L}}_{B:hhhr} &= \dfrac{1}{4}\sqrt{\dfrac{3}{2}} \bigg[ -\hat{h}\ltr\hat{h}^{\prime}\rtr^{2} + 4\ltr\hat{h}^{\prime}\rtr \ltr\hat{h}\hat{h}^{\prime}\rtr + \hat{h}\ltr \hat{h}^{\prime}\hat{h}^{\prime} \rtr - 4 \ltr \hat{h}\hat{h}^{\prime}\hat{h}^{\prime} \rtr \bigg]\hat{r}
\end{align}

\begin{align}
    \overline{\mathcal{L}}_{A:hhrr} &= -\dfrac{1}{12} (\partial\hat{h})_{\mu} (\partial^{\mu}\hat{h})\hat{r}^{2} +\dfrac{1}{24} (\partial_{\mu}\hat{h})^{2}\hat{r}^{2} -\dfrac{1}{24}(\partial_{\mu}\hat{h}_{\nu\rho})^{2} \hat{r}^{2} + \dfrac{1}{12} (\partial_{\mu}\hat{h}_{\nu\rho})(\partial^{\nu}\hat{h}^{\mu\rho}) \hat{r}^{2}\nonumber\\
    &\hspace{10 pt}-\dfrac{1}{6}\hat{h}(\partial\hat{h})_{\mu} \hat{r}(\partial^{\mu}\hat{r}) + \dfrac{1}{3}\hat{h}_{\mu\nu} (\partial\hat{h})^{\mu}\hat{r}(\partial^{\nu}\hat{r}) + \dfrac{1}{6}\hat{h}(\partial_{\mu}\hat{h})\hat{r}(\partial^{\mu}\hat{r})-\dfrac{1}{3}\hat{h}_{\mu\nu} (\partial_{\rho}\hat{h}^{\mu\nu})(\partial^{\rho}\hat{h}) \nonumber\\
    &\hspace{10 pt}-\dfrac{1}{3}\hat{h}_{\mu\nu} (\partial^{\mu}\hat{h})\hat{r}(\partial^{\nu}\hat{r}) + \dfrac{1}{3}\hat{h}_{\nu\rho}(\partial^{\rho}\hat{h}^{\mu\nu})\hat{r}(\partial_{\mu}\hat{r}) + \dfrac{1}{16}\hat{h}^{2} (\partial_{\mu}\hat{r})^{2} - \dfrac{1}{8}\ltr\hat{h}\hat{h}\rtr (\partial_{\mu}\hat{r})^{2}\nonumber\\
    &\hspace{10 pt}- \dfrac{1}{4}\hat{h}\hat{h}_{\mu\nu} (\partial^{\mu}\hat{r})(\partial^{\nu}\hat{r}) + \dfrac{1}{2} \ltr \hat{h}\hat{h}\rtr_{\mu\nu} (\partial^{\mu}\hat{r})(\partial^{\nu}\hat{r})\\
    \overline{\mathcal{L}}_{B:hhrr} &= \dfrac{5}{12}\bigg[\ltr \hat{h}^{\prime}\rtr^{2} - \ltr \hat{h}^{\prime}\hat{h}^{\prime} \rtr \bigg]\hat{r}^{2}
\end{align}

\begin{align}
    \overline{\mathcal{L}}_{A:hrrr} &= \dfrac{1}{6\sqrt{6}}\bigg[ 2 (\partial\hat{h})_{\mu} \hat{r}^{2} (\partial^{\mu}\hat{r}) - 2 (\partial_{\mu}\hat{h}) \hat{r}^{2} (\partial^{\mu}\hat{r}) - 3\hat{h}\hat{r} (\partial_{\mu}\hat{r})^{2} + 6 \hat{h}_{\mu\nu}\hat{r}(\partial^{\mu}\hat{r})(\partial^{\nu}\hat{r}) \bigg]\\
    \overline{\mathcal{L}}_{B:hrrr} &= 0\\
    &\text{ }\nonumber\\
    \overline{\mathcal{L}}_{A:rrrr} &= \dfrac{1}{8} \hat{r}^{2} (\partial_{\mu}\hat{r})^{2}\\
    \overline{\mathcal{L}}_{B:rrrr} &= 0 \label{LBrrrr}
\end{align}

\section{Appendix: WFE Expressions} \label{GRAppendices - WFE}
This appendix derives formulas for weak field expanding the inverse metric $\tilde{G}^{MN}$ and the covariant spacetime volume factor $\sqrt{G} \equiv \sqrt{|\det G|}$.

\subsection{Inverse Metric}
Consider a metric $G$ on $X$-dimensional spacetime of the form
\begin{align}
    G_{MN} \equiv c_{0}\,\eta_{MN} + H_{MN}
\end{align}
where the real number $c_{0}$ is positive. If $H_{MN}$ is small relative to $c_{0}\,\eta_{MN}$, we may weak field expand $\tilde{G}^{MN}$ with respect to the field $H_{MN}$. Note that the form of this expansion must be, using the twice-squared bracket notation defined in Section \ref{Classical - Minkowski},
\begin{align}
    \tilde{G}^{MN} \equiv \sum_{n=0}^{+\infty} \tilde{c}_{n}\, \ltr H^{n} \rtr^{MN} \label{1202}
\end{align}
We can solve for the unknown coefficients $\tilde{c}$ in Eq. \eqref{1202} by imposing the inversion condition $G_{MN}\tilde{G}^{NP} = \eta^{P}_{M}$ like so:
\begin{align}
    \eta^{P}_{M} &\equiv \bigg[c_{0}\, \eta_{MN} + H_{MN}\bigg]\,\bigg[ \sum_{n=0}^{+\infty} \tilde{c}_{n}\, \ltr H^{n} \rtr^{NP}\bigg]\\
    &= c_{0} \sum_{n=0}^{+\infty} \tilde{c}_{n} \ltr H^{n}\rtr^{P}_{M} + \sum_{n=0}^{+\infty} \tilde{c}_{n} \ltr H^{n+1}\rtr^{P}_{M}\\
    &= c_{0}\,\tilde{c}_{0}\, \eta_{M}^{P} + \sum_{n=1}^{+\infty} (c_{0}\,\tilde{c}_{n} + \tilde{c}_{n-1})\ltr H^{n} \rtr^{P}_{M}
\end{align}
which forces the recursive relations
\begin{align}
    \tilde{c}_{0} = c_{0}^{-1} \hspace{35 pt} \tilde{c}_{n} = -c_{0}^{-1}\tilde{c}_{n-1} = (-1)^{n} c_{0}^{-(n+1)}
\end{align}
such that, when $G_{MN} = c_{0}\,\eta_{M} + H_{MN}$,
\begin{align}
    \tilde{G}^{MN} = \sum_{n=0}^{+\infty} (-1)^{n} c_{0}^{-(n+1)}\ltr H^{n} \rtr^{MN}
\end{align}

\subsection{Covariant Volume Factor}
Next, let us weak field expand the covariant spacetime volume factor $\sqrt{|\det G|}$ for various choices of the metric $G_{MN}$. As usual, $\det G$ here refers to the determinant of the matrix of components $G_{MN}$. We will increase the complexity of $G$ in stages until it is of the form of the RS1 metric.

\subsubsection{Minkowski Spacetime}
The $X$-dimensional Minkowski metric is defined such that $[\eta_{MN}] \equiv \text{Diag}(+1,-1,\dots,-1) = [\eta^{MN}]$, from which we may immediately calculate
\begin{align}
    \sqrt{|\det \eta|} = \sqrt{|(+1)\cdot(-1)^{X-1}|} = 1
\end{align}
To prepare for more complicated cases, let us also calculate this another way. Namely, we may use the formula,
\begin{align}
    \det A = \exp\left\{\text{tr}\left[\text{Log}\,(A)\right]\right\} \label{1209}
\end{align}
to write
\begin{align}
    \sqrt{\pm \det A} = \begin{cases}
    \left|\exp\left[\tfrac{1}{2}\text{tr}[\text{Log}\,(A)]\right]\right|\\[0.75em]
    \left|i \exp\left\{\tfrac{1}{2}\text{tr}[\text{Log}\,(A)]\right\}\right|
    \end{cases}
\end{align}
In order to ensure the LHS equals $\sqrt{|\det A|}$ when applied to $A=\eta$ (and, later, $A=G$), we will take the $+$ case when $X$ is odd and $-$ case when $X$ is even. This allows us to write,
\begin{align}
    \sqrt{|\det A|} = \bigg|i^{X+1} \exp\left\{\tfrac{1}{2}\text{tr}[\text{Log}\,(A)]\right\} \bigg| \label{1210}
\end{align}
The matrix logarithm present on the RHS of Eq. \eqref{1210} is defined via power series,
\begin{align}
    \text{Log}\left(\mathbbm{1}+A\right) \equiv \sum_{n=1}^{+\infty} \dfrac{(-1)^{n-1}}{n} A^{n} = A - \dfrac{1}{2} A^{2} + \dfrac{1}{3} A^{3} - \dots \label{matrixlogdef}
\end{align}
For a diagonal matrix (and using principal values),
\begin{align}
    \text{Log}\left[\text{Diag}\left(A_{1},\dots,A_{N}\right)\right] = \text{Diag}\left[\log(A_{1}),\dots,\log(A_{N})\right]
\end{align}
such that
\begin{align}
    \text{Log}\, \eta &= \text{Diag}\left[\log(+1),\log(-1),\dots,\log(-1)\right]\\
    &= \text{Diag}\left(0,i\pi,i\pi,\cdots,i\pi\right)
\end{align}
and so
\begin{align}
    \exp\left[\tfrac{1}{2}\text{tr}\left(\text{Log}\,\eta\right)\right] = \exp\left[\tfrac{1}{2}(X-1)i\pi \right] = i^{(X-1)}
\end{align}
Thus,
\begin{align}
    \sqrt{|\det \eta|} &= \left|i^{2X}\right| = \left|(-1)^{X}\right|= 1
\end{align}
which is consistent with our first calculation. This second method is excessive for the Minkowski metric. However, it is useful for more complicated metrics whose determinants cannot be calculated directly.

\subsubsection{Perturbing Minkowski Spacetime}
Next, we consider the perturbed metric $G_{MN} \equiv c_{0}\,\eta_{MN} + H_{MN}$ (note this is the metric we used in the previous subsection). Our goal is to weak field expand $\sqrt{|\det G|}$, i.e. calculate $\sqrt{|\det G|}$ as perturbative expansion in $H$ near the background metric $\eta$. Because $\eta = \tilde{\eta}$ and $\eta^{2} = \eta\, \tilde{\eta} = \mathbbm{1}$ when considered as matrices, we can write $G$ as the following product:
\begin{align}
    G = c_{0}\,\eta + H \hspace{15 pt} = \eta\, (c_{0}\, \mathbbm{1} + \eta H)
\end{align}
If $[\overline{A},\overline{B}] = 0$ for matrices $\overline{A}$ and $\overline{B}$, then we can apply $\text{Log}(\overline{A}\overline{B}) = \text{Log}(\overline{A}) + \text{Log}(\overline{B})$. However, this is not the case for the product in the above expression: $\eta \eta H = H = [H_{MN}]$ whereas $\eta H \eta = [H^{MN}]$ such that $[\eta,c_{0}\mathbbm{1}+\eta H] = [\eta,\eta H]$ is nonzero. Thankfully, there is a simplification afforded to us by the Baker-Campbell-Hausdorff (BCH) formula. The BCH formula is of the form,
\begin{align}
    \text{Exp}(\overline{A})\,\text{Exp}(\overline{B}) = \text{Exp}\left(\overline{A} + \overline{B} + \tfrac{1}{2}[\overline{A},\overline{B}] +\dots\right)
\end{align}
This is useful to us after making the replacements $(\overline{A},\overline{B})\rightarrow (\text{Log}\,\overline{A},\text{Log}\,\overline{B})$ and taking the matrix logarithm of both sides. Then the BCH becomes
\begin{align}
    \text{Log}(\overline{A}\overline{B}) =  \text{Log}(\overline{A}) + \text{Log}(\overline{B}) + \tfrac{1}{2}\left[\text{Log}(\overline{A}),\text{Log}(\overline{B})\right] + \dots
\end{align}
To apply the determinant formula Eq. \eqref{1209}, we take the trace of both sides of this equation. Because the trace distributes over addition, we find
\begin{align}
    \text{tr}\, \text{Log}(\overline{A}\overline{B}) = \text{tr}\,\text{Log}(\overline{A}) + \text{tr}\,\text{Log}(\overline{B}) + \tfrac{1}{2}\text{tr}\left[\text{Log}(\overline{A}),\text{Log}(\overline{B})\right] + \dots
\end{align}
where higher-order terms contain traces of increasingly-many commutators. But the trace of any commutator vanishes because $\text{tr}(XY) = \text{tr}(YX)$, such that
\begin{align}
    \text{tr} [X, Y] = \text{tr}(XY) - \text{tr}(YX) = \text{tr}(XY) - \text{tr}(XY) = 0
\end{align}
Therefore, the traces of all commutators in our modified BCH formula vanish, yielding
\begin{align}
    \text{tr}\, \text{Log}(\overline{A}\overline{B}) = \text{tr}\,\text{Log}(\overline{A}) + \text{tr}\,\text{Log}(\overline{B})
\end{align}
which implies
\begin{align}
    \sqrt{|\det \overline{A}\overline{B}|} = \bigg| i^{X+1} \exp\left\{\tfrac{1}{2}\text{tr}[\text{Log}\,(\overline{A})]\right\} \exp\left\{\tfrac{1}{2}\text{tr}[\text{Log}\,(\overline{B})]\right\}\bigg|
\end{align}
and, setting $\overline{A} = \eta$ and $\overline{B} = c_{0}\mathbbm{1} + \eta H$,
\begin{align}
    \sqrt{|\det G|} = \bigg| i^{X+1} \exp\left\{\tfrac{1}{2}\text{tr}[\text{Log}\,(c_{0}\eta)]\right\} \exp\left\{\tfrac{1}{2}\text{tr}[\text{Log}\,(\mathbbm{1} + \eta H)]\right\}\bigg| \label{1225}
\end{align}
The first exponential can be evaluated exactly: because
\begin{align}
    \text{tr}[\text{Log}(c_{0}\eta)] &= \log[\det(c_{0}\eta)]\\
    &= \log\left[c_{0}^{X}(-1)^{X-1}\right]\\
    &= \log(c_{0}^{X}) + (X-1)i\pi
\end{align}
it is the case that
\begin{align}
    \exp\left\{\tfrac{1}{2}\text{tr}\left[\text{Log}\left(c_{0}\eta)\right) \right]\right\} = c_{0}^{X/2}\,\exp\left[\tfrac{1}{2}(X-1)i\pi\right] = i^{X-1}c_{0}^{X}
\end{align}
Substituting this into Eq. \eqref{1225}, we obtain the exact expression
\begin{align}
    \sqrt{|\det G|} &= c_{0}^{X/2} \exp\left\{\tfrac{1}{2} \text{tr}\left[\text{Log}\left(\mathbbm{1} +\eta H \right)\right]\right\}
\end{align}
Finally, using the perturbative expression for the matrix logarithm Eq. \eqref{matrixlogdef}, we obtain
\begin{align}
    \sqrt{|\det (c_{0}\eta + H)|} &= c_{0}^{X/2} \prod_{n=1}^{+\infty} \exp\left(\dfrac{(-1)^{n-1}}{2n} \ltr H^n \rtr \right) \label{1231}
\end{align}
where $\ltr H \rtr = \eta^{MN} \, H_{NM}$, $\ltr H^{2} \rtr = \eta^{MN} \, H_{NP} \, \eta^{PQ} \, \eta_{QM}$, and so-on. To obtain the $\mathcal{O}(H^{n})$ terms in $\sqrt{|\det G|}$, we should expand each exponential in the product to $\mathcal{O}(H^{n})$. For example, to obtain $\mathcal{O}(H^{4})$ results, the relevant exponentials and their expansions are
\begin{align}
    \exp\left(+\tfrac{1}{2}\ltr H\rtr \right) &= 1 + \tfrac{1}{2}\ltr H\rtr + \tfrac{1}{8} \ltr H \rtr^{2} + \tfrac{1}{48}\ltr H\rtr^{3} + \tfrac{1}{384} \ltr H \rtr^{4} + \mathcal{O}(H^{5})\\[0.75 em]
    \exp\left(-\tfrac{1}{4}\ltr H^{2} \rtr\right) &= 1 - \tfrac{1}{4}\ltr H^{2}\rtr + \tfrac{1}{32} \ltr H^{2} \rtr^{2} + \mathcal{O}(H^{6})\\[0.75 em]
    \exp\left(+\tfrac{1}{8}\ltr H^{3} \rtr\right) &= 1 + \tfrac{1}{8}\ltr H^{3}\rtr + \mathcal{O}(H^{6})\\[0.75 em]
    \exp\left(-\tfrac{1}{16}\ltr H^{4} \rtr\right) &= 1 - \tfrac{1}{16}\ltr H^{4}\rtr + \mathcal{O}(H^{8})
\end{align}
which yields
\begin{align}
    \sqrt{|\det G|} &= c_{0}^{X/2}\bigg[ 1 + \dfrac{1}{2} \ltr H\rtr + \dfrac{1}{8} \bigg(\ltr H \rtr^{2} - 2\ltr H^{2}\rtr\bigg)\nonumber\\
    &\hspace{-45 pt}+ \dfrac{1}{48}\bigg(\ltr H \rtr^{3} - 6 \ltr H \rtr \, \ltr H^{2} \rtr + 6 \ltr H^{3} \rtr\bigg)\nonumber\\
    &\hspace{-45 pt}+\dfrac{1}{384} \bigg(\ltr H \rtr^{4} - 12 \ltr H \rtr^{2} \, \ltr H^{2} \rtr + 12 \ltr H^{2} \rtr^{2} + 24 \ltr H \rtr \, \ltr H^{3} \rtr - 24 \ltr H^{4} \rtr \bigg) + \mathcal{O}(H^{5}) \bigg]
\end{align}

\subsubsection{Block Diagonal Extension}
Suppose we expand the metric $G$ even further into an $(X+1)$-dimensional object $G_{(X+1)\text{D}}$, so that
\begin{align}
    G_{(X+1)\text{D}} = \matrixbb{w_{0}\,G}{\vec{0}^{\, T}}{\vec{0}}{-v_{0}^{2}} = \matrixbb{w_{0} \, (c_{0} \, \eta + H)}{\vec{0}^{\, T}}{\vec{0}}{-v_{0}^{2}}
\end{align}
where $w_{0}$ and $v_{0}$ are real and positive. To find $\sqrt{|\det G_{(X+1)\text{D}}|}$, we employ a fact about block diagonal matrices. Let $M$ be a block diagonal matrix $M\equiv \text{Diag}(A,B)$ where $A$ and $B$ are square matrices. We may define additional matrices $A^{\prime} \equiv \text{Diag}(A,\mathbbm{1}_{B})$ and $B^{\prime} \equiv \text{Diag}(\mathbbm{1}_{A},B)$, where $\mathbbm{1}_{A}$ and $\mathbbm{1}_{B}$ are identity matrices of the same dimensionality as $A$ and $B$ respectively. $\overline{A}$ and $\overline{B}$ commute ($[A^{\prime},B^{\prime}] = 0$) and their product recovers $M$ ($M = A^{\prime} B^{\prime}$). Thus, the BCH formula implies $\text{Log}(M) = \text{Log}(A^{\prime}) + \text{Log}(B^{\prime})$, and
\begin{align}
    \det(M) &= \exp\left[\text{tr}(\text{Log}\,M)\right]\\
    &=\exp\left[\text{tr}\left(\text{Log}\,A^{\prime} + \text{Log}\,B^{\prime}\right)\right]\\
    &= \exp\left[\text{tr}(\text{Log}\,A^{\prime}) + \text{tr}(\text{Log}\,B^{\prime})\right]\\
    &= \exp\left[\text{tr}(\text{Log}\,A^{\prime})\right]\,\exp\left[ \text{tr}(\text{Log}\,B^{\prime})\right]\\
    &= \det(A^{\prime})\,\det(B^{\prime})
\end{align}
Because $\det(\mathbbm{1}_{A}) = \det(\mathbbm{1}_{B}) = 1$, this result implies $\det(A^{\prime}) = \det(A)\det(\mathbbm{1}_{A}) = \det(A)$ and $\det(B^{\prime}) = \det(B)\det(\mathbbm{1}_{B}) = \det(B)$, such that
\begin{align}
    \det\matrixbb{A}{0}{0}{B} = \det(A)\,\det(B)
\end{align}
This generalizes to multiple blocks via recursion, i.e. the determinant of a block diagonal matrix $\det[\text{Diag}(M_{1},M_{2},\dots,M_{n})]$ is the product of the determinant of the individual blocks $\det(M_{1})\,\det(M_{2})\,\dots\,\det(M_{n})$. Using this on our extended metric, we find
\begin{align}
    \sqrt{|\det G_{(X+1)\text{D}}|} = \sqrt{\left|\det[w_{0} G]\,\det(-v_{0}^{2})\right|} = v_{0}\, w_{0}^{X/2}\, \sqrt{|\det G|}
\end{align}
or
\begin{align}
    \sqrt{|\det G_{(X+1)\text{D}}|} = v_{0}\, w_{0}^{X/2}\, \sqrt{|\det(c_{0}\,\eta + H)|}
\end{align}
from which we can use the previous perturbative result, Eq. \eqref{1231}. This is the form relevant to the 5D RS1 model.
\chapter{The 4D Effective RS1 Model and its Sum Rules} \label{Chapter5Dto4D}
\section{Chapter Summary}
The principal result of the last chapter was the weak field expansion (WFE) of the 5D RS1 Lagrangian, as summarized in Eqs. \eqref{LRShHrR}-\eqref{LBrrrr}. Up to quartic order in the fields, we derived each term $\mathcal{L}_{h^{H}r^{R}}^{(\text{RS})}$ containing $H$ instances of the field $\hat{h}_{\mu\nu}(x,y)$ and $R$ instances of the field $\hat{r}(x)$, and partitioned them into A-type and B-type terms according to Eq. \eqref{LRShHrR}:
\begin{align}
    \mathcal{L}^{(\text{RS})}_{h^{H}r^{R}} &= \kappa^{H+R-2} \left[e^{-\pi kr_{c}}\,\varepsilon^{+2}\right]^{R}\bigg[\varepsilon^{-2}\overline{\mathcal{L}}_{A:h^{H}r^{R}}+\varepsilon^{-4}\overline{\mathcal{L}}_{B:h^{H}r^{R}}\bigg]
\end{align}
where $\kappa \equiv \kappa_{\text{5D}}$. This chapter demonstrates how the 5D fields $\hat{h}_{\mu\nu}(x,y)$ and $\hat{r}(x)$ in the 5D WFE RS1 Lagrangian encode information about 4D spin-2 and spin-0 fields respectively. For example, consider the quadratic terms obtained via this process, as recorded in Eqs. \eqref{LAhh}-\eqref{LBrr},
\begin{align}
    \mathcal{L}_{hh}^{(\text{RS})} &= \vep^{-2}\bigg[(\partial \hat{h})_{\mu} (\partial^{\mu} \hat{h})  - (\partial\hat{h})_{\mu}^{2}  + \dfrac{1}{2} (\partial_{\mu}\hat{h}_{\nu\rho})^{2}- \dfrac{1}{2} (\partial_{\mu}\hat{h})^{2} \bigg]+ \vep^{-4}\bigg[\dfrac{1}{2} \ltr \hat{h}^\prime \rtr^2 - \dfrac{1}{2} \ltr \hat{h}^\prime \hat{h}^\prime \rtr\bigg] \label{LRShh}\\
    \mathcal{L}_{rr}^{(\text{RS})} &= \left[e^{-\pi kr_{c}}\, \vep^{+2} \right]^{2}\cdot\bigg[\dfrac{1}{2}(\partial_{\mu}\hat{r})^{2}\bigg] \label{LRSrr}
\end{align}
These are structurally similar to the 4D Lagrangians from Eqs. \eqref{Ls0massless1}, \eqref{Ls2massless1}, and \eqref{Ls2massive1}:
\begin{align}
    \mathcal{L}^{(s=2)}_{\text{massless}} &\equiv (\partial \hat{h})_{\mu} (\partial^{\mu} \hat{h}) - (\partial\hat{h})_{\mu}^{2} +\dfrac{1}{2} (\partial_{\mu} \hat{h}_{\nu\rho})^2 - \dfrac{1}{2} (\partial_{\mu} \hat{h})^2 \label{Ls2massless}\\
    \mathcal{L}^{(s=2)}_{\text{massive}} &\equiv \mathcal{L}^{(s=2)}_{\text{massless}} + m^{2}\bigg[\dfrac{1}{2} \hat{h}^{2} - \dfrac{1}{2} \ltr \hat{h}\hat{h} \rtr \bigg] \label{Ls2massive}\\
    \mathcal{L}^{(s=0)}_{\text{massless}} &\equiv \dfrac{1}{2} (\partial_{\mu} \hat{r})^{2} \label{Ls0massless}
\end{align}
which are the canonical massless spin-2, massive spin-2, and massless spin-0 Lagrangians respectively. Specifically, if $\hat{h}_{\mu\nu}(x,y)$ is momentarily assumed $y$-independent, then the terms proportional to $\vep^{-4}$ in $\mathcal{L}^{(\text{RS})}_{hh}$ from Eq. \eqref{LRShh} vanish. The remaining terms are proportional to $\vep^{-2}$ and exactly mimic the Lorentz structures of the massless spin-2 Lagrangian (Eq. \eqref{Ls2massless}). Furthermore, if we restore the $y$-dependence of $\hat{h}_{\mu\nu}(x,y)$, the Lorentz structures of the newly-revived $\vep^{-4}$ terms mimic the Fierz-Pauli mass terms of the massive spin-2 Lagrangian (Eq. \eqref{Ls2massive}). This hints (correctly) that the 5D field $\hat{h}_{\mu\nu}(x,y)$ contains information about 4D spin-2 particle excitations, with its $y$-dependence specifically encoding information about 4D particle masses. Meanwhile, the Lorentz structure of $\mathcal{L}_{rr}^{(\text{RS})}$ in Eq. \eqref{LRSrr} directly mimics the massless spin-0 Lagrangian (Eq. \eqref{Ls0massless}). The absence of a massive spin-0 structure for the $y$-independent $\hat{r}(x)$ field synergizes well with our existing observation that massive spin-2 structures arose from the $y$-dependence of $h_{\mu\nu}(x,y)$: in all, the $y$-independent field $\hat{r}$ seemingly only contains information about a massless spin-0 particle excitation.

This chapter formalizes how the 5D fields $\hat{h}_{\mu\nu}(x,y)$ and $\hat{r}(x)$ generate 4D fields and thus 4D particle content. The key technique is Kaluza-Klein (KK) decomposition, which allows the 5D fields to be written as sums of 4D fields weighted by extra-dimensional wavefunctions, e.g.
\begin{align}
    \hat{h}_{\mu\nu}(x,y) &= \dfrac{1}{\sqrt{\pi r_c}} \sum_{n=0}^{+\infty} \hat{h}^{(n)}_{\mu\nu}(x)\psi_n(\varphi) \hspace{35 pt} \hat{r}(x) = \dfrac{1}{\sqrt{\pi r_c}}\,
   \hat{r}^{(0)}(x) \psi_0 \label{46}
\end{align}
where $\{\psi_{n}(\varphi)\}$ are the aforementioned wavefunctions and $\varphi = y/r_{c} \in [-\pi,+\pi]$ parameterizes the extra dimension. The zero mode wavefunction $\psi_{0}$ present in both decompositions is independent of $\varphi$ and thus constant across the extra dimension. The wavefunctions $\psi_{n}$ solve a Sturm-Liouville (SL) equation, and thereby form a complete basis for orbifolded-even continuous functions $f(\varphi)$:
\begin{align}
    f(\varphi) = \dfrac{1}{\sqrt{\pi r_{c}}}\, f_{n}\, \psi_{n}(\varphi)\hspace{15 pt}\implies\hspace{15 pt} f_{n} = \sqrt{\dfrac{r_{c}}{\pi}} \int_{-\pi}^{+\pi} d\varphi\hspace{10 pt}\vep^{-2} f(\varphi)\psi_{n}(\varphi)
\end{align}
where $\vep\equiv \exp(kr_{c}|\varphi|)$. Although this decomposition appears more symmetric when expressed in terms of $y = \varphi r_{c}$,
\begin{align}
    f(y) = \dfrac{1}{\sqrt{\pi r_{c}}} f_{n}\, \, \psi_{n}\hspace{-3 pt}\left(\tfrac{y}{r_{c}}\right)\hspace{15 pt}\implies\hspace{15 pt} f_{n} = \dfrac{1}{\sqrt{\pi r_{c}}} \int_{-\pi r_{c}}^{+\pi r_{c}} dy\hspace{10 pt}\vep^{-2} f(y) \,\psi_{n}\hspace{-3 pt}\left(\tfrac{y}{r_{c}}\right)
\end{align}
working in terms of $\varphi$ makes manifest the fact that the wavefunctions and mass spectrum $\{\mu_{n}\} = \{m_{n}r_{c}\}$ depend only on the parameter combination $kr_{c}$ (as opposed to $k$ and $r_{c}$ independently). Thus, we favor the use of $\varphi$ during KK decomposition and the subsequent investigation of important integrals. Such integrals over $\varphi$ are generated when the KK decomposition ansatz is utilized while determining the 4D effective Lagrangian,
\begin{align}
    \mathcal{L}^{(\text{eff})}_{\text{4D}}(x) \equiv \int_{-\pi r_{c}}^{+\pi r_{c}} dy\hspace{5 pt}\mathcal{L}_{\text{5D}}(x,y) \label{L4Deff}
\end{align}
In this way, the 4D effective theory bundles all extra-dimensional dependence into various integrals of products of wavefunctions.

The rest of this chapter proceeds as follows. Footnotes detail how results in this chapter relate to our published works.
\begin{itemize}
    \item[$\bullet$] Section \ref{WfxnDerivation} introduces KK decomposition and derives the wavefunctions necessary for KK decomposition to yield canonical 4D particle content. Because of its importance for future work, the derivation is performed under slightly more general circumstances than is required for this dissertation.
    \item[$\bullet$] Section \ref{4D Effective RS1 Model} then applies KK decomposition to the quadratic 5D Lagrangians, thereby demonstrating that $\hat{h}_{\mu\nu}(x,y)$ embeds a massless spin-2 field $\hat{h}^{(0)}_{\mu\nu}(x)$ (the graviton) and a tower of massive spin-2 fields $\hat{h}^{(n)}_{\mu\nu}(x)$ (massive KK modes) whereas $\hat{r}(x)$ only embeds a massless spin-0 field $\hat{r}^{(0)}(x)$ (the radion). KK decomposition is then applied to the more general weak field expanded 5D Lagrangian. This requires integrating over the extra dimension, which results in interactions weighted by integrals of products of KK wavefunctions. These integrals define A-type and B-type couplings, which originate from A-type and B-type terms respectively. The $kr_{c}$ dependence of these coupling integrals in the large $kr_{c}$ limit is briefly considered.\footnote{A-type and B-type couplings were originally defined in \cite{Chivukula:2019zkt}. The decomposition and derivation of the 4D effective RS1 Lagrangian was originally published in \cite{Chivukula:2020hvi}. The generalized coupling structure $x^{(p)}$ is new to this dissertation, as are the generalizations of the A-type and B-type couplings that it implies.}
    \item[$\bullet$] Section \ref{DerivingSumRules} derives relations (sum rules) between those coupling integrals and the spin-2 KK mode masses.\footnote{Most of the elastic sum rules derived in this chapter were originally published in \cite{Chivukula:2019zkt} and later proved in \cite{Chivukula:2020hvi}; this section significantly generalizes the proofs in \cite{Chivukula:2020hvi}, and the inelastic results are entirely new to this dissertation.}
\end{itemize}
The results of Sections \ref{4D Effective RS1 Model} and \ref{DerivingSumRules} are essential building blocks for the main outcomes of this dissertation. In the next and final chapter of this dissertation, the 4D effective Lagrangian derived in Section \ref{4D Effective RS1 Model} will be used to calculate scattering amplitudes. The sum rules derived in Section \ref{DerivingSumRules} will prove vital for ensuring cancellations in the most divergent high-energy growth of those amplitudes.

\section[Wavefunction Derivation]{Wavefunction Derivation\footnote{This section was originally published as Appendix B of \cite{Chivukula:2020hvi}. In addition to some changes in wording, new content connects the section to the rest of the dissertation and certain points have been elaborated on.}} \label{WfxnDerivation}

Let us now elaborate on the connection between 5D and 4D fields that was established in the chapter summary, and in doing so derive explicit expressions for the wavefunctions that will be utilized in the Kaluza-Klein (KK) decomposition procedure. To demonstrate that the KK decomposition is generically possible, we assume a quadratic 5D Lagrangian $\mathcal{L}_{\text{5D}}$ can be decomposed into a sum of quadratic 4D Lagrangians, derive constraints that are necessary for that assumption to hold true, demonstrate all constraints can be satisfied by solving a certain Sturm-Liouville problem, and then reveal we could have used that problem's solution set to begin with. However, rather than work with Eq. \eqref{LRShh} directly, let us generalize somewhat. This generalization is excessive for our present goals, but is important when considering (for example) natural extensions of this work, including the addition of 5D bulk scalar matter or when constructing models of radion stabilization.

Thus, instead of the massless 5D field $\hat{h}_{\mu\nu}(x,y)$, we consider a {\it massive} 5D field $\Phi_{\vec{\alpha}}(x,y)$ defined over the 5D bulk by a Lagrangian
\begin{align}
    \mathcal{L}_{\text{5D}} &=
    Q_{A}^{\mu\vec{\alpha}\nu\vec{\beta}} e^{-2 k|y|} (\partial_\mu \Phi_{\vec{\alpha}}) (\partial_\nu \Phi_{\vec{\beta}}) + Q_{B}^{\vec{\alpha}\vec{\beta}}
        \left\{
            e^{-4k|y|} (\partial_y \Phi_{\vec{\alpha}}) (\partial_y \Phi_{\vec{\beta}})
            + m_{\Phi}^2 e^{-4k|y|}  \Phi_{\vec{\alpha}} \Phi_{\vec{\beta}}
        \right\} \label{L5DappB}
\end{align}
where the index $\vec{\alpha}$ is a list of Lorentz indices and $m_{\Phi}$ is the 5D mass of the field. The Lorentz tensors $Q_{A}^{\mu\vec{\alpha}\nu\vec{\beta}}$ and $Q_{B}^{\vec{\alpha}\vec{\beta}}$ will eventually be chosen to ensure this procedure yields KK modes with canonical kinetic terms. Note that this Lagrangian can be written equivalently as
\begin{align}
    \mathcal{L}_{\text{5D}}& \cong
    Q_{A}^{\mu\vec{\alpha}\nu\vec{\beta}} e^{-2 k|y|} (\partial_\mu \Phi_{\vec{\alpha}}) (\partial_\nu \Phi_{\vec{\beta}})+ Q_{B}^{\vec{\alpha}\vec{\beta}}
        \left\{
           -\Phi_{\vec{\alpha}}\cdot\partial_y \left[ e^{-4k|y|} (\partial_y \Phi_{\vec{\beta}}) \right]
            + m_{\Phi}^2 e^{-4k|y|} \Phi_{\vec{\alpha}} \Phi_{\vec{\beta}}
        \right\} \label{a414}
\end{align}
via integration by parts. By performing a mode expansion (KK decomposition) on Eq. \eqref{a414} according to the ansatz
\begin{align}
    \Phi_{\vec{\alpha}}(x,y) = \dfrac{1}{\sqrt{\pi r_c}}\sum_{n=0}^{+\infty}  \Phi_{\vec{\alpha}}^{(n)}(x)\psi_{n}(y)~,
\end{align}
we obtain
\begin{align}
    \mathcal{L}_{\text{5D}} & \cong
    \dfrac{1}{\pi r_c} \hspace{-2 pt}\sum_{m,n=0}^{+\infty}\hspace{-5 pt}
    Q_{A}^{\mu\vec{\alpha}\nu\vec{\beta}} (\partial_\mu \Phi^{(m)}_{\vec{\alpha}}) (\partial_\nu \Phi^{(n)}_{\vec{\beta}})\hspace{2 pt}e^{-2 k|y|} \psi^{(m)}\psi^{(n)}\nonumber\\
    &\hspace{35 pt}+ Q_{B}^{\vec{\alpha}\vec{\beta}} \Phi^{(m)}_{\vec{\alpha}}\Phi^{(n)}_{\vec{\beta}}\hspace{2 pt}
        \psi^{(m)}\bigg\{
            -\partial_y \bigg[ e^{-4k|y|} (\partial_y \psi_{n}) \bigg] + m_{\Phi}^2 e^{-4k|y|} \psi_{n}
        \bigg\}~.
\end{align}
Integrating over the extra dimension as in Eq. \eqref{L4Deff} then yields the following effective 4D Lagrangian:
\begin{align}
    \mathcal{L}_{\text{4D}}^{(\text{eff})} &= \sum_{m,n=0}^{+\infty} Q_{A}^{\mu\vec{\alpha}\nu\vec{\beta}} (\partial_\mu \Phi^{(m)}_{\vec{\alpha}}) (\partial_\nu \Phi^{(n)}_{\vec{\beta}}) \cdot N_{A}^{(m,n)} + Q_{B}^{\vec{\alpha}\vec{\beta}} \Phi^{(m)}_{\vec{\alpha}}\Phi^{(n)}_{\vec{\beta}}\cdot N_{B}^{(m,n)}~,\label{L4Dgenmass}
\end{align}
where $N_{A}^{(m,n)}$ and $N_{B}^{(m,n)}$ equal
\begin{align}
    N_{A}^{(m,n)} &= \dfrac{1}{\pi r_c}\int_{-\pi r_c}^{+\pi r_c}dy\hspace{5 pt} e^{-2 k|y|} \psi_{m}\psi_{n}~,\label{NAmn}\\
    N_{B}^{(m,n)} &=  \dfrac{1}{\pi r_c}\int_{-\pi r_c}^{+\pi r_c}dy\hspace{5 pt} \psi_{m}\left\{
            - \partial_y \bigg[ e^{-4k|y|} (\partial_y \psi_{n}) \right] + m_{\Phi}^2 e^{-4k|y|} \psi_{n}\bigg\}~.\label{NBmn}
\end{align}
We desire that this process yields a particle spectrum described by canonical 4D Lagrangians for particles of definite spins and masses. Specifically, we desire that a (bosonic) mode field $\phi_{\vec{\alpha}}(x)$ in the KK spectrum is described by a Lagrangian
\begin{align}
q_{A}^{\mu\vec{\alpha}\nu\vec{\beta}}(\partial_\mu \phi_{\vec{\alpha}})(\partial_\nu \phi_{\vec{\beta}}) + m^2 q_{B}^{\vec{\alpha}\vec{\beta}} \phi_{\vec{\alpha}}\phi_{\vec{\beta}}~,\label{L4DeffappB}
\end{align}
where $m$ is the mass of the KK mode, and the quantities $q_{A}$ and $q_{B}$ are Lorentz tensor structures that reproduce the canonical quadratic Lagrangian appropriate for the internal spin of $\phi_{\vec{\alpha}}$. For example, a massive spin-2 field $\hat{h}_{\mu\nu}$ has the canonical quadratic Lagrangian Eq. \eqref{Ls2massive}, such that $\phi_{\vec{\alpha}}(x) = \hat{h}_{\alpha_{1}\alpha_{2}}(x)$ and we may choose
\begin{align}
    q_{A}^{\mu\alpha_{1}\alpha_{2}\nu\beta_{1}\beta_{2}} &= \eta^{\mu \alpha_{1}} \eta^{\alpha_{2}\nu} \eta^{\beta_{1}\beta_{2}} - \eta^{\mu\nu} \eta^{\alpha_{1}\alpha_{2}}  \eta^{\beta_{1}\beta_{2}} + \dfrac{1}{2} \eta^{\mu\nu} \eta^{\alpha_{1}\beta_{1}}  \eta^{\alpha_{2}\beta_{2}} - \dfrac{1}{2} \eta^{\mu\nu} \eta^{\alpha_{1}\alpha_{2}}  \eta^{\beta_{1}\beta_{2}}\\
    q_{B}^{\alpha_{1}\alpha_{2}\beta_{1}\beta_{2}} &= \dfrac{1}{2} \eta^{\alpha_{1}\alpha_{2}}\eta^{\beta_{1}\beta_{2}} - \dfrac{1}{2}\eta^{\alpha_{1}\beta_{1}}\eta^{\alpha_{2}\beta_{2}}
\end{align}
For a full KK tower, the corresponding canonical quadratic Lagrangian equals (indexing KK number by $n$),\footnote{Restricting the KK decomposition sum to positive KK indices $n$ is inspired by the RS1 model's orbifold symmetry. For example, if we instead considered a (non-orbifolded) torus, we would sum over all integer $n$, with the sign of $n$ describing the rotational direction of the particle's extra-dimensional momentum around the circular extra dimension. From this perspective, imposing an orbifold symmetry causes the $+n$ and $-n$ non-orbifolded states to be combined into an even superposition which we then call the $n$th KK mode of the orbifolded theory.}
\begin{align}
    \mathcal{L}_{\text{4D}}^{(\text{eff})} = \sum_{n=0}^{+\infty} \hspace{5 pt} q_{A}^{\mu\vec{\alpha}\nu\vec{\beta}}(\partial_\mu \phi^{(n)}_{\vec{\alpha}})(\partial_\nu \phi^{(n)}_{\vec{\beta}}) + m_n^2 q_{B}^{\vec{\alpha}\vec{\beta}} \phi^{(n)}_{\vec{\alpha}}\phi^{(n)}_{\vec{\beta}}~.
\end{align}
Comparing to Eq. \eqref{L4Dgenmass}, one recovers this form for the choices $Q=q$ (i.e. if the 5D quadratic tensor structures mimic the 4D canonical quadratic tensor structures), $\Phi^{(n)} = \phi^{(n)}$, $N_{A}^{(m,n)} = \delta_{m,n}$, and $N_{B}^{(m,n)} = m_n^2 \delta_{m,n}$. Consider this condition on $N_{B}^{(m,n)}$ in more detail. Using Eq. \eqref{NBmn}, $N_{B}^{(m,n)} = m_n^2 \delta_{m,n}$ implies
\begin{align}
    &\dfrac{1}{\pi r_c}\int_{-\pi r_c}^{+\pi r_c}dy\hspace{5 pt} \psi^{(m)}\bigg\{
            -\partial_y \bigg[ e^{-4k|y|} (\partial_y \psi_{n}) \bigg] + m_{\Phi}^2 e^{-4k|y|} \psi_{n}\bigg\} = m_n^2 \delta_{m,n}~.
\end{align}
which then becomes, using the condition $N_{A}^{(m,n)} = \delta_{m,n}$ and Eq. \eqref{NAmn},
\begin{align}
    &\int_{-\pi r_c}^{+\pi r_c}dy\hspace{5 pt} \psi_{m}\bigg\{
            \partial_y \bigg[ e^{-4k|y|} (\partial_y \psi_{n}) \bigg] + \bigg( m_n^2 e^{-2k|y|} - m_{\Phi}^2 e^{-4k|y|} \bigg)\psi_{n}\bigg\} = 0~.\label{DerivingSLEq}
\end{align}
If the collection of wavefunctions $\{\psi_{m}\}$ form a complete set, then Eq. \eqref{DerivingSLEq} implies that they are solutions of the following differential equation
\begin{align}
   \partial_y \bigg[ e^{-4k|y|} (\partial_y \psi_{n}) \bigg]
            + \bigg( m_n^2 e^{-2k|y|} - m_{\Phi}^2 e^{-4k|y|} \bigg)\psi_{n} = 0~,
\end{align}
or, when expressed in unitless combinations,
\begin{align}
    0 = \partial_{\varphi}\bigg[ e^{-4kr_{c}|\varphi|} (\partial_{\varphi} \psi_{n}) \bigg] + \bigg( (m_{n}r_{c})^{2} e^{-2kr_{c}|\varphi|}- (m_{\Phi}r_{c})^2 e^{-4kr_{c}|\varphi|} \bigg)\psi_{n}~.\label{SLeq}
\end{align}
In addition to this differential equation, orbifold symmetry requires that the derivatives of the wavefunctions vanish at the orbifold fixed points, i.e. $(\partial_{\varphi}\psi_{n}) = 0$ for $\varphi\in\{0,\pi\}$, which provides the problem with boundary conditions. Finding the solution set $\{\psi_{n}\}$ (and corresponding values of $\{m_nr_{c}\}$) of Eq. \eqref{SLeq} under these boundary conditions is precisely a Sturm-Liouville (SL) problem, for which there is guaranteed a discrete (complete) basis of real wavefunctions satisfying
\begin{align}
    \dfrac{1}{\pi} \int_{-\pi}^{+\pi} d\varphi\hspace{5 pt}e^{-2kr_{c}|\varphi|}\psi_{m}\psi_{n} = N_{A}^{(m,n)} \equiv \delta_{m,n}~, \label{onA}
\end{align}
as required. Hence, by finding wavefunctions $\psi_{n}$ that solve Eqs. \eqref{SLeq} and \eqref{onA}, we can KK decompose the fields in Eq. \eqref{L5DappB} according to the ansatz and (so long as $Q=q$) obtain a tower of canonical quadratic Lagrangians \eqref{L4DeffappB} as desired.

Eq. \eqref{LRShh} is of the form Eq. \eqref{L5DappB} with $m_{\Phi} = 0$. In general, when the bulk mass $m_{\Phi}$ vanishes, Eq. \eqref{SLeq} admits a massless solution ($\psi_{0}$ with $m_{0}=0$) which is flat in the extra dimension ($\partial_y\psi_{0} = 0$). Therefore, the 5D field $\hat{h}_{\mu\nu}$ gives rise to a massless 4D field $\hat{h}^{(0)}_{\mu\nu}$, which we identify with the usual (4D) graviton. The 5D field $\hat{r}$ yields a massless 4D field $\hat{r}^{(0)}$ which we identify as the radion; however, note that Eq. \eqref{LRSrr} is not of the form Eq. \eqref{L5DappB} because of the additional warp factors introduced alongside $\hat{r}$. Thus, its KK decomposition is derived solely from the $y$-independence of $\hat{r}$.\footnote{To prevent the radion from contributing to long-range gravitational forces and to ensure the extra-dimensional is stable against quantum fluctuations, we must include interactions which make the physical 4D spin-0 field become massive, as occurs during radion stabilization \cite{Goldberger:1999wh}. Radion stabilization will be investigated in future works and is not relevant to the present dissertation.}
Normalization fixes  $\psi_{0}$ to equal
\begin{align}
    \psi_{0}=\sqrt{\dfrac{kr_c\pi}{1-e^{-2kr_c\pi}}}~.\label{psi0gen}
\end{align}
When $m_{\Phi} \neq 0$, this solution does not exist.

By construction, the SL equation combined with Eq. $\eqref{onA}$ implies an additional quadratic integral condition:
\begin{align}
    &\dfrac{1}{\pi} \int_{-\pi }^{+\pi} d\varphi\hspace{5 pt} e^{-4kr_{c}|\varphi|}\bigg[ (\partial_{\varphi}\psi_{m})(\partial_{\varphi}\psi_{n}) + (m_{\Phi} r_{c})^{2} \psi_{m}\psi_{n}\bigg] = (m_{n}r_{c})^{2} \delta_{m,n} ~.\label{onB}
\end{align}
When $m_\Phi = 0$, this becomes an orthonormality condition on the set $\{\partial_{\varphi} \psi_{n}\}$.

The existence of a discrete solution set of wavefunctions is guaranteed by the SL problem. Following the notation and arguments from \cite{Goldberger:1999wh}, we now summarize how to find explicit equations for the non-flat wavefunctions in that solution set. Note that
\begin{align}
\partial_{\varphi}|\varphi| = \text{sign}(\varphi) \hspace{35 pt}\text{ and }\hspace{35 pt}\partial^{2}_{\varphi}|\varphi| = 2\big[\delta(\varphi) - \delta(\varphi - \pi)\big]~,
\end{align}
such that $\partial_{\varphi}^{2}|\varphi|=0$ when $\varphi\neq 0,\pi$ and $(\partial_{\varphi}|\varphi|)^{2} = 1$. Thus, Eq. \eqref{SLeq} may be rewritten in terms of quantities $z_n = (m_n/k)e^{+kr_{c}|\varphi|}$ and $f_{n} = (m_n^2/k^2) \psi_{n}/z_n^2$ as
\begin{align}
    z_n^2 \dfrac{d^2 f_n}{dz_n^2} + z_n \dfrac{df_n}{dz_n} + \left[ z_n^2 - \left( 4 + \dfrac{m_{\Phi}^2}{k^2}\right)\right] f_n = 0~.
\end{align}
away from the orbifold fixed points. When $m_{\Phi} = 0$, this differential equation is solved by $f_n$ equal to Bessel functions $J_2(z_n)$ or $Y_2(z_n)$. When $m_{\Phi} \neq 0$, it is instead solved by Bessel functions $J_\nu(z_n)$ and $Y_\nu(z_n)$ where $\nu^2 \equiv 4 + m_\Phi^2 /k^2$. Taking a superposition of the appropriate Bessel functions yields a generic solution $f_n$, which may then be converted back to $\psi_{n}$. By imposing the SL boundary conditions at the orbifold fixed points ($\partial_{\varphi}\psi_{n} =0$ for $\varphi\in\{0,\pi\}$), the wavefunctions are found to equal
\begin{align}
    \psi_{n} = \dfrac{\varepsilon^{2}}{N_n} \left[J_{\nu} \left(\dfrac{\mu_{n}\varepsilon}{kr_{c}} \right) + b_{n\nu}\s Y_{\nu}\left(\dfrac{\mu_{n}\varepsilon}{kr_{c}}\right)\right]~,
    \label{eq:wavefunction}
\end{align}
where $\varepsilon\equiv e^{+kr_{c}|\varphi|}$ and $\mu_{n}\equiv m_{n}r_{c}$, the normalization $N_{n}$ is determined by Eq. \eqref{onA} (up to a sign that we fix by setting $N_{n}>0$ and which yields $\psi_{n}(0)<0$ for nonzero $n$), and the relative weight $b_{n\nu}$ equals
\begin{align}
    b_{n\nu} = - \dfrac{2J_{\nu}\Big|_{\mu_{n}/kr_{c}} + \dfrac{\mu_{n}}{kr_{c}}(\partial J_{\nu})\Big|_{\mu_{n}/kr_{c}}}{2Y_{\nu}\Big|_{\mu_{n}/kr_{c}} + \dfrac{\mu_{n}}{kr_{c}}(\partial Y_{\nu})\Big|_{\mu_{n}/kr_{c}}}~,\label{bnnu}
\end{align}
where $\partial J_{\nu} \equiv \partial J_{\nu}(z)/\partial z$ and $\partial Y_{\nu} \equiv \partial Y_{\nu}(z)/\partial z$. These wavefunctions satisfy Eq. \eqref{onB} where each $\mu_{n}$ solves
\begin{align}
    \left[2 J_{\nu} + \dfrac{\mu_{n}\varepsilon}{kr_{c}} (\partial J_{\nu})\right]\bigg|_{\varphi = \pi}\left[2 Y_{\nu} + \dfrac{\mu_{n}\varepsilon}{kr_{c}}  (\partial Y_{\nu})\right]\bigg|_{\varphi = 0}\hspace{35 pt}&\nonumber\\
    -\left[2 Y_{\nu} + \dfrac{\mu_{n}\varepsilon}{kr_{c}} (\partial Y_{\nu})\right]\bigg|_{\varphi = \pi}\left[2 J_{\nu} + \dfrac{\mu_{n}\varepsilon}{kr_{c}}  (\partial J_{\nu})\right]\bigg|_{\varphi = 0} &= 0~.\label{mn}
\end{align}
Although these wavefunctions were derived by solving Eq. \eqref{SLeq} away from the orbifold fixed points, they solve the equation across the full extra dimension. 

Finally, note that given a 5D Lagrangian consistent with Eq. \eqref{L5DappB}, the wavefunctions $\psi_{n}$ and spectrum $\{\mu_{n}\}$ are entirely determined by the unitless quantities $kr_c$ and $m_{\Phi}r_{c}$. In the RS1 model, the 5D field $\hat{h}_{\mu\nu}$ lacks a bulk mass ($m_{\Phi}=0$) such that $\nu = 2$ and its KK decomposition is dictated by $kr_{c}$ alone.

\section[4D Effective RS1 Model]{4D Effective RS1 Model\footnote{Subsection \ref{4D Effective RS1 - 4D Particle Content} was originally published as Subsection III.A of \cite{Chivukula:2020hvi}. Subsection \ref{4D Effective RS1 - General Procedure} combines content that was originally published as Subsections III.B and C.2 of \cite{Chivukula:2020hvi}. Subsection \ref{4D Effective RS1 - Summary of Results} was originally published as Appendix C of \cite{Chivukula:2020hvi}. Notations and terminology have been updated, and paragraphs that describe convenient wavefunction properties and the generalized coupling structure $x^{(p)}$ have been added.}} \label{4D Effective RS1 Model}

In this section, we carry out the KK mode expansions of $\hat{h}_{\mu\nu}(x,y)$ and $\hat{r}(x)$, thereby obtaining the 4D particle content of the RS1 model, and discuss the form of the interactions among the 4D fields.

\subsection{4D Particle Content} \label{4D Effective RS1 - 4D Particle Content}
The 4D particle content is determined by employing the KK decomposition ansatz \cite{Kaluza:1984ws,Klein:1926tv,Goldberger:1999wh}:
\begin{align}
    \hat{h}_{\mu\nu}(x,y) &= \dfrac{1}{\sqrt{\pi r_c}} \sum_{n=0}^{+\infty} \hat{h}^{(n)}_{\mu\nu}(x)\psi_n(\varphi) \hspace{35 pt} \hat{r}(x) = \dfrac{1}{\sqrt{\pi r_c}}\,
   \hat{r}^{(0)}(x) \psi_0~,\label{KKreduce}
\end{align}
where we recall that $\varphi=y/r_c$.
The coefficients $\hat{h}^{(n)}_{\mu\nu}$ and $\hat{r}^{(0)}$ are 4D spin-2 and spin-0 fields respectively, while each $\psi_{n}$ is a wavefunction which solves the following Sturm-Liouville equation
\begin{align}
    \partial_{\varphi}\left[\varepsilon^{-4} (\partial_{\varphi} \psi_{n})\right] = - \mu_{n}^{2} \varepsilon^{-2} \psi_{n} \label{SLeqx}
\end{align}
subject to the boundary condition $(\partial_{\varphi} \psi_{n})=0$ at $\varphi = 0$ and $\pi$, where $\varepsilon \equiv e^{k |y|}=e^{kr_{c}|\varphi|}$ \cite{Goldberger:1999wh}. As described in the previous section, there exists a unique solution $\psi_{n}$ (up to normalization) per eigenvalue $\mu_{n}$, each of which we index with a discrete KK number $n \in \{0,1,2,\cdots\}$ such that $\mu_{0}=0 < \mu_{1} < \mu_{2} < \cdots$. Given a KK number $n$, the quantity $\mu_{n}$ and wavefunction $\psi_{n}(\varphi)$ are entirely determined by the value of the unitless nonnegative combination $kr_{c}$. We note that with proper normalization the $\psi_{n}$ satisfy two convenient orthonormality conditions:
\begin{align}
    \dfrac{1}{\pi}\int_{-\pi}^{+\pi} d\varphi\hspace{5 pt} \varepsilon^{-2}\,\psi_{m}\,\psi_{n} &= \delta_{m,n}~,\label{onAx}\\
    \dfrac{1}{\pi}\int_{-\pi}^{+\pi} d\varphi\hspace{5 pt} \varepsilon^{-4}\,(\partial_\varphi\psi_m)\,(\partial_\varphi\psi_n) &= \mu_{n}^{2}\delta_{m,n}~.\label{onBx}
\end{align}
Furthermore, the $\{\psi_{n}\}$ form a complete set, such that the following completeness relation holds:
\begin{align}
    \delta(\varphi_{2}-\varphi_{1}) =  \dfrac{1}{\pi}\, \varepsilon(\varphi_{1})^{-2}\sum_{j=0}^{+\infty} \psi_{j}(\varphi_{1})\, \psi_{j}(\varphi_{2})~.
    \label{eq:completeness}
\end{align}
Because of the assumptions behind its derivation, the completeness relation can only be used to combine or separate orbifold-even integrands. For example, if $f(\varphi)\neq 0$ is an orbifold-odd function (such as $(\partial_{\varphi}\psi_{n})$), then splitting $f(\varphi)^{2}$ into a product $f(\varphi)^{2}\cdot 1$ is fine,
\begin{align}
    0 < \dfrac{1}{\pi} \int_{-\pi}^{+\pi} d\varphi\hspace{10 pt} \vep^{+2} f(\varphi)f(\varphi) = \sum_{j} \left[\dfrac{1}{\pi} \int d\varphi \hspace{10 pt} f(\varphi)f(\varphi)\psi_{j}(\varphi)\right] \, \left[\dfrac{1}{\pi} \int d\varphi \hspace{10 pt} \psi_{j}(\varphi)\right]
\end{align}
whereas trying to apply completeness to separate $f(\varphi)^{2}$ into $f(\varphi)\cdot f(\varphi)$ yields a contradiction
\begin{align}
    0 < \dfrac{1}{\pi} \int_{-\pi}^{+\pi} d\varphi\hspace{10 pt} \vep^{+2}f(\varphi)f(\varphi) \neq \sum_{j} \left[\dfrac{1}{\pi} \int d\varphi \hspace{10 pt} f(\varphi)\psi_{j}(\varphi)\right]^{2} = 0
\end{align}
The completeness relation will be vital to relating different coupling structures present in the 4D effective WFE RS1 Lagrangian.

The KK number $n=0$ corresponds to $\mu_{n} = 0$, for which Eq. $\eqref{SLeqx}$ admits a flat wavefunction solution $\psi_0$ corresponding to the massless 4D graviton. Upon normalization via Eq. \eqref{onAx}, this wavefunction equals
\begin{align}
    1 = \dfrac{1}{\pi} \psi_{0}^{2} \int_{-\pi}^{+\pi} \vep^{-2} = \dfrac{1}{\pi kr_{c}}\left[1 - e^{-2\pi kr_{c}}\right] \hspace{20 pt}\implies\hspace{20 pt} \psi_0 = \sqrt{\dfrac{\pi k r_c}{1-e^{-2\pi kr_c}}}
\end{align}
up to a phase that we set to $+1$ by convention. This is the wavefunction that Eq. $\eqref{KKreduce}$ associates with the fields $\hat{h}^{(0)}$ and $\hat{r}^{(0)}$. The lack of higher modes in the KK decomposition of $\hat{r}$ reflects its $y$-independence. In this sense, choosing to associate $\psi_{0}$ with $\hat{r}^{(0)}$ in Eq. $\eqref{KKreduce}$ is merely done for convenience.

Before we compute the interactions between 4D states, let us first apply the ansatz to the simpler quadratic terms. This will illustrate how the KK decomposition procedure typically works, and why the interaction terms are more complicated. 
The 5D quadratic $\hat{h}_{\mu\nu}(x,y)$ Lagrangian equals (from Section \ref{AppendixWFE})
\begin{align}
    \LRSf{hh} = \hspace{3 pt} \varepsilon^{-2}\s\LA{hh} + \varepsilon^{-4}\s\LB{hh}~, \label{L5DRS hh}
\end{align}
where
\begin{align}
     \LA{hh} =& - \hat{h}_{\mu\nu}(\partial^\mu \partial^\nu \hat{h}) + \hat{h}_{\mu\nu} (\partial^\mu \partial_\rho \hat{h}^{\rho\nu})- \dfrac{1}{2} \hat{h}_{\mu\nu}(\square \hat{h}^{\mu\nu}) + \dfrac{1}{2} \hat{h}(\square \hat{h})~,\label{5DhhA}\\
    \LB{hh} =& - \dfrac{1}{2} \ltr \hat{h}^\prime \hat{h}^\prime \rtr + \dfrac{1}{2} \ltr \hat{h}^\prime \rtr^2~,\label{5DhhB}
\end{align}
A prime indicates differentiation with respect to $y$ and a twice-squared bracket indicates a cyclic contraction of Lorentz indices. Similarly, the quadratic 5D $\hat{r}(x)$ Lagrangian equals,
\begin{align}
    \LRSf{rr} = e^{-2 \pi kr_{c}}\s\varepsilon^{+2}\s \LA{rr}~,\label{Lrr RS}
\end{align}
where
\begin{align}
    \LA{rr} = \dfrac{1}{2}(\partial_\mu\hat{r})(\partial^\mu \hat{r})~.\label{LrrA RS}
\end{align}
To obtain the 4D effective equivalents of the above 5D expressions, we must integrate over the extra dimension and employ the KK decomposition ansatz.

First, the quadratic $\hat{h}_{\mu\nu}$ Lagrangian: the first term in Eq. \eqref{L5DRS hh} becomes
\begin{align}
    \LAeff{hh} &\equiv \int_{-\pi r_c}^{+\pi r_c}dy\hspace{5 pt} \varepsilon^{-2}\s\LA{hh}\nonumber\\
    &\hspace{-25 pt}=\int_{-\pi r_c}^{+\pi r_c} dy\hspace{5 pt} \varepsilon^{-2}\left[-\hat{h}_{\mu\nu} (\partial^\mu \partial^\nu \hat{h}) + \hat{h}_{\mu\nu} (\partial^\mu \partial_\rho \hat{h}^{\rho\nu})- \dfrac{1}{2} \hat{h}_{\mu\nu} (\square \hat{h}^{\mu\nu}) + \dfrac{1}{2} \hat{h} (\square\hat{h})\right]\nonumber\\
    &\hspace{-25 pt}= \sum_{m,n=0}^{+\infty} \left[-\hat{h}^{(m)}_{\mu\nu} (\partial^\mu \partial^\nu \hat{h}^{(n)}) + \hat{h}^{(m)}_{\mu\nu} (\partial^\mu \partial_\rho \hat{h}^{(n)\rho\nu}) - \dfrac{1}{2} \hat{h}^{(m)}_{\mu\nu} (\square \hat{h}^{(n)\mu\nu}) + \dfrac{1}{2} \hat{h}^{(m)} (\square\hat{h}^{(n)})\right]\nonumber\\
    &\times \dfrac{1}{\pi}\int_{-\pi}^{+\pi} d\varphi\hspace{5 pt} \varepsilon^{-2}\psi_{m} \psi_{n} ~,
\end{align}
whereas its second term becomes
\begin{align}
    \LBeff{hh} &\equiv \int_{-\pi r_c}^{+\pi r_c} dy\hspace{5 pt}\varepsilon^{-4}\s\LB{hh}\nonumber\\
    &\hspace{-25 pt}= \int_{-\pi r_c}^{+\pi r_c} dy\hspace{5 pt} \varepsilon^{-4}\left[- \dfrac{1}{2}\ltr \hat{h}^\prime \hat{h}^\prime\rtr + \dfrac{1}{2} \ltr \hat{h}^\prime\rtr^2\right]\nonumber\\
    &\hspace{-25 pt}= \sum_{m,n=0}^{+\infty}\left[- \dfrac{1}{2}\ltr \hat{h}^{(m)}\hat{h}^{(n)}\rtr +\dfrac{1}{2} \ltr \hat{h}^{(m)}\rtr \ltr \hat{h}^{(n)} \rtr \right] \, \dfrac{1}{\pi r_c^2} \int_{-\pi}^{+\pi} d\varphi\hspace{5 pt} \varepsilon^{-4}(\partial_\varphi \psi_m)(\partial_\varphi \psi_n)~.
\end{align}
These are simplified via the orthonormality relations Eqs. \eqref{onAx} and \eqref{onBx}, such that the 4D effective Lagrangian resulting from $\LRSf{hh}$ equals, using Eqs. \eqref{Ls2massless} and \eqref{Ls2massive},
\begin{align}
    \LRSfeff{hh} &= \LAeff{hh}+ \LBeff{hh}\nonumber\\
    &= \mathcal{L}_{\text{massless}}^{(s=2)}(\hat{h}^{(0)})+ \sum_{n=1}^{+\infty} \mathcal{L}^{(s=2)}_{\text{massive}}(m_n,\hat{h}^{(n)})~,
\end{align}
wherein $m_{n}\equiv \mu_{n}/r_{c}$. Therefore, KK decomposition of the 5D field $\hat{h}_{\mu\nu}$ results in the following 4D particle content: a single massless spin-2 mode $\hat{h}^{(0)}$, and countably many massive spin-2 modes $\hat{h}^{(n)}$ with $n\in\{1,2,\cdots\}$. The zero mode $\hat{h}^{(0)}$ is consistent with the usual 4D graviton, and will be identified as such. As will be argued in the next subsection, the 4D graviton has dimensionful coupling constant $\kDD = 2/M_{\text{Pl}} = \psi_{0} \kappa/\sqrt{\pi r_c}$ where $M_{\text{Pl}}$ is the reduced 4D Planck mass. In terms of the reduced 4D Planck mass, the full 4D Planck mass equals $\sqrt{8\pi}M_{\text{Pl}}$.

Meanwhile, the 4D effective equivalent of $\mathcal{L}^{(\text{RS})}_{rr}$ from Eq. \eqref{Lrr RS} equals, using Eq. \eqref{Ls0massless},
\begin{align}
    \LRSfeff{rr} &= \int_{-\pi r_c}^{+\pi r_c} dy\hspace{5 pt} \LRSf{rr}\nonumber\\
    &= \int_{-\pi r_c}^{+\pi r_c} dy\hspace{5 pt} e^{-2\pi kr_{c}}\s\varepsilon^{+2}\left[\dfrac{1}{2}  (\partial_\mu \hat{r})(\partial^\mu\hat{r})\right]\nonumber \\
    &=\dfrac{1}{2}(\partial_\mu \hat{r}^{(0)})(\partial^\mu \hat{r}^{(0)}) \cdot \dfrac{{\psi_{0}}^2}{\pi r_c} \int_{-\pi r_c}^{+\pi r_c} dy\hspace{5 pt} e^{+2k(|y|-\pi r_c)} \nonumber \\
    &= \mathcal{L}^{S=0}_{\text{massless}}(\hat{r}^{(0)})~.
\end{align}
Therefore, KK decomposing the 5D $\hat{r}$ field yields only a single massless spin-0 mode $\hat{r}^{(0)}$, which is called the radion. Note the exponential factor in Eq. \eqref{Lrr RS} is inconsistent with the orthonormality equation \eqref{onAx}, so we had to calculate the integral explicitly. Thankfully, the $y$-independent radion field must possess a flat extra-dimensional wavefunction and so the exponential factor can at most affect its normalization. This would not be the case if the radion field's $y$-dependence was unable to be gauged away in Subsection \ref{SS - Perturbing The Vacuum}.

The RS1 model has three independent parameters according to the above construction: the extra-dimensional radius $r_{c}$, the warping parameter $k$, and the 5D coupling strength $\kappa$. However, we use a more convenient set of independent parameters in practice: the unitless extra-dimensional combination $kr_{c}$, the mass $m_{1}$ of the first massive KK mode $\hat{h}^{(1)}$, and the reduced 4D Planck mass $M_{\text{Pl}}$. These sets are related according to the following relations:
\begin{align}
    m_{1} &\equiv\dfrac{1}{r_{c}}\mu_{1}(kr_{c})\hspace{10 pt}\text{via Eq. \eqref{SLeqx}}~,\\
    M_{\text{Pl}} &\equiv \dfrac{2}{\kappa\sqrt{k}} \sqrt{1 - e^{-2kr_{c}\pi}}~.
\end{align}
In our numerical analyses, we will choose $kr_{c} \in [0,10]$, $m_{1} = 1\text{ TeV}$, and $M_{\text{Pl}}= 2.435\times 10^{15}\text{ TeV}$.

When converting the quadratic terms of the 5D RS1 Lagrangian into their 4D effective equivalents, we were able to perform all integrals exactly. This is because all wavefunctions with a nonzero KK number were present in pairs and thus subject to orthonormality relations. Such simplifications are seldom possible when dealing with a product of three or more 5D $\hat{h}_{\mu\nu}$ fields, and instead the integrals lack closed form solutions. As a result, the RS1 model possesses many nonzero couplings between KK modes and calculating a matrix element for $2$-to-$2$ scattering of massive KK modes typically requires a sum over infinitely many diagrams, each of which is mediated by a different massive KK mode and contains various products of these overlap integrals. The next section details the 4D effective Lagrangian and the origin of those integrals. The final section details relations involving these integrals and the KK mode masses.

\subsection{General Procedure} \label{4D Effective RS1 - General Procedure}

The 5D WFE RS1 Lagrangian derived in Section \ref{AppendixWFE} equals a sum of terms, wherein each term contains some number of 5D fields and exactly two derivatives. Each derivative is either a 4D spatial derivative $\partial_\mu$ or an extra-dimension derivative $\partial_y$, and each field is either an $\hat{r}$ or an $\hat{h}_{\mu\nu}$ field. Because the Lagrangian requires an even number of Lorentz indices in order to form a Lorentz scalar, each derivative pair must consist of two copies of the same kind of derivative, i.e. each term in $\mathcal{L}^{(\text{RS})}_{\text{5D}}$ can be classified into one of two categories:
\begin{itemize}
    \item[$\bullet$] {\bf A-Type:} The term has two spatial derivatives $\partial_\mu \cdot \partial_\nu$, or
    \item[$\bullet$] {\bf B-Type:} The term has two extra-dimensional derivatives $\partial_y \cdot \partial_y$~.
\end{itemize}
In addition to fields and derivatives, every term in $\mathcal{L}^{(\text{RS})}_{\text{5D}}$ has an exponential prefactor. That exponential's specific form is entirely determined by its type (whether A- or B-type) and the number of instances of $\hat{r}$ in the term. Each A-type term is associated with a factor $\varepsilon^{-2} = e^{-2kr_c|\varphi|}$ whereas each B-type term is associated with a factor $\varepsilon^{-4} = e^{-4kr_c|\varphi|}$, and every instance of a radion field provides an additional $e^{-\pi kr_{c}}\,\varepsilon^{+2}$ factor. These assignments correctly reproduce the prefactors found in Section \ref{AppendixWFE} via explicit weak field expansion of the 5D RS1 Lagrangian.

Consider a generic A-type term with $H$ instances of $\hat{h}$ and $R$ instances of $\hat{r}$. Schematically, it will be of the form,
\begin{align}
    X_A&\equiv \kappa^{(H+R-2)}\,\left[ \varepsilon^{-2}\right] \,\left[ e^{-\pi kr_{c}}\,\varepsilon^{+2} \right]^R  \,(\partial_\mu^2,\hat{h}^{H}, \hat{r}^R) \nonumber\\
    &= \kappa^{(H+R-2)}\,e^{-R\pi kr_c}\,\varepsilon^{2(R-1)} \,\overline{X}_A~,
\end{align}
where the combination $\overline{X}_A \equiv (\partial_\mu^2,\hat{h}^{H}, \hat{r}^R)$ refers to a fully contracted product of two 4D derivatives, $H$ gravitons, and $R$ radions. The $\mu$ label on $\partial_\mu^2$ above is only schematic and not literal. Similarly, an equivalent B-type term would be of the form,
\begin{align}
    X_B&\equiv \kappa^{(H+R-2)}\,\left[ \varepsilon^{-4}\right]\, \left[ e^{-\pi kr_{c}}\,\varepsilon^{+2}\right]^R\,(\partial_y^2,\hat{h}^{H},\hat{r}^R)\nonumber\\
    &= \kappa^{(H+R-2)}\,e^{-R\pi kr_c}\,\varepsilon^{2(R-2)} \,\overline{X}_B~,
\end{align}
where the combination $\overline{X}_B \equiv (\partial_y^2,\hat{h}^{H}, \hat{r}^R)$ refers to a fully contracted product of two extra-dimensional derivatives, $H$ instances of $\hat{h}_{\mu\nu}$, and $R$ instances of $\hat{r}$. Because we included $\Delta\mathcal{L}$ in $\mathcal{L}^{(\text{RS})}_{\text{5D}}$, each B-type term we consider has each of its $\partial_{y}$ derivatives acting on a different field, and so we assume $\overline{X}_B$ also satisfies this property.

We form a 4D effective Lagrangian by first KK decomposing our 5D fields into states of definite mass (Eq. \eqref{KKreduce}) and then integrating over the extra dimension (Eq. \eqref{L4Deff}). For the schematic A-type term, this procedure yields,
\begin{align}
    X_A^{(\text{eff})} &= \dfrac{r_c}{(\pi r_c)^{(H+R)/2}}\,\kappa^{(H+R-2)}\sum_{n_1,\cdots,n_H=0}^{+\infty} \left(\partial_\mu^2,\hspace{6 pt} \hat{h}^{(n_1)}\cdots \hat{h}^{(n_H)},\hspace{6 pt} \left[\hat{r}^{(0)}\right]^R \right)\nonumber\\
    &\hspace{35 pt}\times e^{-R\pi kr_c}\int_{-\pi}^{+\pi} d\varphi\hspace{5 pt} \varepsilon^{2(R-1)}\psi_{n_1}\cdots\psi_{n_H}\left[\psi_{0}\right]^R~.\nonumber
\end{align}
Define a unitless combination ${a}$ that contains the extra-dimensional overlap integral:
\begin{align}
    {a}_{({R}|\vec{n})} &\equiv {a}_{r\cdots r n_1\cdots n_H} \equiv \dfrac{1}{\pi}\, e^{-R\pi kr_c}\int_{-\pi}^{+\pi}d\varphi\hspace{5 pt} \varepsilon^{2(R-1)}\psi_{n_1}\cdots \psi_{n_H} \left[\psi_{0}\right]^R~,\label{abar}
\end{align}
where $\vec{n} \equiv (n_{1},\cdots ,n_{H})$, there are $R$ instances of the label $r$ are present in ${a}_{r\cdots r n_{1}\cdots n_{H}}$ (e.g. $a_{(2|n_{1}n_{2})} = a_{rrn_{1}n_{2}}$), and ${a}_{r\cdots r n_{1}\cdots n_{H}}$ is fully symmetric in the subscript (e.g. $a_{nnr} = a_{nrn} = a_{rnn}$). Using this, we may now write
\begin{align}
    X_A^{(\text{eff})} &= \left[\dfrac{\kappa}{\sqrt{\pi r_c}}\right]^{H+R-2}\sum_{n_1,\cdots,n_H=0}^{+\infty} {a}_{(R|n_1\cdots n_H)}\left(\partial_\mu^2,\hspace{6 pt} \hat{h}^{(n_1)}\cdots \hat{h}^{(n_H)},\hspace{6 pt} \left[\hat{r}^{(0)}\right]^R\right)~.\label{KKdecompop}
\end{align}

To simplify this expression further, we define a  KK decomposition operator $\mathcal{X}_{(\vec{n})}[\bullet]$. The KK decomposition operator maps a product of $\hat{h}_{\mu\nu}$ and $\hat{r}$ fields to an analogous product of 4D spin-2 fields $\hat{h}^{(n_{i})}_{\mu\nu}$ labeled by KK numbers $\vec{n}=(n_1,\cdots,n_H)$ and 4D radion fields $\hat{r}^{(0)}$. More specifically, $\mathcal{X}$ maps all $\hat{r}$ in its argument to $\hat{r}^{(0)}$ and applies the specified KK labels to the $\hat{h}_{\mu\nu}$ fields ($\hat{h}_{\mu\nu}\mapsto \hat{h}^{(n_i)}_{\mu\nu}$) per term according to the following prescription: the labels are applied left to right in the order that they occur in $\vec{n}$, and are applied to $\hat{h}_{\mu\nu}$ fields of the form $(\partial_y \hat{h})$ before being applied to all other $\hat{h}_{\mu\nu}$ fields. (This prescription ensures we correctly keep track of KK number relative to the soon-to-be-defined quantity ${b}$.) After KK number assignment, any 4D derivatives $\partial_\mu$ in the argument of $\mathcal{X}$ are kept as is, while each extra-dimensional derivative $\partial_y$ is replaced by $1/r_c$.

Using $\mathcal{X}$, we rewrite the A-type expression:
\begin{align}
    X_A^{(\text{eff})} =&  \left[\dfrac{\kappa}{\sqrt{\pi r_c}}\right]^{H+R-2} \sum_{n_1,\cdots,n_H=0}^{+\infty} {a}_{(R|n_1\cdots n_H)}\cdot \mathcal{X}_{(n_1\cdots n_H)}\hspace{-1 pt}\left[\hspace{+2 pt}\overline{X}_A\right]~.
\end{align}
This completes the schematic A-type procedure. B-type terms admit a similar reorganization. First, we KK decompose and integrate $X_B$ to obtain
\begin{align}
    X_B^{(\text{eff})} = &\dfrac{r_c}{(\pi r_c)^{(H+R)/2}}\,\kappa^{(H+R-2)} \sum_{n_1,\cdots,n_H=0}^{+\infty} \left(1,\hspace{6 pt}\hat{h}^{(n_1)}\cdots \hat{h}^{(n_H)},\hspace{6 pt} \left[\hat{r}^{(0)}\right]^R \right)\nonumber\\
    &\hspace{35 pt}\times e^{-R\pi kr_c}\int d\varphi\hspace{5 pt} \varepsilon^{2(R-2)}(\partial_{\varphi}\psi_{n_1})(\partial_{\varphi}\psi_{n_2})\psi_{n_3}\cdots\psi_{n_H}\left[\psi_{0}\right]^R~.
\end{align}
Then we summarize the extra-dimensional overlap integral as a unitless quantity ${b}$:
\begin{align}
    {b}_{(R|\vec{n})} &\equiv {b}_{r\cdots r n_{1}^{\prime}n_{2}^{\prime}n_{3}\cdots n_{H}}~,\nonumber\\
    &\equiv \dfrac{1}{\pi}\, e^{-R\pi kr_c}\int_{-\pi}^{+\pi}d\varphi\hspace{5 pt} \varepsilon^{2(R-2)} (\partial_{\varphi} \psi_{n_1})(\partial_{\varphi} \psi_{n_2}) \psi_{n_3}\cdots \psi_{n_H}\left[\psi_{0}\right]^R~,\label{bbar}
\end{align}
where primes on a KK index in the subscript of $b_{r\cdots r n^{\prime}_{1} n^{\prime}_{2} n_{3} \cdots n_{H}}$ indicates differentiation of the corresponding wavefunction and $b_{r\cdots r n^{\prime}_{1} n^{\prime}_{2} n_{3} \cdots n_{H}}$ is symmetric in its subscript (e.g. $b_{rn^{\prime}n^{\prime} n } = b_{nn^{\prime} r n^{\prime}}$ and so-on). The first two indices listed in the KK number list $\vec{n}$ when expressed in $b_{(R|\vec{n})}$ form will be primed when expressed in $b_{r\cdots r n^{\prime}_{1} n^{\prime}_{2} n_{3} \cdots n_{H}}$ form. With this definition,
\begin{align}
    X_B^{(\text{eff})} =& \left[\dfrac{\kappa}{\sqrt{\pi r_c}}\right]^{H+R-2} \sum_{n_1,\cdots,n_H=0}^{+\infty} {b}_{(R|n_1n_2n_3\cdots n_H)}\dfrac{1}{r_c^2}\left(1,\hspace{6 pt} \hat{h}^{(n_1)}\cdots \hat{h}^{(n_H)},\hspace{6 pt} \left[\hat{r}^{(0)}\right]^R\right)~,
\end{align}
and, via the KK decomposition operator $\mathcal{X}$,
\begin{align}
    X_B^{(\text{eff})} &= \left[\dfrac{\kappa}{\sqrt{\pi r_c}}\right]^{H+R-2} \sum_{n_1,\cdots,n_H=0}^{+\infty} {b}_{(R|n_1n_2n_3\cdots n_H)}\cdot \mathcal{X}_{(n_1\cdots n_H)}\hspace{-1 pt}\left[\hspace{2 pt}\overline{X}_B\right]~,
\end{align}
where we recall that $\mathcal{X}$ maps $\partial_y$ to $1/r_c$ after KK number assignment. This completes the schematic B-type procedure.

We now connect these procedures to the 4D effective RS1 Lagrangian $\mathcal{L}^{(\text{RS,eff})}_{4D}$, following the arrangement of the 5D Lagrangian described in Sec. \ref{AppendixWFE}. Suppose we collect all terms from the WFE RS1 Lagrangian $\mathcal{L}_{5D}^{(\text{RS})}$ that contain $H$ $\hat{h}_{\mu\nu}$ fields and $R$ $\hat{r}$ fields. Label this collection $\mathcal{L}^{(\text{RS})}_{h^H r^R}$. In general, we can subdivide those terms into two sets based on their derivative content, i.e. whether they are A-type or B-type.
\begin{align}
    \mathcal{L}^{(\text{RS})}_{h^H r^R} = \mathcal{L}^{(\text{RS})}_{A:h^H r^R} + \mathcal{L}_{B:h^H r^R}^{(\text{RS})}~.
\end{align}
We may go a step further by using our existing knowledge to preemptively extract powers of the expansion parameter $\kappa$ and any exponential coefficients:
\begin{align}
    \mathcal{L}^{(\text{RS})}_{h^H r^R} &= \kappa^{(H+R-2)}\bigg[e^{-R\pi kr_c}\, \varepsilon^{2(R-1)}\hspace{3 pt}\overline{\mathcal{L}}_{A:h^H r^R} + e^{-R\pi kr_c}\, \varepsilon^{2(R-2)}\hspace{3 pt}\overline{\mathcal{L}}_{B:h^H r^R}\bigg]~.
\end{align}
Finally, we can apply the schematic procedures described above to obtain a succinct expression (a ``5D-to-4D formula") for the effective Lagrangian with $H$ $\hat{h}_{\mu\nu}$ fields and $R$ $\hat{r}$ fields:
\begin{align}
    \boxed{\mathcal{L}^{(\text{RS,eff})}_{h^H r^R} = \left[\dfrac{\kappa}{\sqrt{\pi r_c}}\right]^{(H+R-2)}\sum_{\vec{n}=\vec{0}}^{+\infty}\bigg\{ {a}_{(R|\vec{n})}\cdot \mathcal{X}_{(\vec{n})}\hspace{-3 pt}\left[\hspace{3 pt}\overline{\mathcal{L}}_{A:h^H r^R}\right] + {b}_{(R|\vec{n})}\cdot \mathcal{X}_{(\vec{n})}\hspace{-3 pt}\left[\hspace{3 pt}\overline{\mathcal{L}}_{B:h^H r^R}\right]\bigg\}}~.\label{LRSeffPerTerm}
\end{align}
Consider all terms in this 5D-to-4D formula which only contain the 4D graviton field $\hat{h}^{(0)}_{\mu\nu}(x)$. Because the zero mode wavefunction $\psi_{0}$ does not depend on $y$, these terms must all be A-type, such that we can write
\begin{align}
    \mathcal{L}^{(\text{RS,eff})}_{h^H r^R}\hspace{15 pt} \mathrel{\mathop{\supset}^{\text{only 4D}}_{\text{gravitons}}}\hspace{15 pt} \sum_{H=2}^{+\infty} \left[\dfrac{\kappa}{\sqrt{\pi r_c}}\right]^{(H-2)}\bigg\{ {a}_{(\vec{0}_{H})}\cdot \mathcal{X}_{(\vec{0}_{H})}\hspace{-3 pt}\left[\hspace{3 pt}\overline{\mathcal{L}}_{A:h^H}\right] \bigg\} \label{465zz}
\end{align}
where $\vec{0}_{H}$ is the $H$-dimensional zero vector. Furthermore, the coupling integral $a_{(\vec{0}_{H})}$ is exactly calculable via orthonormality of the wavefunctions,
\begin{align}
    a_{(\vec{0}_{H})} \hspace{10 pt} = \hspace{10 pt} \dfrac{1}{\pi} \int_{-\pi r_{c}}^{+\pi r_{c}}d\varphi\hspace{5 pt} \vep^{-2} \psi_{0}^{H} \hspace{10 pt} = \hspace{10 pt} \psi_{0}^{H-2}\,\cdot \, \dfrac{1}{\pi} \int_{-\pi r_{c}}^{+\pi r_{c}}d\varphi\hspace{5 pt}\vep^{-2} \psi_{0}^{2} \hspace{10 pt} = \hspace{10 pt} \psi_{0}^{H-2}
\end{align}
for all integer $H\geq 2$, such that Eq. \eqref{465zz} becomes
\begin{align}
    \mathcal{L}^{(\text{RS,eff})}_{h^H r^R}\hspace{15 pt} \mathrel{\mathop{\supset}^{\text{only 4D}}_{\text{gravitons}}}\hspace{15 pt} \sum_{H=2}^{+\infty} \left[\dfrac{\kappa \psi_{0}}{\sqrt{\pi r_c}}\right]^{(H-2)} \mathcal{X}_{(\vec{0}_{H})}\hspace{-3 pt}\left[\hspace{3 pt}\overline{\mathcal{L}}_{A:h^H}\right]
\end{align}
Because an $N$-point 4D graviton interaction will generally carry a coupling factor $\kappa_{\text{4D}}^{N-2}$, comparison with the above expression reveals $\kappa_{\text{4D}} = \kappa\psi_{0}/\sqrt{\pi r_{c}}$.

Computationally, a key feature of the 4D effective Lagrangian Eq. \eqref{LRSeffPerTerm} is how the dependence on the physical variables arrange themselves. Consider the set $\{M_{\text{Pl}}, kr_{c}, m_{1}\}$. The parameter $kr_{c}$ determines the wavefunctions $\{\psi_{n}\}$ and spectrum $\{\mu_{n}\}\equiv\{m_{n}r_{c}\}$, and thus $\{a_{(R|\vec{n})},b_{(R|\vec{n})}\}$ as well. Additionally fixing the value of $m_{1}$ determines $r_{c} = \mu_{1}/m_{1}$ and $k = (kr_{c})m_{1}/\mu_{1}$. Finally, fixing $M_{\text{Pl}}$ determines the prefactor $\kappa/\sqrt{\pi r_{c}} = \kappa_{\text{4D}}/\psi_{0}$ $= 2/(M_{\text{Pl}}\psi_{0})$. Therefore, referring back to the specific form of Eq. \eqref{LRSeffPerTerm}, once $kr_{c}$ is fixed, changing $m_{1}$ only affects the relative importance of A-type vs. B-type terms via factors of $r_{c}$ introduced by $\mathcal{X}_{(\vec{n})}[\bullet]$ and changing $M_{\text{Pl}}$ only affects the interaction's overall strength via $[\kappa/\sqrt{\pi r_{c}}]^{(H+R-2)}$. Alternatively, by fixing $\kappa$ and $r_{c}$ instead, the couplings $\{a_{(R|\vec{n})},b_{(R|\vec{n})}\}$ encapsulate the effect of varying $k$.

While the $a_{(R|\vec{n})}$ and $b_{(R|\vec{n})}$ forms are useful when deriving Eq. \eqref{LRSeffPerTerm}, the alternate notations introduced in Eqs. \eqref{abar} and \eqref{bbar} are more useful in practice. They are special instances of a more general structure $x$, which we define as:
\begin{align}
    x^{(p)}_{r \cdots m^{\prime} \cdots n} &\equiv \dfrac{1}{\pi} \int_{-\pi}^{+\pi} d\varphi\hspace{10 pt}\vep^{p} (\partial_{\varphi}\psi_{m})\cdots \psi_{n} \left[ e^{-\pi kr_{c}}\vep^{+2}\psi_{0}\right]^{R} \label{generalizedx}
\end{align}
to which we add an additional factor of $(\partial_{\varphi}|\varphi|)=(\partial_{\varphi}\vep)/kr_{c}$ if there is an odd number of primed labels (without a factor like this, the quantity would automatically vanish because of orbifold symmetry). In terms of $x$, the A-type and B-type couplings equal
\begin{align}
    a_{r^{R}\, m^{\prime} \cdots n} &\equiv x^{(-2)}_{r^{R}m^{\prime}\cdots n} \hspace{35 pt} b_{r^{R}\, m^{\prime} \cdots n} \equiv x^{(-4)}_{r^{R}m^{\prime}\cdots n} \label{AandBintermsofgeneralizedx}
\end{align}
where this generalization now allows A-type and B-type couplings to contain any number of differentiated wavefunctions in principle. This dissertation concerns tree-level massive KK mode scattering, which is calculated from diagrams of the forms described in Section \ref{S - 2to2}. Consequently, the relevant couplings are cubic and quartic couplings of the forms
\begin{align}
    a_{lmn}\hspace{35 pt}b_{l^{\prime}m^{\prime}n}\hspace{35 pt}b_{m^{\prime}n^{\prime}r}\hspace{35 pt}a_{klmn}\hspace{35 pt}b_{k^{\prime}l^{\prime}mn} \label{atoxandbtox}
\end{align}
A cubic A-type radion coupling is not listed because it does not occur in the RS1 model: the existence of an $a_{mnr}$ coupling would violate the gauge symmetries of the 4D graviton (or, in other words, 4D diffeomorphism invariance of the Lagrangian) by necessarily implying the existence of non-diagonal graviton couplings, e.g. $a_{0nr}$.

Pictorially, we indicate the vertices associated with these couplings as small filled circles attached to the appropriate number of particle lines, e.g. the relevant spin-2 exclusive interactions are drawn as
\begin{align}
    \raisebox{-0.45\height}{\verhhh{0.20}} \hspace{10 pt}&\supset\hspace{35 pt} a_{n_{1}n_{2}n_{3}}\hspace{35 pt} b_{n^{\prime}_{\pi[1]}n^{\prime}_{\pi[2]}n_{\pi[3]}}\\
    \raisebox{-0.45\height}{\verhhhh{0.20}} \hspace{10 pt}&\supset\hspace{35 pt} a_{n_{1}n_{2}n_{3}n_{4}}\hspace{35 pt} b_{n^{\prime}_{\pi[1]}n^{\prime}_{\pi[2]}n_{\pi[3]}n_{\pi[4]}}
\end{align}
where overlapping straight and wavy lines indicate a spin-2 particle, and $\pi$ is a generic permutation of the indices.\footnote{Our use of the symbols `$a$' and `$b$' as labels for the coupling integrals $a_{(R|\vec{n})}$ and $b_{(R|\vec{n})}$ was inspired by the integrals $\alpha$ and $\beta(m,n)$ defined in \cite{Davoudiasl:2001uj} which are specifically associated with spin-2 exclusive cubic interactions in the large $kr_{c}$ limit.} If we set $n_{3}=0$ in the triple spin-2 coupling, the corresponding wavefunction $\psi_{0}$ is flat; either $\psi_{0}$ is differentiated (in which case the integral vanishes) or it can be factored out of the $y$-integral thereby allowing us to invoke the wavefunction orthogonality relations on the remaining wavefunction pair. In this way, the triple spin-2 couplings imply that the massless 4D graviton couples diagonally to the other spin-2 states, as required by 4D general covariance:
\begin{align}
    a_{n_1n_20} &= \psi_0\s \delta_{n_1,n_2}~,\\
    b_{n_{1}^{\prime}n_{2}^{\prime}0} &= \mu_{n_1}^{2} \psi_{0} \s \delta_{n_1,n_2}~,\nonumber\\
    b_{0^{\prime}n_{1}^{\prime}n_{2}} &= 0~.\nonumber
\end{align}
The Sturm-Liouville problem Eq. \eqref{SLeqx} that defines the wavefunctions $\{\psi_{n}\}$ also relates various spin-2 exclusive A-type and B-type couplings to each other, which we be explored further in the next section.

When calculating matrix elements of massive KK mode scattering, we must also consider radion-mediated diagrams. As mentioned previously, the RS1 model lacks a cubic A-type (KK mode)-(KK mode)-radion coupling. Furthermore, note that the additional $\vep^{+2}$ exponential factor in the integrand of $b_{n_{1}^{\prime}n_{2}^{\prime}r}$ due to the radion field (as in Eq. \eqref{generalizedx}) prevents the use of the orthonormality relations Eqs. \eqref{onAx} and \eqref{onBx}; therefore, the radion typically couples non-diagonally to massive spin-2 modes. Pictorially,
\begin{align}
    \raisebox{-0.45\height}{\verhhr{0.20}}\hspace{10 pt} &\supset\hspace{10 pt} b_{n_{1}^{\prime}n_{2}^{\prime}r}
\end{align}
where unadorned straight lines indicate a radion.

\subsection{Summary of Results} \label{4D Effective RS1 - Summary of Results}

Section \ref{AppendixWFE} summarized all terms in the 5D WFE RS1 Lagrangian $\mathcal{L}^{(\text{RS})}_{\text{5D}}$ that contain four or fewer fields. In particular, it listed explicit expressions for all relevant $\overline{\mathcal{L}}_{A}$ and $\overline{\mathcal{L}}_{B}$. Application of the 5D-to-4D formula Eq. \eqref{LRSeffPerTerm} to these terms yields a 4D effective WFE RS1 Lagrangian of the following form: 
\begin{align}
    \mathcal{L}^{(\text{RS,eff})}_{4D} =& \mathcal{L}^{(\text{eff})}_{hh} + \mathcal{L}^{(\text{eff})}_{rr} + \mathcal{L}^{(\text{eff})}_{hhh} + \cdots + \mathcal{L}^{(\text{eff})}_{rrr} + \mathcal{L}^{(\text{eff})}_{hhhh} + \cdots + \mathcal{L}^{(\text{eff})}_{rrrr} + \mathcal{O}(\kappa^3)~.
\end{align}
Explicitly, we find, at quadratic order,
\begin{align}
    \mathcal{L}^{(\text{eff})}_{hh} =& \sum_{n=0}^{+\infty} \left[-\hat{h}^{(n)}_{\mu\nu} (\partial^\mu \partial^\nu \hat{h}^{(n)}) + \hat{h}^{(n)}_{\mu\nu} (\partial^\mu \partial_\rho \hat{h}^{(n)\rho\nu})- \dfrac{1}{2} \hat{h}^{(n)}_{\mu\nu} (\square \hat{h}^{(n)\mu\nu}) + \dfrac{1}{2} \hat{h}^{(n)} (\square\hat{h}^{(n)})\right]\nonumber\\
    &\hspace{35 pt}+ m_n^2\left[-\dfrac{1}{2}\ltr \hat{h}^{(n)}\hat{h}^{(n)}\rtr +\dfrac{1}{2} \ltr \hat{h}^{(n)}\rtr \ltr \hat{h}^{(n)} \rtr \right] ~,\\
    \mathcal{L}^{(\text{eff})}_{rr} =& \dfrac{1}{2}(\partial_\mu \hat{r}^{(0)})(\partial^\mu \hat{r}^{(0)})~,
\end{align}
and, at cubic order,
\begin{align}
    \mathcal{L}^{(\text{eff})}_{hhh} =& \dfrac{\kappa}{\sqrt{\pi r_c}}\sum_{l,m,n=0}^{+\infty} \bigg\{{a}_{(0|lmn)}\cdot \mathcal{X}_{(lmn)}\hspace{-3 pt}\left[\hspace{1 pt}\overline{\mathcal{L}}_{A:hhh}\right] +{b}_{(0|lmn)}\cdot\mathcal{X}_{(lmn)}\hspace{-3 pt} \left[\hspace{+1 pt}\overline{\mathcal{L}}_{B:hhh}\right]\bigg\}~,\\
    \mathcal{L}^{(\text{eff})}_{hhr}=& \dfrac{\kappa}{\sqrt{\pi r_c}}\sum_{m,n=0}^{+\infty}\bigg\{ {b}_{(1|mn)}\cdot\mathcal{X}_{(mn)}\hspace{-3 pt} \left[\hspace{1 pt}\overline{\mathcal{L}}_{B:hhr}\right]\bigg\}~,\\
    \mathcal{L}^{(\text{eff})}_{hrr}=& \dfrac{\kappa}{\sqrt{\pi r_c}}\sum_{n=0}^{+\infty}\bigg\{ {a}_{(2|n)}\cdot \mathcal{X}_{(n)}\hspace{-3 pt} \left[\hspace{1 pt}\overline{\mathcal{L}}_{A:hrr}\right]\bigg\}~,\label{L4Deffhhr}\\
    \mathcal{L}^{(\text{eff})}_{rrr}=& \dfrac{\kappa}{\sqrt{\pi r_c}}\bigg\{ {a}_{(3)}\cdot \mathcal{X}\hspace{-3 pt}\left[\hspace{1 pt}\overline{\mathcal{L}}_{A:rrr}\right]\bigg\}~,
\end{align}
and, at quartic order,
\begin{align}
    \mathcal{L}^{(\text{eff})}_{hhhh} =& \left[\dfrac{\kappa}{\sqrt{\pi r_c}}\right]^2\sum_{k,l,m,n=0}^{+\infty} \bigg\{{a}_{(klmn)}\cdot \mathcal{X}_{(klmn)}\hspace{-3 pt} \left[\hspace{1 pt}\overline{\mathcal{L}}_{A:hhhh}\right]+{b}_{(klmn)}\cdot\mathcal{X}_{(klmn)}\hspace{-3 pt} \left[\hspace{1 pt}\overline{\mathcal{L}}_{B:hhhh}\right]\bigg\}~,\\
    \mathcal{L}^{(\text{eff})}_{hhhr} =& \left[\dfrac{\kappa}{\sqrt{\pi r_c}}\right]^2\sum_{l,m,n=0}^{+\infty}\bigg\{ {b}_{(1|lmn)}\cdot\mathcal{X}_{(lmn)}\hspace{-3 pt} \left[\hspace{1 pt}\overline{\mathcal{L}}_{B:hhhr}\right]\bigg\}~,\\
    \mathcal{L}^{\text{(eff)}}_{hhrr} =& \left[\dfrac{\kappa}{\sqrt{\pi r_c}}\right]^2\sum_{m,n=0}^{+\infty} \bigg\{{a}_{(2|mn)}\cdot\mathcal{X}_{(mn)}\hspace{-3 pt} \left[\hspace{1 pt}\overline{\mathcal{L}}_{A:hhrr}\right]+{b}_{(2|mn)}\cdot\mathcal{X}_{(mn)}\hspace{-3 pt} \left[\hspace{1 pt}\overline{\mathcal{L}}_{B:hhrr}\right]\bigg\}~,\\
    \mathcal{L}^{(\text{eff})}_{hrrr} =& \left[\dfrac{\kappa}{\sqrt{\pi r_c}}\right]^2\sum_{n=0}^{+\infty} \bigg\{{a}_{(3|n)} \cdot\mathcal{X}_{(n)}\hspace{-3 pt} \left[\hspace{1 pt}\overline{\mathcal{L}}_{A:hrrr}\right]\bigg\}~,\\
    \mathcal{L}^{(\text{eff})}_{rrrr} =& \left[\dfrac{\kappa}{\sqrt{\pi r_c}}\right]^2 \bigg\{ {a}_{(4)}\cdot\mathcal{X}\hspace{-3 pt} \left[\hspace{1 pt}\overline{\mathcal{L}}_{A:rrrr}\right]\bigg\}~.
\end{align}
The quantity $a_{(R|\vec{n})}$ is defined in Eq. \eqref{abar}, $b_{(R|\vec{n})}$ is defined in Eq. \eqref{bbar}, and the KK decomposition operator $\mathcal{X}$ is introduced below Eq. \eqref{KKdecompop}.

\subsection{Interaction Vertices}
The 4D effective interaction Lagrangians $\mathcal{L}_{h^{H}r^{R}}^{(\text{eff})}$ of the previous subsection imply interaction vertices $v_{h^{H}r^{R}}$. When deriving those vertices, we apply functional derivatives to the interaction Lagrangians, which should in principle be performed according to the definitions
\begin{align}
    \dfrac{\delta}{\delta \hat{r}^{(0)}} \bigg[\hat{r}^{(0)}\bigg] = 1
    \hspace{70 pt}
    \dfrac{\delta}{\delta \hat{h}^{(n_{1})}_{\alpha_{1}\beta_{1}}} \bigg[\hat{h}^{(n)}_{\alpha\beta}\bigg] = \dfrac{1}{2}\left(\eta^{\alpha_{1}}_{\alpha} \eta^{\beta_{1}}_{\beta} + \eta^{\alpha_{1}}_{\beta} \eta^{\beta_{1}}_{\alpha}\right)\delta_{n,n_{1}} \label{FDdef}
\end{align}
However, in practice each pair of spin-2 Lorentz indices in these vertices will end up projected onto either a polarization tensor or a propagator, all of which have already had their Lorentz indices symmetrized. Therefore, we need not additionally symmetrize the indices in Eq. \eqref{FDdef} and in doing so can avoid introducing terms that will otherwise complicate algebraic manipulations. That is, effectively,
\begin{align}
    \dfrac{\delta}{\delta \hat{h}^{(n_{1})}_{\alpha_{1}\beta_{1}}} \bigg[\hat{h}^{(n)}_{\alpha\beta}\bigg] \hspace{5 pt} \stackrel{\text{in practice}}{=} \hspace{5 pt} \eta^{\alpha_{1}}_{\alpha} \eta^{\beta_{1}}_{\beta} \delta_{n,n_{1}}
\end{align}
Furthermore, each 4D derivative $\partial_{\mu}$ acting on the field being differentiated is replaced by $-i\alpha p_{\mu}$, where $\alpha = \pm1$ if the corresponding $4$-momentum is entering (leaving) the vertex. In order to keep track of which $4$-momenta are associated with which fields, we introduce labels on the functional derivative fields. For the spin-2 fields $\hat{h}^{(n)}_{\mu\nu}$, this can be accomplished via the subscripts we already utilized in Eq. \eqref{FDdef}. For the radion fields $\hat{r}^{(0)}$, we add an additional subscript, e.g. $\hat{r}_{1}^{(0)}$. As long as the subscripts are chosen so that they uniquely label fields connected to a given vertex, all is well.

The conversion of a typical term of the 4D effective Lagrangian into the corresponding interaction vertex proceeds like so:
\begin{align}
    v_{hrr} &= i\dfrac{\delta}{\delta \hat{r}_{1}^{(0)}}\dfrac{\delta}{\delta \hat{r}_{2}^{(0)}}\dfrac{\delta}{\delta \hat{h}_{\alpha_{3}\beta_{3}}^{(n_{3})}}\bigg[\mathcal{L}_{hrr}^{(\text{eff})}\bigg] \\
    &\supset i\dfrac{\delta}{\delta \hat{r}_{1}^{(0)}}\dfrac{\delta}{\delta \hat{r}_{2}^{(0)}}\dfrac{\delta}{\delta \hat{h}_{\alpha_{3}\beta_{3}}^{(n_{3})}}\bigg[ \dfrac{\kappa}{\sqrt{\pi r_{c}}} \sum_{n=0}^{+\infty} a_{nrr} \, \hat{h}^{(n)}_{\mu\nu}(\partial^{\mu}\hat{r}^{(0)})(\partial^{\nu}\hat{r}^{(0)})\bigg]\\
    &= i \dfrac{\kappa}{\sqrt{\pi r_{c}}} \, a_{n_{3}rr} \, \dfrac{\delta}{\delta \hat{r}_{1}^{(0)}}\dfrac{\delta}{\delta \hat{r}_{2}^{(0)}} \bigg[ 
    \eta_{\mu}^{\alpha_{3}}\eta_{\nu}^{\beta_{3}}(\partial^{\mu}\hat{r}^{(0)})(\partial^{\nu}\hat{r}^{(0)}) \bigg]\\
    &= i \dfrac{\kappa}{\sqrt{\pi r_{c}}} \, a_{n_{3}rr} \, \dfrac{\delta}{\delta \hat{r}_{1}^{(0)}} \bigg[  \eta_{\mu}^{\alpha_{3}}\eta_{\nu}^{\beta_{3}}(-i\alpha_{2} p_{2}^{\mu})(\partial^{\nu}\hat{r}^{(0)}) + \eta_{\mu}^{\alpha_{3}}\eta_{\nu}^{\beta_{3}}(\partial^{\mu}\hat{r}^{(0)})(-i\alpha_{2} p_{2}^{\nu}) \bigg]\\
    &= i \dfrac{\kappa}{\sqrt{\pi r_{c}}} \, a_{n_{3}rr} \, \bigg[  \eta_{\mu}^{\alpha_{3}}\eta_{\nu}^{\beta_{3}}(-i\alpha_{2} p_{2}^{\mu})(-i\alpha_{1} p_{1}^{\nu}) + \eta_{\mu}^{\alpha_{3}}\eta_{\nu}^{\beta_{3}}(-i\alpha_{1} p_{1}^{\mu})(-i\alpha_{2} p_{2}^{\nu}) \bigg]\\
    &= -i \dfrac{\kappa}{\sqrt{\pi r_{c}}} \, a_{n_{3}rr} \, \alpha_{1}\alpha_{2} (p_{1}^{\alpha_{3}}p_{2}^{\beta_{3}} + p_{1}^{\beta{3}}p_{2}^{\alpha_{3}})
\end{align}
where $1$ and $2$ label attached radion lines and $3$ labels an attached $n_{3}$th spin-2 KK mode.

\subsection[Wavefunctions and Couplings in the Large $kr_{c}$ Limit]{The Large $kr_{c}$ Limit\footnote{This subsection was originally published as Appendices F.1-2 of \cite{Chivukula:2020hvi}.}} \label{Large krc}

Consider how the aforementioned wavefunctions and couplings behave in the limit that $kr_{c}$ is large. In this limit, the behavior of the irregular Bessel functions $Y_\nu$ causes the coefficients $b_{n\nu}$ in Eq. (\ref{bnnu}) to be small, such that the wavefunctions of Eq. \eqref{eq:wavefunction} (having nonzero KK mode number $n$) can be approximated as
\begin{align}
    \psi_{n}(\varphi) \approx \dfrac{1}{N_{n}}e^{+2 kr_{c} |\phi|} J_{2}\left[x_{n} e^{kr_{c}(|\phi|-\pi)}\right]~,
    \label{eq:asymptotic-wf}
\end{align}
where $x_{n}$ is the $n$th root of $J_{1}$ and
\begin{align}
    N_{n} \approx \dfrac{e^{\pi kr_{c}}}{\sqrt{\pi kr_{c}}} J_{0}(x_{n})~.
\end{align}
This approximation of the wavefunction $\psi_{n}$ corresponds to a state with mass
\begin{align}
    m_{n} \approx x_{n} k\, e^{-\pi kr_{c}}~.
\end{align}
This limit---called the ``large $kr_{c}$ limit"---is a good approximation when $kr_{c}\gtrsim 3$ and is popular in the literature.

The above expressions can be further simplified by replacing $\varphi$ with the quantity $u_{n} \equiv x_{n} e^{kr_{c}(\varphi-\pi)}$. In terms of $u_{n}$, the $n\neq 0$ wavefunction factorizes into separate $u_{n}$ and $kr_{c}$-dependent pieces,
\begin{align}
    \psi_{n}(u_{n}) \approx  \dfrac{\sqrt{\pi}}{x_{n}^{2}\,|J_{0}(x_{n})|} \, \left[ u_{n}^{2}\, J_{2}(u_{n}) \right]\, \cdot \,\sqrt{kr_{c}}\,e^{\pi kr_{c}}~.
\end{align}
More generally, for any nonzero KK mode $j\neq n$,
\begin{align}
    \psi_{j}(u_{n}) \approx  \dfrac{\sqrt{\pi}}{x_{n}^{2}\,|J_{0}(x_{j})|}\, \left[ u_{n}^{2}\, J_{2}\left(\tfrac{x_{j}}{x_{n}}u_{n}\right) \right]\, \cdot\,\sqrt{kr_{c}}\,e^{\pi kr_{c}}~,
\end{align}
and, 
\begin{align}
    (\partial_{\varphi} \psi_{j})(u_{n}) \approx  \dfrac{\sqrt{\pi}\, x_{j}}{x_{n}^{3}\,|J_{0}(x_{j})|}\, \left[ u^{3}\, J_{1}\left(\tfrac{x_{j}}{x_{n}}u_{n}\right) \right]\,(kr_{c})^{3/2}\,e^{\pi kr_{c}}~.
\end{align}
Meanwhile, the large $kr_{c}$ approximation of the zero mode wavefunction is
\begin{align}
    \psi_{0} \approx \sqrt{\pi kr_{c}}~.
\end{align}
We can also rewrite coupling integrals as integrals over $u_{n}$ instead of $\varphi$ and (in doing so) factor any $kr_{c}$-dependence from the integral. Specifically, we can convert $\varphi$ integrals of the form
\begin{align}
    \int_{-\pi}^{+\pi} d\varphi \hspace{5 pt}e^{-A kr_{c}|\varphi|} \, f(|\varphi|) = 2 \int_{0}^{+\pi} d\varphi \hspace{5 pt}e^{-A kr_{c}|\varphi|} \,  f(\varphi)~,
\end{align}
to $u_{n} \equiv x_{n} e^{kr_{c}(\varphi-\pi)}$ integrals (using $d\varphi = du_{n}/(kr_{c} u_{n})$)
\begin{align}
    &\int_{-\pi}^{+\pi} d\varphi \hspace{5 pt}e^{-A kr_{c}|\varphi|} \, f(|\varphi|) = \dfrac{2x_{n}^{A}e^{-A kr_{c}\pi}}{kr_{c}}\,\cdot\,\left[ \int_{u_{n}(0)}^{u_{n}(\pi)} \dfrac{du_{n}}{u_{n}^{A+1}}\hspace{5 pt}f\left(\varphi(u_{n})\right)\right]~, 
\end{align}
where we note that the integration limits become independent of $kr_{c}$ in the large $kr_{c}$ limit:
\begin{align}
    u_{n}(0) = e^{-kr_{c} \pi} x_{n} \rightarrow 0\hspace{50 pt}u_{n}(\pi) = x_{n}~.
\end{align}
and thus the integral over $u_{n}$ does not depend on $kr_{c}$. By combining all of the preceding elements, we can factor all $kr_{c}$-dependence out of the coupling integrals in the large $kr_{c}$ limit, and we find
\begin{align}
    a_{nnnn} &\approx C_{nnnn}\, (kr_{c})\, e^{2\pi kr_{c}}~, \label{largekrc1}\\
    a_{nn0} &\approx C_{nn0}\, \sqrt{kr_{c}}~,\\
    b_{n^{\prime}n^{\prime}r} &\approx C_{nnr}\, (kr_{c})^{5/2}\, e^{-\pi kr_{c}}~,\\
    a_{nnj} &\approx C_{nnj}\, \sqrt{kr_{c}}\, e^{\pi kr_{c}}~,
\end{align}
where the coefficients $C$ are given by the following $kr_c$-independent integrals:
\begin{align}
    C_{nnnn} &\equiv \left[\dfrac{2\pi }{x_{n}^{6}\, J_{0}(x_{n})^{4}}\, \int_0^{x_n} du_{n}\hspace{5 pt}u_{n}^{5}\, J_{2}(u_{n})^{4}\right]~,\\
    C_{nn0} &\equiv \left[\dfrac{2\sqrt{\pi }}{x_{n}^{2}\,J_{0}(x_{n})^{2}}\, \int_0^{x_n} du_{n}\hspace{5 pt}u_{n}\, J_{2}(u_{n})^{2}\right]~,\\
    C_{nnr} &\equiv \left[\dfrac{2 \sqrt{\pi} }{x_{n}^{2}\,J_{0}(x_{n})^{2} }\, \int_0^{x_n} du_{n}\hspace{5 pt}u_{n}^{3}\, J_{1}(u_{n})^{2}\right]~,\\
    C_{nnj} &\equiv \left[\dfrac{2\sqrt{\pi}}{x_{n}^{4}\,|J_{0}(x_{j})|\, J_{0}(x_{n})^{2}}\int_0^{x_n} du_{n}\hspace{5 pt}u_{n}^{3}\, J_{2}(u_{n})^{2}\, J_{2}\left(\tfrac{x_{j}}{x_{n}}u_{n}\right)\right] \label{largekrc2}
\end{align}
Although we utilize exact expressions when investigating the high-energy behavior of matrix elements, the approximate expressions derived in this subsection will be useful when we consider the strong coupling scale of the RS1 model in the next chapter.

\section[Sum Rules Between Couplings and Masses]{Sum Rules Between Couplings and Masses\footnote{The material of this section is entirely new to this dissertation. It generalizes results first published in \cite{Chivukula:2019zkt} and later generalized (but not to the same extent) in \cite{Chivukula:2020hvi}. An error in the first version of this thesis led to an incorrect B-to-A formula for the quartic coupling which has since been corrected.}} \label{DerivingSumRules}
This section derives relationships between the spin-2 exclusive couplings (i.e. RS1 couplings not involving the radion) and spin-2 KK spectrum $\{\mu_{n}\}$ that are relevant to tree-level 2-to-2 massive KK mode scattering. We briefly consider the implications of completeness before deriving a means of expressing all cubic and quartic (spin-2 exclusive) B-type couplings in terms of A-type couplings and special objects $B_{(kl)(mn)}$. These B-to-A formulas reduce the problem of finding amplitude-relevant formulas to the problem of simplifying sums of the form $\sum_{j} \mu_{j}^{2p}\, a_{klj}\, a_{mnj}$. The relevant (inelastic, permutation-symmetric inelastic, and elastic) sum rules are derived and then summarized in the final three subsections.

\subsection{Applications of Completeness}
The completeness relation Eq. \eqref{eq:completeness} allows us to collapse certain sums of cubic coupling products into a single quartic coupling. For example, a pair of cubic A-type couplings can be combined into a quartic A-type coupling:
\begin{align}
    \sum_{j} a_{jkl} a_{jmn} &= \sum_{j} \bigg[\dfrac{1}{\pi}\int d\varphi_{1}\hspace{5 pt}\vep(\varphi_{1})^{-2}\,\psi_{j}(\varphi_{1})\, \psi_{k}(\varphi_{1})\, \psi_{l}(\varphi_{1}) \bigg]\nonumber\\
    &\hspace{35 pt}\times \bigg[\dfrac{1}{\pi}\int d\varphi_{2} \hspace{5 pt} \vep(\varphi_{2})^{-2}\, \psi_{j}(\varphi_{2})\, \psi_{m}(\varphi_{2})\, \psi_{n}(\varphi_{2}) \bigg]\\
    &\hspace{-60 pt}= \dfrac{1}{\pi^{2}}\int d\varphi_{1}\,d\varphi_{2}\hspace{5 pt}\vep(\varphi_{1})^{-2}\,\vep(\varphi_{2})^{-2}\, \psi_{k}(\varphi_{1})\, \psi_{l}(\varphi_{1}) \, \psi_{m}(\varphi_{2})\, \psi_{n}(\varphi_{2}) \bigg[ \sum_{j} \psi_{j}(\varphi_{1})\, \psi_{j}(\varphi_{2}) \bigg]\\
    &\hspace{-60 pt}= \dfrac{1}{\pi^{2}}\int d\varphi_{1}\,d\varphi_{2}\hspace{5 pt}\vep(\varphi_{1})^{-2}\,\vep(\varphi_{2})^{-2}\,  \psi_{k}(\varphi_{1})\, \psi_{l}(\varphi_{1}) \, \psi_{m}(\varphi_{2})\, \psi_{n}(\varphi_{2}) \bigg[ \pi \vep(\varphi_{2})^{+2}\, \delta(\varphi_{2}-\varphi_{1})\bigg]\\
    &\hspace{-60 pt}= \dfrac{1}{\pi} \int d\varphi_{1}\hspace{5 pt}\vep(\varphi_{1})^{-2}\, \psi_{k}(\varphi_{1}) \, \psi_{l}(\varphi_{1}) \, \psi_{m}(\varphi_{1}) \, \psi_{n}(\varphi_{1})\\
    &\hspace{-60 pt}= a_{klmn}
\end{align}
By applying this same procedure to other A-type and B-type couplings, we find
\begin{align}
    a_{klmn} &= \sum_{j} a_{jkl}\, a_{jmn} = \sum_{j} a_{jkm}\,a_{jln} = \sum_{j} a_{jkn}\,a_{jlm} \label{acompleteness}\\
    b_{k^{\prime}l^{\prime} mn} &= \sum_{j} b_{k^{\prime} l^{\prime} j}\, a_{jmn}
\end{align}
Furthermore, by combining cubic B-type couplings in this same way, we define an important new integral that will be present in many of our derivations:
\begin{align}
    c_{k^{\prime}l^{\prime}m^{\prime}n^{\prime}} & \equiv x^{(-6)}_{k^{\prime}l^{\prime}m^{\prime}n^{\prime}} = \dfrac{1}{\pi} \int d\varphi\hspace{5 pt} \vep^{-6}(\partial_{\varphi}\psi_{k})(\partial_{\varphi}\psi_{l})(\partial_{\varphi}\psi_{m})(\partial_{\varphi}\psi_{n}) \label{cklmnDEF}\\
    &= \sum_{j} b_{k^{\prime}l^{\prime}j}\,b_{jm^{\prime}n^{\prime}} = \sum_{j} b_{k^{\prime}m^{\prime}j}\,b_{jl^{\prime}n^{\prime}} = \sum_{j} b_{k^{\prime}n^{\prime}j}\,b_{jl^{\prime}m^{\prime}}\nonumber
\end{align}
where the generic coupling integral $x$ is defined in Eq. \eqref{generalizedx}. Another object that will be useful throughout the rest of the chapter is the symbol $\mathcal{D} \equiv \vep^{-4}\partial_{\varphi}$, which is a combination of quantities that is often present as a result of the Sturm-Liouville equation. We will ultimately derive sum rules that allow us to rewrite certain useful sums of intermediate masses and couplings in terms of just the quartic A-type coupling $a_{klmn}$, three $B_{(kl)(mn)}$ objects (of which any two fix the value of the third), and the integral $c_{k^{\prime}l^{\prime}m^{\prime}n^{\prime}}$.

\subsection{B-to-A Formulas}
This subsection details how to eliminate all B-type couplings (e.g. $b_{l^{\prime}m^{\prime}n}$ and $b_{k^{\prime}l^{\prime}mn}$) in favor of A-type couplings (e.g. $a_{lmn}$ and $a_{klmn}$) and new structures $B_{(kl)(mn)}$. To begin, we note we can absorb a factor of $\mu^{2}$ into A-type couplings with help from the Sturm-Liouville equation. A standard application of this technique proceeds as follows:
\begin{align}
    \mu_{n}^{2}\, a_{lmn} &= \dfrac{1}{\pi}\int d\varphi\hspace{5 pt} \vep^{-2} \psi_{l} \psi_{m} \left[\mu_{n}^{2}\psi_{n}\right]\\
    &= \dfrac{1}{\pi} \int d\varphi\hspace{5 pt}\vep^{-2}  \psi_{l}\psi_{m}\bigg[-\vep^{+2}\partial_{\varphi}(\mathcal{D}\psi_{n})\bigg]\label{286}\\
    &= \dfrac{1}{\pi} \int d\varphi\hspace{5 pt}\partial_{\varphi}\left[\psi_{l}\psi_{m}\right] (\mathcal{D}\psi_{n}) \label{287}\\
    &= \dfrac{1}{\pi} \int d\varphi\hspace{5 pt} \vep^{-4} (\partial_{\varphi}\psi_{l})\psi_{m} (\partial_{\varphi} \psi_{n}) + \dfrac{1}{\pi} \int d\varphi\hspace{5 pt} \vep^{-4} \psi_{l}(\partial_{\varphi}\psi_{m}) (\partial_{\varphi} \psi_{n})\\
    &= b_{l^{\prime}mn^{\prime}} + b_{lm^{\prime}n^{\prime}}
\end{align}
where integration by parts was utilized between Eqs. \eqref{286} and \eqref{287}; because $(\mathcal{D}\psi_{n})$ vanishes on the boundaries, there is no surface term. This and the equivalent calculation with the quartic A-type coupling yield
\begin{align}
    \mu_{n}^{2}\, a_{lmn} &= b_{l^{\prime}mn^{\prime}} + b_{lm^{\prime}n^{\prime}} \label{AIMS3}\\
    \mu_{n}^{2}\, a_{klmn} &= b_{k^{\prime}lmn^{\prime}} + b_{kl^{\prime}mn^{\prime}} + b_{klm^{\prime}n^{\prime}} \label{AIMS4}
\end{align}
By considering different permutations of KK indices, each of these equations corresponds to three and four unique constraints respectively. Because there are only three unique cubic B-type couplings with KK indices $l$, $m$, and $n$ (specifically, $b_{l^{\prime}m^{\prime}n}$, $b_{l^{\prime}mn^{\prime}}$, and $b_{lm^{\prime}n^{\prime}}$), Eq. \eqref{AIMS3} can be inverted to yield
\begin{align}
    \boxed{b_{l^{\prime} m^{\prime} n} = \dfrac{1}{2}\left[\mu_{l}^{2} + \mu_{m}^{2} - \mu_{n}^{2} \right] a_{lmn}} \label{blmnRED}
\end{align}
with which we can eliminate all cubic B-type couplings in favor of the cubic A-type coupling.

There are {\it six} unique quartic B-type couplings with KK indices $k$, $l$, $m$, and $n$. We first halve this set by rewriting each quartic B-type coupling $b_{k^{\prime}l^{\prime}mn}$ in terms of new objects $B_{(kl)(mn)}$. These new objects are motivated as follows: note that Eq. \eqref{AIMS4} implies
\begin{align}
    \dfrac{1}{2} \bigg[ \mu_{k}^{2} + \mu_{l}^{2} - \mu_{m}^{2} - \mu_{n}^{2} \bigg]\, a_{klmn} &= b_{k^{\prime}l^{\prime}mn} - b_{klm^{\prime}n^{\prime}}
\end{align}
Equivalently, we may write this as
\begin{align}
    b_{k^{\prime}l^{\prime}mn} + \dfrac{1}{2} \bigg[\mu_{m}^{2} + \mu_{n}^{2}\bigg]\, a_{klmn} &= b_{klm^{\prime}n^{\prime}} + \dfrac{1}{2}\bigg[\mu_{k}^{2} + \mu_{l}^{2} \bigg]\, a_{klmn} \label{BMotivation}
\end{align}
In other words, the quantity on the LHS possesses a symmetry under the pair swap $(k,l)\leftrightarrow (m,n)$. Furthermore, this symmetry is maintained under the addition of any quantity $\tilde{B}_{(kl)(mn)}$ which is also symmetric under this pair swap. Inspired by Eq. \eqref{BMotivation}, we define
\begin{align}
    B_{(kl)(mn)} \equiv b_{k^{\prime}l^{\prime}mn} + \dfrac{1}{2} \bigg[\mu_{m}^{2} + \mu_{n}^{2}\bigg]\, a_{klmn} + \tilde{B}_{(kl)(mn)}
\end{align}
We will fix the quantity $\tilde{B}_{(kl)(mn)}$ momentarily. Because the B-type couplings satisfy Eq. \eqref{AIMS4}, the sum of all unique $B$ objects satisfies
\begin{align}
    B_{(kl)(mn)} + B_{(km)(ln)} + B_{(kn)(lm)} = \vec{\mu}^{\,2} \, a_{klmn} + \tilde{B}_{(kl)(mn)} + \tilde{B}_{(km)(ln)} + \tilde{B}_{(kn)(lm)} 
\end{align}
where $\vec{\mu}^{\,2} \equiv \mu_{k}^2+\mu_{l}^2+\mu_{m}^{2}+\mu_{n}^{2}$. That is, we can ensure the convenient property
\begin{align}
    B_{(kl)(mn)} + B_{(km)(ln)} + B_{(kn)(lm)} \dot{=} 0
\end{align}
as long as we choose $\tilde{B}_{(kl)(mn)}$ such that
\begin{align}
     \tilde{B}_{(kl)(mn)} + \tilde{B}_{(km)(ln)} + \tilde{B}_{(kn)(lm)} = -\vec{\mu}^{\,2} \, a_{klmn}
\end{align}
One immediate choice (and the choice we take now) is to set each $\tilde{B}$ equal to one third of $-\vec{\mu}^{\,2}\, a_{klmn}$
\begin{align}
    \tilde{B}_{(kl)(mn)} \dot{=} -\dfrac{1}{3}\, a_{klmn}
\end{align}
This yields (as a replacement rule for $b_{k^{\prime}l^{\prime}mn}$ and definition of $B_{(kl)(mn)}$)
\begin{align}
    \boxed{b_{k^{\prime}l^{\prime}mn} = B_{(kl)(mn)}+\dfrac{1}{6}\bigg[2(\mu_{k}^{2}+\mu_{l}^{2})-(\mu_{m}^{2}+\mu_{n}^{2})\bigg] a_{klmn}} \label{bklmnRED}
\end{align}
where $B$ is symmetric within each pair and between pairs:
\begin{align}
    B_{(kl)(mn)} = B_{(mn)(kl)} = B_{(mn)(lk)}
\end{align}
and satisfies the additional constraint
\begin{align}
    \boxed{B_{(kl)(mn)} + B_{(km)(ln)} + B_{(kn)(lm)} = 0}\label{SumOfB}
\end{align}
such that only two among $\{B_{(kl)(mn)},B_{(km)(ln)},B_{(kn)(lm)}\}$ are linearly independent. Note that $B_{(kl)(mn)}$ has the same symmetry properties as $\sum_{j} \mu_{j}^{2p}\,a_{klj}\, a_{mnj}$. Because Eq. \eqref{bklmnRED} reduces B-type couplings to A-type couplings as much is as possible, we refer to it as the quartic B-to-A rule. This and Eq. \eqref{blmnRED} comprise the desired B-to-A formulas.

The above rules are sufficient as-is for reducing the sum $\sum_{j} \mu_{j}^{2}\, a_{klj}\, a_{mnj}$ and yielding the first non-trivial sum rule. Using the cubic coupling equation Eq. \eqref{AIMS3} with completeness yields,
\begin{align}
    b_{k^{\prime}l^{\prime}mn} &= \sum_{j} b_{k^{\prime}l^{\prime}j}\,a_{jmn}\\
    &= \dfrac{1}{2}\sum_{j} \left[\mu_{k}^{2} + \mu_{l}^{2} - \mu_{j}^{2} \right] a_{jkl}\, a_{jmn}\\
    &= \dfrac{1}{2}\left[\mu_{k}^{2} + \mu_{l}^{2}\right]a_{klmn} - \dfrac{1}{2} \sum_{j=0} \mu_{j}^{2}\, a_{jkl}\, a_{jmn}\label{ElimB4Intermediate}
\end{align}
Meanwhile, the LHS can be simplified via  Eq. \eqref{bklmnRED}. Solving for the undetermined sum then gives us,
\begin{align}
    \boxed{\sum_{j=0} \mu_{j}^{2} \, a_{klj}\, a_{mnj} = -2\, B_{(kl)(mn)} + \dfrac{1}{3}\,\vec{\mu}^{\, 2}\, a_{klmn}} \label{A2B4}
\end{align}
where $\vec{\mu}^{\,2} \equiv \mu_{k}^2+\mu_{l}^2+\mu_{m}^{2}+\mu_{n}^{2}$.

The B-to-A formulas greatly reduce the number of relations we must consider. For example, when calculating a tree-level 2-to-2 KK mode scattering amplitude, we encounter quantities such as
\begin{align}
    \sum_{j} b_{k^{\prime}l^{\prime}j}\, a_{jmn} \hspace{35 pt} \sum_{j} b_{k^{\prime}l^{\prime}j}\, b_{m^{\prime}n^{\prime}j} \hspace{35 pt} \sum_{j} \mu_{j}^{2}\, b_{k^{\prime}l^{\prime}j}\, b_{m^{\prime}n^{\prime}j}
\end{align}
where the indices $\{k,l,m,n\}$ are associated with external KK modes and the index $j$ labels an intermediate KK mode that must be summed over in the course of summing over all diagrams. However, by converting all B-type couplings to A-type couplings and $B_{(kl)(mn)}$ objects, the quantities can be evaluated so long as we know instead how to evaluate
\begin{align}
    \sum_{j} a_{jkl}\, a_{jmn} \hspace{35 pt} \sum_{j} \mu_{j}^{2}\, a_{jkl}\, a_{jmn} \hspace{35 pt} \sum_{j} \mu_{j}^{4}\, a_{jkl} \,a_{jmn} \hspace{35 pt} \sum_{j} \mu_{j}^{6}\, a_{jkl} \,a_{jmn} 
\end{align}
Indeed, these are precisely the sums that are relevant to cancelling the high-energy growth of the KK mode scattering amplitudes, which is the goal of this dissertation. The remainder of this chapter is dedicated to rewriting these quantities in terms of the quartic A-type coupling, the $B_{(kl)(mn)}$ objects, and the integral $c_{k^{\prime}l^{\prime}m^{\prime}n^{\prime}}$ of Eq. \eqref{cklmnDEF}. The first two of these rewrites were achieved in Eqs. \eqref{acompleteness} and \eqref{A2B4} respectively. Therefore, we turn our focus to $\sum_{j} \mu_{j}^{4}\, a_{jkl}\, a_{jmn}$ and then $\sum_{j} \mu_{j}^{6}\, a_{jkl} \,a_{jmn}$.

\subsection{The $\mu_{j}^{4}$ Sum Rule}
The $\sum_{j} \mu_{j}^{4}\,a_{jkl}\, a_{jmn}$ relation is relatively straightforward. As defined in Eq. \eqref{cklmnDEF}, we can rewrite $c_{k^{\prime}l^{\prime}m^{\prime}n^{\prime}}$ in terms of B-type cubic couplings, to which we can then apply the B-to-A formulas:
\begin{align}
    c_{k^{\prime}l^{\prime}m^{\prime}n^{\prime}} &= \sum_{j=0} b_{k^{\prime}l^{\prime}j}b_{m^{\prime}n^{\prime}j}\\
    &\hspace{-35 pt}= \dfrac{1}{4}\sum_{j}\left[\mu_{k}^{2} + \mu_{l}^{2} -\mu_{j}^{2}\right]\left[\mu_{m}^{2} + \mu_{n}^{2} -\mu_{j}^{2} \right] a_{jkl}a_{jmn}\\
    &\hspace{-35 pt}= \dfrac{1}{4}(\mu_{k}^{2} + \mu_{l}^{2})(\mu_{m}^{2} + \mu_{n}^{2}) a_{klmn} - \dfrac{1}{4}(\vec{\mu}^{\,2}) \sum_{j} \mu_{j}^{2} a_{jkl} a_{jmn} + \dfrac{1}{4}\sum_{j} \mu_{j}^{4} a_{jkl} a_{jmn}
\end{align}
such that, using Eq. \eqref{A2B4} and solving for the undetermined sum $\sum_{j}\mu_{j}^{4}\,a_{jkl}\,a_{jmn}$,
\begin{align}
    \sum_{j} \mu_{j}^{4} a_{jkl} a_{jmn} = 4\, c_{k^{\prime}l^{\prime}m^{\prime}n^{\prime}} - 2\, (\vec{\mu}^{\,2})\, B_{(kl)(mn)} + \left[\dfrac{1}{3} (\vec{\mu}^{\,2})^{2} - (\mu_{k}^{2} + \mu_{l}^{2})(\mu_{m}^{2} + \mu_{n}^{2}) \right] a_{klmn} \label{muj4sumruleX}
\end{align}
as desired. Deriving the $\sum_{j} \mu_{j}^{6}\,a_{jkl}\, a_{jmn}$ relation requires significantly more work. 

\subsection{The $\mu_{j}^{6}$ Sum Rule}
Before beginning the derivation of the next sum rule, it is advantageous to define a well-organized polynomial basis with which we can write our results succinctly. In particular, because the sums we wish to simplify ($\sum_{j} \mu^{2p}\,a_{klj}\,a_{mnj}$) and relevant quartic degrees of freedom all have (at least) the symmetries of $B_{(kl)(mn)}$, it is useful to define a symmetrization operation that forms quantities with symmetries identical to $B_{(kl)(mn)}$:
\begin{align}
    \langle f_{klmn} \rangle &\equiv \bigg\{ \bigg[ f_{klmn} + (k\leftrightarrow l) \bigg] + (m\leftrightarrow n) \bigg\} + (kl\leftrightarrow mn)\\
    &\hspace{-35 pt} = f_{klmn} + f_{lkmn} + f_{klnm} + f_{lknm} + f_{mnkl} + f_{mnlk} + f_{nmkl} + f_{nmlk}
\end{align}
This allows us to quickly construct a finite basis for polynomials of $\mu^{2}\in \{\mu_{k}^{2},\mu_{l}^{2},\mu_{m}^{2},\mu_{n}^{2}\}$ having the aforementioned symmetry structures. For a single power of $\mu^{2}$, there is only one basis element:
\begin{align}
    \alpha_{(kl)(mn)}^{(1,1)} \equiv \langle \mu_{k}^{2} \rangle = 2 \vec{\mu}^{\,2} \equiv 2(\mu_{k}^{2} + \mu_{l}^{2} + \mu_{m}^{2} + \mu_{n}^{2})
\end{align}
For two powers of $\mu^{2}$, there is three:
\begin{align}
    \alpha_{(kl)(mn)}^{(2,1)} \equiv \langle \mu_{k}^{4} \rangle \hspace{35 pt} \alpha_{(kl)(mn)}^{(2,2)} \equiv \langle \mu_{l}^{2}\mu_{k}^{2} \rangle \hspace{35 pt} \alpha_{(kl)(mn)}^{(2,3)} \equiv \langle \mu_{m}^{2}\mu_{k}^{2} \rangle
\end{align}
and for three powers of $\mu^{3}$, there is four:
\begin{align}
    \alpha_{(kl)(mn)}^{(3,1)} \equiv \langle \mu_{k}^{6} \rangle \hspace{18 pt}&\hspace{18 pt} \alpha_{(kl)(mn)}^{(3,2)} \equiv \langle \mu_{l}^{2}\mu_{k}^{4} \rangle\nonumber\\
    \alpha_{(kl)(mn)}^{(3,3)} \equiv \langle \mu_{m}^{2}\mu_{k}^{4} \rangle \hspace{18 pt}&\hspace{18 pt} \alpha_{(kl)(mn)}^{(3,4)} \equiv \langle \mu_{m}^{2}\mu_{l}^{2}\mu_{k}^{2} \rangle
\end{align}
With these, we can generically construct any polynomial of the squared masses (up to cubic degree) having the aforementioned symmetry properties:
\begin{align}
    M_{(kl)(mn)}^{(1)}(c_{1}) &= 2 c_{1}\, \vec{\mu}^{\,2}\\
    M_{(kl)(mn)}^{(2)}(c_{1},c_{2},c_{3}) &= \sum_{i=1}^{3} c_{i}\, \alpha_{(kl)(mn)}^{(2,i)}\\
    M_{(kl)(mn)}^{(3)}(c_{1},c_{2},c_{3},c_{4}) &= \sum_{i=1}^{4} c_{i}\, \alpha_{(kl)(mn)}^{(3,i)} \label{M3Def}
\end{align}
Note that these symbols are intentionally linear in their $c_{i}$ arguments. In this language, Eq. \eqref{muj4sumruleX} may be rewritten as
\begin{align}
    \boxed{\sum_{j} \mu_{j}^{4} \, a_{klj}\, a_{mnj} = 4\, c_{k^{\prime}l^{\prime}m^{\prime}n^{\prime}} -2\, \vec{\mu}^{\, 2} \, B_{(kl)(mn)} + \dfrac{1}{6}\, M^{(2)}_{(kl)(mn)}(1,1,-1) \, a_{klmn}}
\end{align}
We now proceed to the $\sum_{j} \mu_{j}^{6} \, a_{klj} \, a_{mnj}$ rule.

As in the previous subsection, we begin our derivation by applying the B-to-A formulas to a sum of cubic B-type couplings:
\begin{align}
    \sum_{j} \mu_{j}^{2}\, b_{jk^{\prime}l^{\prime}}\,b_{jm^{\prime}n^{\prime}} &= \dfrac{1}{4}\sum_{j}\left[\mu_{k}^{2} + \mu_{l}^{2} -\mu_{j}^{2}\right]\left[\mu_{m}^{2} + \mu_{n}^{2} -\mu_{j}^{2} \right] \mu_{j}^{2} a_{jkl}a_{jmn}\\
    &\hspace{-70 pt}= \dfrac{1}{4}\sum_{j}\mu_{j}^{6} \, a_{klj} \, a_{jmn} - \dfrac{1}{4} \vec{\mu}^{\,2} \sum_{j} \mu_{j}^{4} \, a_{klj}\, a_{mnj} + \dfrac{1}{4} (\mu_{k}^{2} + \mu_{l}^{2})(\mu_{m}^{2} + \mu_{n}^{2})\sum_{j} \mu_{j}^{2} \, a_{klj}\, a_{mnj}\\
    &\hspace{-70 pt}= \dfrac{1}{4}\sum_{j}\mu_{j}^{6}\, a_{klj}\,a_{jmn} -\vec{\mu}^{\,2}\, c_{k^{\prime}l^{\prime}m^{\prime}n^{\prime}} - \dfrac{1}{12}\,\vec{\mu}^{\,2}\,(\mu_{k}^{2}+\mu_{l}^{2} -\mu_{m}^{2} -\mu_{n}^{2})^{2}\, a_{klmn}\nonumber\\
    &+ \dfrac{1}{2}\, \bigg[(\vec{\mu}^{\,2})^{2} - (\mu_{k}^{2}+\mu_{l}^{2})\,(\mu_{m}^{2}+\mu_{n}^{2})\bigg] B_{(kl)(mn)} \label{mj2bjklbjmnRHS}
\end{align}
On the RHS, only the desired sum remains undetermined. However, unlike the previous subsection, we do not yet have a simplification of the LHS of this expression. To obtain such a simplification, we would like to absorb the $\mu_{j}^{2}$ factor into $b_{jm^{\prime}n^{\prime}}$, and thus we next consider:
\begin{align}
    \mu_{j}^{2}\, b_{jm^{\prime}n^{\prime}} &= \dfrac{1}{\pi} \int d\varphi\hspace{5 pt} \vep^{-2} \left[\mu_{j}^{2} \vep^{-2} \psi_{j} \right] (\partial_{\varphi} \psi_{m}) (\partial_{\varphi} \psi_{n}) \\
    &= \dfrac{1}{\pi} \int d\varphi\hspace{5 pt} \vep^{-4} (\partial_{\varphi}\psi_{j})\, \partial_{\varphi}\left[\vep^{-2}(\partial_{\varphi} \psi_{m})(\partial_{\varphi}\psi_{n}) \right]\\
    &= \dfrac{1}{\pi} \int d\varphi\hspace{5 pt} \vep^{-4} (\partial_{\varphi}\psi_{j})\, \partial_{\varphi}\left[\vep^{+6}(\mathcal{D} \psi_{m})(\mathcal{D}\psi_{n}) \right]\\
    &= \dfrac{1}{\pi}\int d\varphi\hspace{5 pt}\vep^{+2}(\partial_{\varphi} \psi_{j})\left[+6(kr_{c})(\partial_{\varphi}|\varphi|) (\mathcal{D}\psi_{m})(\mathcal{D}\psi_{n})\right.\nonumber\\
    &\hspace{35 pt}\left.- \mu_{m}^{2}\vep^{-2} \psi_{m}(\mathcal{D}\psi_{n}) - \mu_{n}^{2}\vep^{-2} (\mathcal{D}\psi_{m}) \psi_{n} \right]\\
    &=\dfrac{6kr_{c}}{\pi} \int d\varphi\hspace{5 pt}(\partial_{\varphi}|\varphi|) \vep^{+6}(\mathcal{D}\psi_{j})(\mathcal{D}\psi_{m})(\mathcal{D}\psi_{n}) - \mu_{m}^{2}\, b_{j^{\prime}mn^{\prime}} - \mu_{n}^{2}\, b_{j^{\prime}m^{\prime}n}
\end{align}
where all integrals cover $\varphi\in[-\pi,+\pi)$, and from which
\begin{align}
    \mu_{j}^{2} b_{jm^{\prime}n^{\prime}} &= 6(kr_{c})x^{(-6)}_{j^{\prime}m^{\prime}n^{\prime}} - \mu_{m}^{2} b_{j^{\prime}m n^{\prime}} - \mu_{n}^{2} b_{j^{\prime}m^{\prime} n} \label{xxxEq4165}
\end{align}
After applying the cubic B-to-A formulas, we can solve for the new integral $x^{(-6)}_{j^{\prime}m^{\prime}n^{\prime}}$:
\begin{align}
    6(kr_{c})x^{(-6)}_{j^{\prime}m^{\prime}n^{\prime}} &= -\dfrac{1}{2}\bigg[ \mu_{j}^{4} - 2\,(\mu_{m}^{2}+\mu_{n}^{2})\,\mu_{j}^{2} + (\mu_{m}^{2}-\mu_{n}^{2})^{2}\bigg] a_{jmn} \label{2114}\\
     &\hspace{-50 pt}= -\dfrac{1}{2}\, (\mu_{j}-\mu_{m}-\mu_{n})\,(\mu_{j}+\mu_{m}-\mu_{n})\,(\mu_{j}-\mu_{m}+\mu_{n})\,(\mu_{j}+\mu_{m}+\mu_{n})\,a_{jmn}
\end{align}
where the generalized coupling $x$ is defined in Eq. \eqref{generalizedx}. Eq. \eqref{xxxEq4165} still does not allow us to evaluate the LHS of Eq. \eqref{mj2bjklbjmnRHS} because it was derived only using the B-to-A relations and, thus, if applied to the LHS will merely reproduce the RHS of Eq. \eqref{mj2bjklbjmnRHS}. We must find another route. Ideally, we will find a way of using completeness to perform the sum over the index $j$ on the LHS of Eq. \eqref{mj2bjklbjmnRHS}, which cannot be accomplished so long as all wavefunctions are differentiated as in $x^{(-6)}_{j^{\prime}m^{\prime}n^{\prime}}$. Therefore, we can continue making progress by using integration by parts to remove the derivative from $\psi_{j}$ in the integral $x^{(-6)}_{j^{\prime}m^{\prime}n^{\prime}}$:
\begin{align}
    x^{(-6)}_{j^{\prime}m^{\prime}n^{\prime}} &= \dfrac{1}{\pi}\int d\varphi\hspace{5 pt}\vep^{+2}(\partial_{\varphi}|\varphi|) (\partial_{\varphi}\psi_{j})(\mathcal{D}\psi_{m})(\mathcal{D}\psi_{n})\\
    &= -\dfrac{1}{\pi} \int d\varphi\hspace{5 pt} \psi_{j} \partial_{\varphi}\bigg[\vep^{+2}(\partial_{\varphi}|\varphi|) (\mathcal{D}\psi_{m})(\mathcal{D}\psi_{n})\bigg]
\end{align}
The distribution of $\partial_{\varphi}$ on the quantity in square brackets will yield, among other terms,
\begin{align}
    \dfrac{1}{\pi} \int d\varphi\hspace{5 pt} \vep^{+2}\, (\partial_{\varphi}^{2}|\varphi|)\, \psi_{j} (\mathcal{D}\psi_{m})(\mathcal{D}\psi_{n})
\end{align}
which vanishes because $(\partial_{\varphi}^{2}|\varphi|) = 2(\delta_{0} - \delta_{\pi r_{c}})$ and $(\partial_{\varphi}\psi_{n}) = 0$ at the branes. Keeping this in mind, the remaining terms are
\begin{align}
    x^{(-6)}_{j^{\prime}m^{\prime}n^{\prime}} &= -2\,(kr_{c})\,x^{(-6)}_{jm^{\prime}n^{\prime}} + \mu_{m}^{2}\, x^{(-4)}_{jmn^{\prime}} + \mu_{n}^{2}\, x^{(-4)}_{jm^{\prime}n}
\end{align}
Combining this with Eq. \eqref{xxxEq4165}, we derive
\begin{align}
    \mu_{j}^{2} b_{jm^{\prime}n^{\prime}} = -12\,(kr_{c})^{2}\,x^{(-6)}_{jm^{\prime}n^{\prime}} + 6\,(kr_{c})\bigg[\mu_{m}^{2}\, x^{(-4)}_{jmn^{\prime}} + \mu_{n}^{2}\, x^{(-4)}_{jm^{\prime}n}\bigg] - \mu_{m}^{2}\, b_{j^{\prime}mn^{\prime}} - \mu_{n}^{2}\, b_{j^{\prime}m^{\prime}n} \label{mul2blmn}
\end{align}
All terms on the RHS of this equation either lack derivatives on $\psi_{j}$ or have fewer than four powers of $\mu_{j}$ after applying the B-to-A formulas and hence all sums can be handled via existing sum rules. Therefore, we proceed:
\begin{align}
    \sum_{j} \mu_{j}^{2}\, b_{k^{\prime}l^{\prime}j}b_{jm^{\prime}n^{\prime}} &= \sum_{j} b_{k^{\prime}l^{\prime}j}\bigg\{-12(kr_{c})^{2}x^{(-6)}_{jm^{\prime}n^{\prime}} + 6(kr_{c})\bigg[\mu_{m}^{2} x^{(-4)}_{jmn^{\prime}} + \mu_{n}^{2} x^{(-4)}_{jm^{\prime}n}\bigg]\nonumber\\
    &\hspace{35 pt}- \mu_{m}^{2} b_{j^{\prime}mn^{\prime}} - \mu_{n}^{2} b_{j^{\prime}m^{\prime}n}\bigg\}\\
    &= -12(kr_{c})^{2} d_{klmn} +6(kr_{c})\bigg[\mu_{m}^{2}x^{(-6)}_{k^{\prime}l^{\prime}mn^{\prime}} + \mu_{n}^{2}x^{(-6)}_{k^{\prime}l^{\prime}m^{\prime}n}\bigg]\nonumber\\
    &\hspace{35 pt}- \bigg[\mu_{m}^{2} \sum_{j} b_{k^{\prime}l^{\prime}j}b_{j^{\prime}mn^{\prime}} + \mu_{n}^{2} \sum_{j} b_{k^{\prime}l^{\prime}j}b_{j^{\prime}m^{\prime}n}\bigg] \label{2121}
\end{align}
where
\begin{align}
    d_{klmn} &\equiv x^{(-8)}_{k^{\prime}l^{\prime}m^{\prime}n^{\prime}} = \dfrac{1}{\pi} \int d\varphi\hspace{5 pt}\vep^{+8}\,(\mathcal{D}\psi_{k})  (\mathcal{D}\psi_{l})  (\mathcal{D}\psi_{m}) (\mathcal{D}\psi_{n}) 
\end{align}
The latter sums each satisfy, via the A-to-B formulas and existing sum rules,
\begin{align}
    \sum_{j} b_{k^{\prime}l^{\prime}j}\,b_{j^{\prime}m^{\prime}n} = -c_{k^{\prime}l^{\prime}m^{\prime}n^{\prime}} + \mu_{m}^{2}\, B_{(kl)(mn)} + \dfrac{1}{6}\,\mu_{m}^{2}\,(2\mu_{k}^{2} + 2\mu_{l}^{2} -\mu_{m}^{2} - \mu_{n}^{2})\,a_{klmn}
\end{align}
so that together we find,
\begin{align}
    \mu_{m}^{2} \sum_{j} b_{k^{\prime}l^{\prime}j}\,b_{j^{\prime}n^{\prime}m}+\mu_{n}^{2} \sum_{j} b_{k^{\prime}l^{\prime}j}\,b_{j^{\prime}m^{\prime}n} &= -(\mu_{m}^{2}+\mu_{n}^{2})\,c_{k^{\prime}l^{\prime}m^{\prime}n^{\prime}} + 2\, \mu_{m}^{2}\,\mu_{n}^{2}\, B_{(kl)(mn)}\nonumber\\
    &\hspace{15 pt}+ \dfrac{1}{3}\,\mu_{m}^{2}\,\mu_{n}^{2}\,(2\mu_{k}^{2} + 2\mu_{l}^{2} -\mu_{m}^{2} - \mu_{n}^{2})\,a_{klmn}
\end{align}
Next, we must determine how to rewrite $d_{klmn}$, $x^{(-6)}_{k^{\prime}l^{\prime}m^{\prime}n}$, and $x^{(-6)}_{k^{\prime}l^{\prime}mn^{\prime}}$ from Eq. \eqref{2121} in terms of $a_{klmn}$, the $B_{(kl)(mn)}$, and $c_{k^{\prime}l^{\prime}m^{\prime}n^{\prime}}$. The necessary equation for the latter two quantities can be derived by absorbing a factor of $\mu^{2}$ into the quartic B-type coupling:
\begin{align}
    \mu_{n}^{2}\, b_{k^{\prime}l^{\prime}mn} &= \dfrac{1}{\pi} \int d\varphi\hspace{5 pt} \vep^{-2} (\partial_{\varphi} \psi_{k}) (\partial_{\varphi} \psi_{l}) \psi_{m} \left[\mu_{n}^{2}\, \vep^{-2}\, \psi_{n} \right]\\
    &\hspace{-25 pt}= \dfrac{1}{\pi} \int d\varphi\hspace{5 pt} \vep^{-4} (\partial_{\varphi}\psi_{n})\, \partial_{\varphi}\left[\vep^{-2}\,(\partial_{\varphi} \psi_{k})(\partial_{\varphi}\psi_{l})\psi_{m} \right]\\
    &\hspace{-25 pt}= \dfrac{1}{\pi} \int d\varphi\hspace{5 pt} \vep^{-4} (\partial_{\varphi}\psi_{n})\, \partial_{\varphi}\left[\vep^{+6}\,(\mathcal{D} \psi_{k})(\mathcal{D}\psi_{l}) \psi_{m} \right]\\
    &\hspace{-25 pt}= \dfrac{1}{\pi}\int d\varphi\hspace{5 pt}\vep^{+2}\,(\partial_{\varphi} \psi_{n})\,\Big[+6\,(kr_{c})\,(\partial_{\varphi}|\varphi|)\, (\mathcal{D}\psi_{k})(\mathcal{D}\psi_{l})\psi_{m}\nonumber\\
    &\hspace{-10 pt}- \mu_{k}^{2}\,\vep^{-2}\, \psi_{k}(\mathcal{D}\psi_{l})\psi_{m} - \mu_{l}^{2}\,\vep^{-2}\, (\mathcal{D}\psi_{k}) \psi_{l} \psi_{m} + \vep^{+4}\,(\mathcal{D}\psi_{k})(\mathcal{D}\psi_{l})(\mathcal{D}\psi_{m}) \Big]\\
    &\hspace{-25 pt}=\dfrac{6}{\pi}\,(kr_{c}) \int d\varphi\hspace{5 pt}(\partial_{\varphi}|\varphi|)\, \vep^{+6}\,(\mathcal{D}\psi_{k})(\mathcal{D}\psi_{l})\psi_{m}(\mathcal{D}\psi_{n}) - \mu_{k}^{2}\, b_{kl^{\prime}mn^{\prime}} - \mu_{l}^{2}\, b_{k^{\prime}lmn^{\prime}}\nonumber\\
    &\hspace{-10 pt} + \dfrac{1}{\pi} \int d\varphi\hspace{5 pt} \vep^{+10}\,(\mathcal{D}\psi_{k})(\mathcal{D}\psi_{l})(\mathcal{D}\psi_{m})(\mathcal{D}\psi_{n})
\end{align}
which implies
\begin{align}
     6\,(kr_{c})\, x^{(-6)}_{k^{\prime}l^{\prime}mn^{\prime}} = \mu_{k}^{2}\, b_{kl^{\prime} m n^{\prime}} + \mu_{l}^{2}\, b_{k^{\prime} l m n^{\prime}} + \mu_{n}^{2}\, b_{k^{\prime} l^{\prime} m n} - c_{k^{\prime}l^{\prime}m^{\prime}n^{\prime}} \label{mu2bklmn}
\end{align}
Note the special role of the label $m$ on both sides of this equation.  By applying the B-to-A formulas, this yields a formula for $x^{(-6)}_{k^{\prime}l^{\prime}mn^{\prime}}$ and its various permutations in terms of $a_{klmn}$, the $B_{(kl)(mn)}$, and $c_{k^{\prime}l^{\prime}m^{\prime}n^{\prime}}$ as desired.

Now to do the same for $d_{klmn}$. Consider absorbing $\mu_{n}^{2}$ into $x^{(-6)}_{k^{\prime}l^{\prime}m^{\prime}n}$:
\begin{align}
    \mu_{n}^{2} x^{(-6)}_{k^{\prime} l^{\prime} m^{\prime} n} &= \dfrac{1}{\pi} \int d\varphi\hspace{5 pt} (\partial_{\varphi}|\varphi|)\, \vep^{+8}\, (\mathcal{D}\psi_{k}) (\mathcal{D}\psi_{l}) (\mathcal{D}\psi_{m})\bigg[\mu_{n}^{2}\,\vep^{-2}\,\psi_{n}\bigg]\\
    &= \dfrac{1}{\pi} \int d\varphi\hspace{5 pt} (\mathcal{D}\psi_{n})\, \partial_{\varphi}\bigg[(\partial_{\varphi}|\varphi|)\, \vep^{+8} \, (\mathcal{D}\psi_{k})  (\mathcal{D}\psi_{l})  (\mathcal{D}\psi_{m})\bigg] \label{2131}
\end{align}
Because $(\partial_{\varphi}^{2}|\varphi|) = 2(\delta_{0} - \delta_{\pi r_{c}})$ and $(\partial_{\varphi}\psi_{n}) = 0$ at the branes,
\begin{align}
    \dfrac{1}{\pi}\int d\varphi\hspace{5 pt}(\partial^{2}_{\varphi}|\varphi|) \vep^{+8} (\mathcal{D}\psi_{k}) (\mathcal{D}\psi_{l}) (\mathcal{D}\psi_{m}) (\mathcal{D}\psi_{n}) = 0
\end{align}
such that Eq. \eqref{2131} implies, after multiplying both sides by $6\,kr_{c}$,
\begin{align}
    48\,(kr_{c})^{2} \,d_{klmn} = 6\,(kr_{c})\bigg[\mu_{k}^{2} x^{(-6)}_{kl^{\prime}m^{\prime}n^{\prime}} + \mu_{l}^{2} x^{(-6)}_{k^{\prime}lm^{\prime}n^{\prime}} + \mu_{m}^{2} x^{(-6)}_{k^{\prime}l^{\prime}mn^{\prime}} + \mu_{n}^{2} x^{(-6)}_{k^{\prime}l^{\prime}m^{\prime}n}\bigg]
\end{align}
We multiplied by $6\,kr_{c}$ to enable the use of Eq. \eqref{mu2bklmn} on every term of the RHS, which now yields
\begin{align}
     48\,(kr_{c})^{2} \,d_{klmn} &= -\vec{\mu}^{\,2}\, c_{k^{\prime}l^{\prime}m^{\prime}n^{\prime}} + 2\, \mu_{k}^{2}\mu_{l}^{2}\,b_{klm^{\prime}n^{\prime}} + 2\, \mu_{k}^{2}\mu_{m}^{2}\,b_{kl^{\prime}mn^{\prime}} + 2\, \mu_{k}^{2}\mu_{n}^{2}\,b_{kl^{\prime}m^{\prime}n}  \nonumber\\
     &\hspace{35 pt}+ 2\, \mu_{l}^{2}\mu_{m}^{2}\,b_{k^{\prime}lmn^{\prime}} + 2\, \mu_{l}^{2}\mu_{n}^{2}\,b_{k^{\prime}lm^{\prime}n} + 2\, \mu_{m}^{2}\mu_{n}^{2}\,b_{k^{\prime}l^{\prime}mn}\label{xxx4185}
\end{align}
with which we can use the B-to-A formulas to obtain a formula for $d_{klmn}$ in terms of $a_{klmn}$, the $B_{(kl)(mn)}$ objects, and $c_{k^{\prime}l^{\prime}m^{\prime}n^{\prime}}$ as desired. Now every term on the RHS of our alternate expression for $\sum_{j} \mu_{j}^{2}\,b_{k^{\prime}l^{\prime}j}b_{jm^{\prime}n^{\prime}}$ from Eq. \eqref{2121} can be expressed in terms of $c_{k^{\prime}l^{\prime}m^{\prime}n^{\prime}}$, the $B_{(kl)(mn)}$, or $a_{klmn}$. In this way, we obtain
\begin{align}
    \sum_{j} \mu_{j}^{2}\, b_{k^{\prime}l^{\prime}j}\,b_{jm^{\prime}n^{\prime}} &= \dfrac{1}{4}\,\vec{\mu}^{\,2}\, c_{k^{\prime}l^{\prime}m^{\prime}n^{\prime}} - \dfrac{1}{2}\,(\mu_{k}^{2}\mu_{l}^{2} + \mu_{m}^{2}\mu_{n}^{2}) \, B_{(kl)(mn)}\nonumber\\
    &\hspace{15 pt} + \dfrac{1}{2}\,(\mu_{k}^{2}\mu_{m}^{2} + \mu_{l}^{2}\mu_{n}^{2}) \, B_{(km)(ln)} + \dfrac{1}{2}\,(\mu_{k}^{2}\mu_{n}^{2} + \mu_{l}^{2}\mu_{m}^{2}) \, B_{(kn)(lm)}\nonumber\\
    &\hspace{15 pt}+\dfrac{1}{24}\, M_{(kl)(mn)}^{(3)}(0,1,-2,2) \, a_{klmn} \label{RHS2formuj2bb}
\end{align}
where $M_{(kl)(mn)}^{(3)}$ was defined in Eq. \eqref{M3Def}. Note this form has manifest $(k,l)\leftrightarrow(m,n)$ pair swap symmetry, and is not unique because $B_{(km)(ln)} + B_{(kn)(lm)} +  B_{(kl)(mn)} = 0$.

Finally, by combining Eqs. \eqref{mj2bjklbjmnRHS} and \eqref{RHS2formuj2bb}, we arrive at last at the $\sum_{j} \mu_{j}^{6}\,a_{klj}\,a_{jmn}$ sum rule:
\begin{align}
    \sum_{j} \mu_{j}^{6}\,a_{klj}\,a_{jmn} &= 5\,\vec{\mu}^{\,2}\, c_{k^{\prime}l^{\prime}m^{\prime}n^{\prime}} + \dfrac{1}{2}\, \bigg[ M^{(2)}_{(km)(ln)}(0,1,0)\,B_{(km)(ln)}  + (m\leftrightarrow n) \bigg] \nonumber\\
    &- \dfrac{1}{2}\, M^{(2)}_{(kl)(mn)}(2,3,2)\, B_{(kl)(mn)}+\dfrac{1}{6}\,M^{(3)}_{(kl)(mn)}(1,4,-4,0)\, a_{klmn}
\end{align}
We can utilize the constraint $B_{(km)(ln)} + B_{(kn)(lm)} +  B_{(kl)(mn)} = 0$ to rewrite this into a form with fewer fractional coefficients:
\begin{align}
    \sum_{j} \mu_{j}^{6}\,a_{klj}\,a_{jmn} &= 5\,\vec{\mu}^{\,2} c_{k^{\prime}l^{\prime}m^{\prime}n^{\prime}} +  M^{(2)}_{(km)(ln)}(1,1,1)\,B_{(km)(ln)}  +  M^{(2)}_{(kn)(lm)}(1,1,1)\,B_{(kn)(lm)} \nonumber\\
    &-  M^{(2)}_{(kl)(mn)}(0,1,0)\, B_{(kl)(mn)} +\dfrac{1}{6}\,M^{(3)}_{(kl)(mn)}(1,4,-4,0)\, a_{klmn}
\end{align}
In the next few subsections, we summarize the principal results of this section in increasingly specific cases: inelastic, permutation-symmetric inelastic, and finally elastic.

\subsection{Summary of Sum Rules (Inelastic)}
All B-type couplings $\{b_{l^{\prime}m^{\prime}n},b_{k^{\prime}l^{\prime}mn}\}$ can be eliminated in favor of A-type couplings $\{a_{lmn},a_{klmn}\}$ and new $B_{(kl)(mn)}$ objects via the B-to-A formulas
\begin{align}
    b_{l^{\prime} m^{\prime} n} &= \dfrac{1}{2}\left[\mu_{l}^{2} + \mu_{m}^{2} - \mu_{n}^{2} \right] a_{lmn} \label{eq4152}\\ b_{k^{\prime}l^{\prime}mn} &= B_{(kl)(mn)}+\dfrac{1}{6}\bigg[2(\mu_{k}^{2}+\mu_{l}^{2})-(\mu_{m}^{2}+\mu_{n}^{2})\bigg] a_{klmn}\nonumber
\end{align}
where the $B_{(kl)(mn)}$ are constrained such that $B_{(km)(ln)} + B_{(kn)(lm)} +  B_{(kl)(mn)} = 0$, and are symmetric in each individual pair $(k,l)$ and $(m,n)$ as well as with respect to the pair swap replacement $(k,l)\leftrightarrow (m,n)$. Applying the B-to-A formulas reduces the number of sums relevant to the cancellations we examine in the next chapter. These sums are
\begin{align}
    \sum_{j=0} a_{klj}\, a_{mnj} &= a_{klmn}\\
    \sum_{j=0} \mu_{j}^{2} \, a_{klj}\, a_{mnj} &= -2\, B_{(kl)(mn)} + \dfrac{1}{3}\,\vec{\mu}^{\, 2}\, a_{klmn}\\
    \sum_{j=0} \mu_{j}^{4} \, a_{klj}\, a_{mnj} &= 4\, c_{k^{\prime}l^{\prime}m^{\prime}n^{\prime}} -2\, \vec{\mu}^{\, 2} \, B_{(kl)(mn)} + \dfrac{1}{6}\, M^{(2)}_{(kl)(mn)}(1,1,-1) \, a_{klmn}\\
    \sum_{j=0} \mu_{j}^{6} \, a_{klj}\, a_{mnj}  &= 5\,\vec{\mu}^{\,2} c_{k^{\prime}l^{\prime}m^{\prime}n^{\prime}} +  M^{(2)}_{(km)(ln)}(1,1,1)\,B_{(km)(ln)}  +  M^{(2)}_{(kn)(lm)}(1,1,1)\,B_{(kn)(lm)} \nonumber\\
    &-  M^{(2)}_{(kl)(mn)}(0,1,0)\, B_{(kl)(mn)} +\dfrac{1}{6}\,M^{(3)}_{(kl)(mn)}(1,4,-4,0)\, a_{klmn}
\end{align}
where $\vec{\mu}^{\,2}\equiv \mu_{k}^{2}+\mu_{l}^{2}+\mu_{m}^{2}+\mu_{n}^{2}$, and
\begin{align}
    c_{k^{\prime}l^{\prime}m^{\prime}n^{\prime}} \equiv \dfrac{1}{\pi} \int d\varphi\hspace{5 pt}\vep^{-6}(\partial_{\varphi}\psi_{k})(\partial_{\varphi}\psi_{l})(\partial_{\varphi}\psi_{m})(\partial_{\varphi}\psi_{n})
\end{align}
The last two sum rules can be combined as to cancel all factors of $c_{k^{\prime}l^{\prime}m^{\prime}n^{\prime}}$, and thereby yields
\begin{align}
    \sum_{j=0} \mu_{j}^{4} \, \bigg(\mu_{j}^{2} - \dfrac{5}{4}\, \vec{\mu}^{\,2} \bigg) \, a_{klj}\, a_{mnj} &= \dfrac{1}{4}\, M^{(2)}_{(kl)(mn)}(5,1,10) \, B_{(kl)(mn)} \nonumber\\
    &\hspace{-105 pt} + M^{(2)}_{(km)(ln)}(1,1,1)\, B_{(km)(ln)} + M^{(2)}_{(kn)(lm)}(1,1,1) \, B_{(kn)(lm)} \nonumber\\
    &\hspace{-70 pt} - \frac{1}{24}\, M^{(3)}_{(kl)(mn)}(1,-1,16,0) \, a_{klmn}
\end{align}
These equations extend and generalize the sum rules derived in \cite{Chivukula:2020hvi}.

One might wonder whether we can express $B_{(kl)(mn)}$ or $c_{k^{\prime}l^{\prime}m^{\prime}n^{\prime}}$ in terms of the eigenvalues $\{\mu^{2}_{k},\mu^{2}_{l},\mu^{2}_{m},\mu^{2}_{n}\}$ and $a_{klmn}$, i.e. if we may further reduce how many degrees of freedom we consider. We believe this to be unlikely. For example, because $B_{(kl)(mn)}$ is simply related to $b_{k^{\prime}l^{\prime}mn}$ which possesses two derivatives, we expect such a relation for $B_{(kl)(mn)}$ would involve some $\mathcal{O}(\mu^{2})$ polynomial times $a_{klmn}$. However, only one such $\mathcal{O}(\mu^{2})$ polynomial possesses the symmetries of $B_{(kl)(mn)}$: $\vec{\mu}^{\,2}$, such that we'd have $B_{(kl)(mn)} \propto \vec{\mu}^{\,2}\, a_{klmn}$. This form violates the constraint $B_{(kl)(mn)} + B_{(km)(ln)} + B_{(kn)(lm)} = 0$, and is thus forbidden. Similarly, if $c_{k^{\prime}l^{\prime}m^{\prime}n^{\prime}}$ could be reduced in terms of $a_{klmn}$, then by derivative counting we would expect $c_{k^{\prime}l^{\prime}m^{\prime}n^{\prime}} \propto \mathcal{O}(\mu^{4})\, a_{klmn}$ for some $\mathcal{O}(\mu^{4})$ polynomial fully symmetric in $(k,l,m,n)$ (this symmetry also makes including the $B_{(kl)(mn)}$ objects pointless). In the elastic limit ($k=l=m=n$), this would necessarily imply that the ratio $c_{n^{\prime}n^{\prime}n^{\prime}n^{\prime}} / (\mu_{n}^{4} a_{nnnn})$ is independent of KK number $n$, but numerical checks confirm this is not the case. Therefore, if there exist relations for rewriting $B_{(kl)(mn)}$ and $c_{k^{\prime}l^{\prime}m^{\prime}n^{\prime}}$ in terms of eigenvalues and $a_{klmn}$, then they are more complicated than the sorts of relations we presently consider.

\subsection{Summary of Sum Rules (Permutation-Symmetric Inelastic)}
Define the fully symmetric quantity
\begin{align}
    (aa)_{klmn;j} \equiv a_{klj} a_{mnj} + a_{kmj} a_{lnj} + a_{knj} a_{lmj} 
\end{align}
Using the constraint that the $B$'s sum to zero, we immediately derive
\begin{align}
    \sum_{j=0} (aa)_{klmn;j} &= 3\, a_{klmn}\\
    \sum_{j=0} \mu_{j}^{2} \, (aa)_{klmn;j} &= \vec{\mu}^{\, 2}\, a_{klmn}\\
    \sum_{j=0} \mu_{j}^{4} \, (aa)_{klmn;j} &= 12\, c_{k^{\prime}l^{\prime}m^{\prime}n^{\prime}} + \vec{\mu}^{\,4} \, a_{klmn}\\
    \sum_{j=0} \mu_{j}^{6} \, (aa)_{klmn;j} &= 15\, \vec{\mu}^{\,2}\, c_{k^{\prime}l^{\prime}m^{\prime}n^{\prime}} + \vec{\mu}^{\,6} \, a_{klmn}
\end{align}
from the inelastic sum rules of the previous subsection, where $\vec{\mu}^{\,p} \equiv \mu_{k}^{p} + \mu_{l}^{p} + \mu_{m}^{p} + \mu_{n}^{p}$, and
\begin{align}
    \sum_{j=0} \mu_{j}^{4} \, \bigg(\mu_{j}^{2} - \dfrac{5}{4}\, \vec{\mu}^{\,2} \bigg) \, (aa)_{klmn;j} &= \bigg[ \vec{\mu}^{\, 6} - \dfrac{5}{4}\, \vec{\mu}^{\,2}\, \vec{\mu}^{\,4} \bigg]\, a_{klmn}
\end{align}
These expressions are particularly useful because the matrix element $\mathcal{M}$ describing the process $(k,l)\rightarrow (m,n)$ necessarily possesses crossing symmetry.
 
\subsection{Summary of Sum Rules (Elastic)}\label{SummaryOfSumRulesElastic}For the purposes of this dissertation, we are particularly interested in the elastic massive KK mode scattering process, wherein $k=l=m=n\,(\neq0)$ and relations of the previous subsections simplify. Consider, for example, the $B$'s constraint in this context:
\begin{align}
    B_{(km)(ln)} + B_{(kn)(lm)} +  B_{(kl)(mn)} = 0 \hspace{15 pt}\stackrel{\text{elastic}}{\longrightarrow}\hspace{15 pt} B_{(nn)(nn)} = 0
\end{align}
such that all the $B$'s become identical and vanish. The relevant B-to-A formulas become
\begin{align}
    b_{n^{\prime}n^{\prime}j} &= \dfrac{1}{2} \left[ \mu_{n}^{2} - \mu_{j}^{2} \right] a_{nnj}\hspace{35 pt}b_{j^{\prime}n^{\prime}n} = \dfrac{1}{2}\, \mu_{j}^{2}\, a_{nnj}\hspace{35 pt}b_{n^{\prime}n^{\prime}nn} = \dfrac{1}{3}\, \mu_{n}^{2}\, a_{nnnn}
\end{align}
whereas the sum rules reduce to
\begin{align}
    \sum_{j} a_{jnn}^{2} &= a_{nnnn} \label{SumRuleO5}\\
    \sum_{j} \mu_{j}^{2}\, a_{jnn}^{2} &= \dfrac{4}{3}\, \mu_{n}^{2}\, a_{nnnn} \label{SumRuleO4}\\
    \sum_{j} \mu_{j}^{4}\, a_{jnn}^{2} &= 4\, c_{n^{\prime}n^{\prime}n^{\prime}n^{\prime}} + \dfrac{4}{3}\, \mu_{n}^{4}\, a_{nnnn} \label{SumRuleO3x}\\
    \sum_{j} \mu_{j}^{6} a_{jnn}^{2} &= 20\, \mu_{n}^{2} \,c_{n^{\prime}n^{\prime}n^{\prime}n^{\prime}} + \dfrac{4}{3} \,\mu_{n}^{6} \,a_{nnnn} \label{SumRuleO2x}
\end{align}
with the last two expressions combining to yield
\begin{align}
    \sum_{j} \bigg[\mu_{j}^{2} -5 \mu_{n}^{2}\bigg] \mu_{j}^{4}\,a_{jnn}^{2} &= -\dfrac{16}{3}\, \mu_{n}^{6}\, a_{nnnn} \label{SumRule02xO3x}
\end{align}
We now have all the elements necessary to begin calculating and analyzing amplitudes, which is the focus on the next chapter.
\chapter{Massive Spin-2 KK Mode Scattering in the RS1 Model}

\section{Chapter Summary}
We will now apply the original material from chapters 3 and 4 to achieve the main theoretical results of this dissertation. In the last chapter, we used weak field expansion (WFE) and Kaluza-Klein (KK) decomposition to rewrite the 5D fields of the 5D RS1 model in terms of the following 4D field content: a massless spin-2 graviton $\hat{h}^{(0)}_{\mu\nu}$, a tower of massive spin-2 states $\hat{h}^{(n)}_{\mu\nu}$ with KK numbers $n\in \{1,2,\dots\}$, and a massless spin-0 radion $\hat{r}^{(0)}$. We also derived the interactions between these 4D states by integrating the 5D WFE RS1 Lagrangian (which we derived in Chapter \ref{C - 5D RS1} and summarized in Eqs. \eqref{LRShHrR}-\eqref{LBrrrr}) over the extra dimension, thereby obtaining the 4D effective RS1 Lagrangian $\mathcal{L}_{\text{4D}}^{(\text{eff})}$ up to quartic order in the fields. The 5D and 4D effective theories were found to be related via the 5D-to-4D formula, Eq. \eqref{LRSeffPerTerm}:
\begin{align}
    \mathcal{L}^{(\text{RS,eff})}_{h^H r^R} = \left[\dfrac{\kappa}{\sqrt{\pi r_c}}\right]^{(H+R-2)}\sum_{\vec{n}=\vec{0}}^{+\infty}\bigg\{ {a}_{(R|\vec{n})}\cdot \mathcal{X}_{(\vec{n})}\hspace{-3 pt}\left[\hspace{3 pt}\overline{\mathcal{L}}_{A:h^H r^R}\right] + {b}_{(R|\vec{n})}\cdot \mathcal{X}_{(\vec{n})}\hspace{-3 pt}\left[\hspace{3 pt}\overline{\mathcal{L}}_{B:h^H r^R}\right]\bigg\}~.
\end{align}
where $a_{(R|\vec{n})}$ and $b_{(R|\vec{n})}$ are integrals of products of KK wavefunctions which depend on the number of radions $R$ and the KK numbers $\vec{n} = (n_{1},\cdots,n_{H})$ of the $H$ spin-2 modes in each given term. Specifically, these integrals were defined in Eqs. \eqref{abar} and \eqref{bbar} (and later generalized in Eq. \eqref{generalizedx}):
\begin{align}
    {a}_{({R}|\vec{n})} &\equiv {a}_{r\cdots r n_1\cdots n_H} \equiv \dfrac{1}{\pi} e^{-R\pi kr_c}\int_{-\pi}^{+\pi}d\varphi\hspace{5 pt} \varepsilon^{2(R-1)}\psi_{n_1}\cdots \psi_{n_H} \left[\psi_{0}\right]^R~,\\
    {b}_{(R|\vec{n})} &\equiv {b}_{r\cdots r n_{1}^{\prime}n_{2}^{\prime}n_{3}\cdots n_{H}}~,\nonumber\\
    &\equiv \dfrac{1}{\pi} e^{-R\pi kr_c}\int_{-\pi}^{+\pi}d\varphi\hspace{5 pt} \varepsilon^{2(R-2)} (\partial_{\varphi} \psi_{n_1})(\partial_{\varphi} \psi_{n_2}) \psi_{n_3}\cdots \psi_{n_H}\left[\psi_{0}\right]^R~,
\end{align}
where $\vep \equiv e^{+kr_{c}|\varphi|}$, which define the A-type and B-type couplings respectively. Using the fact that the wavefunctions $\psi_{n}$ satisfy a Sturm-Liouville problem, Eq. \eqref{SLeqx},
\begin{align}
    \partial_{\varphi}\left[\varepsilon^{-4} (\partial_{\varphi} \psi_{n})\right] = - \mu_{n}^{2} \varepsilon^{-2} \psi_{n}
\end{align}
with $(\partial_{\varphi}\psi_{n}) = 0$ at the branes ($\varphi\in\{0,\pi\}$), various relations between the couplings and mass spectrum $\{\mu_{n}\} =\{m_{n}r_{c}\}$ were derived (Eqs. \eqref{eq4152}-\eqref{SumRule02xO3x}). This included formulas for rewriting certain B-type couplings in terms of A-type couplings, Eq. \eqref{eq4152},
\begin{align}
    b_{l^{\prime} m^{\prime} n} = \dfrac{1}{2}\left[\mu_{l}^{2} + \mu_{m}^{2} - \mu_{n}^{2} \right] a_{lmn}\hspace{35 pt} b_{n^{\prime}n^{\prime}nn} = \dfrac{1}{3}\, a_{nnnn}
\end{align}
and certain elastic sum rules
\begin{align}
    \sum_{j} a_{jnn}^{2} &= a_{nnnn} \\
    \sum_{j} \mu_{j}^{2}\, a_{jnn}^{2} &= \dfrac{4}{3}\, \mu_{n}^{2} \,a_{nnnn} \\
    \sum_{j} \mu_{j}^{4}\, a_{jnn}^{2} &= 4 \,c_{n^{\prime}n^{\prime}n^{\prime}n^{\prime}} + \dfrac{4}{3} \,\mu_{n}^{4} \,a_{nnnn} \\
    \sum_{j} \mu_{j}^{6}\, a_{jnn}^{2} &= 20\, \mu_{n}^{2} \,c_{n^{\prime}n^{\prime}n^{\prime}n^{\prime}} + \dfrac{4}{3} \,\mu_{n}^{6} \,a_{nnnn}
\end{align}
where $c_{k^{\prime}l^{\prime}m^{\prime}n^{\prime}} \equiv \frac{1}{\pi} \int d\varphi\hspace{5 pt}\vep^{-6}(\partial_{\varphi}\psi_{k})(\partial_{\varphi}\psi_{l})(\partial_{\varphi}\psi_{m})(\partial_{\varphi}\psi_{n})$, with the last two expressions combining to yield
\begin{align}
    \sum_{j} \bigg[\mu_{j}^{2} -5\,\mu_{n}^{2}\bigg] \mu_{j}^{4}\,a_{jnn}^{2} &= -\dfrac{16}{3}\, \mu_{n}^{6}\, a_{nnnn}
\end{align}
This chapter uses all of these results to calculate and then analyze matrix elements.

Recall our analogy between the Standard Model and the RS1 model from Chapter \ref{C - Introduction}, which we originally laid out in Table \ref{Table - Analogy SM to RS1} and have repeated in Table \ref{Table - Analogy SM to RS1 - Again} for convenience. In this chapter, we finally confirm several elements of this table and draw the major conclusions of this dissertation:\footnote{These conclusions have been published across several papers: the high-energy scaling behaviors of the helicity-zero spin-2 KK mode scattering matrix element and each of its channels were published in \cite{Chivukula:2019rij}; the four sum rules which make those scaling behaviors possible for the elastic process were published in \cite{Chivukula:2019zkt}, which also included proofs for two of the sum rules; all of these results were then elaborated on and generalized in \cite{Chivukula:2020hvi}. That most recent paper also provides explicit versions of the 5D and 4D effective WFE RS1 Lagrangians (which we recounted and updated in Chapters 3 and 4), proves another sum rule (which we generalized to inelastic processes in Chapter 4 alongside the other sum rules), analyzes how truncation of the KK tower affects the total matrix element, and calculates the strong coupling scale from the 4D effective theory.}
\begin{itemize}
    \item[$\bullet$] Scattering of massive spin-2 KK modes in the RS1 model has a matrix element that grows like $\mathcal{O}(s)$ at large energies, regardless of helicity combination. Thus, scattering of the massive spin-2 KK modes behaves just like the scattering of 4D gravitons at high energies.
    \item[$\bullet$] Truncating the tower of massive spin-2 states (i.e. ignoring KK modes with KK numbers greater than some value $N$) generates a matrix element that grows like $\mathcal{O}(s^{5})$, which replicates the bad high-energy behavior of, for example, massive spin-2 scattering in Fierz-Pauli gravity.
    \item[$\bullet$] Eliminating the radion from the matrix element calculation causes the matrix element to grow like $\mathcal{O}(s^{3})$, which still reflects the explicit breaking of the underlying symmetry group but is more mild energy growth than the growth we attained by eliminating massive KK modes.
\end{itemize}
The rest of the chapter proceeds as follows:
\begin{itemize}
    \item[$\bullet$] Section \ref{S - Motivation and Definitions} establishes the definitions and conventions necessary to calculate the tree-level 2-to-2 scattering matrix element for massive spin-2 KK modes in the center-of-momentum frame. The section ends with some considerations regarding numerical analysis of the RS1 model.
    \item[$\bullet$] Section \ref{sec:level4} considers the scattering of helicity-zero massive spin-2 states in the 5D orbifolded torus (5DOT) model, the limit of the RS1 model in which $kr_{c}$ vanishes. The 5DOT model exhibits discrete KK momentum conservation: this allows all coupling integrals to be evaluated analytically and ensures only a finite number of diagrams contribute to the matrix element. The helicity-zero matrix element is found to grow like $\mathcal{O}(s)$ for any combination of external KK numbers that conserves discrete KK momentum (and otherwise vanishes). The helicity-zero process $(1,4)\rightarrow(2,3)$ lacks any massless intermediate states because of KK momentum conservation and thus the partial wave amplitudes of its matrix element can be calculated without running into massless poles. We calculate its leading partial wave amplitude $a^{0}$ and find via the partial wave amplitude constraints that the 5DOT strong coupling scale is $\Lambda_{\text{strong}}^{(\text{5DOT})} = \sqrt{8\pi}\,M_{\text{Pl}}$.
    \item[$\bullet$] Section \ref{sec:level5} considers the elastic scattering of massive spin-2 states in the RS1 model in which all external KK modes have equal KK number $n$, beginning with helicity-zero elastic scattering. The $\mathcal{O}(s^{\sigma})$ contributions to the helicity-zero matrix element are demonstrated to cancel via certain sum rules for $\sigma =5$, $4$, $3$, and finally $2$. Of the sum rules obtained, only one linear combination was not proved in the previous chapter: this combination involves the radion coupling, and its validity is instead demonstrated numerically. An analytic expression for the residual $\mathcal{O}(s)$ amplitude is provided. Lastly, it is noted that the aforementioned helicity-zero sum rules are sufficient to ensure all elastic massive spin-2 KK mode scattering matrix elements grow at most like $\mathcal{O}(s)$, regardless of helicity combination. 
    \item[$\bullet$] Section \ref{sec:level6} is devoted to several numerical investigations. Subsection \ref{sec:level6a} demonstrates cancellations down to $\mathcal{O}(s)$ in helicity-zero elastic and inelastic scattering matrix elements. Subsection \ref{sec:level6b} investigates how truncation affects the accuracy of the matrix element and its leading $\mathcal{O}(s^{\sigma})$ contributions ($\sigma\in\{1,2,3,4,5\}$) relative to the full matrix element without truncation. Subsection \ref{sec:level6c} calculates the RS1 strong coupling scale $\Lambda_{\text{strong}}^{(\text{RS1})}$ using the 4D effective RS1 theory. Massless poles in RS1 matrix elements are avoided by comparing the leading $\mathcal{O}(s)$ matrix element growth in the RS1 model to the exactly calculable equivalent in the 5DOT model. This yields $\Lambda_{\text{strong}}^{(\text{RS1})}\sim \Lambda_{\pi}$ as expected based on the 5D RS1 theory.
\end{itemize}
This completes the major results this dissertation intended to present. The next chapter provides a brief conclusion that summarizes our original results as well as directions for future work.

\begin{table}
\bgroup
\begin{center}
    \begin{tabular}{| c || c | c |}
        \hline
         & & \\[-0.75 em]
         & {\bf Standard Model} & {\bf Randall-Sundrum 1}\\[0.25 em]
         \hline
         \hline
         & & \\[-0.75 em]
         The fundamental symmetry group... & $\mathbf{SU(2)_{W}}\times \mathbf{U(1)_{Y}}$ & 5D diffeomorphisms\\[0.25 em]
         \hline
         & & \\[-0.75 em]
         ... w/ unitarity-violation scale... & N/A & $\Lambda_{\pi} = M_{\text{Pl}}\, e^{-kr_{c}\pi}$\\[0.25 em]
         \hline
         & & \\[-0.75 em]
         ... and gauged by the... & electroweak bosons & 5D RS1 graviton\\[0.25 em]
         \hline
         & & \\[-0.75 em]
         ... is spontaneously broken by... & the Higgs vev & background geometry\\[0.25 em]
         \hline
         & & \\[-0.75 em]
         ... to a new symmetry group... & $\mathbf{U(1)_{Q}}$ & 4D diffeomorphisms*\\[0.25 em]
         \hline
         & & \\[-0.75 em]
         ... gauged by the... & photon, $\gamma$ & 4D graviton, $h^{(0)}$\\[0.25 em]
         \hline
         & & \\[-0.75 em]
         ... resulting in a spin-0 state... & Higgs boson, $H$ & radion, $r^{(0)}$\\[0.25 em]
         \hline
         & & \\[-0.75 em]
         ... as well as massive states & $W$-bosons, $W^{\pm}$ & spin-2 KK modes, $h^{(n)}$\\[0.25 em]
         built from fund. gauge bosons... & and $Z$-boson, $Z$ & for $n\in\{1,2,\dots\}$\\[0.25 em]
         \hline
         \hline
         & & \\[-0.75 em]
         The $2$-to-$2$ gauge boson process... & $\gamma\gamma \rightarrow \gamma\gamma$ & $h^{(0)}h^{(0)}\rightarrow h^{(0)}h^{(0)}$\\[0.25 em]
         \hline
         & & \\[-0.75 em]
         ... has $\mathcal{M}$ w/ high-energy growth $\sim$ & $\mathcal{O}(s^{0})$ & $\mathcal{O}(s)$\\[0.25 em]
         \hline
         & & \\[-0.75 em]
         ... or, if naively given mass, ... & $\mathcal{O}(s^{2})$ & $\mathcal{O}(s^{5})$\\[0.25 em]
         \hline
         & & \\[-0.75 em]
         ... yet $2$-to-$2$ massive state process & \multirow{2}{*}{$W^{+}W^{-}\rightarrow W^{+}W^{-}$} & \multirow{2}{*}{$h^{(n_{1})}h^{(n_{2})}\rightarrow h^{(n_{3})}h^{(n_{4})}$}\\[0.25 em]
         where mass arises via sym. break... &  & \\[0.25 em]
         \hline
         & & \\[-0.75 em]
         ... has $\mathcal{M}$ w/ high-energy growth $\sim$ & $\mathcal{O}(s^{0})$ & \textcolor{red}{$\mathbf{O(s)}$}\\[0.25 em]
         \hline
         \hline
         & & \\[-0.75 em]
         Breaking the fund. symmetry by... & elim. $Z$ & KK tower truncation \\[0.25 em]
         \hline
         & & \\[-0.75 em]
         ... makes massive states scatter like & \multirow{2}{*}{$\mathcal{O}(s^{2})$} & \multirow{2}{*}{\textcolor{red}{$\mathbf{O(s^{5})}$}}  \\[0.25 em]
         naively-massive gauge bosons, $\mathcal{M}\sim$ & & \\[0.25 em]
         \hline
         \hline
         & & \\[-0.75 em]
         Breaking the fund. symmetry by... & elim. the Higgs & elim. the radion \\[0.25 em]
         \hline
         & & \\[-0.75 em]
         ... makes massive states scatter $\sim$ & $\mathcal{O}(s)$ & \textcolor{red}{$\mathbf{O(s^{3})}$} \\[0.25 em]
         \hline
    \end{tabular}
\end{center}
\egroup
\caption{The Standard Model (SM) and the Randall-Sundrum 1 (RS1) model share a chain of conceptual similarities with respect to the scattering of particles made massive by spontaneous symmetry breaking. The Mandelstam variable $s\equiv E^{2}$, where $E$ is the incoming center-of-momentum energy. Original results presented in this dissertation are indicated in bold and red. (* - Technically, the new symmetry group is the Cartan subgroup of the 5D diffeomorphisms that contains the 4D diffeomorphisms.)}
\label{Table - Analogy SM to RS1 - Again}
\end{table}

\section{Motivation and Definitions} \label{S - Motivation and Definitions}
\subsection{Restating the Problem} \label{SS - Defs - 5D vs 4D}

From the perspective of the 5D Lagrangian, the only excitation in the RS1 model is a massless 5D graviton $H$, which (when using the appropriate five-dimensional generalization of the helicity operator) has five helicity eigenstates. Because each term of $\mathcal{L}_{\text{5D}}$ contains two derivatives, each interaction vertex contains at most two powers of 4D momentum per term. Consequently, the cubic and quartic couplings grow like $\mathcal{O}(s)$ at high energies
\begin{align}
    \raisebox{-0.45\height}{\verHHH{0.3}} \sim \kappa_{\text{5D}}\, s\hspace{35 pt}\raisebox{-0.45\height}{\verHHHH{0.3}} \sim \kappa_{\text{5D}}^{2}\,s
\end{align}
whereas the propagator falls like $\mathcal{O}(s^{-1})$
\begin{align}
    \propH{0.6} \sim \dfrac{1}{s}
\end{align}
and the external polarizations are independent of $s$. The total tree-level matrix element for 2-to-2 scattering of 5D gravitons is the sum of four diagrams:
\begin{align}
    \raisebox{-0.45\height}{\sgDHHHH{0.3}} + \raisebox{-0.45\height}{\sDHHHHH{0.3}} +  \raisebox{-0.45\height}{\tDHHHHH{0.3}} +  \raisebox{-0.45\height}{\uDHHHHH{0.3}}
\end{align}
By combining the existing scaling arguments for each piece of each diagram, we find the overall matrix element must grow at most like $\mathcal{O}(s)$. We can arrive at this same conclusion by considering each graviton at energies so large that it can be localized with a width significantly less than the compactification radius $r_{c}$ and inverse warping parameter $1/k$. At these energies, the only dimensionful parameter remaining is the coupling strength $\kappa_{\text{5D}}$. Therefore, because a 5D 2-to-2 scattering matrix element has units $[\text{Energy}]^{-1}$ and $\kappa_{\text{5D}}$ has units $[\text{Energy}]^{-3/2}$, the 5D matrix element must scale at high energies like
\begin{align}
    \mathcal{M}_{HH\rightarrow HH} \sim \kappa_{\text{5D}}^{2} s
\end{align}
up to dimensionless multiplicative constants. This scaling provides a strict constraint on the high-energy behavior of the 4D matrix elements, which we now consider.

Consider the same argument from the 4D perspective. Instead of perturbing the metric $G$ to yield a 5D graviton field $\hat{H}_{MN}(x,y)$, it is perturbed by 5D fields $\hat{h}_{\mu\nu}(x,y)$ and $\hat{r}(x)$ which transform covariantly under the 4D Lorentz group. As detailed in Section \ref{4D Effective RS1 Model}, $\hat{h}_{\mu\nu}$ embeds a Kaluza-Klein (KK) tower of 4D spin-2 fields $\hat{h}_{\mu\nu}^{(n)}(x)$, where $n=0$ corresponds to the massless 4D graviton, and $\hat{r}(x)$ embeds a massless 4D spin-0 state $\hat{r}^{(0)}(x)$ called the radion. This dissertation focuses on tree-level 2-to-2 scattering of massive KK modes with (nonzero) KK indices $n_{1}$, $n_{2}$, $n_{3}$, and $n_{4}$. The matrix element $\mathcal{M}_{n_{1}n_{2}\rightarrow n_{3}n_{4}}$ for this process is calculated from infinitely-many diagrams, which we categorize into subsets for ease of writing and discussion. All together, for any combination of external helicities the matrix element equals
\begin{align}
    \mathcal{M}_{n_{1}n_{2}\rightarrow n_{3}n_{4}} \equiv \mathcal{M}_{\text{c}} + \mathcal{M}_{\text{r}} + \sum_{j=0}^{+\infty} \mathcal{M}_j~,
    \label{eq:matrix-element}
\end{align}
within which
\begin{align}
    \mathcal{M}_{\text{c}} & \equiv \raisebox{-0.45\height}{\sgDklmn{0.20}}\nonumber \\
    \mathcal{M}_{\text{r}} &\equiv \raisebox{-0.45\height}{\sDklrmn{0.20}} +  \raisebox{-0.45\height}{\tDklrmn{0.20}} +  \raisebox{-0.45\height}{\uDklrmn{0.20}}\\
    \mathcal{M}_{j} &\equiv \raisebox{-0.45\height}{\sDkljmn{0.20}} +  \raisebox{-0.45\height}{\tDkljmn{0.20}} +  \raisebox{-0.45\height}{\uDkljmn{0.20}}\nonumber
\end{align}
where subscript ``c" denotes the contact diagram,  ``r" denotes the sum of diagrams mediated by the radion $\hat{r}^{(0)}$, and ``$j$" denotes sum of diagrams mediated by the $j$th spin-2 KK mode $\hat{h}^{(j)}$. The relevant vertices scale like
\begin{align}
    \raisebox{-0.45\height}{\verhhr{0.3}} \sim \dfrac{\kappa_{\text{5D}}}{\sqrt{\pi r_{c}}} \hspace{35 pt}\raisebox{-0.45\height}{\verhhh{0.3}} \sim \dfrac{\kappa_{\text{5D}}}{\sqrt{\pi r_{c}}} s \hspace{35 pt}\raisebox{-0.45\height}{\verhhhh{0.3}} \sim \left[\dfrac{\kappa_{\text{5D}}}{\sqrt{\pi r_{c}}}\right]^{2} s
\end{align}
where the $hhr$ interaction does not grow in energy because the corresponding interaction Lagrangian (Eq. \eqref{L4Deffhhr}) contains no 4D derivatives, and the relevant propagators scale like
\begin{center}
    \raisebox{-0.35\height}{\begin{tikzpicture}
    \begin{feynman}[medium]
        \vertex (n1) at (-1,0) {};
        \vertex (a1) at (-0.7,0.35) {};
        \vertex (b1) at (0.7,0.35) {};
        \vertex (n2) at (1,0) {};
        
        \diagram* {
            (n1) -- [edge label = $r$] (n2)
        };
    \end{feynman}
    \end{tikzpicture}} $\sim \dfrac{1}{s}$ \hspace{35 pt} \raisebox{-0.35\height}{\begin{tikzpicture}
    \begin{feynman}[medium]
        \vertex (n1) at (-1,0) {$\mu\nu$};
        \vertex (a1) at (-0.7,0.35) {};
        \vertex (b1) at (0.7,0.35) {};
        \vertex (n2) at (1,0) {$\rho\sigma$};
        
        \diagram* {
            (n1) -- [photon, edge label = $0$] (n2),
            (n1) -- [] (n2),
        };
    \end{feynman}
    \end{tikzpicture}} $\sim \dfrac{1}{s}$ \hspace{35 pt} \raisebox{-0.35\height}{\begin{tikzpicture}
    \begin{feynman}[medium]
        \vertex (n1) at (-1,0) {$\mu\nu$};
        \vertex (a1) at (-0.7,0.35) {};
        \vertex (b1) at (0.7,0.35) {};
        \vertex (n2) at (1,0) {$\rho\sigma$};
        
        \diagram* {
            (n1) -- [photon, edge label = $n\neq 0$] (n2),
            (n1) -- [] (n2),
        };
    \end{feynman}
    \end{tikzpicture}} $\sim s$
\end{center}
according to Eqs. \eqref{masslessspin0propagator1}-\eqref{Bmunu}. The external massive spin-2 states can take on any one of five possible helicities $\lambda \in\{-2,-1,0,1,2\}$, and are described by polarization tensors $\epsilon_{\lambda}^{\mu\nu}$ which have leading $\mathcal{O}(s^{2-|\lambda|})$ high-energy behavior (Eqs. \eqref{C1Spin1Polarizations} and \eqref{C1Spin2Polarizations}). In order to maximize energy growth, we focus on the helicity-zero process wherein $\lambda_{1}=\lambda_{2}=\lambda_{3}=\lambda_{4}=0$. Under this assumption, if we combine these diagrammatic elements we find the diagrams seemingly scale like
\begin{align}
    \mathcal{M}_{j > 0} &\sim \mathcal{O}(s^{7})\\
    \mathcal{M}_{0} \text{ and } \mathcal{M}_{c} &\sim \mathcal{O}(s^{5})\\
    \mathcal{M}_{r} &\sim \mathcal{O}(s^{3})
\end{align}
such that naively we expect the matrix element $\mathcal{M}_{n_{1}n_{2}\rightarrow n_{3}n_{4}}$ to grow like $\mathcal{O}(s^{7})$ when all external massive spin-2 states have vanishing helicity. Explicit evaluation reveals that the scaling is slightly more mild in practice: per diagram,
\begin{align}
    \mathcal{M}_{j} \text{ and } \mathcal{M}_{c} &\sim \mathcal{O}(s^{5})\\
    \mathcal{M}_{r} &\sim \mathcal{O}(s^{3})
\end{align}
where cancellations occur such that each diagram in $\mathcal{M}_{j>0}$ only grows like $\mathcal{O}(s^{5})$. This suggests that $\mathcal{M}_{n_{1}n_{2}\rightarrow n_{3}n_{4}}$ might grow as fast as $\mathcal{O}(s^{5})$. However, such rapid energy growth would starkly contrast the high-energy growth of the 4D graviton, whose own 2-to-2 scattering matrix element only grows as fast as $\mathcal{O}(s)$. Inspired by the analogy with the Standard Model in Table \ref{Table - Analogy SM to RS1 - Again}, wherein the massive $W$-bosons scatter with the same high-energy behavior as photons due to the underlying electroweak symmetry $\mathbf{SU(2)_{W}}\times\mathbf{U(1)_{Y}}$, we expect that the matrix elements for scattering massive KK modes (which are generated by the 5D RS1 graviton just like the 4D graviton) should exhibit the same high-energy growth as graviton scattering, and indeed: this chapter demonstrates that cancellations occur between the diagrams in Eq. \eqref{eq:matrix-element} which reduce the naive $\mathcal{O}(s^{5})$ growth down to $\mathcal{O}(s)$ growth. These cancellations require precise relationships between the KK mode mass spectra and coupling integrals.

This chapter isolates those relationships and demonstrates they hold true in the 4D effective field theory. After this, the strong coupling scale $\Lambda_{\pi}$ is calculated directly from the 4D effective theory, and the effects of KK mode truncation are investigated.

\subsection[Definitions]{Definitions\footnote{This subsection was originally published as Subsection IV.A of \cite{Chivukula:2020hvi}, up to minor changes in wording.}} \label{sec:level4a}

The preceding chapters detailed how to determine the vertices relevant to tree-level $2$-to-$2$ scattering of massive spin-2 helicity eigenstates in the center-of-momentum frame. This subsection recounts the other diagrammatic pieces which go into calculating the diagrams relevant to those matrix elements. For scattering of nonzero KK modes $(n_{1},n_{2})\rightarrow(n_{3},n_{4})$ with helicities $(\lambda_{1},\lambda_{2})\rightarrow(\lambda_{3},\lambda_{4})$, we choose coordinates such that the initial particle pair have 4-momenta satisfying
\begin{align}
    p^\mu_{1} &= (E_{1}, + |\vec{p}_{i}| \hat{z})\hspace{25 pt}p_{1}^2 = m_{n_{1}}^2\\
    p^\mu_{2} &= (E_{2}, - |\vec{p}_{i}| \hat{z})\hspace{25 pt}p_{2}^2 = m_{n_{2}}^2
\end{align}
and the final particle pair have 4-momenta satisfying
\begin{align}
    p^\mu_{3} &= (E_{3}, + \vec{p}_f)\hspace{25 pt}p_{3}^2 = m_{n_{3}}^2\\
    p^\mu_{4} &= (E_{4}, - \vec{p}_f)\hspace{25 pt}p_{4}^2 = m_{n_{4}}^2
\end{align}
where $\vec{p}_f \equiv |\vec{p}_f| (\sin\theta\cos\phi,\sin\theta\sin\phi,\cos\theta)$. That is, the initial pair approach along the $z$-axis and the final pair separate along the line described by the angles $(\phi,\theta)$. The helicity-$\lambda$ spin-2 polarization tensor $\epsilon_{\lambda}^{\mu\nu}(p)$ for a particle with 4-momentum $p$ is defined according to
\begin{align}
    \epsilon_{\pm 2}^{\mu\nu} &= \epsilon_{\pm1}^{\mu} \epsilon_{\pm 1}^{\nu}~,\\
    \epsilon_{\pm 1}^{\mu\nu} &= \dfrac{1}{\sqrt{2}}\left[\epsilon_{\pm 1}^{\mu}\epsilon_{0}^{\nu} + \epsilon_{0}^{\mu} \epsilon_{\pm 1}^{\nu}\right]~\\
    \epsilon_{0}^{\mu\nu} &= \dfrac{1}{\sqrt{6}}\bigg[\epsilon^{\mu}_{+1} \epsilon^{\nu}_{-1} + \epsilon^{\mu}_{-1}\epsilon^{\nu}_{+1} + 2\epsilon^{\mu}_{0}\epsilon^{\nu}_{0}\bigg]~,
\end{align}
where the $\epsilon_{s}^\mu$ are the (particle-direction dependent) spin-1 polarization vectors
\begin{align}
    \epsilon_{\pm 1}^{\mu} &= \pm\dfrac{e^{\pm i\phi}}{\sqrt{2}}\bigg(0,- c_{\theta}c_{\phi} \pm i s_{\phi}, -c_{\theta}s_{\phi} \mp ic_{\phi},s_{\theta}\bigg)~,\\
    \epsilon_{0}^{\mu} &= \dfrac{E}{m}\bigg(\sqrt{1-\dfrac{m^2}{E^2}}, \hat{p}\bigg)~,\label{ep10}
\end{align}
$(c_{x},s_{x})\equiv(\cos x,\sin x)$, and $\hat{p}$ is a unit vector in the direction of the momentum \cite{Han:1998sg}. We use the Jacob-Wick second particle convention, which adds a phase $(-1)^{2-\lambda}e^{-2\lambda i\phi}$ to $\epsilon^{\mu\nu}_{\lambda}$ when the polarization tensor describes $h^{(n_{2})}$ or $h^{(n_{4})}$  \cite{Jacob:1959at}. Due to rotational invariance, we may set the azimuthal angle $\phi$ to $0$ without loss of generality. Meanwhile, the propagators for virtual spin-0 and spin-2 particles of mass $M$ and 4-momentum $P$ are, respectively,
\begin{align}
    \raisebox{2 pt}{\propSc{0.5}}\hspace{18 pt} &= \dfrac{i}{P^2-M^2}\\
    \raisebox{-0.35\height}{\propGr{0.5}} &= \dfrac{i B^{\mu\nu,\rho\sigma}}{P^2-M^2}
\end{align}
where
\begin{align}
    &B^{\mu\nu,\rho\sigma} \equiv \dfrac{1}{2}\bigg[\overline{B}^{\mu\rho}\overline{B}^{\nu\sigma} + \overline{B}^{\nu\rho}\overline{B}^{\mu\sigma} - \dfrac{1}{3}(2+\delta_{0,M}) \overline{B}^{\mu\nu}\overline{B}^{\rho\sigma}\bigg]\nonumber\\
    &\left.\overline{B}^{\alpha\beta}\right|_{M=0} = \eta^{\alpha\beta}\hspace{50 pt}\left.\overline{B}^{\alpha\beta}\right|_{M\neq 0} \equiv \eta^{\alpha\beta} - \dfrac{P^\alpha P^\beta}{M^2}
\end{align}
and $\eta^{\mu\nu} = \text{Diag}(+1,-1,-1,-1)$ is the flat 4D metric \cite{Han:1998sg}. The massless spin-2 propagator is derived in the de Donder gauge by adding a gauge-fixing term
\begin{align}
    \mathcal{L}_{\text{gf}} &\equiv -\bigg(\partial^{\mu}\hat{h}_{\mu\nu} - \dfrac{1}{2} \partial_{\nu} \hat{h} \bigg)^{2}
\end{align}
to the Lagrangian. The Mandelstam variable $s \equiv (p_{1}+p_{2})^2 = (E_{1}+E_{2})^2$ provides a convenient frame-invariant measure of collision energy. The minimum value of $s$ that is kinematically allowed equals $s_{\text{min}}\equiv \max[(m_{n_{1}}+m_{n_{2}})^{2},(m_{n_{3}}+m_{n_{4}})^{2}]$. When dealing with explicit full matrix elements, we will replace $s\in[s_{\text{min}},+\infty)$ with the unitless $\mathfrak{s} \in [0,+\infty)$ which is defined according to $s \equiv s_{\text{min}}(1+\mathfrak{s})$.

As discussed in Subsection \ref{SS - Defs - 5D vs 4D}, any tree-level massive spin-2 scattering amplitude can be written as
\begin{align}
    \mathcal{M}_{n_{1}n_{2}\rightarrow n_{3}n_{4}} \equiv \mathcal{M}_{\text{c}} + \mathcal{M}_{\text{r}} + \sum_{j=0}^{+\infty} \mathcal{M}_j~,
    \label{eq:Mtot-def}
\end{align}
where $\mathcal{M}_{n_{1}n_{2}\rightarrow n_{3}n_{4}}$ will be abbreviated to $\mathcal{M}$ when the process can be understood from context, and we separate the contributions arising from contact interactions, radion exchange, and a sum over the exchanged intermediate KK states $j$ (and where ``0" labels the massless graviton). In practice, this sum cannot be completed in entirety and must instead be truncated. Therefore, we also define the truncated matrix element
\begin{align}
    \mathcal{M}^{[N]} \equiv \mathcal{M}_{\text{c}} + \mathcal{M}_{\text{r}} + \sum_{j=0}^{N} \mathcal{M}_j~,
    \label{eq:M-N-def}
\end{align}
which includes the contact diagram, the radion-mediated diagrams, and all KK mode-mediated diagrams with intermediate KK number less than or equal to $N$.

We are concerned with the high-energy behavior of these matrix elements, and will therefore examine the high-energy behavior of each of the contributions discussed.
Because the polarization tensors $\epsilon^{\mu\nu}_{\pm1}$ introduce odd powers of energy, $\sqrt{s}$ is a more appropriate expansion parameter for generic helicity combinations. Thus, we series expand the diagrams and total matrix element in $\sqrt{s}$ and label the coefficients like so:
\begin{align}
    \mathcal{M}(s,\theta) \equiv \sum_{\sigma \in \tfrac{1}{2}\mathbb{Z}} \overline{\mathcal{M}}^{(\sigma)}(\theta)\cdot s^{\sigma}
\end{align}
for half-integer $\sigma$. We also define $\mathcal{M}^{(\sigma)}\equiv \overline{\mathcal{M}}^{(\sigma)}\cdot s^{\sigma}$. In what follows, we demonstrate that $\mathcal{M}^{(\sigma)}$ vanishes for $\sigma > 1$ regardless of helicity combination and we present the residual linear term in $s$ for helicity-zero elastic scattering.

\subsection{Comments on Numerical Evaluation}
The previous chapter detailed how to manipulate integrals of products of wavefunctions from a purely analytic perspective, so let us take a moment to consider the numerical perspective. In those cases where it is desirable to numerically evaluate matrix elements, it can be difficult to achieve a desired numerical accuracy for a variety of reasons. For example, the determination of the massive spin-2 KK mode spectrum via
\begin{align}
    \left[2 J_{2} + \dfrac{\mu_{n}\varepsilon}{kr_{c}} (\partial J_{2})\right]\bigg|_{\varphi = \pi}\left[2 Y_{2} + \dfrac{\mu_{n}\varepsilon}{kr_{c}}  (\partial Y_{2})\right]\bigg|_{\varphi = 0}\hspace{35 pt}&\nonumber\\
    -\left[2 Y_{2} + \dfrac{\mu_{n}\varepsilon}{kr_{c}} (\partial Y_{2})\right]\bigg|_{\varphi = \pi}\left[2 J_{2} + \dfrac{\mu_{n}\varepsilon}{kr_{c}}  (\partial J_{2})\right]\bigg|_{\varphi = 0}& = 0\label{massequation}
\end{align}
(which is Eq. \eqref{mn} when $\nu = 2$) amounts to solving for the roots of the LHS to some desired accuracy. However, the spacing of those roots can vary dramatically depending on the value of $kr_{c}$, which means (depending on your root-solving method) there is the possibility to inadvertently skip roots. To avoid this, we can use our exact knowledge of the eigenvalue spectrum when $kr_{c} = 0$ (considered in the next section) and when $kr_{c}$ is large (Subsection \ref{Large krc}) to reparameterize Eq. \eqref{massequation} in terms of a variable wherein the roots are more evenly spaced. For this purpose, we use
\begin{align}
    \mu_{n} \equiv \dfrac{c_{n}}{n} \bigg[ (kr_{c})\, x_{n} \, e^{-kr_{c}\pi} + n\, e^{-kr_{c\pi}} \bigg]
\end{align}
and solve Eq. \eqref{massequation} for $c_{n}$ instead. Having obtained a sufficiently-accurate eigenvalue spectrum, it is then useful to rewrite $\psi_{n}$ into the form
\begin{align}
    \psi_{n} = \dfrac{\varepsilon^{2}}{N_n} \left[b^{(\text{den})}_{n2}\, J_{2} \left(\dfrac{\mu_{n}\varepsilon}{kr_{c}} \right) - b^{(\text{num})}_{n2}\s Y_{2}\left(\dfrac{\mu_{n}\varepsilon}{kr_{c}}\right)\right] \label{PSI2}
\end{align}
rather than Eq. \eqref{eq:wavefunction}, where $b^{(\text{num})}_{n2}$ and $b^{(\text{den})}_{n2}$ indicate the numerator and denominator of Eq. \eqref{bnnu} respectively. (The value of $N_{n}$ must change to accommodate this new form but is still determined by Eq. \eqref{onA}.) This new form helps avoid numerical instability during the occasions when $b^{(\text{den})}_{n2}$ is close to zero. Furthermore, it is worthwhile to directly utilize the analytic form of derivatives wherever possible. Specifically, this means using
\begin{align}
    \partial J_{\nu} \equiv \dfrac{1}{2}\left[J_{\nu-1} - J_{\nu+1}\right] \hspace{35 pt} \partial Y_{\nu} \equiv \dfrac{1}{2}\left[Y_{\nu-1} - Y_{\nu+1}\right]
\end{align}
and
\begin{align}
    (\partial_{\varphi} \psi_{n}) = \dfrac{\vep^{3}}{N_{n}}\mu_{n}\left[b^{(\text{den})}_{n2}\, J_{1} \left(\dfrac{\mu_{n}\varepsilon}{kr_{c}} \right) - b^{(\text{num})}_{n2}\s Y_{1}\left(\dfrac{\mu_{n}\varepsilon}{kr_{c}}\right)\right]  (\partial_{\varphi}|\varphi|)
\end{align}
which uses the same $N_{n}$ derived when normalizing $\psi_{n}$ in Eq. \eqref{PSI2}. These changes all help in gaining as much numerical accuracy as possible before calculating coupling integrals. As detailed in Section \ref{4D Effective RS1 Model}, interaction vertices in the effective theory are proportional to integrals of products of wavefunctions and their derivatives. Each wavefunction $\psi_{n}$ oscillates through zero $n$ times over the (half) domain $\varphi\in [0,\pi]$ and is exponentially distorted towards $\varphi = \pm \pi$ by an amount determined by the specific value of $kr_{c}$ selected. Consequently, interaction vertices involving even relatively modest mode numbers ($n \sim 10$) generate integrands that are highly oscillatory. Those dramatic oscillations in the integrand lead to cancellations between large positive and large negative values in the integral, which can eliminate many significant digits worth of numerical confidence. The number of significant digits retained following these cancellations depends on just how accurately the different maxima and minima cancel one-another, which varies dramatically from integral to integral. In this sense, the integrals required for investigations of the 4D effective RS1 model are numerically unstable. This results in a time-consuming feedback loop: the numerical accuracy of the spectrum and wavefunctions must be increased until the coupling integrals are sufficiently accurate, which can not be known until those integrals are attempted. Furthermore, because we are interesting in demonstrating cancellations between diagrams in the matrix element, we are often evaluating perturbative expressions in an attempt to ``measure zero": because higher-order terms in those expansions contribute less than lower-order terms, their effects are only evident if the lower-order terms are evaluated to sufficient accuracy, further increasing the need for highly-accurate results. We can only be confident we have calculated all elements of the calculation to sufficient accuracy once all evidence of numerical noise is absent from certain cross-checks (such as the sum rules analytically proved in Section \ref{DerivingSumRules}). Unfortunately, there seems to be no means of avoiding this time-consuming complication.

\section[Elastic Scattering in the 5D Orbifolded Torus]{Elastic Scattering in the 5D Orbifolded Torus\footnote{The first paragraph of this section originates from Section IV of \cite{Chivukula:2020hvi}. The rest of this section's content was original published as Subsection IV.B of \cite{Chivukula:2020hvi} up to minor changes.}} \label{sec:level4}

In this section, we begin our analysis of the scattering amplitudes of the massive spin-2 KK modes. As described above, the full tree-level scattering amplitudes will require summing over the exchange of all intermediate states, and we will find that the cancellations needed to reduce the growth of RS1 scattering amplitudes from ${\cal O}(s^5)$ to ${\cal O}(s)$ will only completely occur once all states are included. In the present section, we analyze KK mode scattering in a limit that only has finitely many nonzero diagrams per matrix element: the 5D Orbifolded Torus model.

The 5D Orbifolded Torus (5DOT) model is obtained by taking the limit of the RS1 metric Eq. \eqref{GMNwvform} as $k r_c$ vanishes, while simultaneously maintaining a nonzero finite first mass $m_1$ (or, equivalently, a nonzero finite $r_c$). Consequently, the 5DOT metric lacks explicit dependence on $y$,
\begin{align}
    \GEDOT = \matrixbb{e^{\tfrac{-\kappa\hat{r}}{\sqrt{6}}}(\eta_{\mu\nu} + \kappa \hat{h}_{\mu\nu})}{0}{0}{-\left(1+\tfrac{\kappa\hat{r}}{\sqrt{6}}\right)^2}~, \label{GMN5DOT}
\end{align}
and as $kr_{c}\rightarrow 0$ the massive wavefunctions go from exponentially-distorted Bessel functions to simple cosines:
\begin{align}
    \psi_n = \begin{cases}
        \psi_0 = \tfrac{1}{\sqrt{2}}\\
        \psi_n = -\cos(n|\varphi|)\hspace{15 pt} 0<n\in\mathbb{Z}
    \end{cases}\label{RSwfxn}
\end{align}
with masses given by $\mu_{n} = m_{n} r_{c} = n$ and 5D gravitational coupling $\kappa = \sqrt{2\pi r_c}\hspace{1 pt}\kDD = \sqrt{8\pi r_{c}}/M_{\text{Pl}}$. In the absence of warp factors, the radion couples diagonally and spin-2 interactions display discrete KK momentum conservation. Explicitly, an $H$-point vertex $\hat{h}^{(n_1)}\cdots \hat{h}^{(n_H)}$ in the 4D effective 5DOT model has vanishing coupling if there exists no choice of $c_i \in \{-1,+1\}$ such that $c_1 n_1 +\cdots + c_H n_H =0$. For example, the three-point couplings $\ahhh{n_{1}}{n_{2}}{n_{3}}$ and $\bhhh{n_{1}^{\prime}}{n_{2}^{\prime}}{n_{3}}$ are nonzero only when $n_{1}=|n_{2}\pm n_{3}|$. Therefore, unlike when $kr_c$ is nonzero, the 5DOT matrix element $\mathcal{M}^{(\text{5DOT})}$ for a process $(n_{1},n_{2})\rightarrow (n_{3},n_{4})$ consists of only finitely many nonzero diagrams.

\begin{footnotesize}
\begin{table*}[t]
\begin{tabular}{|c|c|c|c|c|}
\hline
& & & & \\[-0.75 em]
& $ s^{5}$ & $s^{4}$ & $s^{3}$ & $s^{2}$  \\
& & & & \\[-0.75 em]
\hline
& & & & \\[-0.75 em]
$\frac{1}{\kappa^{2}}\mathcal{M}_{\text{c}}$
& $ - \frac{  r_{c}^{7} [7 + c_{2\theta}] s_{\theta}^2 }{ 3072 n^{8} \pi } $
& $ \frac{  r_{c}^{5} [63 - 196\hspace{1 pt}c_{2\theta} + 5\hspace{1 pt}c_{\theta}] }{ 9216 n^{6} \pi } $
& $ \frac{  r_{c}^{3} [-185 + 692\hspace{1 pt}c_{2\theta}  +5\hspace{1 pt}c_{\theta}] }{ 4608 n^{4} \pi } $
& $ - \frac{  r_{c} [5 + 47\hspace{1 pt}c_{2\theta}] }{ 72 n^{2} \pi } $  \\
& & & & \\[-0.75 em]
\hline
& & & & \\[-0.75 em]
$ \frac{1}{\kappa^{2}}\mathcal{M}_{2n}$
& $ \frac{  r_{c}^{7} [7 + c_{2\theta}] s_{\theta}^{2} }{ 9216 n^{8} \pi } $
& $ \frac{  r_{c}^{5} [-13+ c_{2\theta}] s_{\theta}^{2} }{ 1152 n^{6} \pi } $
& $ \frac{  r_{c}^{3} [97 + 3\hspace{1 pt}c_{2\theta}] s_{\theta}^{2} }{ 1152 n^{4} \pi } $
& $ \frac{  r_{c} [-179 + 116\hspace{1 pt}c_{2\theta} - c_{\theta}] }{ 1152 n^{2} \pi } $   \\
& & & & \\[-0.75 em]
\hline 
& & & & \\[-0.75 em]
$ \frac{1}{\kappa^{2}}\mathcal{M}_{0}$
& $ \frac{  r_{c}^{7} [7 + c_{2\theta}] s_{\theta}^{2} }{ 4608 n^{8} \pi } $
& $ \frac{  r_{c}^{5} [-9 + 140\hspace{1 pt}c_{2\theta} - 3 c_{\theta}] }{ 9216 n^{6} \pi } $
& $ \frac{  r_{c}^{3} [15 - 270\hspace{1 pt}c_{2\theta} - c_{\theta}] }{ 2304 n^{4} \pi } $
& $ \frac{  r_{c} [175 +624\hspace{1 pt}c_{2\theta}  + c_{\theta}] }{ 1152 n^{2} \pi } $   \\
& & & & \\[-0.75 em]
\hline
& & & & \\[-0.75 em]
$\frac{1}{\kappa^{2}}\mathcal{M}_{\text{r}}$
& $ 0 $
& $ 0 $
& $ - \frac{  r_{c}^{3} s_{\theta}^{2} }{ 64 n^{4} \pi } $
& $ \frac{  r_{c} [7 + c_{2\theta}] }{ 96 n^{2} \pi } $ \\
& & & & \\[-0.75 em]
\hline
\hline
& & & & \\[-0.75 em]
Sum & $0$ & $0$ & $0$ &  $0$ \\[0.25 em]
\hline 
\end{tabular} 
\caption{Cancellations in the $(n,n) \to (n,n)$ 5DOT amplitude, where $(c_{\theta},s_{\theta}) = (\cos\theta,\sin\theta)$. Originally published in \cite{Chivukula:2019rij} \label{tab:nn2nn}}
\end{table*}
\end{footnotesize}

For $(n,n)\rightarrow (n,n)$, the 5DOT matrix element arises from four types of diagrams:
\begin{align}
    \mathcal{M}^{(\text{5DOT})}_{(n,n)\rightarrow (n,n)} = \mathcal{M}_{\text{c}} + \mathcal{M}_{\text{r}} +  \mathcal{M}_{0} + \mathcal{M}_{2n}~.
\end{align}
Using Eq. \eqref{AandBintermsofgeneralizedx} and the 5DOT wavefunctions, we find:
\begin{align}
    n^{2} a_{nnnn} \hspace{5 pt} &= \hspace{5 pt} 3b_{n^{\prime}n^{\prime}nn} \hspace{5 pt} = \hspace{5 pt} \dfrac{3}{4}n^{2}~,\nonumber \\
    n^{2} a_{nn0} \hspace{5 pt} &= \hspace{5 pt} b_{n^{\prime}n^{\prime}0} \hspace{5 pt} = \hspace{5 pt} b_{n^{\prime}n^{\prime}r} \hspace{5 pt} =\hspace{5 pt}  \dfrac{1}{\sqrt{2}}n^{2}~,\label{eq:orbifold-couplings}\\
    n^{2} a_{nn(2n)} \hspace{5 pt} &= \hspace{5 pt} - b_{n^{\prime}n^{\prime}(2n)} \hspace{5 pt} = \hspace{5 pt} \dfrac{1}{2} b_{(2n)^{\prime}n^{\prime}n} \hspace{5 pt} = \hspace{5 pt} -\dfrac{1}{2}n^{2}~, \nonumber 
\end{align}
where here again the subscript ``0" refers to the massless 4D graviton. We focus first on the scattering of helicity-zero states, which have the most divergent high-energy behavior (we return to consider other helicity combinations in Sec. \ref{sec:nonlongitudinal}). Figure \ref{tab:nn2nn} lists $\mathcal{M}_{\text{c}}^{(\sigma)}$, $\mathcal{M}_{\text{r}}^{(\sigma)}$, $\mathcal{M}_{0}^{(\sigma)}$, and $\mathcal{M}_{2n}^{(\sigma)}$ for $\sigma\geq 1$, and demonstrates how cancellations occur among them such that $ \overline{\mathcal{M}}^{(\sigma)} = 0$  for $\sigma > 1$. The leading contribution in incoming energy is
\begin{align}
    \overline{\mathcal{M}}^{(1)} = \dfrac{3\kappa^2}{256\pi r_c} \left[7+\cos(2\theta)\right]^2\csc^{2}\theta\,.\label{M15DOTnnnn}
\end{align}
More generally, for a generic helicity-zero 5DOT process $(n_{1},n_{2})\rightarrow (n_{3},n_{4})$, the leading high-energy contribution to the matrix element equals
\begin{align}
    \overline{\mathcal{M}}^{(1)} = \dfrac{\kappa^2}{256\pi r_c}x_{n_{1}n_{2}n_{3}n_{4}} \left[7+\cos(2\theta)\right]^2\csc^{2}\theta~, \label{M15DOT}
\end{align}
where $x$ is fully symmetric in its indices, and satisfies
\begin{align}
    x_{aaaa} = 3,\hspace{15 pt}x_{aabb} = 2,\hspace{5 pt}\text{ otherwise}\hspace{5 pt}
    x_{abcd} = 1~,\nonumber
\end{align}
when discrete KK momentum is conserved (and, of course, vanishes when the process does not conserve KK momentum). We next discuss the full calculation, including subleading terms. 

The complete (tree-level) matrix element for the elastic helicity-zero 5DOT process equals
\begin{align}
    \mathcal{M}^{(\text{5DOT})} = \dfrac{\kappa^{2} n^{2}\left[P_{0} + P_{2} c_{2\theta} + P_{4} c_{4\theta} + P_{6}c_{6\theta} \right] \csc^{2}\theta}{256\pi r_{c}^{3} \mathfrak{s}(\mathfrak{s}+1)(\mathfrak{s}^{2} + 8 \mathfrak{s} + 8 - \mathfrak{s}^{2} c_{2\theta})}~,\label{M5DOTtotal}
\end{align}
where
\begin{align}
    P_{0} &= 510\s \mathfrak{s}^{5} + 3962\s\mathfrak{s}^{4} + 8256\s \mathfrak{s}^{3} + 7344\s \mathfrak{s}^{2} + 3216\s \mathfrak{s} + 704~,\\
    P_{2} &= -429 \s\mathfrak{s}^{5} + 393 \s\mathfrak{s}^{4} + 3936 \s\mathfrak{s}^{3} + 5584 \s\mathfrak{s}^{2} + 3272 \s\mathfrak{s} + 768~,\\
    P_{4} &= -78 \s\mathfrak{s}^{5} - 234 \s\mathfrak{s}^{4} + 192 \s\mathfrak{s}^{3} + 1552 \s\mathfrak{s}^{2} + 1776 \s\mathfrak{s} + 576~,\\
    P_{6} &= -3 \s\mathfrak{s}^{5} - 25 \s\mathfrak{s}^{4} - 96 \s\mathfrak{s}^{3} - 144\s\mathfrak{s}^{2} - 72 \s\mathfrak{s}~,
\end{align}
and $\mathfrak{s}$ is defined such that $s \equiv s_{\text{min}} (1 + \mathfrak{s})$. In this case, $s_{\text{min}} = 4 m_{n}^{2} = 4n^{2}/r_{c}^{2}$. The multiplicative $\csc^{2}\theta$ factor in Eq. \eqref{M5DOTtotal} is indicative of $t$- and $u$-channel divergences from the exchange of the massless graviton and radion, which introduces divergences at $\theta = 0,\pi$. Such IR divergences prevent us from directly using a partial wave analysis to determine the strong coupling scale of this theory. In order to characterize the strong coupling scale of this theory, we must instead investigate a nonelastic scattering channel for which  KK momentum conservation implies that no massless states can contribute, $\mathcal{M}_{0} = \mathcal{M}_{r} = 0$. (In this case, the $\csc^{2}\theta$ factor present in Eq. $\eqref{M15DOT}$ is an artifact of the high-energy expansion and is absent from the full matrix element.) 

Consider for example the helicity-zero 5DOT process $(1,4)\rightarrow (2,3)$.  The total matrix element is computed from four diagrams
\begin{align}
    \raisebox{-0.45\height}{\sgDA{0.20}}+\raisebox{-0.45\height}{\sDA{0.20}}+\raisebox{-0.45\height}{\tDA{0.20}}+\raisebox{-0.45\height}{\uDA{0.20}}
\end{align}
which together yield, after explicit computation,
\begin{align}
    \mathcal{M}= \dfrac{ \kappa^{2} \mathfrak{s} }{12800 \pi r_{c}^{3} (\mathfrak{s} + 1)^{2} Q_{+} Q_{-}} \sum_{i=0}^{4} Q_{i} c_{i\theta}~,
\end{align}
where
\begin{align}
    Q_{\pm} &= 25(\mathfrak{s} + 1) \pm \left[3 + \sqrt{(25\s\mathfrak{s} + 16)(25\s\mathfrak{s}+24)} \cos\theta\right]~,\\
    Q_{0} &= 15\left(2578125\s \mathfrak{s}^{4} + 9437500\s\mathfrak{s}^{3} + 12990000\s \mathfrak{s}^{2} + 7971000\s \mathfrak{s} + 1840564\right)~,\\
    Q_{1} &= 72 \sqrt{(25\s\mathfrak{s}+16)(25\s\mathfrak{s}+24)}(50\s\mathfrak{s}+43)(50\s\mathfrak{s}+47)~,\\
    Q_{2} &= 4\left(2734375\s \mathfrak{s}^{4} + 11562500\s\mathfrak{s}^{3} + 18047500\s \mathfrak{s}^{2} + 12340500\s \mathfrak{s} + 3121692\right)~,\\
    Q_{3} &= 24 \sqrt{(25\s\mathfrak{s}+16)(25\s\mathfrak{s}+24)}(50\s\mathfrak{s}+51)(50\mathfrak{s}+59)~,\\
    Q_{4} &= 390625\s\mathfrak{s}^{4} + 2187500\s\mathfrak{s}^{3} + 4360000\s\mathfrak{s}^{2} +3729000\s\mathfrak{s} + 1165956~,
\end{align}
and $s_{\text{min}} = 25/r_{c}^{2}$. As expected, unlike the elastic 5DOT matrix element (\ref{M5DOTtotal}), the $(1,4)\rightarrow(2,3)$ 5DOT matrix element is finite at $\theta =0,\pi$.

Given a 2-to-2 scattering process with helicities $(\lambda_{1},\lambda_{2})$ $\rightarrow$ $(\lambda_{3},\lambda_{4})$, the corresponding partial wave amplitudes $a^{J}$ are defined as \cite{Jacob:1959at}
\begin{align}
a^J(s) = \dfrac{1}{64 \pi^2} \int d\Omega\hspace{10 pt} {\mathcal{D}}_{\lambda_i,\lambda_f}^{J}(\phi,\theta)\, \mathcal{M}(s,\phi,\theta)~, \label{563zzzz}
\end{align}
where $\lambda_{i} = \lambda_{1} - \lambda_{2}$ and $\lambda_{f} = \lambda_{3} - \lambda_{4}$, $d\Omega = d(\cos\theta)\s d\phi$, and the Wigner D-matrix $\mathcal{D}^{J}_{\lambda_{a},\lambda_{b}}$ is normalized according to
\begin{align}
    \int d\Omega \hspace{10 pt} \mathcal{D}^J_{\lambda_{a},\lambda_{b}}(\phi,\theta) \cdot \mathcal{D}^{J^\prime *}_{\lambda_{a}^{\prime},\lambda_{b}}(\phi,\theta) = \dfrac{4\pi}{2J+1}\cdot  \delta_{J,J^{\prime}}\cdot \delta_{\lambda_{a},\lambda^{\prime}_{a}}~.
\end{align}
Because $(1,4)\rightarrow(2,3)$ is an inelastic process, its partial wave amplitude is constrained by unitarity to satisfy
\begin{align}
    \beta_{14}\beta_{23}\hspace{5 pt} |a^{J}(s)|^{2} \leq \frac{1}{4}~,
\end{align}
where
\begin{align}
    \beta_{jk} \equiv \dfrac{1}{s} \sqrt{\bigg[s-(m_{j}-m_{k})^{2}\bigg]\,\bigg[s-(m_{j}+m_{k})^{2}\bigg]}
\end{align}
The leading partial wave amplitude of the $(1,4)\rightarrow(2,3)$ helicity-zero  5DOT matrix element corresponds to $J=0$, and has leading term
\begin{align}
    a^{0} \simeq \dfrac{s}{16\pi M_{\text{Pl}}^{2}}\s \ln\left(\dfrac{s}{s_{\text{min}}}\right)~.
\end{align}
Hence, this matrix element violates unitarity when $|a^0| \simeq 1/2$, or equivalently when the value of $E\equiv \sqrt{s}$ is near or greater than $\Lambda_{\text{strong}}^{(\text{5DOT})} \equiv \sqrt{8\pi} M_{\text{Pl}}$.\footnote{In \cite{Chivukula:2020hvi}, there was a missing relative factor of two between the definition of $a^{J}(s)$ and the partial wave amplitude unitarity constraints, thus yielding $\Lambda_{\text{strong}}^{(\text{5DOT})} \equiv \sqrt{4\pi} M_{\text{Pl}}$ instead of $\Lambda_{\text{strong}}^{(\text{5DOT})} \equiv \sqrt{8\pi} M_{\text{Pl}}$. This has been corrected in this dissertation.} Because $M_{\text{Pl}}$ labels the reduced Planck mass, $\Lambda_{\text{strong}}^{(\text{5DOT})}$ equals the conventional Planck mass. We will use this inelastic calculation as a benchmark for estimating the strong coupling scale associated with other processes. 

We now consider the behavior of scattering amplitudes in the RS1 model.

\section[Elastic Scattering in the Randall-Sundrum 1 Model]{Elastic Scattering in the Randall-Sundrum 1 Model\footnote{The section description comes from Section V of \cite{Chivukula:2020hvi}. The section content comprises Subsections V.B through V.G of \cite{Chivukula:2020hvi} with some modification to update notation and utilize the new expressions of the sum rules from the previous chapter.}} \label{sec:level5}

This section discusses the computation of the elastic scattering amplitudes of massive spin-2 KK modes in the RS1 model, for arbitrary values of the curvature of the internal space. For any nonzero curvature, every KK mode in the infinite tower contributes to each scattering process and the cancellation from $\mathcal{O}(s^5)$ to $\mathcal{O}(s)$ energy growth only occurs when all of these states are included.

In the subsequent subsections, we apply the sum rules to determine the leading high-energy behavior of the amplitudes for helicity-zero $(n,n)\rightarrow (n,n)$ scattering of KK modes.  Finally, Sec. \ref{sec:nonlongitudinal} extends this analysis to all other helicity combinations of $(n,n)\rightarrow (n,n)$ KK mode scattering.

\subsection{Cancellations at $\mathcal{O}(s^{5})$ in RS1} \label{ElasticCancellationsOs5}
We will now go order by order in powers of $s$ through the contributions to the helicity-zero $(n,n)\rightarrow (n,n)$ scattering amplitude in the RS1 model, and apply the sum rules derived in the previous chapter. When reporting contributions, all spin-2 exclusive B-type couplings $b_{\vec{n}}$ couplings have already been converted into spin-2 exclusive A-type $a_{\vec{n}}$ couplings via Eq. \eqref{eq4152}.

As described in Subsection \ref{SS - Defs - 5D vs 4D}, the contact diagram and spin-2-mediated diagrams individually diverge like $\mathcal{O}(s^{5})$. Their contributions to the elastic helicity-zero RS1 matrix element equal
\begin{align}
    \overline{\mathcal{M}}_{c}^{(5)} &=-\dfrac{\kappa^{2}\s a_{nnnn}}{2304\s\pi r_{c}\s m_{n}^{8}}\left[7+\cos(2\theta)\right]\sin^{2}\theta~,\\
    \overline{\mathcal{M}}_{j}^{(5)} &= \dfrac{\kappa^{2} \s a_{nnj}^{2}}{2304\s\pi r_{c}\s m_{n}^{8}}\left[7+\cos(2\theta)\right]\sin^{2}\theta~,
\end{align}
such that they sum to
\begin{align}
    \overline{\mathcal{M}}^{(5)}&= \dfrac{\kappa^{2}\left[7+\cos(2\theta)\right]\sin^{2}\theta}{2304\s \pi r_{c} \s m_{n}^{8}}\left\{ \sum_{j=0}^{+\infty} a_{nnj}^{2} - a_{nnnn}\right\}~. \label{ELM5RS1}
\end{align}
This vanishes via Eq. \eqref{SumRuleO5}, which we will, henceforth, refer to as the $\mathcal{O}(s^{5})$ sum rule.

\subsection{Cancellations at $\mathcal{O}(s^{4})$ in RS1} \label{ElasticCancellationsOs4}

The $\mathcal{O}(s^{4})$ contributions to the elastic helicity-zero RS1 matrix element equal
\begin{align}
    \overline{\mathcal{M}}_{c}^{(4)} &=\dfrac{\kappa^{2}\s a_{nnnn}}{6912\s\pi r_{c}\s m_{n}^{6}}\left[63-196\cos(2\theta)+5\cos(4\theta)\right]~,\\
    \overline{\mathcal{M}}_{j}^{(4)} &= -\dfrac{\kappa^{2}\s a_{nnj}^{2}}{9216\s\pi r_{c}\s m_{n}^{6}}\bigg\{\left[7+\cos(2\theta)\right]^{2}\dfrac{m_{j}^{2}}{m_{n}^{2}} +2\left[9-140\cos(2\theta)+3\cos(4\theta)\right]\bigg\}~.
\end{align}
Using the $\mathcal{O}(s^{5})$ sum rule, $\overline{\mathcal{M}}^{(4)}$ equals
\begin{align}
    \overline{\mathcal{M}}^{(4)}&= \dfrac{\kappa^{2}\left[7+\cos(2\theta)\right]^{2}}{9216\s \pi r_{c} \s m_{n}^{6}}\left\{ \dfrac{4}{3}\, a_{nnnn} - \sum_{j} \dfrac{m_{j}^{2}}{m_{n}^{2}} \, a_{nnj}^{2}\right\}~. \label{ELM4RS1}
\end{align}
This vanishes via Eq. \eqref{SumRuleO4}, which we shall refer to as the $\mathcal{O}(s^{4})$ sum rule.

\subsection{Cancellations at $\mathcal{O}(s^{3})$ in RS1} \label{ElasticCancellationsOs3}

Once the $\mathcal{O}(s^{5})$ and $\mathcal{O}(s^{4})$ contributions are cancelled, the radion-mediated diagrams, which diverge like $\mathcal{O}(s^{3})$, become relevant to the leading behavior of the elastic helicity-zero RS1 matrix element. Furthermore, because of differences between the massless and massive spin-2 propagators, $\overline{\mathcal{M}}_{0}$ and $\overline{\mathcal{M}}_{j>0}$ differ from one another at this order and lower. The full set of relevant contributions is therefore
\begin{align}
    \overline{\mathcal{M}}^{(3)}_{c} &= \dfrac{\kappa^{2}\s a_{nnnn}}{3456\s \pi r_{c}\s m_{n}^{4}} \left[-185+692\cos(2\theta)+5\cos(4\theta)\right]~,\\
    \overline{\mathcal{M}}^{(3)}_{r} &= -\dfrac{\kappa^{2}}{32 \s \pi r_{c} \s m_{n}^{4}}\left[\dfrac{b_{n^{\prime}n^{\prime}r}^{2}}{(m_{n}r_{c})^{4}}\right] \sin^{2}\theta~,\\
    \overline{\mathcal{M}}^{(3)}_{0} &= \dfrac{\kappa^{2}\s a_{nn0}^{2}}{1152\s \pi r_{c}\s m_{n}^{4}} \left[15-270\cos(2\theta)-\cos(4\theta)\right]~,\\
    \overline{\mathcal{M}}^{(3)}_{j>0} &= \dfrac{\kappa^{2}\s a_{nnj}^{2}}{2304\s \pi r_{c}\s m_{n}^{4}} \bigg\{5\left[1-\cos(2\theta)\right] \dfrac{m_{j}^{4}}{m_{n}^{4}}+\left[69+60\cos(2\theta)-\cos(4\theta)\right]\dfrac{m_{j}^{2}}{m_{n}^{2}}\nonumber\\
    &\hspace{10 pt}+2\left[13-268\cos(2\theta)-\cos(4\theta)\right]\bigg\}~,
\end{align}
After applying the $\mathcal{O}(s^{5})$ and $\mathcal{O}(s^{4})$ sum rules, $\overline{\mathcal{M}}^{(3)}$ equals
\begin{align}
    \overline{\mathcal{M}}^{(3)} &= \dfrac{5\s\kappa^{2}\s \sin^{2}\theta}{1152 \s \pi r_{c} \s m_{n}^{4}}\bigg\{ \sum_{j} \dfrac{m_{j}^{4}}{m_{n}^{4}} \, a_{nnj}^{2} - \dfrac{16}{15} \, a_{nnnn} -\dfrac{4}{5}\left[ \dfrac{9\s b_{n^{\prime}n^{\prime}r}^{2}}{(m_{n}r_{c})^{4}} - a_{nn0}^{2}\right]\bigg\}~. \label{ELM3RS1}
\end{align}
These contributions cancel if and only if the following $\mathcal{O}(s^{3})$ sum rule holds true:
\begin{align}
    \sum_{j=0}^{+\infty} \mu_{j}^{4}\,a_{nnj}^{2} = \dfrac{16}{15}\,\mu_{n}^{4}\, a_{nnnn} + \dfrac{4}{5}\bigg[9\,b_{n^{\prime}n^{\prime}r}^{2} - \mu_{n}^{4}\, a_{nn0}^{2}\bigg]\label{SumRuleOs3} 
\end{align}
We do not yet have an analytic proof of this sum rule; however we have verified that the right-hand side numerically approaches the left-hand side as the maximum intermediate KK number $N_{\text{max}}$ is increased to $100$ for a wide range of values of $kr_{c}$, including $kr_{c} \in \{10^{-3},10^{-2},10^{-1},1,2,\ldots,10\}$.\footnote{The cancellations implied by this sum rule correspond to the vanishing of ${\mathcal R}^{[N](3)}$ in Fig. \ref{fig:cancel1111} as $N$ increases.}

The $\mathcal{O}(s^{3})$ sum rule may also be written as
\begin{align}
    3\bigg[9b_{n^{\prime}n^{\prime}r}^{2} - \mu_{n}^{4} \,a_{nn0}^{2}\bigg] &= 15 \,c_{n^{\prime}n^{\prime}n^{\prime}n^{\prime}} + \mu_{n}^{4}\,a_{nnnn} \label{SumRuleOs3b} 
\end{align}
by applying Eq. \eqref{SumRuleO3x} to Eq. \eqref{SumRuleOs3}.

\subsection{Cancellations at $\mathcal{O}(s^{2})$ in RS1} \label{ElasticCancellationsOs2}
The contributions to the elastic helicity-zero matrix element at $\mathcal{O}(s^{2})$ equal
\begin{align}
    \overline{\mathcal{M}}_{c}^{(2)} &= -\dfrac{\kappa^{2}\s a_{nnnn}}{54\s \pi r_{c} \s m_{n}^{2}}\left[5 +47 \cos(2\theta)\right]~,\\
    \overline{\mathcal{M}}_{r}^{(2)} &= \dfrac{\kappa^{2}}{48\s \pi r_{c} \s m_{n}^{2}}\left[\dfrac{b_{n^{\prime}n^{\prime}r}^{2}}{(m_{n}r_{c})^{4}}\right]\left[7 + \cos(2\theta)\right]~,\\
    \overline{\mathcal{M}}_{0}^{(2)} &= \dfrac{\kappa^{2}\s a_{nn0}^{2}}{576 \s \pi r_{c} \s m_{n}^{2}} \left[175 + 624 \cos(2\theta) + \cos(4\theta)\right]~,\\
    \overline{\mathcal{M}}_{j>0}^{(2)} &= \dfrac{\kappa^{2}\s a_{nnj}^{2}}{6912 \s \pi r_{c}\s m_{n}^{2}} \bigg\{4\left[7+\cos(2\theta)\right]\left[5 - 2\dfrac{m_{j}^{2}}{m_{n}^{2}}\right] \dfrac{m_{j}^{4}}{m_{n}^{4}}\nonumber\\
    &\hspace{10 pt}-\left[1291+1132\cos(2\theta)+9\cos(4\theta)\right] \dfrac{m_{j}^{2}}{m_{n}^{2}}\nonumber\\
    &\hspace{10 pt}+4\left[553+1876\cos(2\theta)+3\cos(4\theta)\right]\bigg\}~.
\end{align}
By applying the $\mathcal{O}(s^{5})$ and $\mathcal{O}(s^{4})$ sum rules (but {\it not} the $\mathcal{O}(s^{3})$ sum rule), the total $\mathcal{O}(s^{2})$ contribution equals
\begin{align}
    \overline{\mathcal{M}}^{(2)} &= \dfrac{\kappa^{2}\s \left[7+\cos(2\theta)\right]}{864\s \pi r_{c} \s m_{n}^{2}}\bigg\{\sum_{j}\left[\dfrac{m_{j}^{2}}{m_{n}^{2}} - \dfrac{5}{2}\right]\dfrac{m_{j}^{4}}{m_{n}^{4}} a_{nnj}^{2}+\dfrac{8}{3}a_{nnnn} -2\left[ \dfrac{9\s b_{n^{\prime}n^{\prime}r}^{2}}{(m_{n}r_{c})^{4}} - a_{nn0}^{2}\right]\bigg\}~, \label{ELM2RS1}
\end{align}
which vanishes if and only if the following $\mathcal{O}(s^{2})$ sum rule holds:
\begin{align}
    \sum_{j=0}^{+\infty} \left[ \mu_{j}^{2} - \dfrac{5}{2}\mu_{n}^{2}\right]\mu_{j}^{4}\, a_{nnj}^{2} &= -\dfrac{8}{3}\,\mu_{n}^{6}\, a_{nnnn} + 2\,\mu_{n}^{2}\bigg[9b_{n^{\prime}n^{\prime}r}^{2} - \mu_{n}^{4}\, a_{nn0}^{2}\bigg]~.\label{SumRuleOs2}
\end{align}
Again, we do not yet have a proof for this sum rule, despite strong numerical evidence that it is correct (refer to Sec. \ref{sec:level6}). However, combining the $\mathcal{O}(s^{3})$ and $\mathcal{O}(s^{2})$ sum rules (Eqs. \eqref{SumRuleOs3} and \eqref{SumRuleOs2}), yields an equivalent set of rules
\begin{align}
    \sum_{j=0}^{+\infty}\left[\mu_{j}^{2} - 5 \mu_{n}^{2} \right] \mu_{j}^{4} \,a_{nnj}^{2} &= -\dfrac{16}{3}\,\mu_{n}^{6} \,a_{nnnn}~,\label{SumRuleOsA}\\
    3\bigg[9b_{n^{\prime}n^{\prime}r}^{2} - \mu_{n}^{4}\, a_{nn0}^{2}\bigg] &= 15\, c_{n^{\prime}n^{\prime}n^{\prime}n^{\prime}} + \mu_{n}^{4}\, a_{nnnn}~.\label{SumRuleOsB}
\end{align}
where Eq. \eqref{SumRuleOsA} is precisely Eq. \eqref{SumRule02xO3x} (which we proved in Section \ref{DerivingSumRules}) and Eq. \eqref{SumRuleOsB} is Eq.\eqref{SumRuleOs3b} again. Therefore, if the $\mathcal{O}(s^{3})$ sum rule holds true, then the $\mathcal{O}(s^{2})$ sum rule must also hold true, and vice versa. Of the relations necessary to ensure cancellations, only Eq. \eqref{SumRuleOsB} remains unproven.

Finally, we note that the sum rules we have derived for the RS1 model in Eqs. (\ref{SumRuleO5}), (\ref{SumRuleO4}), (\ref{SumRuleOs3}), and (\ref{SumRuleOs2}),
are consistent with those inferred by the authors of \cite{Bonifacio:2019ioc} when they assumed helicity-zero massive spin-2 mode scattering amplitudes in KK theories will ultimately grow like $\mathcal{O}(s)$. A description of the correspondence of our results with theirs is given in Appendix E of \cite{Chivukula:2020hvi}.

\subsection{The Residual $\mathcal{O}(s)$ Amplitude in RS1} \label{ElasticCancellationsOs1}

After applying all the sum rules above\footnote{The elastic 5D Orbifolded Torus couplings (\ref{eq:orbifold-couplings}) directly satisfy all of these sum rules.} (including Eq. \eqref{SumRuleOsB}, which lacks an analytic proof), the leading contribution to the elastic helicity-zero matrix element is found to be $\mathcal{O}(s)$. The relevant contributions, sorted by diagram type, equal
\begin{align}
    \overline{\mathcal{M}}_{c}^{(1)} &= \dfrac{\kappa^{2}\s a_{nnnn}}{1728\s \pi r_{c}}\left[1505 + 3108 \cos(2\theta) -5 \cos(4\theta)\right]~,\\
    \overline{\mathcal{M}}_{r}^{(1)} &= -\dfrac{\kappa^{2}}{24\s \pi r_{c}}\left[\dfrac{b_{n^{\prime}n^{\prime}r}^{2}}{(m_{n}r_{c})^{4}}\right]\left[9 + 7\cos(2\theta)\right]~,\\
    \overline{\mathcal{M}}_{0}^{(1)} &= \dfrac{\kappa^{2}\s a_{nn0}^{2}\s \csc^{2}\theta}{2304 \s \pi r_{c}} \left[748 + 427 \cos(2\theta)+ 1132\cos(4\theta)-3\cos(6\theta)\right]~,\\
    \overline{\mathcal{M}}_{j>0}^{(1)} &= \dfrac{\kappa^{2}\s a_{nnj}^{2} \s \csc^{2}\theta}{6912 \s \pi r_{c}} \bigg\{3\left[7+\cos(2\theta)\right]^{2}\dfrac{m_{j}^{8}}{m_{n}^{8}}-4\left[241+148\cos(2\theta)-5\cos(4\theta)\right]\dfrac{m_{j}^{6}}{m_{n}^{6}}\nonumber\\
    &\hspace{35 pt}+4\left[787+604\cos(2\theta)-47\cos(4\theta)\right]\dfrac{m_{j}^{4}}{m_{n}^{4}}\nonumber\\
    &\hspace{35 pt}-\left[3854+5267\cos(2\theta)+98\cos(4\theta)-3\cos(6\theta)\right]\dfrac{m_{j}^{2}}{m_{n}^{2}}\nonumber\\
    &\hspace{35 pt}+\left[2156+1313\cos(2\theta)+3452\cos(4\theta)-9\cos(6\theta)\right]\bigg\}~.
\end{align}
Combining them according to Eq. \eqref{eq:Mtot-def} yields
\begin{align}
    \overline{\mathcal{M}}^{(1)} &= \dfrac{\kappa^{2}\s \left[7+\cos(2\theta)\right]^{2}\csc^{2}\theta}{2304\s \pi r_{c}}\bigg\{\sum_{j}\dfrac{m_{j}^{8}}{m_{n}^{8}} a_{nnj}^{2}+\dfrac{28}{15} a_{nnnn} - \dfrac{48}{5}\left[ \dfrac{9\s b_{n^{\prime}n^{\prime}r}^{2}}{(m_{n}r_{c})^{4}} - a_{nn0}^{2}\right]\bigg\}~.\label{ELM1RS1}
\end{align}
This is generically nonzero, and thus represents the true leading high-energy behavior of the elastic helicity-zero RS1 matrix element.

\begin{figure*}[t]
\includegraphics[scale=0.144]{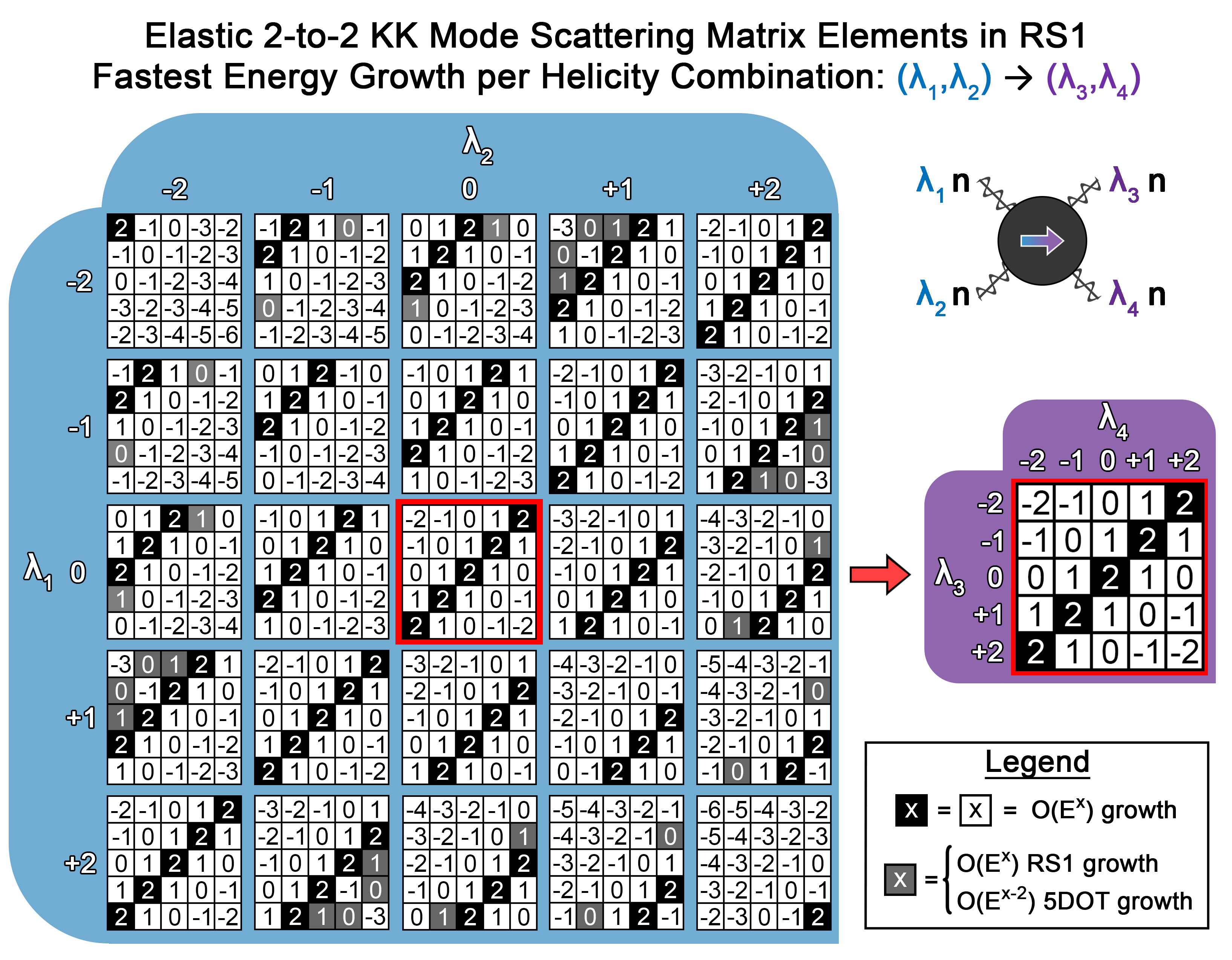}
 \caption{This table-of-tables gives the leading order (in energy) growth of elastic $(n,n) \to (n,n)$ scattering for different incoming ($\lambda_{1,2}$) and outgoing ($\lambda_{3,4}$) helicity combinations in RS1. In the cases listed in grey, the leading order behavior is softer in the orbifolded torus limit (by two powers of center-of-mass energy).}
\label{fig:tableoftables}
\end{figure*}

\subsection{\label{sec:nonlongitudinal}Other Helicity Combinations} \label{ElasticCancellationsNonlongitudinal}

The sum rules of the previous subsections were derived by considering what cancellations were necessary to ensure the elastic helicity-zero RS1 matrix element grew no faster than $\mathcal{O}(s)$, a constraint which in turn comes from considering the spontaneous symmetry breaking of extra-dimensional physics. This bound on high-energy growth must hold for scattering of all helicities, and---indeed---upon studying the nonzero-helicity scattering amplitudes, we find that the sum rules derived in the helicity-zero case are sufficient to ensure {\it all} elastic RS1 matrix elements grow at most like $\mathcal{O}(s)$.

Figure \ref{fig:tableoftables} lists the leading high-energy behavior of the elastic RS1 matrix element for each helicity combination after  the sum rules have been applied. These results are expressed in terms of the leading exponent of incoming energy $E\equiv \sqrt{s}$. For example, the elastic helicity-zero matrix element diverges like $\mathcal{O}(s)=\mathcal{O}(E^{2})$ and so its growth is recorded as ``2" in the table. As expected, no elastic RS1 matrix element grows faster than $\mathcal{O}(E^{2})$.

Some matrix elements grow more slowly with energy in the 5DOT model than they do in the more general RS1 model; they are indicated by the gray boxes in Fig. \ref{fig:tableoftables}. For these instances, the leading $\mathcal{M}^{(\sigma)}$ contribution in the RS1 model is always proportional to the same combination of couplings
\begin{align}
    \left[3\, a_{nn0}^{2} + 16\, a_{nnnn}\right] \mu_{n}^{4} - 27\s b_{n^{\prime}n^{\prime}r}^{2}~,
\end{align}
which vanishes exactly when $kr_{c}$ vanishes. Regardless of the specific helicity combination considered, no full matrix element vanishes.

\begin{figure*}[t]
\includegraphics[width=\linewidth]{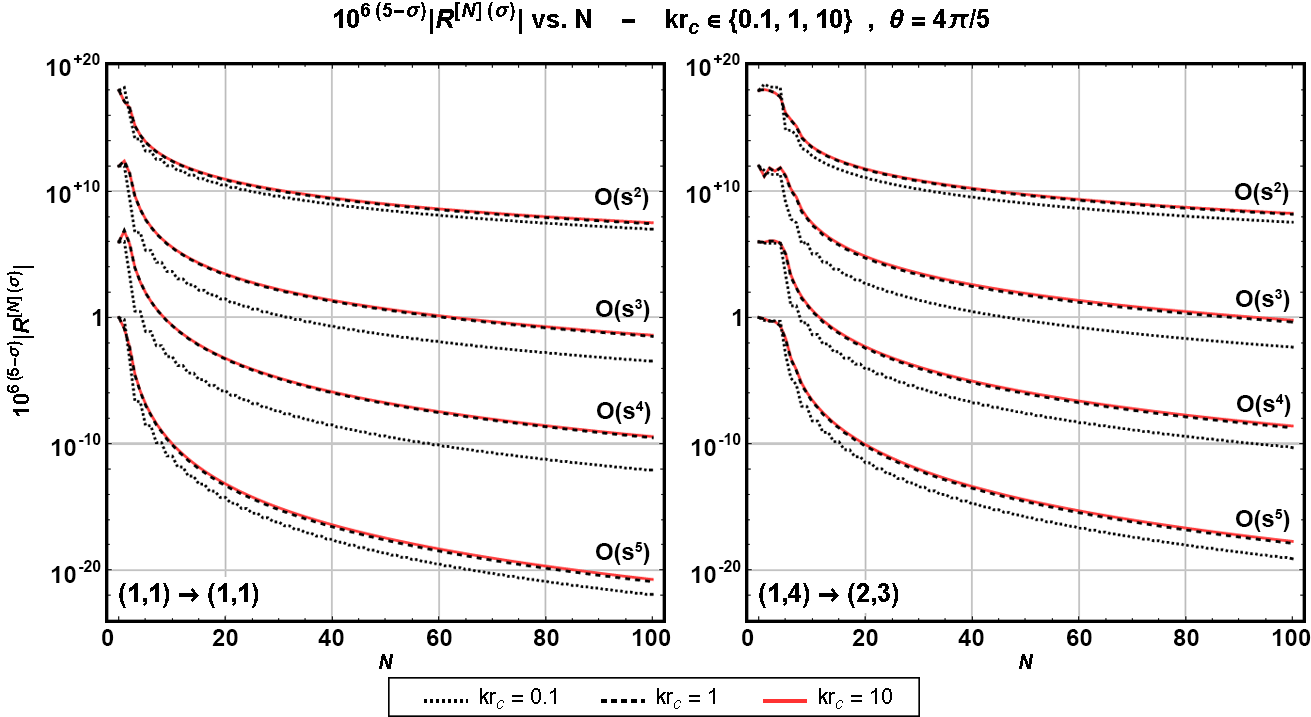}
\caption{This figure plots the ratio $\mathcal{R}^{[N](\sigma)}(kr_{c},\theta) = \mathcal{M}^{[N](\sigma)}/\mathcal{M}^{[0](\sigma)}$ (defined in Eq. \eqref{eq:defR}), where $\mathcal{M}^{[N](\sigma)}$ is the $\mathcal{O}(s^{\sigma})$ contribution to the matrix element describing helicity-zero scattering of KK modes $(1,1) \to (1,1)$ (left) and $(1,4) \to (2,3)$ (right) as a function of the number of KK intermediate states included in the calculation ($N$). The curves are drawn for $kr_c=0.1,\, 1,\, 10$ at fixed $\theta=4\pi/5$. In all cases, the remaining matrix element falls rapidly with the addition of more intermediate states, thereby demonstrating the cancellation of all high-energy growth faster than $\mathcal{O}(s)$. To visually separate the different curves, the value of the ratio at $N=0$ has been artificially normalized to $(1,10^6,10^{12},10^{18})$ for $\sigma=5,4,3,2$ respectively.}
\label{fig:cancel1111}
\end{figure*}

\section[Numerical Study of Scattering Amplitudes in the Randall-Sundrum 1 Model]{Numerical Study of Scattering Amplitudes in the Randall-Sundrum 1 Model \footnote{The content of this section was originally published as Section VI and Appendix F.3 of \cite{Chivukula:2020hvi}, up to minor changes in wording and notation.}} \label{sec:level6}

This section presents a detailed numerical analysis of the scattering in the RS1 model. In Sec. \ref{sec:level6a} we demonstrate that the cancellations demonstrated for elastic scattering occur for inelastic scattering channels as well, with the cancellations becoming exact as the number of included intermediate KK modes increases. In Sec. \ref{sec:level6b} we examine the truncation error arising from keeping only a finite number of intermediate KK mode states. We then return in Sec. \ref{sec:level6c} to the question of the validity of the KK mode EFT. In particular, we demonstrate directly from the scattering amplitudes that the cutoff scale is proportional to the RS1 emergent scale \cite{ArkaniHamed:2000ds,Rattazzi:2000hs}
\begin{align}
    \Lambda_\pi = M_{\text{Pl}}\, e^{-k\pi r_c}~,
\end{align}
which is related to the relative locations of branes \cite{Randall:1999ee,Randall:1999vf}.

\subsection{\label{sec:level6a}Numerical Analysis of Cancellations in Elastic \& Inelastic Scattering Amplitudes}

We have demonstrated that the elastic scattering amplitudes in the Randall-Sundrum model grow only as ${\cal O}(s)$ at high energies, and have analytically derived the sum rules which enforce these cancellations. Physically, we expect similar cancellations and sum rules apply for arbitrary inelastic scattering amplitudes as well. However, we have not yet found an analytic derivation of this property.\footnote{This is to be contrasted with the situation for KK compactifications on Ricci-flat manifolds, where an analytic demonstration of the needed cancellations has been found \cite{Bonifacio:2019ioc}.}

Instead, we demonstrate here numerical checks with which we observe behavior consistent with the expected cancellations. To do so, we must first rewrite our expressions so we may vary $kr_c$ while keeping $M_{\text{Pl}}$ and $m_1$ fixed. We do so by noting that we may rewrite the common matrix element prefactor as
\begin{align}
    \dfrac{\kappa^{2}}{\pi r_{c}} = \dfrac{\kappa_{\text{4D}}^{2}}{{\psi_{0}}^{2}} = \dfrac{1}{\pi kr_{c}}\left[1- e^{-2kr_{c}\pi}\right]\dfrac{4}{M_{\text{Pl}}^{2}}~,
\end{align}
and that $r_{c} = \mu_{1}/m_{1}$, such that $\mathcal{M}^{(\sigma)}$ can be factorized for any process (and any helicity combination) into three unitless pieces, each of which depends on a different independent parameter:
\begin{align}
    \mathcal{M}^{(\sigma)} \equiv   \left[\mathcal{K}^{(\sigma)}(kr_{c},\theta)\right] \cdot \left[\dfrac{s}{M_{\text{Pl}}^{2}}\right] \cdot \left[\dfrac{\sqrt{s}}{m_{1}}\right]^{2(\sigma-1)}~. \label{MmsE}
\end{align}
This defines the dimensionless quantity $\mathcal{K}^{(\sigma)}$ (in the first square brackets) characterizing the residual growth of order $(\sqrt{s})^{2\sigma}$ in any scattering amplitude. We can apply this decomposition to the truncated matrix element contribution $\mathcal{M}^{[N](\sigma)}$ defined in Eq. (\ref{eq:M-N-def}) as well. By comparing $\mathcal{M}^{[N](\sigma)}$ to $\mathcal{M}^{[0](\sigma)}$ and increasing $N$ when $\sigma > 1$, we can measure how cancellations are improved by including more KK states in the calculation and do so in a way that depends only on $kr_{c}$ and $\theta$. Therefore, we define
\begin{align}
    \mathcal{R}^{[N](\sigma)}(kr_{c},\theta) \equiv \dfrac{\mathcal{M}^{[N](\sigma)}}{\mathcal{M}^{[0](\sigma)}} = \dfrac{\mathcal{K}^{[N](\sigma)}}{\mathcal{K}^{[0](\sigma)}}~,
    \label{eq:defR}
\end{align}
which vanishes as $N\rightarrow +\infty$ if and only if $\mathcal{M}^{[N](\sigma)}$ vanishes as $N\rightarrow +\infty$. Because $\mathcal{R}^{[N](\sigma)}$ depends continuously on $\theta$, we expect that so long as we choose a value of $\theta$ such that $\mathcal{K}^{[N](\sigma)}\neq 0$, its exact value is unimportant to confirming cancellations. Figure \ref{fig:cancel1111} plots $10^{6(5-\sigma)}\mathcal{R}^{[N_{\text{max}}](\sigma)}$ for the helicity-zero processes $(1,1)\rightarrow(1,1)$ and $(1,4)\rightarrow(2,3)$ as functions of $N_{\text{max}}\rightarrow 100$  for $kr_{c}\in \{10^{-1},1,10\}$ and $\theta=4\pi/5$. The factor of $10^{6(5-\sigma)}$ only serves to vertically separate the curves for the reader's visual convenience; without this factor, the curves would all begin at $\mathcal{R}^{[0](\sigma)}=1$ and thus overlap substantially.

We find that, both for the case of elastic scattering $(1,1) \to (1,1)$ where we have an analytic demonstration of the cancellations and for the inelastic case $(1,4) \to (2,3)$ where we do not, $\mathcal{M}^{[N](\sigma)} \to 0$ as $N\to \infty$. Furthermore, we find that the rate of convergence is similar in the two cases. In addition, and perhaps more surprisingly, the rate of convergence is relatively independent of the value of $kr_c$ for values between $1/10$ and $10$.

\begin{figure*}[t]
\center
\includegraphics[width=\linewidth]{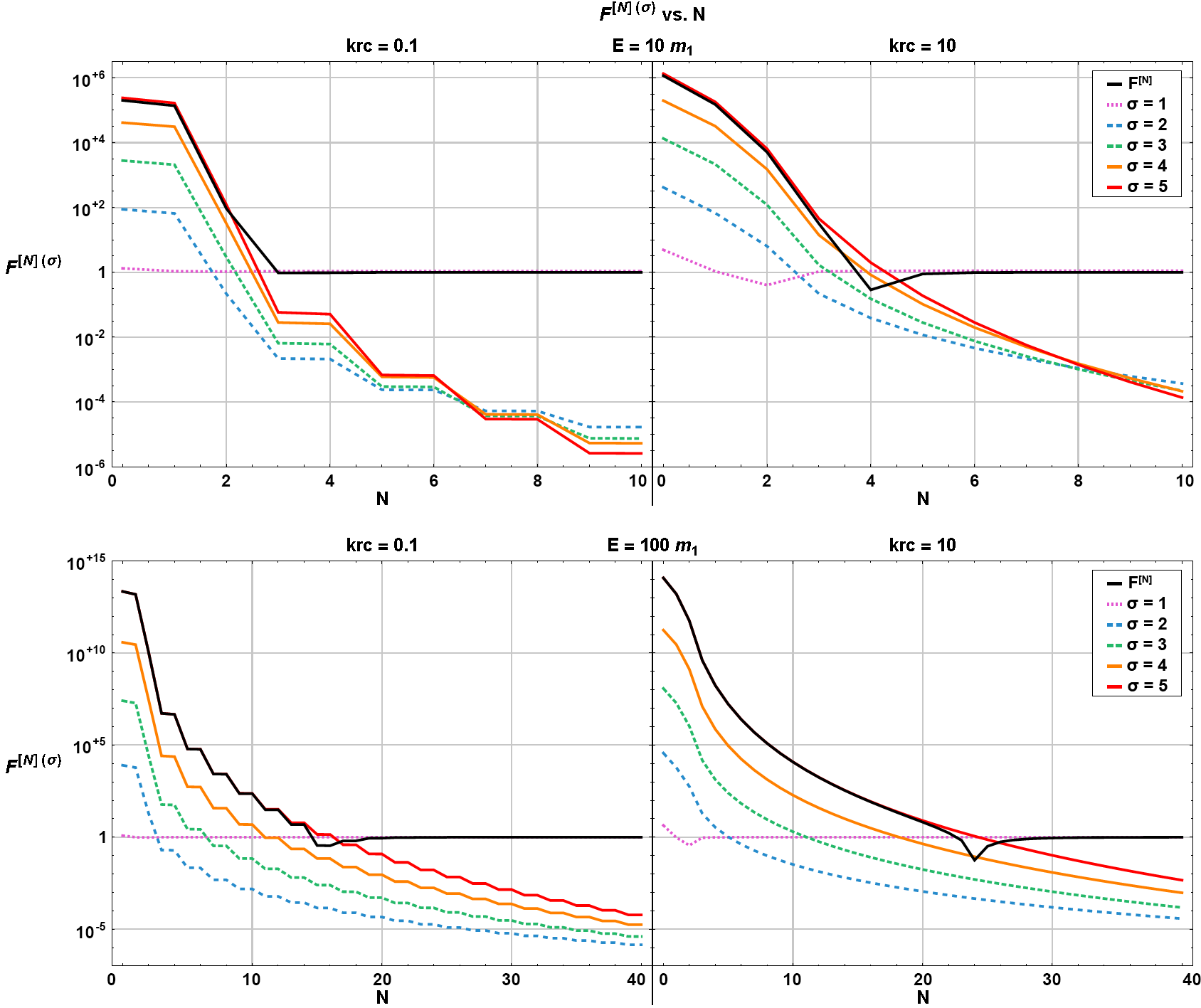}
\caption{This figure plots an upper bound on the size of the residual truncation error relative to the size of the full matrix element for the process $(1,1)\to (1,1)$ as a function of the number of included KK modes $N$, for $E=10m_1$ (upper pair) and $E=100m_1$ (lower pair), and $kr_c=0.1$ (left pair) and $kr_c=10$ (right pair). $\mathcal{F}^{[N](\sigma)}(kr_c,s)$ from Eq. (\ref{eq:defF1}) is drawn in color, for $\sigma=1$ - $5$, and $\mathcal{F}^{[N]}(kr_c,s)$ from Eq. (\ref{eq:defF2}) is drawn in black. We find that the size of the truncation error falls rapidly as the number of included intermediate states $N$ increases. We also find that, for $E \gg m_1$ and with a sufficient number of intermediate states included, $\mathcal{M}^{[N](1)}$ is a good approximation of the full matrix element. Note that if an insufficient number of intermediate KK modes is included, the truncation error is large and $\mathcal{M}^{[N](5)}$ dominates.}
\label{fig:truncation}
\end{figure*}

\subsection{\label{sec:level6b}Truncation Error}

In the RS1 model, the exact tree-level matrix element for any scattering amplitude requires summing over the entire tower of KK states. In practice, of course, any specific calculation will only include a finite number of intermediate states $N$. In this subsection we investigate the size of the ``truncation error" of such a calculation. For simplicity, in this section we will focus on the helicity-zero elastic scattering amplitude $(1,1) \to (1,1)$ and investigate the size of the truncation error for different values of $kr_c$ and center-of-mass scattering energy. 

For $\sigma>1$, consider the ratio
\begin{align}
    {\mathcal F}^{[N](\sigma)}(kr_{c},s)\equiv \max_{\theta \in [0,\pi]}\left|\dfrac{\mathcal{M}^{[N](\sigma)}(kr_{c},s,\theta)}{\mathcal{M}(kr_{c},s,\theta)}\right|~, 
\end{align}

which measures the size of each truncated matrix element contribution relative to the full amplitude.\footnote{In practice, we approximate the ``full" amplitude by  ${\mathcal{M}^{[N=100]}(kr_{c},s,\theta)}$, which we have checked provides ample sufficient numerical accuracy for the quantities reported here.} For sufficiently large $N$ and $\sigma > 1$ we have confirmed numerically that  the ratio $|\mathcal{M}^{[N](\sigma)}/\mathcal{M}^{[N]}|$ reaches a global maximum at $\theta = \pi/2$ for $\sigma > 1$. Therefore
\begin{align}
    {\mathcal F}^{[N](\sigma)}(kr_{c},s) = \left|\dfrac{\mathcal{M}^{[N](\sigma)}(kr_{c},s,\theta)}{\mathcal{M}(kr_{c},s,\theta)}\right|_{\theta = \pi/2}~.
    \label{eq:defF1}
\end{align}

Unlike $\mathcal{M}^{(\sigma)}$ for $\sigma >1$, $\mathcal{M}^{(1)}$ diverges at $\theta \in \{0,\pi\}$ because of a $\csc^{2}\theta$ factor, as indicated in Eq. \eqref{ELM1RS1}, which arises from the $t$- and $u$-channel exchange of light states.\footnote{Formally, the sum over intermediate KK modes in $\overline{\mathcal{M}}^{(1)}$ from Eq. (\ref{ELM1RS1}) extends over all masses, but the couplings $a_{11n}$ vanish as $n$ grows and suppress the contributions from heavy states.} The total elastic RS1 amplitude $\mathcal{M}$, on the other hand, only has such IR divergences due to the exchange of the massless graviton and radion. For this reason, and as confirmed by the numerical evaluation of $\mathcal{M}^{[N](1)}/\mathcal{M}^{[N]}$,  the divergences at $\theta\in\{0,\pi\}$ of $\mathcal{M}^{[N](1)}$  are actually slightly more severe than the corresponding divergences of $\mathcal{M}^{[N]}$, and so the ratio $\mathcal{M}^{[N](1)}/\mathcal{M}^{[N]}$ grows large in the vicinity of $\theta \in \{0,\pi\}$. However, this unphysical divergence is confined to nearly forward or backward scattering; otherwise the ratio is approximately flat. Thus for $\sigma = 1$ we study the analogous quantity
\begin{align}
    \mathcal{F}^{[N](1)}(kr_{c},s) =  \left|\dfrac{\mathcal{M}^{[N](\sigma)}(kr_{c},s,\theta)}{\mathcal{M}(kr_{c},s,\theta)}\right|_{\theta = \pi/2}~.
\end{align}

We also define the overall accuracy of the partial sum over intermediate states using a version of this quantity for which no expansion in powers of energy has been made: 
\begin{align}
    \mathcal{F}^{[N]}(kr_{c},s) \equiv \left|\dfrac{\mathcal{M}^{[N]}(kr_{c},s,\tfrac{\pi}{2})}{\mathcal{M}(kr_{c},s,\tfrac{\pi}{2})}\right|~.
\label{eq:defF2}
\end{align}

Because $\mathcal{F}^{[N](\sigma)}$ ($\mathcal{F}^{[N]}$)  measures the discrepancy between any given contribution $\mathcal{M}^{[N](\sigma)}$ ($\mathcal{M}^{[N]}$) and the full matrix element $\mathcal{M}$, we study these quantities to understand the truncation error. In the upper two panes of Fig. \ref{fig:truncation} we plot these quantities as a function of maximal KK number $N$ for $kr_{c} = 1/10$ and $kr_{c} = 10$ at the representative energy $s=(10 m_{1})^2$, for $m_1=1$ TeV. The lower two panes of Fig. \ref{fig:truncation} plot similar information but at the energy $s=(100 m_{1})^2$. The $kr_{c}=10$ panes contain the more phenomenologically relevant information. In all cases, we find that including sufficiently many modes in the KK tower yields an accurate result for angles away from the forward or backward scattering regime. When including only a small number of modes $N$, the contribution from $\mathcal{M}^{[N](5)}$ (the residual contribution arising from the non-cancellation of the ${\cal O}(s^{5})$ contributions) dominates and the truncation yields an inaccurate result. As one increases the number of included modes, this unphysical ${\cal O}(s^{5})$ contribution to the amplitude falls in size until the full amplitude is dominated by $\mathcal{M}^{[N](1)}$, which is itself a good approximation to the complete tree-level amplitude. For $kr_{c}=1/10$, the number of states $N$ required to reach this ``crossover", however, increases from $3$ to $15$ as $\sqrt{s}$ increases from $10 m_{1}$ to $100 m_{1}$. Consistent with our analysis in the previous subsection, however, the truncation error is less dependent on $kr_{c}$; the number of states required to reach crossover increases by less than a factor of $2$ when moving from $kr_{c}=1/10$ to $kr_{c}=10$ at fixed $\sqrt{s}$. 

Lastly, we note that the vanishing of $\mathcal{F}^{[N](3)}$ as $N$ increases is a numerical test of the $\mathcal{O}(s^3)$ sum rule in Eq. \eqref{SumRuleOs3}. 

\begin{figure*}[t]
\centering
\includegraphics[scale=0.5]{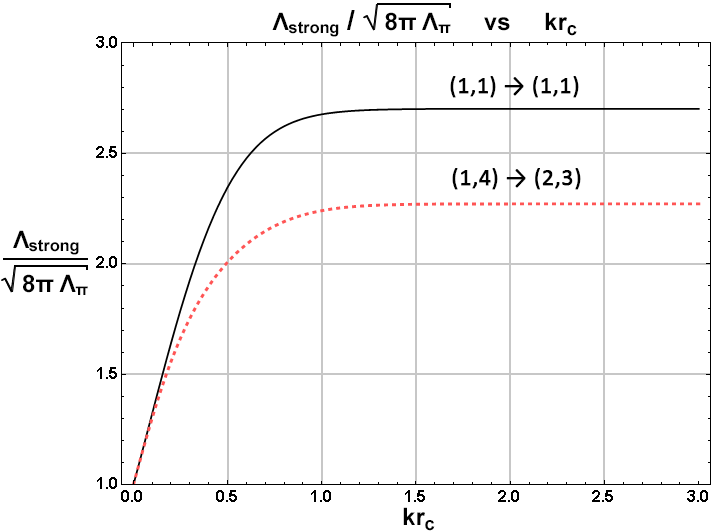}
\caption{The strong coupling scale $\Lambda^{(\text{RS1})}_{\text{strong}}(kr_{c})$, Eq. (\ref{eq:strongscale}), as a function of $kr_c$ for the processes $(1,1)\to (1,1)$ and $(1,4)\to (2,3)$. We find that this scale is comparable to $\sqrt{8\pi}\Lambda_{\pi}$.}
\label{fig:strongcoupling}
\end{figure*}

\subsection{\label{sec:level6c}The Strong Coupling Scale at Large $kr_c$}

In Section \ref{sec:level4} we analyzed the tree-level scattering amplitude $(1,4) \to (2,3)$ and discovered that 5D gravity compactified on a (flat) orbifolded torus becomes strongly coupled at the non-reduced Planck scale, $ \Lambda_{\text{strong}}^{(\text{5DOT})} \equiv \sqrt{8\pi} M_{\text{Pl}}$. In the large $kr_{c}$ limit of the RS1 model, however, we expect that all low-energy mass scales are determined by the emergent scale  \cite{ArkaniHamed:2000ds,Rattazzi:2000hs}
\begin{align}
    \Lambda_\pi = M_{\text{Pl}}\, e^{-\pi k r_{c}}~,
\label{eq:LambdaPi}
\end{align}
which is related to the relative locations of the branes \cite{Randall:1999ee,Randall:1999vf}. In this section we describe how this emergent scale arises from an analysis of the elastic KK scattering
amplitude in the large-$kr_c$ limit.

Consider the helicity-zero  $(n,n) \to (n,n)$ scattering amplitude. As plotted explicitly for $n=1$ in the previous subsection, at energies $s \gg m^2_n$ the scattering amplitude is dominated by the leading term ${\mathcal M}^{(1)}(kr_c,s,\theta)$ given in Eq.  \eqref{ELM1RS1}. The analogous expression in the 5D orbifold torus is given by Eq. (\ref{M15DOTnnnn}). We note that the angular dependence of these two expressions is precisely the same, and therefore we can compare their amplitudes by taking their ratio. This gives
the purely $kr_c$-dependent result\footnote{Formally, as in the case of toroidal compactification, the full amplitude has an infrared (IR) divergence due to the exchange of the massless graviton and radion modes. By taking the ratio of the amplitudes in the RS1 model to that in the 5D orbifolded torus in the high energy limit, the IR divergences cancel exactly and we can relate the strong coupling scale in the RS1 model to that in the case of toroidal compactification.} 
\begin{align}
    \dfrac{\mathcal{M}^{(1)}(kr_{c})}{\mathcal{M}^{(1)}(0)} = \left[\dfrac{1- e^{-2\pi kr_{c}}}{2 \pi kr_{c}}\right]\cdot \overline{\mathcal{K}}_{nnnn}(kr_{c})~,
\end{align}
where
\begin{align}
    \overline{\mathcal{K}}_{nnnn} &=\dfrac{1}{405}\bigg\{15\sum_{j}\dfrac{m_{j}^{8}}{m_{n}^{8}} a_{nnj}^{2}+28 a_{nnnn} - 144\left[ \dfrac{9\s b_{n^{\prime}n^{\prime}r}^{2}}{(m_{n}r_{c})^{4}} - a_{nn0}^{2}\right]\bigg\}~.\label{Kbarnnnn}
\end{align}
From this ratio, we can estimate the strong coupling scale at nonzero $kr_{c}$:
\begin{align}
    \Lambda^{\rm (RS1)}_{\rm strong} (kr_c) &\equiv \Lambda^{(\text{RS1})}_{\text{strong}}(0)\sqrt{\dfrac{\mathcal{M}^{(1)}(0)}{\mathcal{M}^{(1)}(kr_{c})}}~,\nonumber\\
    &= \dfrac{\Lambda^{\rm (5DOT)}_{\rm strong}}{\sqrt{\overline{\mathcal{K}}_{nnnn}}} \sqrt\frac{2 \pi kr_c}{1 - e^{-2\pi kr_{c}}}~.\label{LambdaStrong}
\end{align}
where we can use our earlier $\Lambda_{\text{strong}}^{(\text{5DOT})} = \sqrt{8\pi} M_{\text{Pl}}$ result.

Now let us consider the $kr_c$ dependence of this expression in the large-$kr_c$ limit. At large $kr_{c}$, Eq. \eqref{LambdaStrong} becomes
\begin{align}
    \Lambda^{(\text{RS1})}_{\text{strong}}(kr_{c}) \approx \sqrt{8 \pi} M_{\text{Pl}} \sqrt{\dfrac{2\pi kr_{c}}{\overline{\mathcal{K}}_{nnnn}}}~.
    \label{eq:strongscale}
\end{align}
whereas, using Eqs. \eqref{largekrc1}-\eqref{largekrc2},
\begin{align}
    \dfrac{m_{j}^{8}}{m_{n}^{8}} a_{nnj}^{2} &\approx \dfrac{x_{j}^{8}}{x_{n}^{8}} C_{nnj}^{2}\, (kr_{c})\, e^{2\pi kr_{c}}~,\\\
    a_{nnnn} &\approx C_{nnnn}\, (kr_{c})\, e^{2\pi kr_{c}}~,\\
    \dfrac{b_{n^{\prime}n^{\prime}r}^{2}}{(m_{n}r_{c})^{4}} &\approx \dfrac{1}{x_{n}^{4}} C_{nnr}^{2}\, (kr_{c})\, e^{2 \pi kr_{c}}~,\\
    a_{nn0}^{2} &\approx C_{nn0}\, (kr_{c})~.
\end{align}
such that
\begin{align}
    \dfrac{\overline{\mathcal{K}}_{nnnn}}{2\pi kr_c} &= \dfrac{ e^{2\pi kr_{c}}}{810\pi\, x_{n}^{8}}\bigg\{ 15\sum_{j=1}^{+\infty}  x_{j}^{8} \, C_{nnj}^{2} + 28 \, x_{n}^{8} \, C_{nnnn} - 1296\,  x_{n}^{4} \, C_{nnr}^{2} \bigg\}~,
\end{align}
In this expression, the $x_{j,n}$ are the $j$th and $n$th zeros of the Bessel function $J_{1}$, respectively; the constants $C_{nnj}$, $C_{nnnn}$, and $C_{nnr}$ (defined explicitly in Subsection \ref{Large krc}) are integrals depending only on the Bessel functions themselves. Therefore, focusing on the overall $kr_{c}$ dependence, we find that
\begin{align}
    \Lambda_{\text{strong}}^{(\text{RS1})} \propto \sqrt{8 \pi}M_{\text{Pl}} e^{-\pi kr_{c}} = \sqrt{8\pi} \Lambda_{\pi}
\end{align}
at large $kr_{c}$, as anticipated. The precise value of the proportionality constant
depends weakly on the process considered, and in the large-$kr_c$ limit for the processes $(n,n) \to (n,n)$ we find
\begin{align}
    \begin{tabular}{| c | c c c c c |}
    \hline $n$ & $1$ & $2$ & $3$ & $4$ & $5$\\
    \hline $\Lambda^{\rm (RS1)}_{\rm strong}/\sqrt{8\pi} \Lambda_\pi $ & $2.701$ & $2.793$ & $2.812$ & $2.819$ & $2.822$\\
    \hline
    \end{tabular}~.
\end{align}
Since these results for the elastic scattering amplitudes follow from the form of the wavefunctions in Eq. (\ref{eq:asymptotic-wf}), similar results will follow for the inelastic amplitudes as well---and they will also be controlled by $\Lambda_\pi$.

In addition to the previous analytic large-$kr_{c}$ analysis, we have also numerically examined the dependence for lower values of $kr_c$ via Eq. \eqref{LambdaStrong}. We display the dependence of $\Lambda^{\rm (RS1)}_{\rm strong}$ as a function of $kr_{c}$ for the processes $(1,1) \to (1,1)$ and $(1,4) \to (2,3)$ in Fig. \ref{fig:strongcoupling}. In all cases, we find that the strong coupling scale is roughly $\Lambda_{\pi}$. Therefore,  in the RS1 model (as conjectured under the AdS/CFT correspondence) all low-energy mass scales are controlled by the single emergent scale $\Lambda_{\pi}$.
\chapter{Conclusion}

Between what we published in \cite{Chivukula:2019rij,Chivukula:2019zkt,Chivukula:2020hvi} and additional original work discussed in this dissertation, we have obtained many substantial original results regarding the Randall-Sundrum 1 model:
\begin{itemize}
    \item[$\bullet$] Summary of the 5D weak field expanded RS1 Lagrangian $\mathcal{L}_{\text{5D}}$ and its 4D effective equivalent $\mathcal{L}_{\text{4D}}^{(\text{eff})}$ through $\mathcal{O}(\kappa_{\text{5D}}^{2})$. (Section \ref{AppendixWFE} and Subsection \ref{4D Effective RS1 - Summary of Results}.)
    \item[$\bullet$] Confirmation that all terms containing factors of $(\partial_{\varphi}|\varphi|)$ or $(\partial^{2}_{\varphi}|\varphi|)$ in $\mathcal{L}_{\text{5D}}$ are cancelled to all orders in the 5D coupling $\kappa_{\text{5D}}$. (Section \ref{SS- Cancel CC-Like})
    \item[$\bullet$] A new parameterization of the 4D effective RS1 Lagrangian as summarized in the 5D-to-4D formula, Eq. \eqref{LRSeffPerTerm}, which categorizes all couplings in the RS1 model as ``A-type" or ``B-type." (Section \ref{4D Effective RS1 Model})
    \item[$\bullet$] The demonstration that the matrix element describing massive spin-2 KK mode scattering in the 5D orbifolded torus model yields $\mathcal{O}(s)$ growth for all helicity combinations. (Section \ref{sec:level4})
    \item[$\bullet$] The demonstration that the matrix element describing massive spin-2 KK mode scattering in the RS1 model yields $\mathcal{O}(s)$ growth for all helicity combinations, including the derivation of sum rules that are sufficient for maintaining the cancellations from $\mathcal{O}(s^{5})$ down to $\mathcal{O}(s)$. (Sections \ref{sec:level5} and \ref{sec:level6})
    \item[$\bullet$] Analytic proofs for many of the sum rules, as well as numerical evidence supporting the one rule lacking an analytic proof. (Section \ref{DerivingSumRules} and Figure \ref{fig:cancel1111})
    \item[$\bullet$] Numerical measurements of how KK tower truncation impacts the accuracy of the full matrix element and its $\mathcal{O}(s^{\sigma})$ contributions ($\sigma \in\{1,2,3,4,5\}$) relative to the full matrix element without truncation. (Subsections \ref{sec:level6a} and \ref{sec:level6b})
    \item[$\bullet$] Calculation of the 5D strong coupling scale $\Lambda_{\pi} = M_{\text{Pl}} \, e^{-kr_{c}\pi}$ directly from the 4D effective RS1 theory via partial wave unitarity constraints. (Subsection \ref{sec:level6c})
\end{itemize}
These results point toward several interesting open questions as well as providing a foundation for future work. There are several projects we will be pursuing (including some for which substantial progress has already been made):
\begin{itemize}
    \item[$\bullet$] {\bf The Role of the Radion:} The single sum rule which lacks an analytical proof is the combined $\mathcal{O}(s^{3})$-$\mathcal{O}(s^{2})$ rule, Eq. \eqref{SumRuleOs3b},
    \begin{align}
        3\bigg[9\,b_{n^{\prime}n^{\prime}r}^{2} - \mu_{n}^{4} \,a_{nn0}^{2}\bigg] &= 15 \, c_{n^{\prime}n^{\prime}n^{\prime}n^{\prime}} + \mu_{n}^{4} \, a_{nnnn}
    \end{align}
    Its lack of proof is due to the curious coupling behavior of the radion. For example, the radion is introduced to the metric in the combination $\hat{u}\equiv (\kappa_{\text{5D}}\,\hat{r}/2\sqrt{6})\, \vep^{+2}\,e^{-kr_{c}\pi}$, which means every instance of the 5D field $\hat{r}(x)$ carries with it a warp factor $\vep^{+2}$, which throws a wrench in the otherwise powerful sum rules machinery developed in Section \eqref{DerivingSumRules}. Is an analytic proof of this sum rule possible? And if so, does it elucidate the role of the radion in the RS1 model?
    \item[$\bullet$] {\bf Radion Stabilization:} The massless radion poses a problem for the RS1 model: if left as is, it generates an attractive Casimir force which pulls the branes at either end of the extra dimension  together, thereby driving the extra dimension to smaller and smaller distance scales until the separation enters the quantum gravity regime and the RS1 model is no longer predictive \cite{PhysRevD.28.772,Goldberger:1999uk}. Furthermore, a massless radion would necessarily generate a scalar-tensor theory of long-distance gravitation at low energies contrary to the usual pure tensor theory of 4D gravity. Therefore, phenomenological applications of the RS1 model require that the radion become massive in a process called radion stabilization. Radion stabilization typically involves adding a massive bulk scalar field to the RS1 Lagrangian that generates a radion potential which stabilizes the positions of the branes. However, we have found that adding a mass to the radion by hand causes the matrix elements describing massive spin-2 KK mode scattering to scale like $\mathcal{O}(s^{2})$ instead of $\mathcal{O}(s)$. In a full model of radion stabilization, are cancellations down to $\mathcal{O}(s)$ maintained? If so, how does the introduction of radion stabilization influence the sum rules?
    \item[$\bullet$] {\bf Bulk and Brane Matter:} Phenomenological applications of the RS1 model are not usually restricted to the pure gravity theory that we consider in this dissertation. Instead, physicists typically add either bulk or brane matter to the RS1 model, and investigate scattering of that matter in different circumstances. When adding (scalar, fermionic, vector) matter to the bulk or a brane, how do the new 2-to-2 scattering matrix elements scale at large energies? What new sum rules (if any) are implied?
    
    We have actually already completed the analyses of bulk and brane scalar matter, wherein we find that the process $\phi\phi\rightarrow h^{(n)}h^{(n)}$ for a bulk or brane scalar $\phi$ exhibits cancellations down to $\mathcal{O}(s)$---and derive several new sum rules.
    \item[$\bullet$] {\bf Machinery:} Because of the complexity of diagrams involving multiple massive spin-2 particles, the analytic calculations required for the analyses in this dissertation were nontrivial. They required the development of a program that uses specialized techniques in order to complete the calculation in a timely fashion. It is our goal to generalize and clean up this code as to make it available for use to the wider physics community.
\end{itemize}
Thus, this dissertation presents original results about massive spin-2 KK mode scattering in the 4D effective Randall-Sundrum 1 model, and these results are of existing and future relevance in theoretical and phenomenological contexts.

%

\end{doublespace}
\bibliographystyle{ieeetr}
\bibliography{references}

\end{document}